\documentclass[aps,prd,twocolumn,groupedaddress,nofootinbib]{revtex4-1}
\pdfoutput=1
\usepackage{ifpdf}
\ifpdf
\usepackage[pdftex]{graphicx}
\else
\usepackage{graphicx}
\fi
\ifpdf
\DeclareGraphicsExtensions{.pdf, .jpg, .tif}
\else
\DeclareGraphicsExtensions{.eps, .jpg}
\fi

\usepackage{graphicx,color,amsmath}
\usepackage[caption=false]{subfig}
\usepackage{amsmath,amssymb}
\usepackage{wrapfig}
\usepackage{graphicx}
\usepackage{hyperref}
\usepackage{cleveref}
\usepackage[shortlabels]{enumitem}
\usepackage{graphicx}
\usepackage{comment}
\usepackage[normalem]{ulem}

\usepackage{tikz}
\usetikzlibrary{decorations.pathmorphing}

\tikzset{snake it/.style={decorate, decoration=snake}}

\newcommand{\yr}{{\, {\rm yr}}}
\newcommand{\Myr}{{\, {\rm Myr}}}

\newcommand{\kpc}{{\, {\rm kpc}}}

\newcommand{\cm}{{\, {\rm cm}}}

\newcommand{\eV}{{\, {\rm eV}}}

\newcommand{\MeV}{{\, {\rm MeV}}}
\newcommand{\GeV}{{\, {\rm GeV}}}

\newcommand{\Hz}{{\, {\rm Hz}}}
\newcommand{\kHz}{{\, {\rm kHz}}}

\newcommand{\beq}{\begin{equation}}
\newcommand{\eeq}{\end{equation}}
\newcommand{\bal}{\begin{align}}
\newcommand{\eal}{\end{align}}

\newcommand{\msun}{M_{\odot}}
\newcommand{\ep}{\varepsilon_{211}}

\newcommand{\LL}{{\mathcal{L}}}
\newcommand{\OO}{{\mathcal{O}}}

\newcommand{\MM}{{\mathcal{M}}}
\newcommand{\Mpl}{{M_{\rm pl}}}
\newcommand{\Mbh}{M}
\newcommand{\BH}{\text{BH}}
\newcommand{\SR}{\text{SR}}

\newcommand{\di}{\text{{d}}}
\newcommand{\pare}[1]{\left(#1\right)}
\newcommand{\parea}[1]{\left[#1\right]}

\newcommand{\MBH}{M}
\newcommand{\Mc}{M_\text{c}}
\newcommand{\Mct}{M_\text{c2}}

\newcommand{\kinto}{\kappa^{\lambda}_1}
\newcommand{\kgro}{\kappa^\text{gr}_1}
\newcommand{\kintt}{\kappa^{\lambda}_2}
\newcommand{\kgrt}{\kappa^\text{gr}_2}
\newcommand{\kintop}{\kappa^{\lambda}_3}
\newcommand{\kgrop}{\kappa^\text{gr}_3}
\newcommand{\kinttp}{\kappa^{\lambda}_4}
\newcommand{\kgrtp}{\kappa^\text{gr}_4}

\newcommand{\eone}{\varepsilon_{211}}
\newcommand{\eoned}{\dot{\varepsilon}_{211}}
\newcommand{\etwo}{\varepsilon_{322}}
\newcommand{\etwod}{\dot{\varepsilon}_{322}}
\newcommand{\rinv}{\pare{\frac{\Mpl}{f}}}

\newcommand{\gsro}{\gamma^\text{SR}_{211}}
\newcommand{\gsrt}{\gamma^\text{SR}_{322}}
\newcommand{\gbh}{\gamma_{211\times211}^{322\times\text{BH}}}
\newcommand{\ginf}{\gamma_{322\times322}^{211\times\infty}}

\newcommand{\gsr}{\gamma_\text{SR}}

\newcommand{\bra}[1]{\left<#1\right|}
\newcommand{\ket}[1]{\left|#1\right>}

\definecolor{mypurple}{RGB}{164,64,214}
\definecolor{myorange}{RGB}{255,165,0}

\newcommand{\abs}[1]{\left| #1 \right|} 
\newcommand{\pd}[2]{\frac{\partial #1}{\partial #2}} 
\newcommand{\pdd}[2]{\frac{\partial^2 #1}{\partial #2^2}} 

\makeatletter
\def\l@subsection#1#2{}
\def\l@subsubsection#1#2{}
\makeatother
 
\begin{document}

\title{Black hole superradiance of self-interacting scalar fields}

\author{Masha Baryakhtar}
\email{mbaryakhtar@nyu.edu}
\affiliation{Center for Cosmology and Particle Physics, Department of Physics, New York University, New York, NY 10003, USA}
\author{Marios Galanis}
\email{mgalanis@stanford.edu}
\affiliation{Stanford Institute for Theoretical Physics, Stanford University, Stanford, CA 94305, USA}
\author{Robert Lasenby}
\email{rlasenby@stanford.edu}
\affiliation{Stanford Institute for Theoretical Physics, Stanford University, Stanford, CA 94305, USA}
\author{Olivier Simon}
\email{osimon@stanford.edu}
\affiliation{Stanford Institute for Theoretical Physics, Stanford University, Stanford, CA 94305, USA}

\date{\today}

\begin{abstract}

Black hole superradiance is a powerful probe of light,
	weakly-coupled hidden sector particles. Many candidate
	particles, such as axions, generically have self-interactions
	that can influence the evolution of the superradiant
	instability.	As pointed out in \cite{Gruzinov_2016} in the
	context of a toy model, much of the existing literature on
	spin-0 superradiance does not take into account the most
	important self-interaction-induced processes. These processes
	lead to energy exchange between quasi-bound levels and particle
	emission to infinity; for large self-couplings, superradiant
	growth is saturated at a quasi-equilibrium configuration of
	reduced level occupation numbers. In this paper, we perform a
	detailed analysis of the rich dynamics of spin-0 superradiance
	with self-interactions, and the resulting observational
	signatures. We focus on quartic self-interactions, which
	dominate the evolution for most models of interest. We explore
	multiple distinct regimes of parameter space introduced by a
	non-zero self-interaction, including the simultaneous population
	of two or more bound levels; at large coupling, we confirm the
	basic picture of quasi-equilibrium saturation and provide
	evidence that the ``bosenova'' collapse does not occur in most
	of the astrophysical parameter space. Compared to gravitational
	superradiance, we find that gravitational wave ``annihilation''
	signals and black hole spin-down are parametrically suppressed
	with increasing interactions, while new gravitational wave
	``transition'' signals can take place for moderate interactions.
	The novel phenomenon of scalar wave emission
	is less suppressed at large couplings, and if the particle has Standard Model
	interactions, then coherent,
	monochromatic axion wave signals from black hole superradiance
	may be detectable in proposed axion  dark matter experiments.

\end{abstract}

\maketitle

\tableofcontents


\section{Introduction}

As discovered by Penrose~\cite{Penrose1971}, it is possible
to extract energy and angular momentum from rotating
black holes. While the Penrose thought experiments
were in terms of mechanical scattering,
equivalent processes were developed by the Zeldovich group for bosonic
waves \cite{Zeldovich1971,Misner1972,Starobinskii1973}.
This phenomenon, termed ``superradiance'', is expected
to occur in nature and, for certain initial conditions, amplify photon and graviton waves  passing near rotating black holes.
Moreover, if there exists a new bosonic particle
with a small mass, bound states of this particle
could be exponentially amplified around astrophysical
black holes, forming very high occupation number
``clouds'' that could lead to a range of observational
signatures.

Black hole (BH) superradiance as a probe
of new ultralight particles was first proposed
in~\cite{Arvanitaki:2009fg}, which has given rise to an 
extensive literature.
Superradiance of new particles, including
spin-0~\cite{Arvanitaki:2010sy,Ternov:1978gq,Zouros:1979iw,Detweiler:1980uk,Dolan:2007mj,Yoshino:2013ofa,Arvanitaki:2014wva,Brito:2014wla,Brito:2015oca,Arvanitaki:2016qwi}, spin-1~\cite{Rosa:2011my,Pani:2012bp,Pani:2012vp,East:2017mrj,Baryakhtar:2017ngi,Baumann:2019eav},
and spin-2~\cite{Brito:2013wya,Brito:2020lup}
fields, have been investigated,  with observational signatures including black hole spin-down, gravitational wave
emission, and modified black hole in-spiral dynamics; see the above for further references and \cite{Brito:2015oca} for a review.

Gravitational interactions are all that is necessary for BH superradiance, which makes superradiance a unique window on new particles that are otherwise inaccessible to experimental probes.  However, many beyond-Standard-Model
particle candidates have other interactions. These can include self-interactions,
interactions with Standard Model (SM) states, and interactions with other hidden sector states.
For some new particles, including the well-motivated QCD axion \cite{Peccei+1977,Weinberg1978,Wilczek1978},
both self-interactions and interactions with the SM are required by the model.
Therefore, it is important to understand the
consequences of such interactions for the growth and
behavior of superradiant bound states.

In this paper, we analyze in detail the consequences of a quartic self-interaction for
the superradiance phenomenology of a light scalar around astrophysical black holes. We find
that over a large range of parameter space of interest to light axion models, the addition of
a quartic coupling leads to rich dynamics in the evolution of the superradiant instability, and
new observational consequences. These dynamics include limiting the maximum number of particles in a bound level, populating levels inaccessible through gravitational superradiance alone, saturation to quasi-equilibrium configurations of two or more levels, and emission of  non-relativistic and relativistic scalar waves to infinity.  As we demonstrate, an effective quartic term is generically the most important effect driving the evolution, for much of the astrophysically
relevant parameter space.

BH superradiance of a self-interacting scalar was first introduced in Ref.~\cite{Arvanitaki:2010sy}, which discussed phenomena including
relativistic scalar emission, level mixing, and the possibility of a ``bosenova'' --- a rapid,
non-perturbative collapse of the cloud due to attractive self-interactions.
The bosenova process was studied numerically  in Ref.~\cite{Yoshino:2012kn,Yoshino:2015nsa}, and these results were used in subsequent phenomenological investigations~\cite{Fukuda:2019ewf,2009.07206}. 
However, as we will discuss, these previous analyses did not take into account self-interaction-induced energy transfers between different superradiant levels. This was pointed out (for a toy model)
in~\cite{Gruzinov_2016}, which showed that these energy transfer processes, along with 
scalar emission, can result in saturation to a two-level equilibrium configuration before the cloud has had a chance
to grow large enough for a bosenova.  We provide evidence that during evolution from astrophysical initial conditions, a ``bosenova'' does not occur in much of the phenomenologically-relevant parameter space: scalar field values remain small and the cloud size required for collapse is not reached.

For small enough self-couplings --- including much of the superradiance parameter space
for the QCD axion --- self-interaction effects are unimportant. Superradiance proceeds as in the purely-gravitational case: a non-relativistic bound state of scalars is populated by extracting energy and angular momentum from the rotating black hole, and subsequently annihilates to gravitational radiation. 

Slightly larger self-interactions result in non-relativistic scalar radiation to infinity. This new energy loss mechanism reduces the power emitted over time in gravitational wave ``annihilation'' signals. The interactions also populate higher angular momentum levels; the simultaneous occupation of several bound states can give rise to gravitational wave ``transition'' signals, in which scalars emit lower frequency gravitational waves by transitioning between two occupied levels.

Large enough self-interactions, including those typical
of axion dark matter produced through the misalignment
mechanism, significantly reduce the occupation number of the cloud. Instead of being limited by angular momentum conservation, superradiant growth is cut off  early by self-interactions. The smaller cloud size suppresses the peak gravitational wave signal strains.  For even larger self-couplings, the occupation of the cloud reaches quasi-equilibrium at parametrically smaller occupation values, as found in \cite{Gruzinov_2016}.  
In this regime,  the self-interactions parametrically slow the spin-down of the BH compared to the purely-gravitational case. 

Throughout, a new phenomenon of
almost-monochromatic, non-relativistic scalar wave
emission occurs; for large self-interactions, the
signal amplitude is constant on timescales up to
the age of the universe. If couplings to Standard
Model particles are present in addition to the
self-interaction, then this scalar radiation
may be detectable in proposed axion dark matter
experiments. For a range of
models, the self-interaction and SM interactions
are controlled by the same scale; consequently, the signal in
Earth-based detectors can persist for arbitrarily
small occupation numbers, as long as the classical
scalar field description holds.

Many of our analyses in this paper use hydrogenic approximations
for bound states around BHs. Consequently, they are valid
for scalar Compton wavelengths bigger than a few times the black hole light-crossing time. Understanding the behavior of
more massive scalars would require numerical techniques.
Since some of the most dramatic superradiance signatures
may occur for slightly heavier scalars, further
investigations of this kind are strongly motivated.

We review purely gravitational superradiance of  scalar (spin-0) fields in Sec.~\ref{secspin0},
and discuss the new processes introduced by quartic (and cubic) interactions in Sec.~\ref{sec:quartic}.
In Sec.~\ref{secperturb}, we explore in detail the  evolution of the superradiant cloud in the presence
of quartic self-interactions, which lead to several distinct regions in 
mass-coupling parameter space. In Sec.~\ref{sec:nonperturb}, we discuss
the maximum amplitude reached by the axion field, and whether this is large enough to
cause non-perturbative behavior such as a ``bosenova''.
We study the observable signatures of axion superradiance in the presence
of self-interactions: spin down of astrophysical black holes (Sec.~\ref{sec:spindown}), gravitational wave annihilations and transitions (Sec.~\ref{sec:gw}), and axion waves (Sec.~\ref{sec:scalarwaves}). We provide more detailed calculations related to both self-interactions  and gravitational superradiance in App.~\ref{app:param} --\ref{app:dmestimates}. We conclude  and comment on directions for future investigations in Sec.~\ref{sec:concl}.

\section{Spin-0 superradiance}
\label{secspin0}

In this section, we give a brief review of BH superradiance for a scalar with purely
gravitational interactions. There is a broad
literature on this topic; for a review,
see~\cite{Brito:2015oca}. We take our
signature to be $-+++$, and assume natural units
with $c = \hbar = 1$ unless otherwise indicated.
We use the convention $\Mpl \equiv 1/\sqrt{G}$
throughout.

In the Kerr background, the Killing vector tangent
to the horizon, in static (Boyer-Lindquist)
coordinates, is $\xi = \partial_t +
\Omega_H\partial_\phi$. Here, $\Omega_H= \frac{1}{2
r_g}\left(\frac{a_*}{1 + \sqrt{1 - a_*^2}}
\right)$ is the angular velocity of the horizon
and $a_* = J/GM^2$ is the dimensionless spin of
the BH, where $J$ is the BH's angular momentum, $M$ is its mass,
and $r_g \equiv G M$.
Consequently, a wave with
frequency $\omega$, and angular momentum $m$ about
the BH spin axis, has energy flux $\propto \omega
(\omega - m \Omega_H)$ across the horizon,
relative to distant observers (the energy flux is necessarily
ingoing for local observers near the horizon). For
$\omega < m \Omega_H$, there is energy and angular
momentum extraction from the BH, as measured at
infinity.

Massive bosonic fields have quasi-bound states around a BH. In a Schwarzschild background, all of these
states are unstable to decay. However, in a Kerr background, states with $\omega < m \Omega_H$ are unstable
to \emph{growth} \cite{Detweiler_1980,Zouros:1979iw,Gaina1988,Dolan:2007mj}.\footnote{For complex $\omega$, as appropriate
for an unstable state,  the energy flux across the horizon is negative if $\frac{|\omega|^2}{{\rm Re} \, \omega} < m \Omega_H$~\cite{Dolan:2007mj}}
Exponential growth of these superradiant states,
starting either from a pre-existing astrophysical
population in the field, or from quantum fluctuations,
will occur given enough time. If we start from the vacuum state, 
then ignoring the BH interior gives effectively non-unitary
evolution of the field outside
(due to the absorbing boundary conditions at the horizon),
producing a mixed state. Interactions with external systems will generally decohere this into
an almost-coherent state, with well-defined phase and amplitude. This process is analogous to the growth of a large-occupation-number
laser field from quantum fluctuations~\cite{Mandel_1995}.

The energy flux across the horizon, for a scalar
field $\varphi$, is $\dot E_\infty \sim A_H
|\varphi_H|^2 \omega (\omega - m \Omega_H)$,
where $|\varphi_H|$ is the amplitude of the field
at the horizon (in in-going coordinates, for
which $\varphi$ is smooth at the horizon), and $A_H$
is the area of the BH horizon. This
flux determines the growth rate of a quasi-bound
state. For a scalar of mass $\mu \ll r_g^{-1}$,
the lowest energy states
are analogous to hydrogenic bound states, since the
effect of the BH at large radii is that of a point
source with a $1/r$ potential. The hydrogenic
level with principal quantum number $n$, total angular
momentum $l$, and azimuthal angular momentum $m$ 
(around the BH spin axis)
has frequency $\omega = \omega_r + i \omega_i$,
where
\begin{equation}
	\omega_r \simeq \mu \left(1 - \frac{\alpha^2}{2 n^2} + \mathcal{O}(\alpha^4)\right)
\end{equation}
with $\alpha \equiv G M \mu$ acting as the equivalent
of the fine-structure constant~\cite{Arvanitaki:2014wva,Baumann:2019eav}.
The imaginary part of the frequency is
\begin{equation}
	\omega_i \propto \alpha^{4l + 5}(m \Omega_H - \omega_r)
	(1 + \OO(\alpha))
	\label{eqwi1}
\end{equation}
Strictly speaking, for $m \neq 0$, the leading-$\alpha$
form of this expression is simply $\alpha^{4l + 5} m \Omega_H$.
However, if $m \Omega_H$ is also small relative
to $r_g^{-1}$, then the expression in Eq.~\eqref{eqwi1} is
appropriate (and more generally, changes sign at the correct $\omega_r$).
The 211  ($n=2, l=1, m=1$) level, which has the fastest
growth rate at small $\alpha$, 
has $\omega_i = \frac{a_*}{48} \alpha^8 \mu$ at leading
order in $\alpha$. The ``superradiance rate'', which is usually defined as
 the growth rate of the occupation number, is
 $\Gamma_{\rm SR} \equiv 2 \omega_i$.
The $\alpha^{4l + 5}$ scaling for the growth
rate corresponds to the field amplitude
at the BH horizon --- for higher-$l$ modes, 
the amplitude is suppressed by the angular momentum barrier,
leading to exponentially smaller growth rates
for higher $l$ modes~\cite{Zouros:1979iw,Detweiler:1980uk,Arvanitaki:2010sy,Baumann:2019eav}.

While the expansions above were phrased in terms
of $\alpha$ being small, it is actually
the case that $\alpha/l$ is a good expansion parameter.
Whenever a level is superradiant,
we must have $\alpha < m/2$, so $\alpha/l < 1/2$, 
and the hydrogenic approximation can be used.

If the Compton wavelength of the particle is very large, i.e.\
$\alpha \ll 1$, then all of the superradiance rates
are suppressed by a high power of $\alpha$,
$\Gamma \propto \alpha^{4l + 4} \mu$, so are very small.
Conversely, if
the Compton wavelength of the particle is
significantly smaller than the size of the BH,
i.e.\ $\alpha \gg 1$, then only modes with $m
\gg 1$ can be superradiant; however, these
have exponentially suppressed growth rates.
Consequently, for observationally-relevant
superradiance rates, the Compton wavelength of the
particle should approximately match the size of
the BH. For stellar-mass black holes, $M_{\rm BH}
\sim 10 M_{\odot}$, this corresponds to $\mu \sim
10^{-13} - 10^{-11} \eV$.
While the superradiant growth rates around
such BHs are rather slow on particle physics scales ---  with $e$-folding times a few minutes or longer ---
they can still be much faster than other astrophysical processes
and timescales, allowing superradiance to occur in realistic
astrophysical environments.

Once a Kerr BH is ``born'' , e.g.\ in a binary merger or a supernova,
the superradiant bound states start growing in amplitude.
The fastest-growing level, which usually has the minimum $m$
satisfying the superradiance condition (except
close to the $\omega_r = m \Omega_H$ threshold), is the first to extract
a significant amount of angular momentum from the BH,
spinning it down to $\Omega_H \simeq \omega/m$.\footnote{Strictly
speaking, $a_*$ asymptotes towards
the $\Omega_H = \omega/m$ threshold,
since the superradiance rate is $\propto (m \Omega_H - \omega)$,
so vanishes at the threshold. However, 
we will mostly ignore this small effect
in the rest of the paper, and will refer to the
BH being spun down ``to the spin threshold''.} For modes with the same $m$, the most tightly
bound mode is often (for small $m$) the one with the largest
growth rate, since it has larger amplitude
at the horizon. Consequently, if $\omega = m \Omega_H$
for that mode, then $\omega > m \Omega_H$ for the other
modes, and they are not growing (this is not always
true for $m \ge 3$; see Sec.~\ref{sec:gw}).

Since the angular momentum of an astrophysical
BH is very large, 
\begin{equation}
	J = a_* G M^2 = a_* \frac{ M^2}{ \Mpl^2}
\simeq 10^{78} a_* \left(\frac{M}{10 M_\odot}\right)^2,
	\label{eqjbh}
\end{equation}
it takes $\sim \log(J/m) \sim 180$ e-folds of superradiant growth
to cause $\OO(1)$ BH spin-down. 
Correspondingly, the fully-grown superradiant cloud has an extremely high occupation
number $\sim \OO(1) J$.
This corresponds to an energy density which is 
significantly higher than astrophysical DM
densities (assuming that DM is not in extremely dense clumps), App.~\ref{app:dmestimates}.
Consequently, the presence or absence of 
an astrophysical scalar field abundance makes little
difference to its superradiant growth.

The oscillating scalar field  sources gravitational wave (GW) radiation, at a frequency
$\simeq 2 \mu$ --- on a particle level, this corresponds
to scalars annihilating to gravitons in the black hole background.
The emitted power scales as $P \propto G N^2 \mu^4 \alpha^{16 + 4l}$, where $N$ is the occupation number of the mode~\cite{Yoshino:2013ofa,Arvanitaki:2014wva,Brito:2014wla}.
The smallness of $G$, and the high power of $\alpha$, mean
that this process is slow; in particular, it is always
too slow to disrupt the initial superradiant growth
of the level \cite{Yoshino:2013ofa}.

The superradiant growth of higher $l$ levels will also take place. 
Once lower-$l$ modes have grown to saturation,
higher-$l$ modes can still be superradiant, but
their growth rate is slower, so there is 
a parametric separation between the growth times of
successive levels. The annihilation process generally depletes the majority of the scalar cloud before
the next level grows. Once the next level significantly spins down
the BH, the first mode now has $\omega > m \Omega_H$,
so is decaying with a rate comparable to its
initial growth rate, and its remaining density falls back into the BH.
Over sufficiently long times, a similar process
will repeat for the next level. 

There are a number of observational signatures
of purely gravitational scalar superradiance. The first
is a lack of old, fast-spinning BHs, at masses
for which the scalar would have spun them down
in the time available. There have been $\sim
10$ measurements of stellar-mass BH spins in X-ray
binary systems~\cite{McClintock:2013vwa};
for high-spin BHs, these measurements can be
accurate to a few percent, and have been used to
set constraints the mass of weakly-interacting
scalars~\cite{Arvanitaki:2014wva}. LIGO
observations of binary BH mergers also
enable spin measurements of the pre-merger
BHs~\cite{Arvanitaki:2016qwi,ng2019searching}. While most of these measurements
are currently too imprecise to provide evidence for existence of a scalar~\cite{Arvanitaki:2016qwi,ng2019searching,Ng_2020ruv}, initial bounds are already possible \cite{Ng_2020ruv} (see section~\ref{sec:spindown} for a more detailed discussion).

Another possibility is the observation of gravitational
radiation from the scalar cloud.
For stellar-mass black holes, this radiation
could potentially be observed at LIGO~\cite{Arvanitaki:2014wva,
Arvanitaki:2016qwi,Brito+2017,Brito+2017a,Zhu:2020tht};
for heavier BHs, lower-frequency observatories
such as LISA or atom interferometers~\cite{graham2017midband}
could have sensitivity~\cite{Arvanitaki:2014wva,Brito+2017,Brito+2017a}.
The presence of a scalar cloud during a binary
merger could also change inspiral dynamics, 
yielding further gravitational wave
signatures~\cite{Baumann2020,Zhang2020,Zhang2019,Baumann+2019}. While LIGO only observes the last
few periods of BBH mergers, making such observations
difficult, lower-frequency detectors will
observe many more cycles, which will likely
improve their chances of observing such effects.


\section{Quartic self-interactions}
\label{sec:quartic}

For a spin-0 particle, the simplest
non-gravitational interaction is a quartic self-interaction.
This is generic in the sense that, if we expand
a potential about a symmetric minimum, then the quartic
is the most important interaction term for small
amplitudes. 

More specifically, a naturally small mass for a scalar field, as required for superradiance around astrophysical black holes, can be achieved through the breaking of a shift symmetry at some high energy scale
$f_a$.  A potential of the form  $V(\varphi) = \Lambda^4 g(\varphi/f_a)$ can be generated from
non-perturbative physics, so that $\Lambda \ll f_a$. 
For the case of a generic potential $g$,
expanding around the minimum
of the potential gives a mass scale $\mu^2 = g'' {\Lambda^4}/{f_a^2}$
and a self-interaction term of order $\lambda = g^{(4)} {\Lambda^4}/{f_a^4}$.

A well-known example is the QCD axion;
given a coupling $\LL \supset \frac{\varphi}{f_a}
\frac{g_s^2}{32 \pi^2} G_{\mu\nu}^a \tilde G^{a,\mu\nu}$
of the axion $\varphi$ to the QCD pseudoscalar
field strength, it acquires a potential of the form ~\cite{di_Cortona_2016}
\begin{equation}
V(\varphi) \simeq - m_{\pi}^2 f_{\pi}^2
\sqrt{1 - \frac{4 m_u m_d}{(m_u + m_d)^2} \sin^2(\varphi/(2 f_a))}.
\end{equation}
resulting in a mass $\mu \simeq 6 \times 10^{-12} \eV \frac{10^{18} \GeV}{f_a}$, and quartic self-interaction~\cite{di_Cortona_2016},
\begin{equation}
	\lambda  \simeq 0.3 \mu^2/f_a^2
	\simeq 10^{-80} \left(\frac{\mu}{10^{-12} \eV}\right)^4
	\label{eqlambdaqcd}
\end{equation}

For more general axion-like particles, the natural parametric value of the quartic coupling is
\begin{equation}
	\lambda \sim \frac{\mu^2}{f_a^2} 
	\simeq 10^{-74} \left(\frac{\mu}{10^{-12} \eV}\right)^2
	\left(\frac{10^{16} \GeV}{f_a}\right)^2,
	\label{eqlambda1}
\end{equation}
where we chose the nominal value of $\mu$ to be in the range of interest for stellar-mass BHs, and $f_a$ to be around the Grand Unification (GUT) scale, for illustration. For example, a motivated target model
is an axion-like particle which makes up $\OO(1)$ of the dark matter
abundance. If it is produced in the early universe
by the misalignment mechanism, and starts out with a
field value that is $\sim \OO(1) f_a$, then 
the scale for which we obtain the correct DM abundance
is $f_a \simeq 3 \times 10^{14} \GeV (10^{-12} \eV / \mu)^{1/4}$
(assuming a time-independent potential,
unlike the QCD axion case). This gives a typical quartic
coupling of 
\begin{equation}
	\lambda \sim  10^{-71} \left(\frac{\mu}{10^{-12} \eV}\right)^{5/2}.
	\label{eqlambdaDM}
\end{equation}
We will see that even such tiny self-coupling values
can have important consequences for the dynamics and phenomenology
of spin-0 superradiance.

The Lagrangian for a scalar field $\varphi$ with a quartic
coupling $\lambda$
in a fixed background spacetime is  given by
\begin{equation}
	\LL = - \frac{1}{2} (D_\mu \varphi) (D^\mu \varphi) - \frac{1}{2} \mu^2 \varphi^2
	+ \frac{1}{4!} \lambda \varphi^4,
\end{equation}
where $D^\mu$ is the covariant derivative
and $\mu$ is the mass of $\varphi$.
This gives the equation of motion
\begin{equation}
	(D^2 - \mu^2) \varphi = - \frac{\lambda}{6} \varphi^3.
	\label{eqEom1}
\end{equation}
The quartic interaction strength $\lambda$ can have either sign; $\lambda > 0$ corresponds
to an attractive self-interaction, as is the case for axion-like-particles,
while $\lambda < 0$ is repulsive.  For future convenience, we also define an energy scale $f$ such that the quartic
$\lambda \equiv \mu^2/f^2$; for an axion-like particle, we expect $f \sim f_a$, where $f_a$ is the symmetry-breaking scale.

The states that dominate the evolution of superradiance are generally
non-relativistic, hydrogen-like wavefunctions;
these have the fastest growth rates and
so obtain the largest amplitudes.
Consequently, it is helpful to perform a non-relativistic
reduction, writing
\begin{equation}
\label{eq:nr_ansatz}
	\varphi = \frac{1}{\sqrt{2\mu}} (\psi e^{- i \mu t} + {\rm c.c}).
\end{equation}
Here, the ``wavefunction'' $\psi$ is a complex scalar
field, with $\int dV |\psi|^2 \simeq N$
the occupation number.
The equation of motion is
\begin{equation}
	(D^2 - \mu^2) \psi e^{- i \mu t}
	 + {\rm c.c.} = \frac{-\lambda}{12 \mu}
	 \left(\psi^3 e^{-3 i \mu t} + 3 \psi^2 \psi^* e^{-i \mu t} \right)
	 + {\rm c.c.}
	 \label{eqpsi1}
\end{equation}
If $\psi$ changes slowly with time, compared
to $\mu^{-1}$, then
we can ignore the $\partial_t^2 \psi$ terms,
and extract the $e^{-i \mu t}$ part of the EoM to obtain
the Gross-Pitaevskii equation \cite{Arvanitaki:2010sy},
\begin{equation}
	\left(i \partial_t + \frac{\nabla^2}{2\mu} + \frac{\alpha}{r}\right)
	\psi \simeq \frac{-3}{24 \mu^2} \lambda \psi^2 \psi^*.
	\label{eq:eqgp1}
\end{equation}
The  $\psi^3 e^{- 3 i \mu t}$ term
in Eq.~\eqref{eqpsi1} leads to additional subdominant processes,
such as the emission of relativistic $\varphi$
waves, that are not captured by Eq.~\eqref{eq:eqgp1} (see Sec.~\ref{secrelscalar} and App.~\ref{app:relemission}). 

As a visual aid for understanding the $\lambda$-induced
interactions, we can use a diagrammatic notation for
the terms of 
\begin{equation}
	\frac{\lambda}{4!} \varphi^4 = \frac{\lambda}{96 \mu^2}
	\left(\psi e^{- i \mu t} + \psi^* e^{i \mu t}\right)^4
	\label{eqlterm}
\end{equation}
in close analogy to Feynman diagrams.
If we expand $\psi = \sum \alpha_i \psi_i$ in some basis
$\{\psi_i\}$,
then 
legs on the left-hand-side of the diagram will correspond to 
$\psi_i$ terms in Eq.~\eqref{eqlterm}, while legs on the right-hand-side will
correspond to $\psi_i^*$ terms.
For example, relativistic emission sourced
by the $211$ hydrogenic level
corresponds to the diagram
\begin{center}
	\begin{tikzpicture}[scale=0.7]
	\draw (-1,1) -- (0,0);
	\draw (-1,0) -- (0,0);
	\draw (-1,-1) -- (0,0);
	\draw (0,0) -- (1,0);
	\node[anchor=east] at (-1,1) {211};
	\node[anchor=east] at (-1,0) {211};
	\node[anchor=east] at (-1,-1) {211};
	\node[anchor=west] at (1,0) {$l=3,m=3$};
\end{tikzpicture}
\end{center}
in the sense that the relevant terms in the equation
of motion are obtained from terms involving $\psi_{211}^3$
in the Lagrangian, which source a $l,m=3,3$ relativistic mode.
We will make use of these diagrams throughout this section.

The (typically tiny) values of $\lambda$ introduced in Eq.~\eqref{eqlambda1} have
very little effect on processes involving only a
few $\varphi$ quanta. In particular, if we start
in a vacuum (or near-vacuum) state,
the first process of interest is the superradiant
growth of the most unstable hydrogenic levels, exactly as in the
purely-gravitational case. However, since the occupation
number $N$ of a superradiant level can reach
exponentially large values (Eq.~\eqref{eqjbh}), the large field amplitude can compensate for a small self-interaction, and
the quartic term's effects can qualitatively alter the dynamics of superradiance.  We investigate these effects below.

Higher-dimensional interactions, corresponding
to higher powers of the field, will be present
in general. However, we will see that,
in much of the astrophysically-relevant parameter
space, the field
never reaches large enough amplitudes
for them to be important, for natural 
hierarchies between the mass, quartic, and higher-order terms (see section~\ref{sec:theta}). The case of an additional cubic coupling leads to qualitatively similar dynamics as for the quartic alone, as discussed in section~\ref{sec:cubic}.


In the presence of a quartic interaction,
three types of perturbative processes affect the evolution
of the levels (here, perturbative is meant in the sense
that dynamics can be treated as involving approximately hydrogenic
modes, interacting on timescales long compared to their
oscillation times). 
These are relativistic emission of
axions to infinity (Sec.~\ref{secrelscalar}),
non-relativistic emission of axions to
infinity (Sec.~\ref{secnonrelem}), and
bound-state interactions leading to energy exchange
between levels
(Sec.~\ref{secboundints}). We will see in the
following sections that the latter two processes
will be most important for determining the
dynamics of the scalar cloud.

\subsection{Relativistic scalar emission}
\label{secrelscalar}

One of the simplest kinds of process
arising from the equation of motion (Eq.~\eqref{eqEom1})
is the $3 \rightarrow 1$ process
in which bound-state particles ``annihilate'' into a
relativistic $\varphi$. In terms of the non-relativistic
reduction, the relativistic mode $\varphi_\infty$ is sourced by
\begin{equation}
	(D^2 - \mu^2) \varphi_\infty \simeq \frac{-\lambda/6}{(2 \mu)^{3/2}} \psi^3 e^{-3 i \mu t} + {\rm c.c.}
\end{equation}
This can be solved via Green's function methods,
using the solution of $(D^2 - \mu^2) \varphi = 0$ in the Kerr
background.
For small $\alpha$, when the wavelength $\sim \mu^{-1}$ of the
emitted radiation is much larger than the horizon scale $r_g$,
we can ignore the near-horizon structure of the Kerr metric,
and consider only its $1/r$ behavior.
These calculations are discussed in more detail in App.~\ref{appcalcs}.

For radiation sourced by the 211 hydrogenic level,
which we write as $211 \times 211 \times 211 \rightarrow \infty$,
the emitted power to infinity is
\begin{equation}
	P \simeq  1.5 \times 10^{-8} \alpha^{17}
	\mu^2 \lambda^2 N_{211}^3,
	\label{eqrelpower}
\end{equation}
at leading order in $\alpha$.
The corresponding diagram is
\begin{center}
	\begin{tikzpicture}[scale=0.7]
	\draw (-1,1) -- (0,0);
	\draw (-1,0) -- (0,0);
	\draw (-1,-1) -- (0,0);
	\draw (0,0) -- (1,0);
	\node[anchor=east] at (-1,1) {211};
	\node[anchor=east] at (-1,0) {211};
	\node[anchor=east] at (-1,-1) {211};
	\node[anchor=west] at (1,0) {$\infty$};
\end{tikzpicture}
\end{center}
In principle, the emitted mode
has $\omega < m \Omega_H$ when the 211 level
is superradiant, and so will extract additional energy from the BH. However, like the SR rate of bound states, this horizon flux is suppressed by the small overlap between the BH and the radiation,
and is consequently a subleading effect
in the small-$\alpha$ limit.

Eq.~\eqref{eqrelpower} is $\sim 15$ times larger than the estimate
in~\cite{Arvanitaki:2010sy}. The latter
effectively solved the equation $\partial^2 \varphi_\infty
= \frac{-\lambda/6}{(2 \mu)^{3/2}} \psi^3 e^{- 3 i \mu t} + {\rm c.c.}$;
that is, they approximated the emitted radiation as being
massless, and propagating on a flat-space background.

If there is some occupation number in states
other than 211, then any combination of three initial
states can result in relativistic radiation.
If the 
bound states have orbital angular 
momenta $l,l',l''$, then the emitted power scales
as $P \propto \alpha^{11 + 2(l + l' + l'')} \mu^2 N N' N''$,
where $N,N',N''$ are the respective occupation numbers.
In particular, as we will see below,
populations in multiple superradiant levels
can lead to forced oscillations in the $l=0,m=0$
mode. This might lead us to wonder whether the less severe
$\alpha$ suppression in the 
\begin{center}
	\begin{tikzpicture}[scale=0.7]
	\draw (-1,1) -- (0,0);
	\draw (-1,0) -- (0,0);
	\draw (-1,-1) -- (0,0);
	\draw (0,0) -- (1,0);
		\node[anchor=east] at (-1,1) {(0,0)};
		\node[anchor=east] at (-1,0) {(0,0)};
		\node[anchor=east] at (-1,-1) {(0,0)};
	\node[anchor=west] at (1,0) {$\infty$};
\end{tikzpicture}
\end{center}
process, as compared to $211 \times 211 \times 211
\rightarrow \infty$, can compensate for
the smaller amplitude of the $00$ mode in comparison
to 211. However, for the 211 and 322 occupation numbers
attained in the evolution of the cloud
(see section~\ref{secperturb}), the emitted
power via $211\times211\times 211\rightarrow \infty$, Eq.~\eqref{eqrelpower},
is suppressed by fewer powers of $\alpha$, and numerically always
much larger.

\begin{figure*}
\begin{center}
	\begin{tikzpicture}[scale=0.3]
	\draw (-1,1) -- (1,-1);
	\draw (-1,-1) -- (1,1);
	\node[anchor=east] at (-1,1) {211};
	\node[anchor=east] at (-1,-1) {211};
	\node[anchor=west] at (1,1) {322};
	\node[anchor=west] at (1,-1) {BH};
\end{tikzpicture}
\quad
	\begin{tikzpicture}[scale=0.3]
	\draw (-1,1) -- (1,-1);
	\draw (-1,-1) -- (1,1);
	\node[anchor=east] at (-1,1) {322};
	\node[anchor=east] at (-1,-1) {322};
	\node[anchor=west] at (1,1) {211};
	\node[anchor=west] at (1,-1) {$\infty$};
\end{tikzpicture}
\end{center}
\begin{center}
	\begin{tikzpicture}[scale=0.4]
	\draw (-1,1) -- (0,0);
	\draw (-1,0) -- (0,0);
	\draw (-1,-1) -- (0,0);
	\draw (0,0) -- (1,0);
	\node[anchor=east] at (-1,1) {211};
	\node[anchor=east] at (-1,0) {211};
	\node[anchor=east] at (-1,-1) {211};
	\node[anchor=west] at (1,0) {$\infty$};
\end{tikzpicture}
\quad
	\begin{tikzpicture}[scale=0.4]
	\draw (-1,1) -- (0,0);
	\draw (-1,0) -- (0,0);
	\draw (-1,-1) -- (0,0);
	\draw (0,0) -- (1,0);
	\node[anchor=east] at (-1,1) {211};
	\node[anchor=east] at (-1,0) {211};
	\node[anchor=east] at (-1,-1) {322};
	\node[anchor=west] at (1,0) {$\infty$};
\end{tikzpicture}
\quad
	\begin{tikzpicture}[scale=0.4]
	\draw (-1,1) -- (0,0);
	\draw (-1,0) -- (0,0);
	\draw (-1,-1) -- (0,0);
	\draw (0,0) -- (1,0);
	\node[anchor=east] at (-1,1) {211};
	\node[anchor=east] at (-1,0) {322};
	\node[anchor=east] at (-1,-1) {322};
	\node[anchor=west] at (1,0) {$\infty$};
\end{tikzpicture}
\quad
	\begin{tikzpicture}[scale=0.4]
	\draw (-1,1) -- (0,0);
	\draw (-1,0) -- (0,0);
	\draw (-1,-1) -- (0,0);
	\draw (0,0) -- (1,0);
	\node[anchor=east] at (-1,1) {322};
	\node[anchor=east] at (-1,0) {322};
	\node[anchor=east] at (-1,-1) {322};
	\node[anchor=west] at (1,0) {$\infty$};
\end{tikzpicture}
\end{center}
\begin{center}
	\begin{tikzpicture}[scale=0.5]
	\draw (-1,1) -- (0,0);
	\draw (-1,-1) -- (0,0);
		\draw[snake it] (0,0) -- (1,0);
	\node[anchor=east] at (-1,1) {211};
	\node[anchor=east] at (-1,-1) {211};
	\node[anchor=west] at (1,0) {$\infty$};
\end{tikzpicture}
\quad
	\begin{tikzpicture}[scale=0.5]
	\draw (-1,1) -- (0,0);
	\draw (-1,-1) -- (0,0);
		\draw[snake it] (0,0) -- (1,0);
	\node[anchor=east] at (-1,1) {211};
	\node[anchor=east] at (-1,-1) {322};
	\node[anchor=west] at (1,0) {$\infty$};
\end{tikzpicture}
\quad
	\begin{tikzpicture}[scale=0.5]
	\draw (-1,1) -- (0,0);
	\draw (-1,-1) -- (0,0);
		\draw[snake it] (0,0) -- (1,0);
	\node[anchor=east] at (-1,1) {322};
	\node[anchor=east] at (-1,-1) {322};
	\node[anchor=west] at (1,0) {$\infty$};
\end{tikzpicture}
\quad 
	\begin{tikzpicture}[scale=0.5]
	\draw (-1,0) -- (0,0);
	\draw (0,0) -- (1,1);
		\draw[snake it] (0,0) -- (1,-1);
	\node[anchor=east] at (-1,0) {322};
	\node[anchor=west] at (1,1) {211};
	\node[anchor=west] at (1,-1) {$\infty$};
\end{tikzpicture}
\end{center}
\caption{Processes
relevant to the evolution of the 211 and 322 hydrogenic modes. The first row 
corresponds to the interactions between non-relativistic modes  (Sec.~\ref{secnonrelem} and~\ref{secboundints}) and
the second corresponds to the emission of relativistic scalar
radiation (Sec.~\ref{secrelscalar}), both mediated by the quartic self-interaction. The third row corresponds to the emission
of gravitational radiation (indicated
by wavy legs), also present in gravitational superradiance.}
\label{figtls}
\end{figure*}
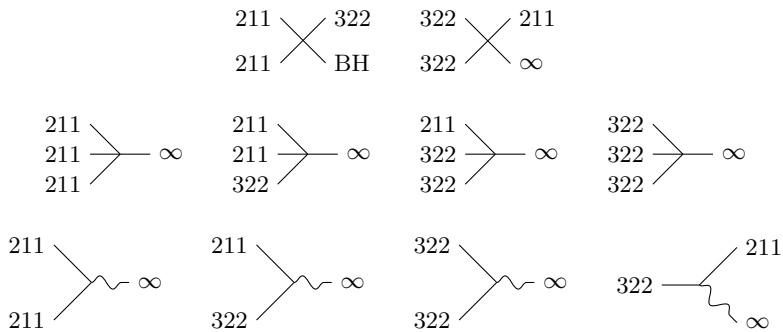

\subsection{Non-relativistic scalar emission}
\label{secnonrelem}

Emission to unbound states can also occur in the non-relativistic
regime.
Suppose that we have bound oscillations $\psi_j(t) = \psi_j e^{-i \tilde \omega_j t}$,
where $j$ labels a particular bound state, with 
frequencies 
$\tilde \omega_{j,j',j''} < 0$ (i.e.\ the physical frequencies
are $\omega = \mu + \tilde \omega < \mu$).
If $\tilde \omega_j + \tilde \omega_{j'} - \tilde \omega_{j''} > 0$, then the $\psi_j \psi_{j'} \psi_{j''}^*$ term in the equation
of motion will source unbound, non-relativistic radiation,
corresponding to the diagram
\begin{center}
	\begin{tikzpicture}[scale=0.7]
	\draw (-1,1) -- (1,-1);
	\draw (-1,-1) -- (1,1);
	\node[anchor=east] at (-1,1) {$j$};
	\node[anchor=east] at (-1,-1) {$j'$};
	\node[anchor=west] at (1,1) {$j''$};
	\node[anchor=west] at (1,-1) {$\infty$};
\end{tikzpicture}
\end{center}
Since the emitted state is also non-relativistic,
we can consistently use the
Gross-Pitaevskii equation (Eq.~\eqref{eq:eqgp1}).
Writing $\psi$ for the radiated wave, we want to solve
\begin{equation}
 \left(\tilde \omega + \frac{\nabla^2}{2 \mu} + \frac{\alpha }{r} \right)
	\psi = \frac{-3}{12 \mu^2} \lambda \psi_j \psi_{j'} \psi_{j''}^*
	\label{eqgpnr}
\end{equation}
(with the appropriate multiplicity factors). 
For each of the different spherical harmonic components
in the right hand side of Eq.~\eqref{eqgpnr}, we can write
a one-dimensional radial equation for the part of $\Psi$
with the corresponding angular dependence. 
These radial equations can be non-dimensionalised~\cite{Gruzinov_2016},
showing that the power
emitted in non-relativistic modes is given by
$P~\propto~\alpha^4 \lambda^2 N_j N_{j'} N_{j''} \mu^2$
at leading order in $\alpha$, where $N_j, N_{j'}, N_{j''}$
are the occupation numbers of the bound modes. 
The constant factors can be found by numerically solving the 
radial equations, as reviewed in App.~\ref{appnonrelem}.

Considering an example which will, in many circumstances,
be very important for the cloud's evolution,
suppose that we have
some population in the 211 and 322 modes. 
Taking $\psi_{j,j'} = \psi_{322}$ and $\psi_{j''} = \psi_{211}$,
we have $2 \tilde \omega_{322} - \tilde \omega_{211} \simeq
\frac{\alpha^2 \mu}{72} > 0$, so emission to infinity is possible.
As reviewed in App.~\ref{appnonrelem}, this emission is dominantly
sourced
at radii $r \sim r_c \equiv r_g / \alpha^2$,
i.e. where most of the cloud's mass sits. Since the dominant
part of the BH potential is $\sim 1/r$ at large distances, which
is spherically symmetric, both the bound modes and the emitted
wave will have have approximately spherical harmonic
angular dependence.
For this particular case,
$Y_{22}^2 Y_{11}^* = \sqrt{\frac{5}{42 \pi^2}} Y_{33}
- \sqrt{\frac{5}{1848 \pi^2}} Y_{53}$, so the emitted quanta
are in the $l=3,m=3$ and $l=5,m=3$ modes.
At leading order in $\alpha$, the total emitted power for
the 
\begin{equation}
	\begin{tikzpicture}[scale=0.7]
	\draw (-1,1) -- (1,-1);
	\draw (-1,-1) -- (1,1);
	\node[anchor=east] at (-1,1) {322};
	\node[anchor=east] at (-1,-1) {322};
	\node[anchor=west] at (1,1) {211};
		\node[anchor=west] at (1,-1) {$(3,3),(5,3)$};
\end{tikzpicture}
	\label{eq2213}
\end{equation}
process is 
\begin{equation}
	P \simeq 10^{-8} \alpha^4 \lambda^2 \mu^2 N_{322}^2 N_{211}
	\label{eqpemit1}
\end{equation}
with the $(l,m) = (3,3)$ radiation dominating the emitted power.\footnote{This expression corresponds to the classical wave equation;
in the quantum case, the final state occupation number
$N_{211}$ should be replaced by $N_{211} + 1$. We use the classical expression for brevity in the remainder of the text,
though the quantum version is important in allowing levels
to grow from vacuum fluctuations.}
This is a factor $4$ smaller than the rate
given in~\cite{Gruzinov_2016}, due to the hydrogenic
wavefunctions used in the latter
having a normalization that is a factor $\sqrt{2}$ too large.
The rates for processes involving different bound
states are discussed in App.~\ref{appnonrelem},
and tabulated in Table~\ref{tablenrem}.

At larger $\alpha$, deviations from the non-relativistic
approximation become more important. 
However, at small enough $\alpha$ such that 211 is still
superradiant, 
the $\psi_{211}$ and $\psi_{322}$ wavefunctions
are still well-approximated by the hydrogenic form,
except near the origin.
Since 
the source term $\psi_{322}^2 \psi_{211}^*$ for the non-relativistic
radiation
is largest at the characteristic radius of the bound states,
$a \sim r_g/\alpha^2$, 
where the potential is dominantly $\sim 1/r$,
we would expect the corrections to the non-relativistic
calculation to be small. This can be confirmed by 
performing a numerical computation in the 
Kerr background, the results of which match the leading-order
formula for the emitted power (Eq.~\eqref{eqpemit1})
at the few percent level.

As well as relativistic effects, there will also
be higher-order effects of $\lambda$; for example,
self-interaction-induced distortions to the bound
state wavefunctions, and to the radiated wave.
For $\varphi/f \ll 1$, these effects will be small.
In much of the astrophysically-relevant parameter
space, this condition holds, as we discuss
in section~\ref{sec:nonperturb}.

\subsection{Bound state interactions}
\label{secboundints}

If we have bound oscillations
$\psi_{j,j',j''}$ for which $\tilde \omega = \tilde \omega_j + \tilde \omega_{j'} 
- \tilde \omega_{j''} < 0$,
then the oscillation that they source is also bound.
For example, the $\psi_{211}^2 \psi_{322}^*$ term has
 frequency $2 \tilde \omega_{211} - \tilde \omega_{322} \simeq 
-\frac{7 \alpha^2 \mu}{36} < 0$.
In general, $\tilde \omega$ will not be very close
to the frequency of any of the hydrogenic bound levels 
(with some exceptions that we review below)
so the oscillation that they source will be forced.

Depending on the angular properties of the driving modes,
the forced oscillation may gain or lose energy from the BH.
If it loses energy to the BH, then for a forcing term
$\psi_j \psi_{j'} \psi_{j''}^*$, this corresponds
to energy loss from the $\psi_j, \psi_{j'}$ modes,
but energy \emph{gain} for the $\psi_{j''}$ mode. The example that
will be the most important for us is when $\psi_j, \psi_{j'} = \psi_{211}$,
and $\psi_{j''} = \psi_{322}$:
\begin{equation}
	\begin{tikzpicture}[scale=0.7]
	\draw (-1,1) -- (1,-1);
	\draw (-1,-1) -- (1,1);
	\node[anchor=east] at (-1,1) {211};
	\node[anchor=east] at (-1,-1) {211};
	\node[anchor=west] at (1,1) {322};
	\node[anchor=west] at (1,-1) {$m=0$};
\end{tikzpicture}
	\label{eq1120}
\end{equation}

The forced oscillation has $m=0$, so loses
energy through the BH horizon. Given some amplitude
in the 211 and 322 modes, each $\sim \mu$ of 
energy lost from the forced oscillation into the BH
corresponds to $\sim 2 \mu$ loss from the 211 mode,
and $\sim \mu$ gain in the 322 mode. The energy loss
rate is proportional to the squared amplitude 
of the forced oscillation, which is $\propto
N_{211}^2 N_{322}$. Consequently, if we have a large
initial occupation number in 211, and a small initial occupation
number in 322, then this process will lead to the
exponential growth of $N_{322}$, at the expense of 211.

This picture makes intuitive sense when the amplitudes
of the ``forcing'' modes (211 and 322 in the above example)
are large. However, if we are interested in e.g.\ the
growth of 322 from quantum fluctuations, we might worry about
the validity of treating it as a forcing for the $m=0$
oscillation. A more systematic approach (reviewed
in App.~\ref{app:param}) is to assume
that we have some large-amplitude $\psi_c$, and treat this
as the source for only two of the ``legs'', i.e.\ to solve
\begin{equation}
	(i \partial_t + \MM)\psi = \frac{-3 \lambda}{24 \mu^2}(\psi_c^2 \psi^*
	+ |\psi_c|^2 \psi)
	\label{eqGPparam}
\end{equation}
(here, $\mathcal{M}$ represents the other terms
in the non-relativistic Hamiltonian, including
an absorbing term corresponding to the BH horizon)
with $\psi_c$ acting as a \emph{parametric} driving
term, rather than a simple forcing.
When the amplitude of this driving term is small,
its effects can be described as perturbations
to the usual modes, ``mixing'' them with others.
The key point is that, if the $\psi_c^2 \psi^*$ term
induces a mixing with a decaying mode,
then this contributes a \emph{growing} term
to the original $\psi$ mode. In our 
$211 \times 211 \rightarrow 322 \times {\rm BH}$ example, 
if we take $\psi_c = \psi_{211}$, then this acts
as a parametric driving, which mixes 322 with decaying
modes such as 100. This results in the same growth rate for
322 as we would calculate from the forced oscillation picture
above. Quantitatively, the energy flux into the BH is,
at leading order in $\alpha$,
\begin{equation}
	P \simeq 4 \times 10^{-7} \alpha^7 \lambda^2 (1 + 
	\sqrt{1 - a_*^2}) \mu^2 N_{211}^2 N_{322}
	\label{eqpcross1}
\end{equation}
More generally, for $\psi_{j,j',j''}$ such that the forced oscillation
has a $m=0$ component, the energy flux
through the BH horizon is 
$P \propto \alpha^7 \lambda^2 (1 + \sqrt{1 - a_*^2}) \mu^2
N_{j} N_{j'} N_{j''}$.

These calculations are discussed in
App.~\ref{appdd}, and rates for different
processes are tabulated in Table~\ref{tablebound}.
The listed processes all correspond to forced
oscillations with $m=0$. Forced oscillations with 
larger $|m|$ have smaller energy fluxes
into (or out of) the horizon, corresponding
to bound state interaction rates that
are suppressed by higher powers of $\alpha$.

At larger $\alpha$, there will be deviations from 
the leading power-law behavior of Eq.~\eqref{eqpcross1}.
Since the energy lost through the forced oscillation
depends on its value at the horizon, i.e.\ on the behavior
at small distances, we would expect these deviations to be
relatively greater than those for non-relativistic
radiation in the previous subsection.
As we discuss in App.~\ref{app:param}, the behavior
is similar to that of the 100 level's decay rate,
with the rate a factor few larger than the leading-order value
at $\alpha \sim 0.2$. While we provide leading-$\alpha$ expressions in the text, the semi-analytic and numerical results from App.~\ref{app:param} are used for our results.

If all
four legs of the interaction are almost on-shell,
then the $\alpha$ scaling of the energy flux
can be different from that of Eq.~\eqref{eqpcross1}.
An example, that will be of interest in
section~\ref{secperturb}, is
\begin{equation}
	\begin{tikzpicture}[scale=0.7]
	\draw (-1,1) -- (1,-1);
	\draw (-1,-1) -- (1,1);
	\node[anchor=east] at (-1,1) {211};
	\node[anchor=east] at (-1,-1) {311};
	\node[anchor=west] at (1,1) {322};
		\node[anchor=west] at (1,-1) {$m=0$};
\end{tikzpicture}
	\label{eq1120res}
\end{equation}
Since $\omega_r = \mu (1 - \alpha^2/(2 n^2) + \OO(\alpha^4))$,
we have
$\omega_{211} + \omega_{311} - \omega_{322} = \omega_{200} + \OO(\alpha^4)$ (whereas for $211 \times 211 \rightarrow 322 \times {\rm BH}$,
$2 \omega_{211} -  \omega_{322}$ is $\OO(\alpha^2)$ away 
from the frequency of any quasi-bound level).
Consequently, the 200 forced oscillation dominates the energy flux
into the BH, and we obtain
\begin{equation}
	P \simeq 3 \times 10^{-10} \alpha^3 \lambda^2 (1 + \sqrt{1 - a_*^2})
	\mu^2 N_{211} N_{311} N_{322}
\end{equation}
This parametrically faster
rate means that any 311 occupation can be quickly depleted by this
process, as we will see in section~\ref{n11}.

\subsection{Cubic couplings}
\label{sec:cubic}

In the above, we assumed that the self-interactions consist of 
a quartic $\lambda \varphi^4$ interaction.
A generic scalar can also have a cubic term,
\begin{equation}
	\LL \supset - \frac{1}{2}\mu^2 \varphi^2 + \frac{g}{3!} \varphi^3 
	+ \frac{\lambda}{4!} \varphi^4.
\end{equation}
If we write $\lambda =   \mu^2 / f^2$,
then a natural value  for the cubic is $g = C \mu^2 / f$,  $C\sim \mathcal{O}(1)$.
For example, if we take a cosine potential and add a slope
\begin{equation}
	V(\varphi) = \mu^2f^2 \left(1-\cos(\varphi/f) - C \varphi/f\right),
\end{equation}
then the expansion of the potential around its minimum is
\begin{equation}
V(\varphi_0+\delta\varphi)	= 
	\frac{\mu^2}{2} \delta \varphi^2
	- \frac{C}{3!} \frac{\mu^2 }{f}\delta \varphi^3
	- \frac{1}{4!}\frac{\mu^2}{f^2}\delta \varphi^4 + \dots
\end{equation}
to leading order in small $C$ and $\delta \varphi$.\par
At leading order in $g$, the only relevant process
is relativistic $2 \rightarrow 1$ emission,
in analogy to the relativistic $3 \rightarrow 1$ emission
discussed in section~\ref{secrelscalar}. For definiteness, consider again the situation for the level with the fastest superradiant rate, 211. The leading order cubic process is
\begin{equation} 
\label{eq211cubic}
	\begin{tikzpicture}[scale=0.5]
	\draw (-1,1) -- (0,0);
	\draw (-1,-1) -- (0,0);
		\draw (0,0) -- (1,0);
	\node[anchor=east] at (-1,1) {211};
	\node[anchor=east] at (-1,-1) {211};
	\node[anchor=west] at (1,0) {$\infty$};
\end{tikzpicture}
\end{equation}
with  power:
\begin{equation}
\label{cubicpower}
P\simeq 10^{-4} \alpha^{12} C^2 (\mu^4/f^2) N_{211}^2 .
\end{equation}
More generally, for radiation sourced by quasi-bound
levels with orbital angular momentum $l$ and $l'$, 
the emitted power scales as $P \propto \alpha^{8 + 2 (l + l')}
C^2 (\mu^4/f^2) N N'$.
Unlike for the case of relativistic $3 \rightarrow 1$ emission
via a quartic coupling (section~\ref{secrelscalar}), the leading-$\alpha$ contribution
can be obtained by treating the radiation
as propagating in flat space,
i.e.\ by solving $(\partial^2 - \mu^2) \varphi_\infty =$ source.

Similarly to the discussion in section~\ref{secrelscalar},
we can ask whether the smaller $\alpha$ suppression
of the 
\begin{center}
\label{cubic}
	\begin{tikzpicture}[scale=0.5]
	\draw (-1,1) -- (0,0);
	\draw (-1,-1) -- (0,0);
		\draw (0,0) -- (1,0);
		\node[anchor=east] at (-1,1) {(0,0)};
		\node[anchor=east] at (-1,-1) {(0,0)};
	\node[anchor=west] at (1,0) {$\infty$};
\end{tikzpicture}
\end{center}
process, sourced by forced oscillations in the $l=0,m=0$
mode, can compensate for its smaller source amplitude
compared to $211 \times 211 \rightarrow \infty$.
For the 211 and 322 occupation numbers
attained (section~\ref{secperturb}), the power from
the latter process is again parametrically and numerically larger.

In the next section, we will show that, at the very least for large parts of parameter space, relativistic processes in general (from cubic or quartic vertices) are less important than quartic self-interactions between non-relativistic states. \par
As well as these leading-order processes, interactions between non-relativistic modes are generated at order $g^2$:
\begin{center}
 	\vspace{0.06cm}
	\begin{tikzpicture}[scale=0.5]
	\draw (-1,1) -- (0,0);
	\draw (-1,-1) -- (0,0);
		\draw (0,0) -- (1,0);
		\draw (1,0) -- (2,1);
		\draw (1,0) -- (2,-1);
\end{tikzpicture}
\quad
	\begin{tikzpicture}[scale=0.5]
	\draw (-1,1) -- (0,0.5);
	\draw (-1,-1) -- (0,-0.5);
	\draw (0,0.5) -- (0,-0.5);
	\draw (0,0.5) -- (1,1);
	\draw (0,-0.5) -- (1,-1);
\end{tikzpicture}
\quad
	\begin{tikzpicture}[scale=0.5]
	\draw (-1,-1) -- (-0.5,0);
	\draw (1,-1) -- (0.5,0);
	\draw (-0.5,0) -- (0.5,0);
	\draw (-1,1) -- (0.5,0);
	\draw (1,1) -- (-0.5,0);
\end{tikzpicture}
	\vspace{0.06cm}
\end{center}
In terms of interactions between 
non-relativistic modes, these are equivalent
to a quartic interaction
$\lambda_{\rm eff} = \frac{5}{3} \frac{g^2}{\mu^2} 
= \frac{5 C^2}{3} \frac{\mu^2}{f}$, which is always attractive.\footnote{We find that the contribution of the cubic coupling to the effective quartic is greater than the one in \cite{Gruzinov_2016} by a factor of 5/4.}
It should be noted that this is only true
for non-relativistic modes; other
processes induced at order $g^2$, such as $3 \rightarrow 1$ emissions,
will not be captured
by the same effective quartic.
Nevertheless, as we will discuss in section~\ref{secperturb}, in many
circumstances, only processes involving non-relativistic
states are important for the evolution of the
field around the BH.

Since the most important behavior can generally be captured
by an effective quartic coupling, we will ignore
the cubic coupling for most of this paper,
setting $C=0$. For $C \neq 0$, 
one can use the replacement rule
\beq
\frac{1}{f^2} \rightarrow \frac{1}{f_\text{eff}^2} = \left(1+\frac{5}{3}C^2\right)\frac{1}{f^2}
\eeq
for processes involving only non-relativistic states.

\subsection{Summary}

In ``gravitational'' superradiance, there are two generic ways for bound states to gain or lose energy and thus particle number: superradiance itself, in which the black hole acts as an energy and angular momentum source, and gravitational radiation, which carries energy and angular momentum to infinity. We have seen that in the presence of quartic self-interactions, three new classes of processes are introduced:  emission of relativistic axion waves to infinity, emission of non-relativistic axion waves to infinity, and excitation of forced oscillations which typically are absorbed back into the black hole. 

A non-zero cubic self-interaction can act as an
additional source of relativistic emission, as
well as contributing to an effective quartic term.
We will see that, unless the cubic coupling is tuned so as to
suppress the effective quartic coupling, or the cubic is rather large compared to its natural value ($|C|\gg1$), relativistic emission
generally does not have an important effect on the dynamics.

The first investigation of  scalar self-interactions in BH superradiance
was in Ref.~\cite{Arvanitaki:2010sy},
which carried out a very similar analysis
to ours; for example,  Eq.~(50) in Ref.~\cite{Arvanitaki:2010sy} corresponds to our Eq.~\eqref{eqGPparam} describing bound-state interactions.
However, in considering whether a perturbation
grows or shrinks,  Ref.~\cite{Arvanitaki:2010sy} focused on the energy flux through
the BH horizon, and did not take into account
energy transfer, through the parametric forcing term, between bound states. 
Since the BH absorbs energy in e.g. the $211 \times 211 \rightarrow 322 \times {\rm BH}$ process,
the conclusion was that interaction between modes suppresses occupation number growth.
This seems to account for the discrepancy between
our analysis and the conclusions of Ref.~\cite{Arvanitaki:2010sy}.

The processes outlined in this section create new energy loss
mechanisms for bound states, thereby typically limiting their
occupation numbers below those of gravitational superradiance. They
also create the ability to exchange particles efficiently between
bound states with different energy and angular momentum, enabling the
growth of high angular momentum states on timescales much faster
than the growth possible through gravitational superradiance alone. In the
following section, we will discuss in detail the new dynamics for a
range of self-interaction strengths.

Finally, similarly to the emission processes discussed
above, there will also be effects
that are higher order in $\lambda$. In particular,
if the amplitude of the cloud becomes too large,
then the attractive self-interactions
will lead to a rapid, non-perturbative collapse, 
the ``bosenova'' \cite{Arvanitaki:2010sy}.
However, we will see that, for most parts of parameter
space, the leading order in $\lambda$ processes that
we have described will prevent the field
from reaching such large amplitudes.
We discuss such non-perturbative behavior in more detail in section~\ref{sec:nonperturb}.


\section{Perturbative evolution}
\label{secperturb}

\begin{table*}
	\begin{center}
		\begin{tabular}{c|l|l}
			process & Rate constant (occupation numbers $N$)  & Rate constant (normalized occupation numbers $\varepsilon$)\\ \hline
			211 superradiance & 
			$\, \, \Gamma_{211}^{\rm SR} \simeq 4 \times 10^{-2} \alpha^8 (a_* - 2 \alpha (1 + \sqrt{1 - a_*^2}) \mu$
			& 
			$\, \, \gamma_{211}^{\rm SR} = \Gamma_{211}^{\rm SR}$ \\
		\begin{tikzpicture}[scale=0.3]
	\draw (-1,1) -- (0,0);
	\draw (-1,-1) -- (0,0);
		\draw[snake it] (0,0) -- (1,0);
	\node[anchor=east] at (-1,1) {211};
	\node[anchor=east] at (-1,-1) {211};
	\node[anchor=west] at (1,0) {$\infty$};
\end{tikzpicture} & 
	  $\, \, \Gamma^\text{GW}_{211\times 211}$  $\simeq 10^{-2} \alpha^{12}\left(\frac{\mu}{\Mpl}\right)^2\,\mu$ & 
			$\, \, \gamma_{211\times 211}^{\rm GW} \simeq  10^{-2} \alpha^{14} \mu $\\
	\begin{tikzpicture}[scale=0.3]
	\draw (-1,1) -- (1,-1);
	\draw (-1,-1) -- (1,1);
	\node[anchor=east] at (-1,1) {211};
	\node[anchor=east] at (-1,-1) {211};
	\node[anchor=west] at (1,1) {322};
	\node[anchor=west] at (1,-1) {BH};
	\end{tikzpicture} & 
			$\, \, \Gamma^{322\times{\rm BH}}_{211\times211} \simeq 4 \times 10^{-7} \alpha^7 \lambda^2 (1 + \sqrt{1 - a_*^2}) \mu$
			&
			$\, \, \gamma^{322\times {\rm BH}}_{211\times 211} \simeq 4 \times 10^{-7} \alpha^{11} \left(\frac{\Mpl}{f}\right)^4 (1 + \sqrt{1 - a_*^2}) \mu$ \\
	\begin{tikzpicture}[scale=0.3]
	\draw (-1,1) -- (1,-1);
	\draw (-1,-1) -- (1,1);
	\node[anchor=east] at (-1,1) {322};
	\node[anchor=east] at (-1,-1) {322};
	\node[anchor=west] at (1,1) {211};
	\node[anchor=west] at (1,-1) {$\infty$};
	\end{tikzpicture} &
			$\, \, \Gamma^{211\times \infty}_{322 \times 322} \simeq 10^{-8} \alpha^4 \lambda^2 \mu$
			&
			$\, \, \gamma^{211 \times \infty}_{322 \times 322} \simeq 10^{-8} \alpha^8
			\left(\frac{\Mpl}{f}\right)^4 \mu$ \\
			& & \\
			322 superradiance & 
			$\, \, \Gamma_{322}^{\rm SR} \simeq 8 \times 10^{-5} \alpha^{12} (a_* - \alpha (1 + \sqrt{1 - a_*^2}) \mu$
			& 
			$\, \, \gamma_{322}^{\rm SR} = \Gamma_{211}^{\rm SR}$ \\
		\end{tabular}
	\end{center}
	\label{tabrates1}
	\caption{Rates for the most important processes involved in 
	the evolution of the 211 and 322 hydrogenic levels,
	at leading order in $\alpha$. The second column shows
	the rate constants appropriate for occupation numbers
	$N_{211}$ etc, as per equations~\eqref{eqdotnj} and~\eqref{eqn221d},
	while the third column shows the rate constants
	for normalized occupation numbers $\varepsilon_{211} \equiv
	N_{211} / (G M_{\rm BH}^2)$ etc, as per Eq.~\eqref{eqepsdot}.}
	\label{tab:ratesummary}
\end{table*}

In this section, we study the evolution 
of the cloud-BH system, when the new dynamics
introduced by self-interactions
can be treated perturbatively. That is,
we treat the cloud as consisting of approximately
hydrogenic levels, interacting on timescales
long compared to their oscillation timescales.
Although the processes are individually simple,
the number of them involved can make the narrative hard to follow.
Accordingly, we have collated some of the most important
information into a number of tables and figures.
Table~\ref{tab:ratesummary} lists the most
important processes affecting level evolution,
and gives their rates.
Fig.~\ref{fig:paramspace} is an important guide
to how our discussion is structured, showing the
four qualitatively different regimes of parameter space
that we analyze.
Table~\ref{tab:paramsummary} gives approximate
expressions for the boundaries of these regions,
and points to their definitions in the text.
Fig.~\ref{fig:nevolution} shows examples of the
time evolution of the cloud-BH system, drawn from the four
different regions.
Table~\ref{tab:paramsummary2} summarizes the
level occupation numbers, observational signatures,
and characteristic timescales associated with each region.

\subsection{Evolution of occupation numbers}
\label{secevol}

The evolution of the scalar field
around the BH is driven by the gravitational processes
discussed in Sec.~\ref{secspin0} ---
superradiant growth or decay, and GW emission ---
and by the interaction-mediated processes discussed in Sec.~\ref{sec:quartic}.
As we have seen, when these processes can be
treated perturbatively, they can be viewed
as transferring energy to and from the 
quasi-bound states of the field (which are themselves
only slightly perturbed from their
hydrogenic forms). Putting everything together,
we can write down a set of coupled differential
equations, governing the evolution of the occupation
numbers of the modes.

Schematically, if we write the occupation
number of level $j$ as $N_j$ (where we index
the different quasi-bound states by a single index $j$),
then
\begin{align}
	\label{eqdotnj}
	\dot N_j &= \Gamma^{\rm SR}_j N_j \\
	&+ \sum_{j'} (- c\Gamma^{\rm GW}_{j\times j'} 
	+ \Gamma^\text{GW}_{j'\rightarrow j} - \Gamma^\text{GW}_{j\rightarrow j'})N_j N_{j'} \nonumber \\
	&+ \sum_{j', j''} (\Gamma^{j\times k}_{j'\times j''} - c \Gamma^{j''\times k}_{j\times j'}
	- c \Gamma^{j'\times k}_{j\times j''} - c\Gamma_{j\times j'\times j''}^\infty) N_j N_{j'} N_{j''} \nonumber
\end{align}
where the $c\Gamma$ notation encodes the appropriate
multiplicity factors, and 
\begin{itemize}
	\item $\Gamma_j^{\rm SR}$ is the growth(/decay) rate
corresponding to the mode's flux across
the BH horizon
\item $\Gamma^{\rm GW}_{j\times j'}$ is
the annihilation rate of $j\times j'$ to gravitational radiation.
\item  $\Gamma^\text{GW}_{j\rightarrow j'}$ is the rate of transitions,
	via gravitational-wave emission, from 
		$j'$ to $j$.
	\item $\Gamma^{j \times k}_{j'\times j''}$ is the rate
		of the 
\begin{center}
	\begin{tikzpicture}[scale=0.3]
	\draw (-1,1) -- (1,-1);
	\draw (-1,-1) -- (1,1);
		\node[anchor=east] at (-1,1) {$j'$};
	\node[anchor=east] at (-1,-1) {$j''$};
	\node[anchor=west] at (1,1) {$j$};
		\node[anchor=west] at (1,-1) {$k$};
\end{tikzpicture}
\end{center}
		process, where the $k$ leg corresponds
		to non-relativistic scalar emission, or to bound forced
		oscillation damped by the BH. For emission to infinity, we will
		sometimes write $\Gamma_{j\times j'}^{j''\times \infty}$,
		while for a bound forced oscillation,
		we will write $\Gamma_{j\times j'}^{j''\times {\rm BH}}$.
	\item $\Gamma_{j\times j'\times j''}^\infty$ is the rate of 
		the
\begin{center}
	\begin{tikzpicture}[scale=0.5]
	\draw (-1,1) -- (0,0);
	\draw (-1,0) -- (0,0);
	\draw (-1,-1) -- (0,0);
	\draw (0,0) -- (1,0);
	\node[anchor=east] at (-1,1) {$j$};
	\node[anchor=east] at (-1,0) {$j'$};
	\node[anchor=east] at (-1,-1) {$j''$};
	\node[anchor=west] at (1,0) {$\infty$};
\end{tikzpicture}
\end{center}
relativistic emission process. Repeated indices will sometimes be abbreviated using an exponential (i.e. $\Gamma_{j\times j \times j}^\infty = \Gamma_{j^3}^\infty $)
\end{itemize}


For example, the evolution of the fastest-growing level is given by
\begin{align}
	\label{eqn221d}
	\dot N_{211} &= \Gamma^{\rm SR}_{211} N_{211} \\
	&- 2 \Gamma^{\rm GW}_{211\times 211} N_{211}^2
	- \Gamma^{\rm GW}_{211\rightarrow 322} N_{211} N_{322} + \dots \nonumber \\
	&- 2 \Gamma^{322\times \BH}_{211\times 211} N_{211}^2 N_{322}
	+ \Gamma^{211\times \infty}_{322\times 322} N_{211} N_{322}^2 + \dots \nonumber \\
	&- 3 \Gamma_{(211)^3}^\infty N_{211}^3 - 2 \Gamma_{(211)^2\times322}^\infty
	N_{211}^2 N_{322}
	+ \dots \nonumber
\end{align}
Some of the key rates, at leading order in $\alpha$, are listed in Table~\ref{tab:ratesummary}.

While, as we observed above, $\lambda$ is often
extremely small, the $N_j$ can become extremely large.
From Eq.~\eqref{eqjbh}, the angular momentum of
a BH is 
$J = a_* G M^2 \simeq 10^{78} a_* \left(\frac{M}{10 M_\odot}\right)^2$.
To spin it down by $\OO(1)$, as is necessary to saturate
the superradiant instability, we need $N_j$ to be of this order.
Consequently, it is often more convenient to 
work in terms of ``normalized'' occupation numbers, 
$\varepsilon_j \equiv N_j / (G M_{\rm BH}^2) < 1$,
and normalized rates $\gamma$ such that
\begin{align}
	\label{eqepsdot}
	\dot \varepsilon_j &= \gamma^{\rm SR}_j \varepsilon_j \\
	&+ \sum_{j'} (- c\gamma^{\rm GW}_{j\times j'} 
	+ \gamma^\text{GW}_{j'\rightarrow j} - \gamma^\text{GW}_{j'\rightarrow j}) \varepsilon_j \varepsilon_{j'}\nonumber\\
	&+ \sum_{j',j''} (\gamma^{j \times k}_{j'\times j''} - c \gamma^{j''\times k}_{j\times j'}
	- c \gamma^{j'\times k}_{j\times j''}
	- c \gamma^\infty_{j\times j'\times j''}) \varepsilon_j \varepsilon_{j'} \varepsilon_{j''}\nonumber
\end{align}

Similarly, it is helpful to write $\lambda \equiv  \mu^2/f^2$,
as motivated around Eq.~\eqref{eqlambda1}.
In terms
of these, the scalings with $\alpha$ and $f$
of the different $\gamma$ are:
\begin{itemize}
	\item For growth (or decay) of a bound oscillation via the BH horizon $\gamma_j^{\rm SR} \propto \alpha^{4l + 4}$
	\item For non-relativistic scalar emissions to infinity, 
		$\gamma^{j \times \infty}_{j'\times j''} \propto \alpha^{8} (\Mpl/f)^4$
	\item For the absorption of energy from a forced bound oscillation with angular momentum $l$ damped by the BH, 
		$\gamma^{j\times \BH}_{j'\times j''} \propto \alpha^{11 + 4l} (\Mpl/f)^4$ (except in the case of ``resonant'' processes,
		as discussed in section~\ref{secboundints}).
	\item For 3-to-1 relativistic scalar emissions to infinity $\gamma^{\infty}_{j\times j'\times j''} \propto \alpha^{2(l+l'+l'')+15}$ 
	\item For annihilation to gravitational waves $\gamma^{\rm GW}_{j\times j'} \propto \alpha^{10 + 2(l + l')}$
	\item For transitions between bounds states with gravitational wave emission, $\gamma^\text{GW}_{j\rightarrow j'}$,
		see section~\ref{sec:trans}.
\end{itemize}

In addition, a non-zero cubic interaction contributes to the evolution equations \eqref{eqepsdot} as
\begin{align}
	\dot \varepsilon_j &= - \sum_{j'} c\gamma_{j \times j'}^{\infty} \varepsilon_j \varepsilon_{j'}\label{eqepsdotcubic} + \dots
\end{align}
with rate  $\gamma_{j\times j'}^{\infty} \propto \alpha^{2(l+l')+10} |C|^2 \left(\Mpl / f\right)^2\mu$. 

The rates that determine the evolution in large
parts of the parameter space are listed in
Table~\ref{tab:ratesummary}, at leading order in
$\alpha$. As discussed above, for some of these
processes, this approximation can be quite poor
at $\alpha$ values of interest, and for
the computations involved in producing our plots,
we use more accurate numerical or semi-analytic expressions.

When all of the $\varepsilon_j$ are very small, then 
only the $\gamma_j^{\rm SR}$ are important,
and evolution proceeds as in the purely-gravitational
case, with the fastest-growing level increasing
exponentially in amplitude. Since
the $\varepsilon_j$ for this level will usually dominate
exponentially over the other $\varepsilon_{j'}$,
other levels can only be built up (faster than their superradiance rates) through\footnote{
	If the occupied level $j$ is higher-frequency
	than some other level $j'$,
	then transitions from $j$ to $j'$ via
	GW emission can also occur.
	However, as discussed in section~\ref{sec:gw},
	the fastest-growing superradiant level is also
	the most tightly bound superradiant
	level, for $l < 3$.
	Consequently, transitions
	from a superradiant level would have to be
	to decaying levels. Since
	the decay rate through the BH horizon is generally
	significantly larger
	than the growth rate
	due to GW transitions,
	this does not give rise to exponential growth
	of $j'$. For example,
	if we consider $322 \rightarrow 200 + $ GW
	transitions,
	the evolution equation for
	the 200 level is
	$\dot \varepsilon_{200}/\mu
	\simeq 4 \times 10^{-6} \alpha^8 \varepsilon_{322} \varepsilon_{200} - 0.5 \alpha^5 \varepsilon_{200}$, so
	the 200 level is still damped
	even for large $\varepsilon_{322}$.}
	\begin{center}
	\begin{tikzpicture}[scale=0.3]
	\draw (-1,1) -- (1,-1);
	\draw (-1,-1) -- (1,1);
	\node[anchor=east] at (-1,1) {$j$};
	\node[anchor=east] at (-1,-1) {$j$};
	\node[anchor=west] at (1,1) {$j'$};
	\node[anchor=west] at (1,-1) {BH};
\end{tikzpicture}
\end{center}
where the BH leg corresponds to a bound oscillation.
If interaction processes are strong enough to 
significantly affect
the evolution, then
the $j'$ for which this growth rate is fastest 
will be the next level to become important.

For small $\alpha$, the fastest superradiant growth
is for the 211 level, and the fastest quartic process,
given a 211 amplitude, is
\begin{center}
	\begin{tikzpicture}[scale=0.3]
	\draw (-1,1) -- (1,-1);
	\draw (-1,-1) -- (1,1);
	\node[anchor=east] at (-1,1) {211};
	\node[anchor=east] at (-1,-1) {211};
	\node[anchor=west] at (1,1) {322};
	\node[anchor=west] at (1,-1) {BH};
\end{tikzpicture}
\end{center}
as discussed in section~\ref{secboundints}. It turns out that,
similarly to the toy model discussed in~\cite{Gruzinov_2016}, there is a large
parameter space for which \emph{only} the (perturbed) 211 and 322 levels
are ever significantly populated.
This regime will be the main focus of our paper.

Situations in which 211 is the first superradiant
level generally lead to the strongest radiative
signals, either in gravitational or scalar waves.
However, superradiance into higher levels
can be important for other phenomenological signatures,
such as BH spin-down. In such circumstances, levels
other than 211 and 322 will be important. For example,
if 322 is the first level to grow through superradiance,
then 544 will generally be the next level to be built
up through self-interactions. Though we do not investigate
such scenarios in detail in this paper, they represent 
an important subject for future work.

\begin{figure*}[t]
	\includegraphics[width=.9\textwidth]{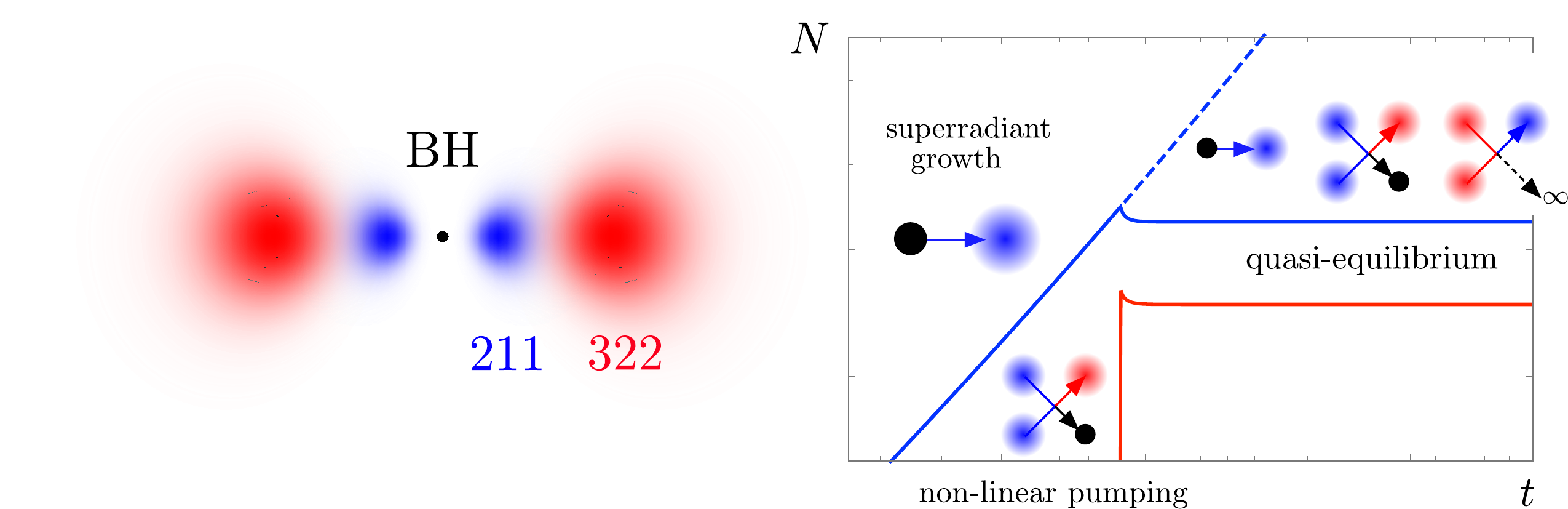}
	\caption{Schematic illustration of the effects
	of a large quartic self-interaction on the growth
	of scalar fields around a spinning BH. The
	left-hand figure shows the energy densities of 
	the 211 (blue) and 322 (red) modes in the
	$(x,z)$ plane, taking the BH spin
	to be in the $z$ direction. The right-hand panel
	shows the evolution of the 211 (blue) and 322 (red) occupation numbers with time (where the $N$ axis is taken to be logarithmic). We assume that the initial BH
	spin is high enough that the first process to occur is superradiant
	growth of 211. In the absence of self-interactions, this growth
	would continue until the BH was spun down to the $m=1$ threshold (as indicated
	by the dashed blue line). When sufficiently large self-interactions
	are present, the 322 mode is built up from the 211 mode,
	via the non-linear pumping process described in section~\ref{secboundints}.
	This stops the growth of 211, and the levels quickly reach
	a quasi-equilibrium configuration, in which the processes
	of 211 superradiance, $211 \times 211 \rightarrow 322 \times$ BH and
	$322 \times 322 \rightarrow 211 \times \infty$ emission
	(section~\ref{secnonrelem}) keep the 211 and 322 occupation numbers
	almost constant.}
	\label{figcartoon}
\end{figure*}

\subsection{Two-level system}
\label{sec:two-level}

If the (suitably perturbed) 211 and 322 modes are
the only ones with significant
occupation numbers, then the relevant processes are
illustrated in Fig.~\ref{figtls}.
Given this multitude of processes,
the behavior of the system seems
potentially very complicated. 
However, we will see that, because the
relativistic emission rates are suppressed by
high powers of $\alpha$ (and the gravitational
radiation rates have an additional relative
suppression of $(f/\Mpl)^4$, which will turn out
to be small when self-interactions are important),
only
the two non-relativistic processes (along with superradiance) are generally significant. 

Assuming that 211 is the fastest-growing mode
at the start of the evolution, these give rise
to fairly simple qualitative behavior, for
large enough couplings $\lambda$. Initially,
the 211 mode grows through superradiance.
Once its occupation number is large enough,
the growth rate of the 322 mode, through the
$211 \times 211 \rightarrow 322 \times {\rm
BH}$ process, becomes significant. This stops
the growth of the 211 mode.
Since 322 is depleted via the $322 \times 322
\rightarrow 211 \times \infty$ process, but
built up via $211 \times 211 \rightarrow 322
\times {\rm BH}$ (and vice versa for 211), the
211 and 322 modes reach a quasi-equilibrium
configuration, in which their occupation numbers
are almost constant. This evolution is illustrated
schematically in Fig.~\ref{figcartoon}, and
is the regime that was studied in the toy model
of~\cite{Gruzinov_2016}.

\begin{figure*}[t!]
	\begin{centering}
		\includegraphics[width=.49\textwidth]{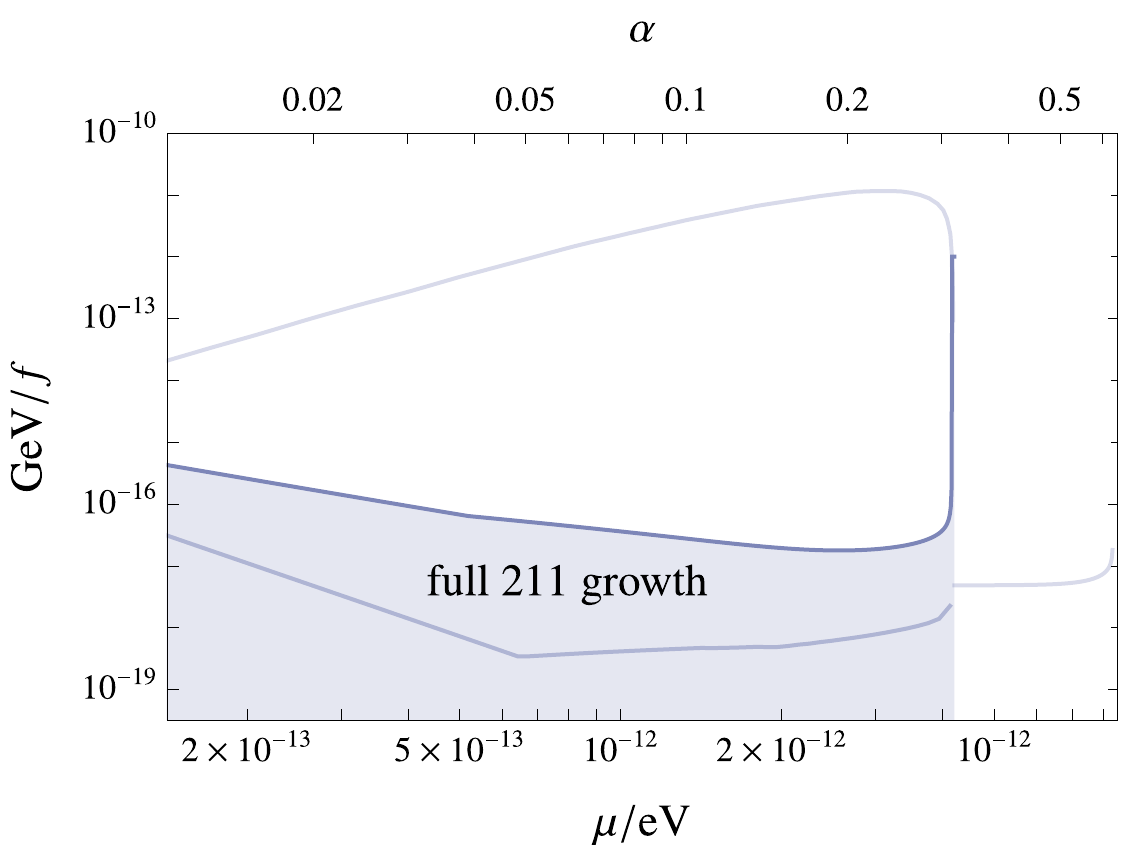}
		\includegraphics[width=.49\textwidth]{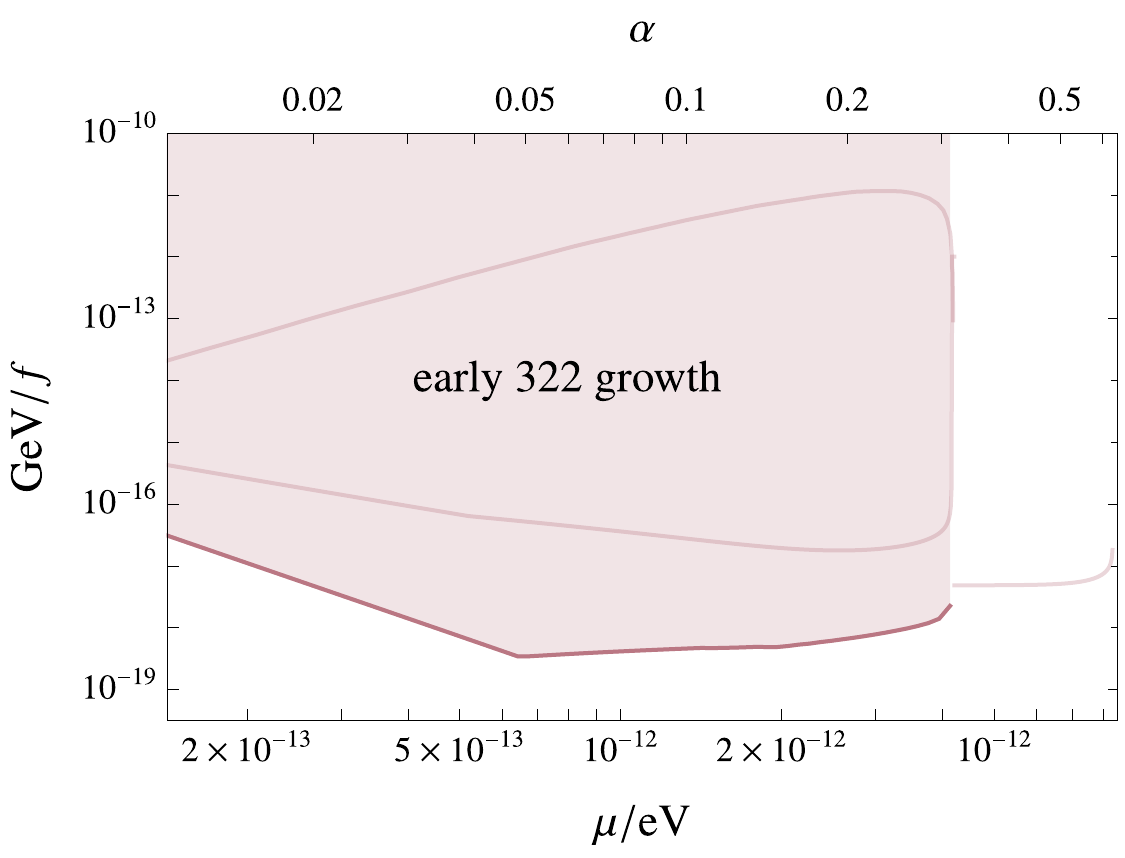}
		\includegraphics[width=.49\textwidth]{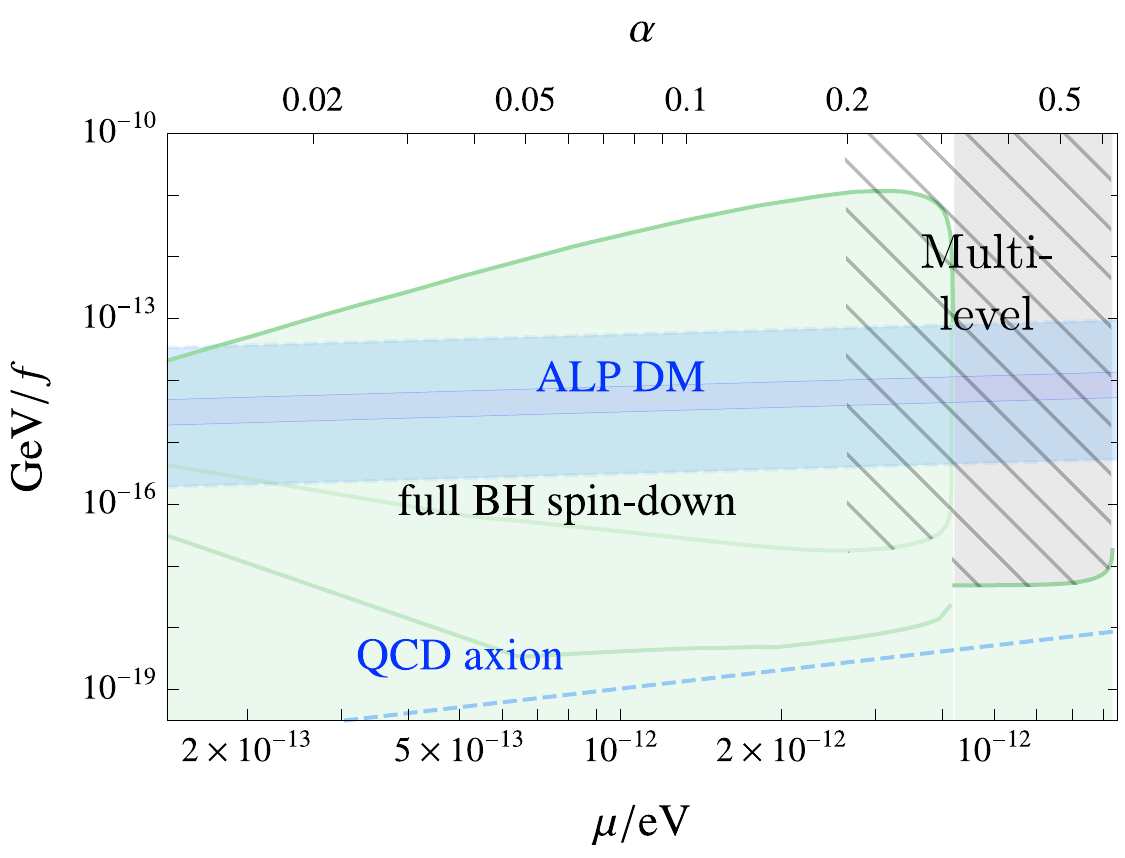}
		\includegraphics[width=.49\textwidth]{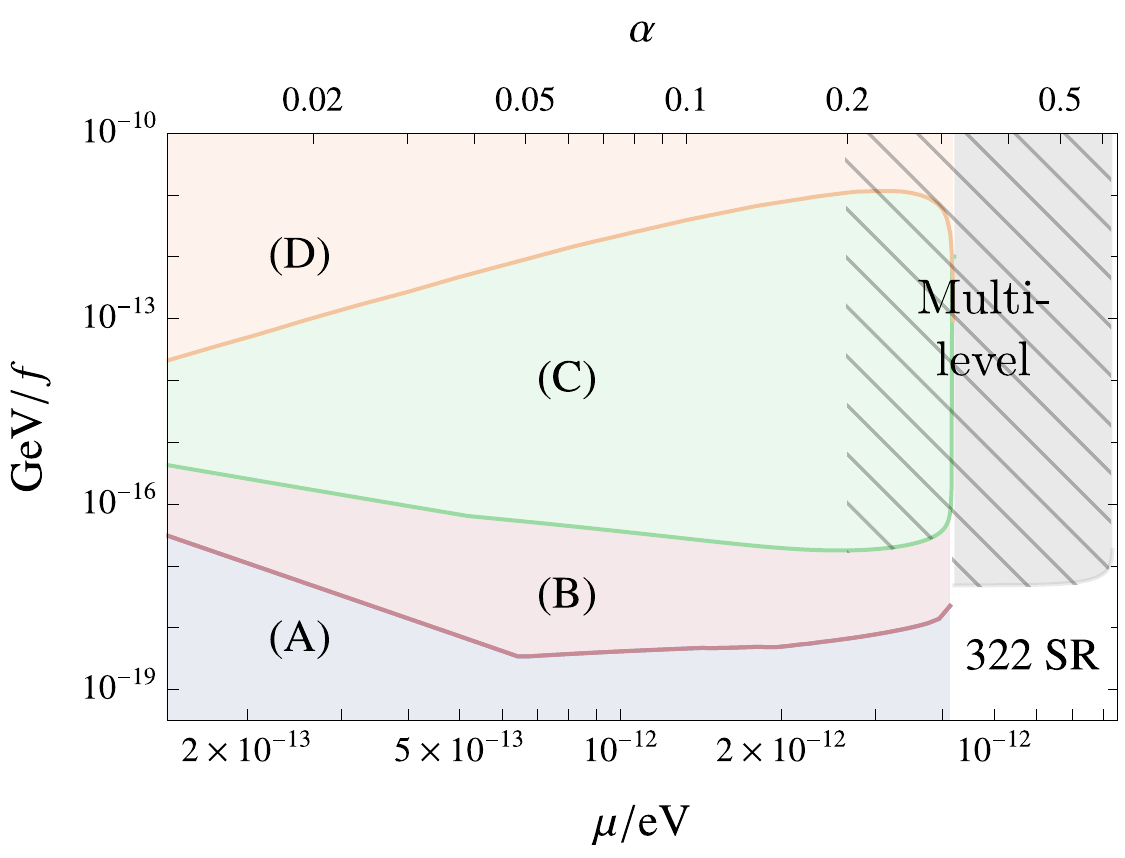}
		\caption{Parameter space for superradiance 
		of a scalar with mass $\mu$ and quartic coupling $\lambda 
		= \mu^2/f^2$, around
		a BH with $M_{\rm BH} = 10 \msun$ and $a_* = 0.9$ (initially),
		given a total evolution time of $10^{10}\yr$.
		\emph{Top-left:} parameter space in which the 211 level
		grows to saturation through superradiance.
		\emph{Top-right:} parameter space in which
		the 322 level grows faster due to self-interactions
		than it would have through superradiance alone.
		\emph{Bottom-left:} parameter space in which
		the BH is spun down to the threshold of 211
		superradiance. For $\mu \gtrsim 4\times 10^{-12} \eV$ (i.e.\
		past the threshold for 211 superradiance),
		we show the parameter space region in which 322
		superradiance is not cut off by self-interactions,
		and we can
		be confident that the BH is spun down to the 
		threshold of 322 superradiance. 
		The gray hatched region corresponds 
		to parameter space in which levels
		other than 211 and 322 are expected
		to grow; we have not fully analyzed the
		behavior in these regimes.
		The blue dashed line corresponds to the quartic
		coupling for the QCD axion. The ``ALP DM'' band
		corresponds to the range of quartic couplings
		that, for an axion with a time-independent cosine potential,
		allow the observed DM abundance to be produced by
		the early-universe misalignment mechanism.
		The darker middle band corresponds to $\OO(1)$ values
		of the initial misalignment angle, while the lighter bands
		above and below correspond to ``tuned'' initial values
		(see Sec.~\ref{secaxionmodels} for details).
		\emph{Bottom-right:} parameter space regions discussed
		in the text. (A) corresponds to the ``small self-coupling''
		regime discussed in section~\ref{seclsmall},
		(B) corresponds to the ``moderate self-coupling''
		regime discussed in section~\ref{seclmoderate},
		(C) corresponds to the ``large self-coupling''
		regime discussed in section~\ref{secllarge},
		and (D) corresponds to the ``lack of BH spindown''
		regime discussed in section~\ref{seclnosd}.
		The ``322 SR'' region is where 322 superradiance is not
		cut off by self-interactions,
		while the gray parameter space above this is when
		this does occur, and further analysis would be required.}
	\label{fig:paramspace}
	\end{centering}
\end{figure*}

The above picture holds for the case of large enough
self-couplings; in the opposite limit of very small
self-couplings, the evolution will be almost the same
as the purely gravitational case. For intermediate values
of $\lambda$, there can be more complicated behaviors.
In the rest of this section, we will make all of these
statements precise, by investigating
in detail the evolution of the cloud, for different
$\mu$ and $f$.
Fig.~\ref{fig:paramspace}, and
Tables~\ref{tab:paramsummary} and~\ref{tab:paramsummary2}, serve as guides to this discussion.
Readers more interested in the observational
effects of superradiance around astrophysical BHs can skip ahead
to sections~\ref{sec:spindown} and~\ref{sec:gw}, referring
back to this section when necessary.

\begin{table*}
	\begin{center}
		\begin{tabular}{l|c|c}
Coupling strength & Fig \ref{fig:paramspace} & Boundary in parameter space \\\hline
			Small (\ref{seclsmall})& A & \begin{tabular}{@{}c@{}}$f > f_\text{AB} \approx \min\Big[3\times10^{16}\GeV\left(\frac{T_\BH}{10^{10} \yr}\right)^\frac{1}{4}\left(\frac{\mu}{10^{-13}\eV}\right)^\frac{1}{4}\left(\frac{\alpha}{0.01}\right)^\frac{11}{4}$, \\ $8\times 10^{18}\GeV \left(\frac{0.01}{\alpha}\right)^\frac{3}{4}
		\left(\frac{a_*}{0.9}\right)^\frac{1}{4}\Big] $ (Eqs. \eqref{eq:f_{AB}}, \eqref{eq:ann_to_graviton})\end{tabular}\\
			Moderate   (\ref{seclmoderate}) & B & $f_\text{AB} > f > f_\text{BC}\approx 2\times10^{16}\GeV\left(\frac{a_*(t_0)}{0.9}\right)^{\frac{1}{4}}\min\left[\left(\frac{\alpha}{0.04}\right)^{\frac{3}{4}},\left(\frac{\alpha}{0.04}\right)^{\frac{3}{2}}\right] 
			$ (Eqs. \eqref{eq:large_interaction_boundary}, \eqref{eq:f_min}, \eqref{eq:f_eq}) \\
Large  (\ref{secllarge}) & C &  $f_\text{BC}>f>f_\text{CD}\approx  3\times 10^{14}\GeV \left(\frac{10^{10} \yr}{T_\BH}\right)^{\frac{1}{2}}\left(\frac{10^{-13} \eV}{\mu}\right)^{\frac{1}{2}}\left(\frac{0.01}{\alpha}\right)^{\frac{5}{2}}\left(\frac{0.9}{a_*(t_0)}\right)^{\frac{3}{4}} $ (Eqs. \eqref{eq:f_CD})  \\
No spindown (\ref{seclnosd}) & D & $ f_\text{CD}>f\gg\mu$  \\\hline\hline
		\end{tabular}
		\end{center}
	\caption{Approximate expressions for the boundaries
	between different regions in $\mu,f$ parameter space,
	as diagrammed in the bottom-right panel of Fig.~\ref{fig:paramspace}. The first column identifies the section in the text
	discussing the particular parameter space region,
	while the third column presents the $f$ range
	(for given $\mu$) corresponding to that region,
	along with references to the relevant equations in the text.
	The expressions given are to leading order in small
	$\alpha$, and numerical coefficients are approximate;
	the reader should refer to the text for more precise
	expressions.}
	\label{tab:paramsummary}
\end{table*}

\begin{table*}
	\begin{center}		
		\begin{tabular}{l|c|c|c|c}
 Coupling strength 
& $\varepsilon^\text{peak}_{211}/\varepsilon_{211}^\text{max}$& $\eta =\varepsilon_{322}/\varepsilon_{211}$  & Signatures & Timescales\\\hline
			Small (\ref{seclsmall}), A& $1$ &$\simeq 0$  & spindown, GW &$ \tau_\text{ann} \approx  10^{5}\yr \left(\frac{0.1}{\alpha}\right)^{14}\left(\frac{10^{-12}\eV}{\mu}\right)$ (Eq.~\eqref{eq:t_ann})\\
Moderate  (\ref{seclmoderate}), B & $1$ & $
   10^{-5}\left(\frac{\alpha}{0.01}\right)^3$&spindown, GW  & $\tau_\text{scalar} \approx 10^{-1}\yr\left(\frac{0.1}{\alpha}\right)^{14} \left(\frac{10^{-12}\eV}{\mu}\right)\left(\frac{f}{10^{17}\GeV}\right)^4$ (Eq.~\eqref{eq:t_scalar})\\
Large (\ref{secllarge}), C & $\left(\frac{f}{f_\text{BC}}\right)^2$&$
   10^{-5}\left(\frac{\alpha}{0.01}\right)^3$ &slow spindown, AW & 
   $\tau_\text{sd} \approx 10^7 \yr \left(\frac{0.01}{\alpha}\right)^5 \left(\frac{10^{-12}\eV}{\mu}\right)\left(\frac{0.9}{a_*}\right)^{\frac{3}{2}}\left(\frac{10^{15}\GeV}{f}\right)^2$ (Eq.~\eqref{eq:spindown_time})\\
No spindown (\ref{seclnosd}), D &$\left(\frac{f}{f_\text{BC}}\right)^2$ &$
    10^{-5}\left(\frac{\alpha}{0.01}\right)^3$ & no spindown, AW& $\tau_\text{sd} \gtrsim T_\BH$ (Eq.~\eqref{eq:no_spindown}) 
\\\hline\hline
		\end{tabular}\\
	\begin{flushleft}
		\vspace{0.2cm}
		$\varepsilon_{211}^\text{eq} \approx \frac{2}{\sqrt3}\frac{\sqrt{\kappa^\infty\kappa^\text{SR}(a_*-2\alpha\tilde r_+)}}{\alpha^3 \kappa^\text{BH}\tilde r_+}\left(\frac{f}{\Mpl}\right)^2 =2.5\times 10^{-1}\left(\frac{0.01}{\alpha}\right)^3\left(\frac{a_*}{0.9}\right)^{1/2}\left(\frac{f}{10^{15}\GeV}\right)^2$ (Eq.~\eqref{eq:quasi_equlibrium_occupations_211}) ; \quad
		\vspace{0.1cm}
		$\varepsilon_{322}^\text{eq} \approx \sqrt{\frac{1}{3}\frac{\kappa^\text{SR}(a_*-2\alpha \tilde r_+)}{\kappa^\infty}}\left(\frac{f}{\Mpl}\right)^2=6.9\times 10^{-6}\left(\frac{a_*}{0.9}\right)^{1/2} \left(\frac{f}{10^{15}\GeV}\right)^2$ (Eq.~\eqref{eq:quasi_equlibrium_occupations_322})
	\end{flushleft}
	\end{center}
	\caption{Summary of important quantities in the parameter
	space regimes A-D (Fig~\ref{fig:paramspace}, Table~\ref{tab:paramsummary}). The second column lists the ratio of the peak value $\varepsilon^\text{peak}_{211}$ attained in the corresponding region to the maximum value attained through gravitational superradiance $\varepsilon_{211}^\text{max}$. The fourth column describes the most important
	observational signatures of superradiance in
	each regime. For regions A and B, these
	are BH spindown (see Sec.~\ref{sec:spindown}), the emission of gravitational
	radiation (see Sec.~\ref{sec:gw}) from $211 \times 211 \rightarrow$ GW annihilations and from $322\rightarrow  211 \times$ GW transitions (only in region B). 
	For regions C and D, gravitational radiation
	is suppressed, but non-relativistic scalar
	radiation (``AW'', for ``axion waves'') from the $322 \times 322 \rightarrow 211 \times \infty$
	process may be detectable, if the
	scalar field couples to SM states (see Sec.~\ref{sec:scalarwaves}). The right-most column gives approximate 
	expressions for the relevant dynamical timescales, which also correspond to typical signal timescales
	of GW radiation (for A and B) and scalar
	radiation (for C and D).
	The expressions given are to leading order in small
	$\alpha$, and numerical coefficients are approximate;
	the reader should refer to the text for more precise
	expressions.}
	\label{tab:paramsummary2}
\end{table*}

\begin{figure*}[t]
\includegraphics[width = .99 \columnwidth]{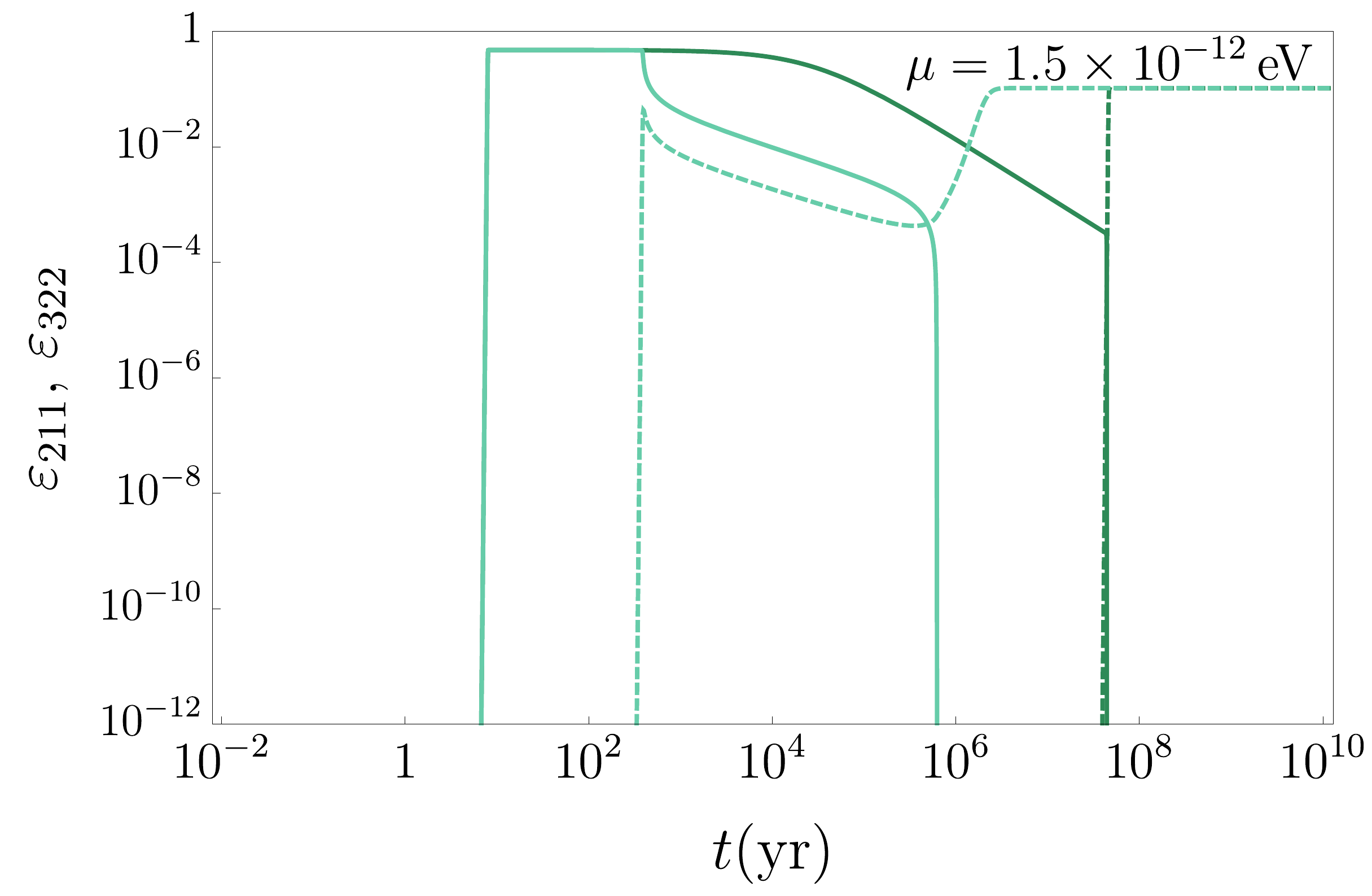}
\includegraphics[width = .964
\columnwidth]{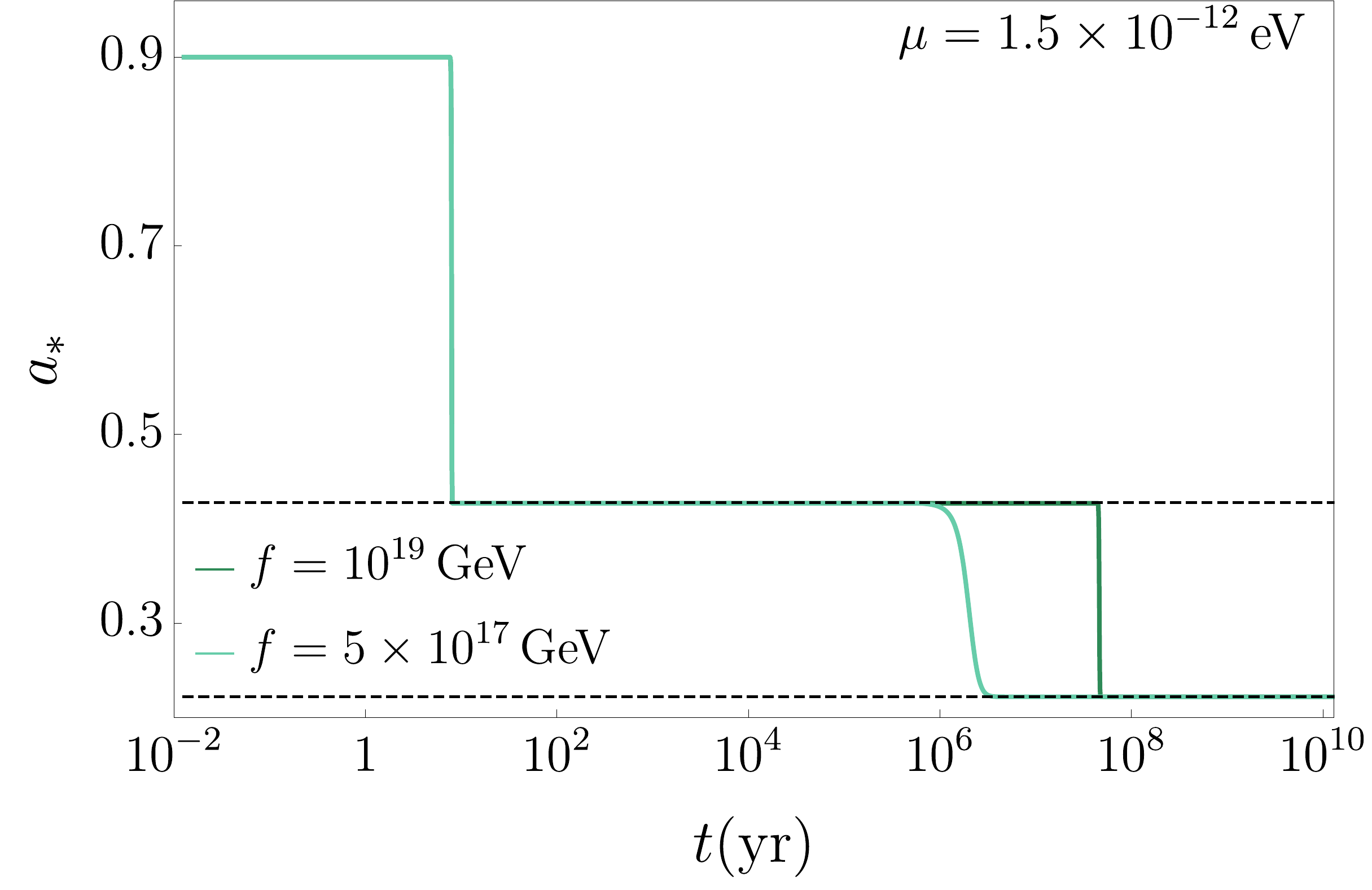}
\includegraphics[width = .99 \columnwidth]{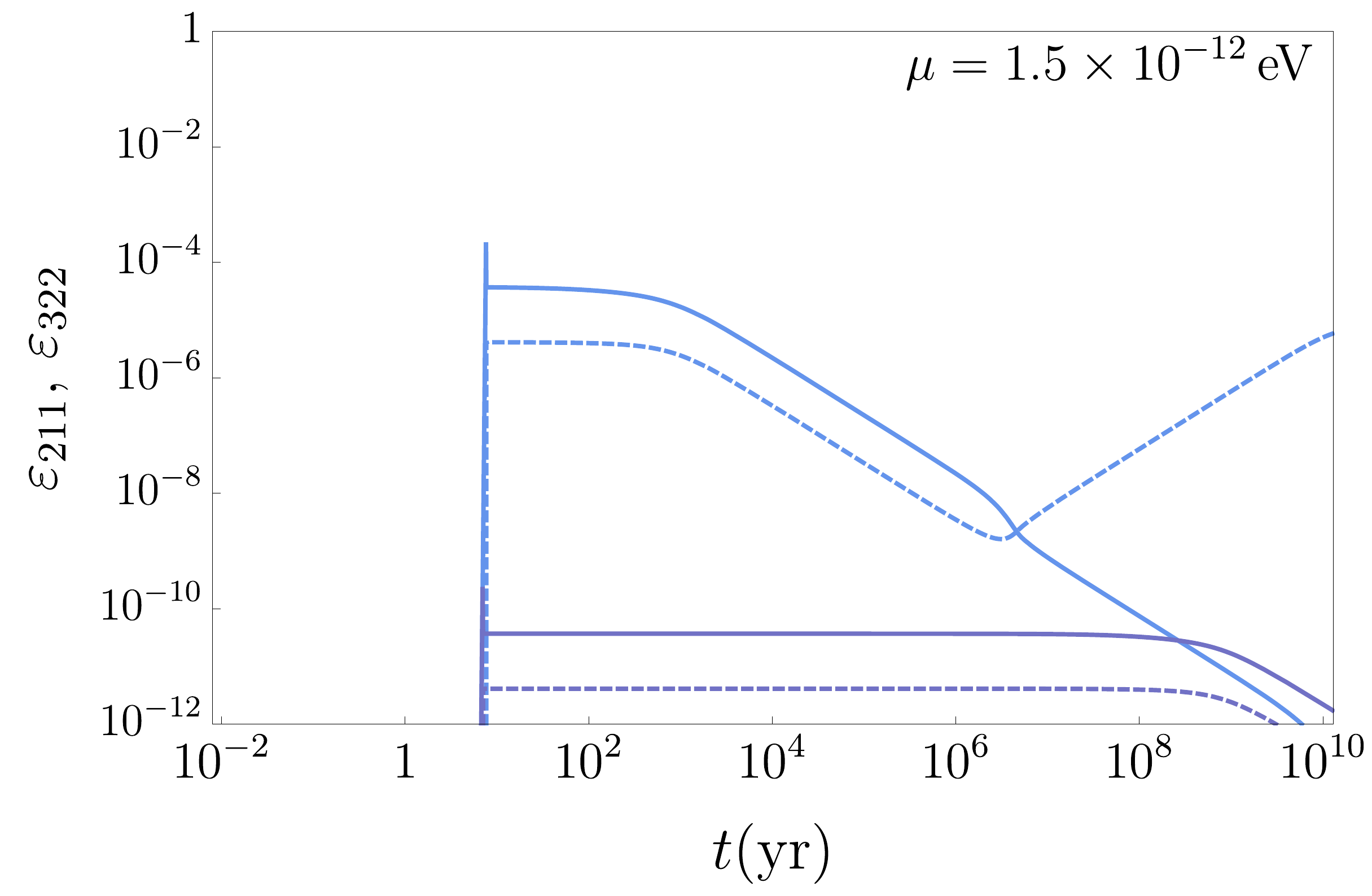}
\includegraphics[width = .964
\columnwidth]{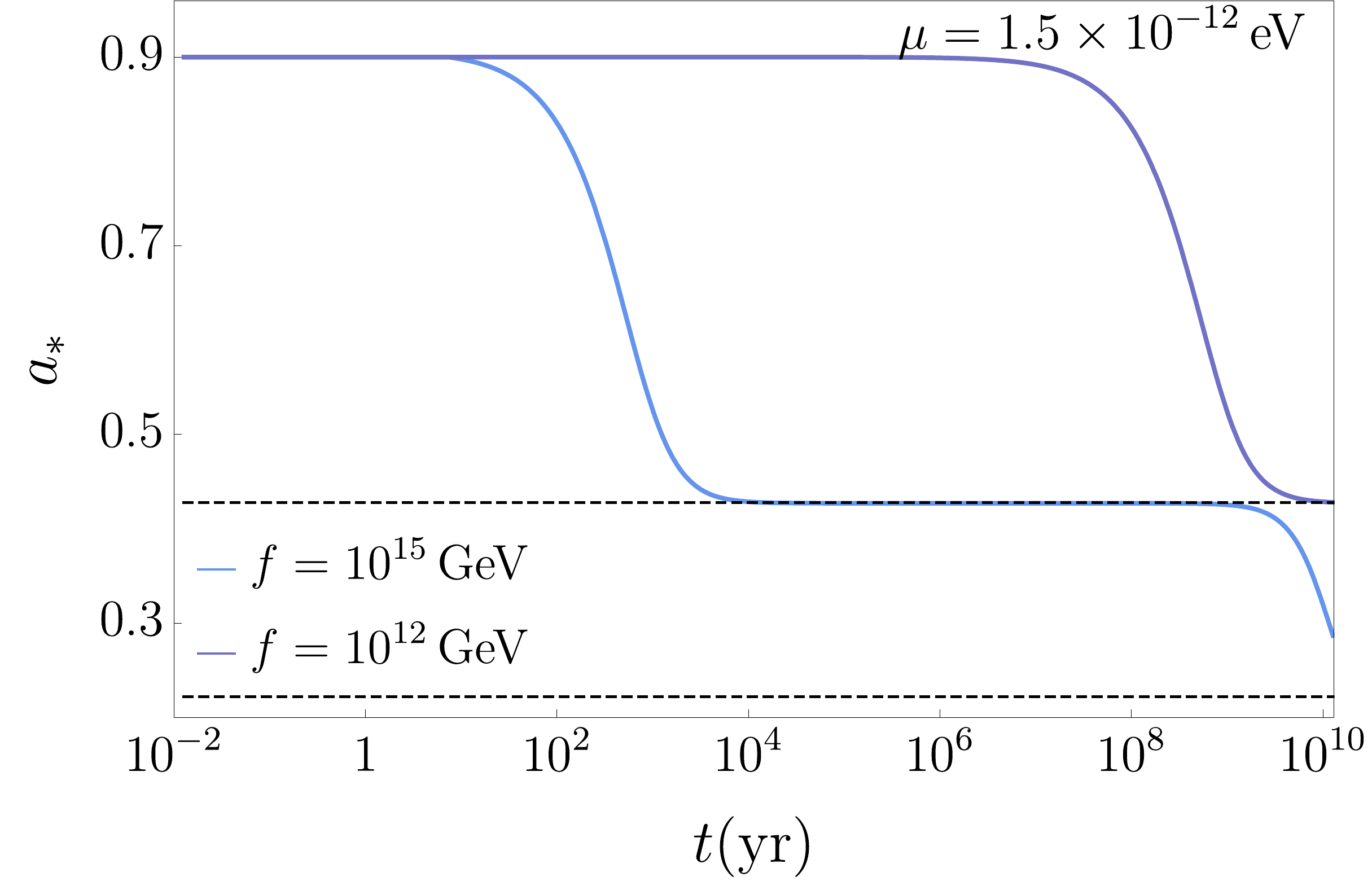}
	\caption{\emph{Left panel:} fractional occupation
	numbers of 211 (solid lines) and 322 (dashed lines) levels,
	and \emph{Right panel:} BH spin,
	as a function of time, for a BH of mass $10 \msun$ and initial spin
	$a_* = 0.9$, given a scalar of mass $\mu =1.5\times 10^{-12} \eV$.
	The different colors correspond to the different self-interaction
	strengths indicated in the right-hand plots (see
	section~\ref{secperturb} for explanations of
	the behaviours at different couplings).}
\label{fig:nevolution}
\end{figure*}

\subsubsection{Evolution equations}

As discussed above, only the processes in Table~\ref{tab:ratesummary}
are generally important in the evolution of the 211/322 system.
We highlight these rates (which are presented outside the parentheses)
in the full evolution equations for the occupation
numbers of the 211 and 322, which are (at leading order in $\alpha$)
\begin{align}
\frac{\dot\varepsilon_{211}}{\mu} 
	={}& \kappa_{211}^\text{SR}\alpha^8(a_*-2\alpha\tilde r_+)\varepsilon_{211}
	\\
	-{}&2\kappa_{211\times 211}^{322\times \text{BH}}\alpha^{11}(\Mpl/f)^4\tilde r_+\varepsilon_{211}^2\varepsilon_{322} \nonumber
	\\
	+{}&\kappa_{322\times 322}^{211\times\infty}\alpha^8(\Mpl/f)^{4} \varepsilon_{322}^2\varepsilon_{211}\nonumber
	\\
	-{}&2\kappa_{211\times 211}^{\rm GW}\alpha^{14}\varepsilon_{211}^2+ \nonumber\\
	\Big(-{}&\kappa_{211\times 322}^{\rm GW}\alpha^{16}\varepsilon_{211}\varepsilon_{322}
+ \kappa_{322\rightarrow 211}^{\rm GW}\alpha^{10}\varepsilon_{211}\varepsilon_{322}\nonumber
	\\
-{}&3\kappa_{(211)^3}^\infty\alpha^{21}(\Mpl/f)^4\varepsilon_{211}^3 \nonumber
	\\
-{}& 2\kappa_{(211)^2\times(322)}^\infty\alpha^{23}(\Mpl/f)^4\varepsilon_{211}^2\varepsilon_{322}\nonumber
	\\
	-{}&\kappa_{(211)\times(322)^2}^\infty\alpha^{25}(\Mpl/f)^4\varepsilon_{211}\varepsilon_{322}^2\Big),\nonumber
\end{align}
\begin{align}
	\frac{\dot\varepsilon_{322}}{\mu} ={}& \kappa^\text{SR}_{322}\alpha^{12}(a_*-\alpha\tilde r_+)\varepsilon_{322}
	\\+{}&\kappa_{211\times 211}^{322\times \text{BH}}\alpha^{11}(\Mpl/f)^{4}\tilde r_+\varepsilon_{211}^2\varepsilon_{322}\nonumber
	\\-{}&2\kappa_{322\times 322}^{211\times\infty} \alpha^8(\Mpl/f)^{4}\varepsilon_{322}^2\varepsilon_{211} + \nonumber
	\\
	\Big(-{}&2\kappa_{322\times322}^{\rm GW}\alpha^{18}\varepsilon_{322}^2 - \kappa_{211\times322}^{\rm GW}\alpha^{16}\varepsilon_{211}\varepsilon_{322} \nonumber
	\\
-{}&\kappa_{322\rightarrow 211}^{\rm GW}\alpha^{10}\varepsilon_{211}\varepsilon_{322}\nonumber
	\\
-{}&3\kappa_{(322)^3}^\infty\alpha^{27}(\Mpl/f)^4\varepsilon_{322}^3 \nonumber
	\\
-{}& \kappa_{(211)^2\times(322)}^\infty\alpha^{23}(\Mpl/f)^4\varepsilon_{211}^2\varepsilon_{322}\nonumber
	\\
	-{}&2\kappa_{(211)\times(322)^2}^\infty\alpha^{25}(\Mpl/f)^4\varepsilon_{211}\varepsilon_{322}^2\Big),\nonumber
\end{align}
where $\tilde{r}_+ \equiv r_+/r_g = 1 +
\sqrt{1 - a_*^2}$, and the $\kappa$ values 
correspond to the $\gamma$ rates, with the leading $\alpha$,
$f$ and $a_*$ dependence factored
out
(e.g.\ $\gamma_{211\times211}^{322\times\text{BH}} = \kappa_{211\times211}^{322\times\text{BH}}
\alpha^{11} (\Mpl/f)^4 \tilde r_+ \mu$, etc).
We also need to keep track of the BH's mass and spin,
for which 
\begin{align}
\label{eqadot}
\frac{\dot a_*}{\mu}={}&-\kappa^\text{SR}_{211}\alpha^8(a_*-2\alpha\tilde r_+) \varepsilon_{211}\\
{}&-2\kappa^\text{SR}_{322} \alpha^{12}(a_*-\alpha\tilde r_+)\varepsilon_{322}, \nonumber
\end{align}
and
\begin{align}
\label{eq:massevol}
	\frac{\dot M}{\mu^2 G M^2} \simeq 
{}&-\kappa^\text{SR}_{211}\alpha^8(a_*-2\alpha\tilde r_+) \varepsilon_{211}\\
{}&-\kappa^\text{SR}_{322} \alpha^{12}(a_*-\alpha\tilde r_+)\varepsilon_{322} \nonumber\\
{}&+\kappa_{211\times211}^{322\times\BH}\alpha^{11}(\Mpl/f)^4\tilde r_+\varepsilon_{211}^2\varepsilon_{322}.\nonumber
\end{align}
A simplifying assumption at small $\alpha$ is to neglect the change in the mass of the black hole; we will often use this approximation in the text. This is equivalent to setting the maximum 211 fractional occupation value attained through purely gravitational evolution, $\varepsilon_{211}^\text{max}$, to $|\Delta a_*| = a_*(t_0) - 4\alpha/(1+4\alpha^2)$. At larger $\alpha$, the mass of the BH changes more significantly and $\varepsilon_{211}^\text{max} > |\Delta a_*|$. Our expressions can still be used, however, with the correct value of $\varepsilon_{211}^\text{max}$, for which we derive good analytic approximations in App. \ref{app:cloudmass}.

\subsubsection{Small self-coupling: gravitational superradiance}
\label{seclsmall}

In the limit of very small coupling, $f
\rightarrow \infty$, the system evolves 
under purely gravitational dynamics, as summarized in
section~\ref{secspin0}.
As long as the fastest and second-fastest growing superradiant
levels have sufficiently different growth rates,
the former will grow first, and attain exponentially larger
occupation numbers than other modes.
For most of this paper, we focus on situations where
the initially fastest-growing mode is the 211 level.
This grows to maximum size, and spins
the BH down to the $m=1$ superradiance threshold, 
in a time 
\begin{equation}
 \frac{\log{GM^2}}{\Gamma_{211}^\text{SR}} 
	\simeq \left(\frac{M_{\rm BH}}{10 M_\odot}\right) \times \begin{cases} 
		9 {\rm \, hour} & \alpha = 0.4 \\
		6 \times 10^3 \yr \left(\frac{0.05}{\alpha}\right)^9 & \alpha \lesssim 0.2 \\
	\end{cases}
\end{equation}
for high spin ($a_* = 0.99$). On a timescale
that,
for small $\alpha$, is parametrically larger, 
the 211 level is depleted
through gravitational wave annihilations,
with a decay time of 
\begin{align}
\label{eq:t_ann}
	\tau_{\rm ann} &\approx \frac{1}{2\Gamma^{\rm GW}_{211 \times 211} N_{211, \max}} \\
	&\simeq \left(\frac{M_{\rm BH}}{10 M_\odot}\right) \times \begin{cases} 
		4 {\rm \, hour} & \alpha = 0.4,  \\
		3 \times 10^9 \yr \left(\frac{0.05}{\alpha}\right)^{15} & \alpha \lesssim 0.2.
	\end{cases}
	\nonumber
\end{align}

On even longer timescales, the fastest-growing
$m=2$ level (i.e.\ 322) spins down the BH via superradiance,
\begin{equation}
	\frac{\log{GM^2}}{\Gamma_{322}^\text{SR}}
	\simeq \left(\frac{M_{\rm BH}}{10 M_\odot}\right) \times \begin{cases} 
		4 {\rm \, yr} & \alpha = 0.4, \\
		10^{11} \yr \left(\frac{0.05}{\alpha}\right)^{13} & \alpha \lesssim 0.5.
	\end{cases}
\end{equation}
By this point, only a small fraction of the initial 211
occupation generally remains (for $\alpha$ large enough
that growth occurs on relevant timescales), so gravitational wave transition
signals from $322 \rightarrow 211 \times {\rm GW}$ events
are small.
The upper panels of Fig.~\ref{fig:nevolution}
illustrate this evolution, for $f \simeq \Mpl$.
For BHs with long enough lifetimes, a similar story
applies to the growth of higher-$m$ levels.

As we discuss below, the purely gravitational
story describes the evolution well if the
self-interaction-induced $211 \times 211
\rightarrow 322 \times$ BH process is always slow
compared to superradiant growth processes.
The parameter space for which this is true
is plotted as region (A) in the bottom-right panel
of Fig.~\ref{fig:paramspace}.

\subsubsection{Moderate self-coupling: early growth of 322 and late equilibrium}
\label{seclmoderate}

If we decrease $f$, while holding other parameters fixed,
the first significant difference from 
purely-gravitational evolution that arises is earlier growth of
the 322 level. 
We label this regime, where 211 still grows to saturation,
but 322 grows sooner than it would have 
if $\lambda = 0$, the ``moderate self-coupling'' regime.
The upper-left panel of Fig.~\ref{fig:nevolution}
illustrates the evolution of the 211 and 322
occupation numbers for an $f$ value in this regime
(as well as for a larger $f$ in the small self-coupling regime).

The parameter space for moderate self-coupling
is plotted
as region (B) in the bottom right-hand panel of 
Fig.~\ref{fig:paramspace},
and corresponds to the intersection of the shaded
regions in the upper two panels.
In this subsection, we will focus on the threshold
between the small self-coupling and moderate
self-coupling regimes, deferring the small-$f$
boundary of the moderate regime (i.e.\ the point
at which 211 no longer grows to saturation)
to the next subsection.

For the $211 \times 211 \rightarrow 322 \times$ BH
process to build up 322 within the lifetime
of the BH, we need
\begin{equation}
\label{eq:f_{AB}}
	\gamma_{211 \times 211}^{322 \times {\rm BH}}
	(\varepsilon_{211}^{\rm max})^2
	\gtrsim \frac{\log (\varepsilon_{322}^{\rm final}/\varepsilon_{322}^{\rm initial})}{T_{\rm BH}}
	\simeq \frac{\log (G M_{\rm BH}^2)}{T_{\rm BH}}
\end{equation}
where $\varepsilon_{211}^{\rm max} \approx a_*(t_0)- a_*^{\rm thresh} \approx a_*(t_0)-4/\alpha(1+4\alpha^2)$
is the occupation number of the saturated 211 level.
Parametrically, if we start from very small fluctuations
in the 322 level,
and $\varepsilon_{322}^{\rm final}$ is not exponentially small,
then $\varepsilon_{322}^{\rm final}/\varepsilon_{322}^{\rm initial}
\sim G M^2$.
For this growth to be faster than 322 superradiance,
we need 
$\gamma_{211 \times 211}^{322 \times {\rm BH}}
	(\varepsilon_{211}^{\rm max})^2 \gtrsim \gamma_{322}^{\rm SR}$.

The condition~\eqref{eq:f_{AB}} is necessary for early 322 growth to occur, but
not sufficient, since annihilations to gravitational
waves may deplete 211 before 322 can grow. In order
for this not to happen, we need 
\begin{align}
    \label{eq:ann_to_graviton}
	\frac{\gamma_{211 \times 211}^{322 \times {\rm BH}} 
	(\varepsilon_{211}^{\rm max})^2}{\log (\varepsilon_{322}^{\rm final}/\varepsilon_{322}^{\rm initial})}  \gtrsim 2 \gamma_{211 \times 211}^{\rm GW} \varepsilon^{\max}_{211}
\end{align}
Replacing the rates by their small-$\alpha$ expansions,
this is equivalent to
\begin{align}
      \frac{\kappa^\BH\tilde r_+ (\Mpl/ f)^4 \varepsilon_{211}^\text{max}}{\log\left(GM^2\right)} \gtrsim 2\kappa_{211\times 211}^\text{GW}\alpha^3.
\end{align}

The combination of the conditions~\eqref{eq:f_{AB}} 
and~\eqref{eq:ann_to_graviton} is responsible for
the shape of the (A)-(B) boundary in Fig.~\ref{fig:paramspace}.
At small $\alpha$, \eqref{eq:f_{AB}} is more constraining,
while at larger $\alpha$, \eqref{eq:ann_to_graviton}
takes over.
The parametric form of this threshold value $f_{\rm AB}$ is given 
in table~\ref{tab:paramsummary}.

\emph{Evolution of levels}:
Unlike in the gravitational scenario, where the
growth of 322 via superradiance is accompanied by a rapid
drop in 211 occupation, here both levels
eventually reach roughly-comparable occupation numbers. 
Subsequently, the joint
cloud is slowly depleted by the
combination of non-relativistic scalar emission
and damping by the BH. Other processes, including
gravitational annihilations and transitions as
well as relativistic scalar emission, are
small perturbations to this overall evolution.

As discussed above, only a few rates drive
the dynamics in the regions of parameter
space for which self-interactions modify the
purely gravitational scenario. These are
$\kappa_{211}^\text{SR}$, $\kappa_{211\times
211}^{322\times\text{BH}}$, and $\kappa_{322\times
322}^{211\times \infty}$ (and $\kappa_{322}^{\rm SR}$,
in some circumstances). To streamline our
notation, 
we will refer to them as $\kappa^\text{SR},
\kappa^\text{BH}$, and $\kappa^{\infty}$
respectively.

In the regime of moderate self-coupling, the
growth of the 211 level occurs as in the purely-gravitational
case; both the occupation number and the
BH angular momentum
change ``suddenly'',
with almost all of the change happening in the last few e-folds of superradiant
growth. This is illustrated in the top 
panels of Fig.~\ref{fig:nevolution}.
The BH spin decreases to $a_* \approx
4\alpha/(1+4\alpha^2)$, and $\varepsilon_{211}$
stays at $\approx\varepsilon_{211}^\text{max}$ for a long time. In the
purely gravitational scenario, the cloud would
then slowly self-annihilate to gravitational waves
until $\sim 200$ e-folds of 322 superradiance have
passed. Here, however,
the quartic process dominates, and 
the 322 growth rate is higher:
\begin{align}
\begin{split}
\label{eq:growth_of_322}
   \frac{\dot\varepsilon_{322}}{\mu} \approx {}& \kappa^\BH\tilde r_+\alpha^{11} (\Mpl/f)^{4}(\varepsilon_{211}^\text{max})^2\varepsilon_{322}.
    \end{split}
\end{align}
Eventually, the 322 occupation number becomes large enough that the
quartic vertex $322\times 322\rightarrow 211\times \infty$ becomes
important and a quasi-equilibrium is established, roughly after time
\begin{align}
\begin{split}
	\label{eqtstar}
t_*\simeq GM\frac{\log(GM^2
)}{\kappa^\text{BH}\tilde r_+\alpha^{12}(\Mpl/f)^4(\varepsilon_{211}^\text{max})^2} 
\end{split}
\end{align}
has passed.

At this point, superradiance to 211 has
effectively shut down, and 322 superradiance
is too slow to be significant. Particles are leaving
the combined cloud, going back to the BH (via
$211 \times 211 \rightarrow 322 \times \rm{BH}$)
and to infinity (via $322\times 322\rightarrow
211\times \infty$). Gravitational and relativistic
scalar processes are suppressed by high powers of
$\alpha$. Accordingly, the coupled dynamics of the
two-level system simplifies to
\begin{subequations}
\label{eq:simplified_evolution}
\begin{align}
\begin{split}
\frac{\dot\varepsilon_{211}}{\mu} \approx{}& -2\kappa^{\text{BH}}\tilde r_+\alpha^{11}(\Mpl/f)^4\varepsilon_{211}^2\varepsilon_{322} \\{}&+\kappa^{\infty}\alpha^8(\Mpl/f)^{4} \varepsilon_{322}^2\varepsilon_{211},
\end{split}
\end{align}
\begin{align}
\begin{split}
\frac{\dot\varepsilon_{322}}{\mu} \approx{}&\kappa^{ \text{BH}}\tilde r_+\alpha^{11}(\Mpl/f)^{4}\varepsilon_{211}^2\varepsilon_{322}\\-{}&2\kappa^{\infty} \alpha^8(\Mpl/f)^{4}\varepsilon_{322}^2\varepsilon_{211},
\end{split}
\end{align}
\begin{align}
\begin{split}
a_* \approx \frac{4\alpha}{1+4\alpha^2}.
\end{split}
\end{align}
\end{subequations}
Since there are no processes (except for
the negligible superradiance of 322) which
contribute particles to the cloud, particles
are only leaving. Accordingly, the system has
no true equilibrium occupations. However,
\eqref{eq:simplified_evolution} still admits
a time-independent equilibrium \emph{ratio} of occupation 
numbers,
$\varepsilon_{322}/\varepsilon_{211} = \eta^B$ to which the system flows,
\begin{align}
\begin{split}
    \label{eq:equilibrium_ratio} \eta^B \simeq  {}& \frac{1}{2}\frac{\kappa^{\text{BH}}\alpha^{3}\tilde r_+}{\kappa^{\infty}}
 \simeq 4\times10^{-5}\left(\frac{\alpha}{0.01}\right)^3.
   \end{split}
\end{align}
For the regime of moderate self-coupling, the scalings in \eqref{eq:equilibrium_ratio} are only representative at leading orders in $\alpha$. A more accurate expression is derived in App. \ref{sec:equilibrium_ratio}.

When the equilibrium ratio is obtained at time $t_*$, the occupations evolve as
\begin{align}
\varepsilon_{211}(t) \simeq \frac{\varepsilon_{211}(t_*)}{\sqrt{1+2\varepsilon_{211}^2(t_*)(t-t_*)/\tau_\text{scalar}}},
\label{eq:scalarevol}
\end{align}
where
\begin{align}
\begin{split}
\label{eq:t_scalar}
\tau_\text{scalar} {}&\equiv \frac{4}{3\mu}\frac{\kappa^\infty}{(\kappa^\BH\tilde r_+)^2\alpha^{14}}\left(\frac{f}{\Mpl}\right)^4\\
\approx{}& 10^{-1}\yr\left(\frac{0.1}{\alpha}\right)^{14} \left(\frac{10^{-12}\eV}{\mu}\right)\left(\frac{f}{10^{17}\GeV}\right)^4 ,
\end{split}
\end{align}
and $\varepsilon_{322}(t)=\varepsilon_{211}(t)\eta^B$.

The joint cloud continues to deplete until the
occupation of 211 has diminished enough that the
superradiance rate of 322 outcompetes the ``stimulated'' emission process $322\times 322 \rightarrow 211\times \infty$, and the cloud starts growing again. A large occupation builds up in 322,
causing rapid 211 depletion via $211\times 211 \rightarrow 322 \times \BH$. Moreover, as superradiance extracts
angular momentum from the BH to 322, the BH's spin
decreases further, making 211 (and other $m=1$ states)
\emph{damped}.
This sequence of events is illustrated in the top
panels of Fig.~\ref{fig:nevolution}
(where the green curves correspond to moderate
self-coupling, and the blue to small self-coupling).

In the
$\lambda=0$ case, $m=2$ superradiance must proceed from zero-point
quantum fluctuations, or from
a small pre-existing astrophysical density. Here, superradiance gets to act on
the \emph{pre-existing} occupation $\varepsilon_{322}$,
since 322 has already been populated by self-interaction-mediated
processes. In this way, self-interactions
``assist" superradiance, sometimes leading to more rapid
saturation of the $m=2$ instability than allowed in the
purely gravitational story.
The $f = 5 \times 10^{17} \GeV$ curves
in the upper panels of Fig.~\ref{fig:nevolution} show
an example of this, with 322 spin-down occurring after
only $\sim {\rm few} \times 10^6 \yr$,
compared to almost $10^8 \yr$ in the purely-gravitational case.

The above discussion summarizes the evolution
of the cloud in the moderate self-coupling
regime.
Before moving on, we will discuss
the effects of processes other than
$211 \times 211 \rightarrow 322 \times$ BH,
$322 \times 322 \rightarrow 211 \times \infty$,
and superradiance, and review why they are (in most cases) subdominant.

\emph{Annihilations to GWs:}
An important point is that, to be in the 
moderate self-coupling regime for astrophysical BH masses, we
need $f \lesssim \Mpl$ (as illustrated in Fig.~\ref{fig:paramspace}).
This is evident from the form
of the threshold $f_{\rm AB}$ given in table~\ref{tab:paramsummary},
$f_{\rm AB} = {\rm min}(f_1, f_2)$.
The first term $f_1$ comes from the condition
$\gamma_{211 \times 211}^{322 \times {\rm BH}} 
(\varepsilon_{211}^{\rm max})^2 \gtrsim \frac{\log (G M_{\rm BH}^2)}{T_{\rm BH}}$;
to make $f_1 \ge \Mpl$, we need to take
$\alpha \gtrsim 0.07$
(for $M_{\rm BH} = \OO(10 M_\odot)$).
Such large values of $\alpha$ make the $f_2$, coming from the
condition that GW annihilations are not too fast
\eqref{eq:ann_to_graviton}, much less than $\Mpl$.
Consequently, 
gravitational wave emission processes suffer
a \emph{suppression} $\sim (f/\Mpl)^4$,
relative to self-interaction-mediated
quartic processes.
This means that, once 322 has reached its
equilibrium ratio with 211 (Eq.~\eqref{eq:equilibrium_ratio}),
even the fastest GW emission process,
$211 \times 211 \rightarrow$ GW,
is generally slower than $211 \times 211 \rightarrow
322 \times$ BH and $322 \times 322 \rightarrow 211 \times \infty$ (at least until the levels have depleted significantly).

\emph{GW transitions:}
From table~\ref{tablegrav}, gravitational wave transitions
$322 \rightarrow 211 + $ GW contribute
a term $\dot \varepsilon_{322} \simeq - 3 \times 10^{-6} \alpha^{10}
\varepsilon_{322} \varepsilon_{211} \mu + \dots$ to the evolution
equations.
If we take $\varepsilon_{322} = \eta^B \varepsilon_{211}$
(Eq.~\eqref{eq:equilibrium_ratio}),
this gives
\begin{equation}
\dot \varepsilon_{322}/\mu \simeq - 3 \alpha^{13}
	\varepsilon_{211}^2 + 0.4 \alpha^{14} \varepsilon_{211}^3  \left(\frac{\Mpl}{f}\right)^4 + \dots
	\label{eq322ev}
\end{equation}
where we have also included the $211 \times 211 \rightarrow 
322 \times$ BH term for comparison.
While the GW transition term is suppressed
by one less power of $\alpha$,
Fig.~\ref{fig:paramspace} illustrates
that, as $\alpha$ decreases, the maximum $f$ for
the moderate self-coupling regime decreases
(from table~\ref{tab:paramsummary}, $f_{AB} \propto \alpha^{11/4}$
for small $\alpha$).
Consequently, the relative $(\Mpl/f)^4$ enhancement
of the quartic self-interaction
terms always wins out.

Even though gravitational wave emission no longer dominates the evolution compared to the small self-interactions regime of gravitational superradiance, GW annihilation signals can still be strong enough for detection in this regime. In addition, the simultaneous occupation of the two levels allows for the possibility of GW signals from transitions. We explore potential signatures in more detail in Sec. \ref{sec:gw}.

\emph{Relativistic $3 \rightarrow 1$ emission:}
As discussed in section~\ref{secrelscalar},
quartic self-interactions also lead to
processes emitting relativistic scalar
waves,
such as $211 \times 211 \times 211 \rightarrow \infty$.
This contributes
\begin{equation}
	\dot \varepsilon_{211}/\mu \simeq - 5 \times 10^{-9}
	\alpha^{21} \left(\frac{\Mpl}{f}\right)^4
	\varepsilon_{211}^3 + \dots
\end{equation}
Because of the high power of $\alpha$ this is suppressed by,
its effect is small compared
to the non-relativistic quartic processes.

\emph{Relativistic cubic emission:}
In section~\ref{sec:cubic}, we discussed how,
in addition to a quartic self-interaction,
there may also be a cubic interaction term,
$\LL \supset \frac{1}{6} C \frac{\mu^2}{f} \varphi^3$,
which can lead to relativistic emission
processes such as $211 \times 211 \rightarrow \infty$.
This contributes
\begin{equation}
	\dot \varepsilon_{211}/\mu \simeq - 2 \times 10^{-4}
	\alpha^{14} |C|^2 \left(\frac{\Mpl}{f}\right)^2
	\varepsilon_{211}^2 + \dots
\end{equation}
Compared to the quartic-induced term
in Eq.~\eqref{eq322ev}, the lower
power of $\Mpl/f$, and the smaller constant factor,
mean that unless $|C| \gg 1$, 
relativistic emission from the cubic coupling will be
a subdominant effect.

\subsubsection{Large self-coupling: early equilibrium and halted extraction of angular momentum}
\label{secllarge}

If we further decrease $f$, we reach a point
where 322 grows large enough, early enough,
that 211 superradiance is disrupted,
and 211 does not reach its saturation value.
We call this the regime of ``large self-coupling''; it corresponds
to regions (C) and (D) in the bottom-right panel
of Fig.~\ref{fig:paramspace},
and to the bottom
panels in Fig.~\ref{fig:nevolution}.

For the $211 \times 211 \rightarrow 322 \times$ BH process to
disrupt 211 superradiance, we need that
$2\gamma_{211\times 211}^{322\times \BH} \varepsilon_{211}\varepsilon_{322}\gtrsim \gamma_{211}^\SR$ before $\varepsilon_{211}$ has grown
to its saturation value.
This does not necessarily preclude 211 reaching
$\varepsilon_{211}^{\rm max}$ ($\varepsilon_{211}$ can
still grow after that point, albeit more slowly than it would have with $\lambda=0$), but it is necessary to have a significant effect.
Parametrically, this condition is approximately equivalent
to 
\begin{align}
\begin{split}
\label{eq:large_interaction_boundary}
\gamma_{211\times 211}^{322\times\BH}(\varepsilon_{211}^\text{max})^2\gtrsim 2\log\left(GM^2\right)\gamma_{211}^\SR,
\end{split}
\end{align}
where we neglect the dependence of the rates on the BH spin (i.e. set $a_*(t)=a_*(t_0)$).
A more precise condition is derived in App. \ref{sec:fa_min_contour}.

The condition \eqref{eq:large_interaction_boundary} can be expressed as
a condition on $f$. 211 superradiance is basically unaffected if $f \gtrsim f_\text{thresh}$, where
\begin{align}
\begin{split}
\label{eq:f_min}
f_\text{thresh} \approx{}& \Mpl\left(\frac{\alpha^{3}}{2\log (GM^2)}\frac{\kappa^\BH \tilde r_+(\varepsilon_{211}^\text{max})^2}{\kappa^\SR a_*(t_0)}\right)^{1/4}\\
\approx{}& 6\times 10^{15}\GeV\left(\frac{\alpha}{0.01}\right)^{3/4}\left(\frac{a_*(t_0)}{0.9}\right)^{1/4}.
\end{split}
\end{align}
The scalings in \eqref{eq:f_min} are only representative when $\alpha
\ll a_*(t_0)$. For larger values of $\alpha$, rates obtained
numerically, and a more precise version of
\eqref{eq:large_interaction_boundary} (App.
\ref{sec:fa_min_contour}), can be used. 

As pointed out in \cite{Gruzinov_2016}, if $a_*$ is held fixed, the system admits equilibrium occupations for which $\dot\varepsilon_{211} = \dot\varepsilon_{322}=0:$
\begin{subequations}
\label{eq:quasi_equlibrium_occupations}
\begin{align}
\begin{split}
\label{eq:quasi_equlibrium_occupations_211}
\varepsilon_{211}^\text{eq}(a_*)
\approx{}& \frac{2}{\sqrt{3}}\frac{\sqrt{\kappa^\infty\kappa^\SR(a_*-2\alpha \tilde r_+) }}{\alpha^3\kappa^\BH \tilde r_+ }\left(\frac{f}{\Mpl}\right)^2\\
\equiv{}&\left(\frac{f}{f_\text{eq}}\right)^2\varepsilon_{211}^\text{max}\\
={}&2.5 \times 10^{-1}\left(\frac{0.01}{\alpha}\right)^3\left(\frac{a_*}{0.9}\right)^{1/2}\left(\frac{f}{10^{15}\GeV}\right)^2,
\end{split}
\end{align}
\begin{align}
\begin{split}
\label{eq:quasi_equlibrium_occupations_322}
\varepsilon_{322}^\text{eq}(a_*)\approx{}&\sqrt{\frac{1}{3}\frac{\kappa^\text{SR}\left(a_*-2\alpha\tilde r_+\right)}{\kappa^\infty}}\left(\frac{f}{\Mpl}\right)^2\\
={}& 6.9\times 10^{-6} \left(\frac{a_*}{0.9}\right)^{1/2} \left(\frac{f}{10^{15}\GeV}\right)^2,
\end{split}
\end{align}
\end{subequations}
where
\begin{align}
\begin{split}
\label{eq:f_eq}
f_\text{eq} \approx{}& \Mpl\left(\frac{\sqrt{3}}{2}\frac{\alpha^3\kappa^\BH \tilde r_+\varepsilon_{211}^\text{max}}{\sqrt{\kappa^\SR \kappa^\infty (a_*-2\alpha\tilde r_+)}}\right)^{1/2}\\
\approx{}& 2\times 10^{15}\GeV \left(\frac{\alpha}{0.01}\right)^{3/2}\left(\frac{a_*(t_0)}{0.9}\right)^{1/4}.
\end{split}
\end{align}
Note that the ratio $\eta^\text{eq} \equiv \varepsilon_{322}^\text{eq}/\varepsilon_{211}^\text{eq}$ is
\beq
\eta^\text{eq} = \frac{\gamma^\BH}{2\gamma^\infty} \approx (\eta^B)_{\text{small }\alpha},
\eeq
according to the approximation \eqref{eq:equilibrium_ratio} valid for small $\alpha$. At larger values of $\alpha$, $\eta^B > \eta^\text{eq}$. See App. \ref{sec:equilibrium_ratio} for more details.

We now consider what happens in the physical case, where
$a_*$ can change. If $\varepsilon_{211}^{\rm eq}$ is much less
than its saturation value, then the timescale to extract
an $\OO(1)$ fraction of the BH's spin is much longer
than the characteristic timescale of the processes
maintaining the equilibrium. Consequently, we expect
the quasi-equilibrium to be maintained to a good
approximation, as $a_*$
undergoes a slow descent. 
The equilibrium occupation numbers
$\varepsilon_{211}^{\rm eq}(a_*)$ and $\varepsilon_{322}^{\rm eq}(a_*)$
stay almost
constant, with the angular momentum extracted from the BH
via 211 superradiance being emitted to infinity
via the $322 \times 322 \rightarrow 211 \times \infty$ process.
This is in contrast to the regimes of small and moderate self-interactions, where the angular momentum lost from the BH builds up in the cloud.

Close to the transition from moderate to large self-interactions, there
is a sliver of parameter space for which the exponential growth of 211 is
maintained for some time and $\mathcal O(1)$ of the maximum spin
extraction occurs, before getting cut short by the equilibrium. Deep
inside the region of small $f$, however, the spin of the BH is
essentially unchanged at the time the equilibrium is established, and
most of the extraction of angular momentum happens adiabatically.  

Although \eqref{eq:quasi_equlibrium_occupations}
is valid at equilibrium, if $\alpha$ is large enough then
$\varepsilon_{211}$ will ``overshoot'' its
equilibrium value before $\varepsilon_{322}$ has
caught up with it.
Before equilibrium, if we neglect the dependence
of $\gamma_{211}^\SR$ on the BH spin,
$\varepsilon_{211}\propto \exp(\gamma_{211}^\SR
t)$. In App. \ref{sec:fa_min_contour}, we
derive an estimate for the value of the exponent
$\gamma_{211}^\SR t$ at the time when $211\times
211\rightarrow 322\times \BH$ is comparable to SR.
To a good approximation
\beq
\varepsilon_{211}^\text{thresh} \approx\sqrt{\frac{2\gamma^\SR_{211}\log(G M^2)}{\gamma_{211\times 211}^{322\times \BH}}}\approx \left(\frac{f}{f_\text{thresh}}\right)^2\varepsilon_{211}^\text{max},
\label{eq:overshoot}
\eeq
where we set $a_*(t)=a_*(t_0)$ in both rates.

Accordingly, the evolution towards equilibrium can happen in two
qualitatively different ways. When $\alpha \gtrsim 0.04$,
$f_\text{thresh}<f_\text{eq}$ and $\varepsilon_{211} ^\text{thresh} >
\varepsilon_{211}^\text{eq}$. In this case, the occupation
$\varepsilon_{211}$ overshoots its equilibrium value and subsequently
evolves toward it from above. This
is illustrated in the bottom-left
panel of Fig.~\ref{fig:nevolution} (for which $\alpha = 0.11$). Conversely, when $\alpha \lesssim 0.04$,
then $\varepsilon_{211} ^\text{thresh} < \varepsilon_{211}^\text{eq}$.
There is no overshoot, and $\varepsilon_{211}$ evolves toward its
equilibrium occupation from below. 

Given this, the boundary between the moderate self-coupling regime,
where $\varepsilon_{211}$ reaches $\varepsilon_{211}^{\max}$,
and large self-coupling, where it does not, is set
by
\beq 
\label{eq:f_{BC}}
f
\lesssim
f_\text{BC}\equiv\min\left[f_\text{thresh},f_\text{eq}\right].\eeq

To review, the evolution of the superradiant cloud, in the regime
of large self-coupling, occurs in different stages:
\begin{enumerate}
	\item An initial stage of exponential 211 growth, during
		which $\varepsilon_{322}$ is too small to significantly
		affect the evolution of $\varepsilon_{211}$.
	\item A ``non-equilibrium'' stage in which $\varepsilon_{211}$
		and $\varepsilon_{322}$ evolve towards their equilibrium values.
		The timescale to approach the equilibrium values is
		at most a logarithmic
		multiple of $1/\gamma_{211}^{\rm SR}$, 
		since the relevant self-interaction processes
		are at least as fast as $\gamma_{211}^{\rm SR}$.
	\item Once $\varepsilon_{211}$ and $\varepsilon_{322}$
		are close to their equilibrium values, there is a
		long period of quasi-adiabatic evolution. 
		The spin-down of the BH due to spin extraction
		through 211 superradiance, which changes
		$a_*$ on a timescale $(\dot a_*/a_*)^{-1}\sim (\varepsilon_{211}^{\rm max}/\varepsilon_{211}^{\rm eq}) / \gamma_{211}^{\rm SR}$, leads to the slow evolution of
		the equilibrium occupation numbers.
	\item If the BH lifetime is long enough
		that spin-down to the $m=1$ threshold occurs, 
		then similar behavior to the moderate self-coupling
		regime will result. The 211 and 322 levels will maintain
		a quasi-equilibrium ratio,
		but with decreasing occupation numbers,
		as scalars are emitted to infinity.
		Eventually, the occupation numbers will become small
		enough that 322 superradiance starts to dominate,
		at which point the 322 occupation number starts
		growing again (e.g.\ the $f=10^{15} \GeV$ curves in the
		bottom-left panel of Fig.~\ref{fig:nevolution}).
\end{enumerate}
Consequently, when $f$ is appreciably smaller than $f_{\rm BC}$,
the first and second stages change $a_*$ by only a small amount, and
the majority of the BH's spin-down to the $m=1$ threshold happens
during the period of
almost adiabatic, quasi-equilibrium evolution.

When the equilibrium occupations
\eqref{eq:quasi_equlibrium_occupations} are obtained, the angular
momentum of the BH decreases according to 
\eqref{eqadot}, with 
$\varepsilon_{211}=\varepsilon_{211}^\text{eq}(a_*)$
(and we can ignore $\kappa_{322}^{\rm SR}$).
The timescale
for spindown is therefore set by
\begin{align}
\begin{split}
\label{eq:spindown_time}
\tau_\text{sd}(a_*) \approx{}&
\frac{\sqrt{3}}{2\alpha^5\mu}\frac{\kappa^\BH \tilde r_+\left(\Mpl/f\right)^2}{\sqrt{\kappa^\infty}\left(\kappa^\SR\left(a_*-2\alpha\tilde r_+\right)\right)^{3/2}}\\
\approx{}& 10^7 \yr \left(\frac{0.01}{\alpha}\right)^5\left(\frac{10^{-12}\eV}{\mu}\right)\\{}&\qquad \times\left(\frac{0.9}{a_*}\right)^\frac{3}{2}\left(\frac{10^{15}\GeV}{f}\right)^2.
\end{split}
\end{align}

While in (slowly-varying) equilibrium,
the cloud emits non-relativistic axion waves
through the $322 \times 322 \rightarrow 211 \times \infty$
process.
These could, in the presence of axion-SM interactions, be
detected by experiments on Earth.
Even though the occupation number of the
cloud decreases $\propto f^2$ for small $f$,
the coupling strength of axion-SM interactions
will generically scale as $\sim 1/f$. Consequently,
the interaction rate of the emitted radiation
with a laboratory target
can be \emph{independent} of $f$ in the small-$f$
regime. This in contrast to gravitational wave
signals, which are suppressed at small $f$.
We discuss this possibility more fully in section~\ref{sec:scalarwaves}.

In the previous subsection
on the moderate self-coupling regime,
we discussed how interaction processes, other than
non-relativistic quartic interactions and superradiance,
are generally subdominant in their effects 
on the evolution of the cloud.
Very similar calculations apply to the large self-coupling
regime; the equilibrium ratio of
$\varepsilon_{322}/\varepsilon_{211}$ is the same,
with the difference being that the equilibrium
occupation numbers are suppressed, scaling $\propto f^2$.

This scaling only makes a difference to comparisons
between processes with different multiplicities. For
annihilation to GWs, the $(f/\Mpl)^2$ scaling of
the occupation number is not enough to make
up for the $(\Mpl/f)^4$ relative enhancement 
of the quartic interaction rates, so GW annihilation processes
are even less important than they are in 
the moderate self-coupling regime.

For relativistic cubic emissions, 
the fastest of which is $211 \times 211 \rightarrow \infty$,
we can compare the contribution to
the evolution rate to that from 
$211 \times 211 \rightarrow 322 \times$ BH:
\begin{align}
	\dot \varepsilon_{211} / \mu &\simeq
	- 2 \times 10^{-4} \alpha^{14} |C|^2
	\left(\frac{\Mpl}{f}\right)^2
	\varepsilon_{211}^2 \nonumber  \\ 
	& \quad - 8 \times 10^{-7} \alpha^{11}
	\left(\frac{\Mpl}{f}\right)^4 \varepsilon_{211}^2 \varepsilon_{322}\nonumber \\
	&\simeq \left(-2 \times 10^{-4} \alpha^{14} |C|^2 - 10^{-3} 
	\alpha^{11}  \right) \nonumber  \\
	& \times \left(\frac{\Mpl}{f}\right)^2 (\varepsilon^{\rm eq}_{211})^2
\end{align}
where the second equality applies 
for the equilibrium occupation numbers
\eqref{eq:quasi_equlibrium_occupations}.
Consequently, if $|C| \lesssim 16 (0.2/\alpha)^{3/2}$,
then the effect of the cubic emission term
is small compared to that of the non-relativistic quartic processes.

For $\alpha \gtrsim 0.04$,
the equilibrium values of $\varepsilon_{211}$ and
$\varepsilon_{322}$ are smaller
than the ``overshoot'' values
at which self-interactions first affect the evolution of 
211. Consequently, if the relativistic cubic processes
are unimportant in equilibrium, then they are
always less important than the quartic 
$211 \times 211 \rightarrow 322 \times$ BH process, whenever
the latter has a significant effect on 211 evolution.

For smaller $\alpha$, the $2 \rightarrow 1$ process
will be relatively most important around the initial 
time at which 211 growth is slowed down, since the equilibrium
occupation numbers are approached from below.
Still, even without calculating the thresholds
carefully, we can see that as long as 
$|C| \lesssim 16 (0.2/0.04)^{3/2} \simeq 180 $, 
cubic emission will be insignificant in that
regime 
(since decreasing $\alpha$ decreases 
the relative importance of cubic emission).
Overall, we can see that, unless $|C| \gg 1$,
relativistic emission through the cubic coupling should always
be a subdominant effect on the evolution
of the 211 level (cubic emission for higher-$l$ levels
is suppressed by higher powers of $\alpha$, so should
generally be less significant again).

\subsubsection{Large self-coupling: lack of BH spindown}
\label{seclnosd}

Since $\varepsilon_{211}^{\rm eq}
\propto f^2$, and the rate of spin extraction
from the BH is $\propto \varepsilon_{211}$, 
the spin-down rate for small enough $f$
will be so slow that the $m=1$ threshold spin
is not reached within the BH lifetime.
The $f=10^{12} \GeV$ curves in
the bottom panels of Fig.~\ref{fig:nevolution}
show an example, if we take the BH
lifetime to be $< 10^{10} \yr$.
This affects BH spin-down signatures of superradiance,
as we discuss in section~\ref{sec:spindown}.

The timescale for spin extraction 
in the large self-coupling regime is set by
$\tau_{\rm sd}$ (Eq.~\eqref{eq:spindown_time}).
Setting this equal to the age $T_{\rm BH}$ 
of the BH gives the threshold value of $f$
\begin{align}
\label{eq:f_CD}
	f_\text{CD}&\approx  3\times 10^{14}\GeV \left(\frac{10^{10} \yr}{T_\BH}\right)^{\frac{1}{2}}\left(\frac{10^{-13} \eV}{\mu}\right)^{\frac{1}{2}}  \\
	& \times	\left(\frac{0.01}{\alpha}\right)^{\frac{5}{2}}\left(\frac{0.9}{a_*(t_0)}\right)^{\frac{3}{4}} \nonumber
\end{align}
i.e.\ if $f \lesssim f_{\rm CD}$, then the BH does
not have time to fully spin down.
The parameter space in which this is the case is
plotted as region (D) in the bottom-right panel of
Fig.~\ref{fig:paramspace},
and is illustrated by
the smallest-$f$ curve
in Fig.~\ref{fig:BHspindown}.
For $f \ll f_{\rm CD}$, which gives $T_{\rm BH} \ll \tau_{\rm sd}$, 
the amount of angular momentum extracted is
\beq
\label{eq:no_spindown}
|\Delta a_*| \simeq \frac{T_\BH}{\tau_\text{sd}(a_*(t_0))}.
\eeq

\subsection{Beyond the two-level system}
\label{sec:higherlevels}

So far, we have focussed on BH-cloud systems
which are dominated by the 211 and 322 hydrogenic levels.
In this subsection, we consider the effect
of other levels on the dynamics, including higher
principal number $n$ and higher angular momentum
numbers $l,m$. We continue to assume that the
initial conditions are such that 211 satisfies the
superradiance condition and is the first level to grow;
this is the regime of fastest black hole spindown
and the largest gravitational and scalar emission rates,
and is thus the most relevant from an observational
perspective. 

We find that, for $\alpha \lesssim 0.2$, 
the two-level picture discussed so far is
probably sufficient, with only 211
and 322 growing to large occupation numbers.
For $\alpha \gtrsim 0.2$, we expect that self-interactions
would cause other levels to grow;
 we leave a full analysis of this regime to future work.

Our analysis in this section focusses
on perturbative processes, assuming that evolution
is well-approximated by a combination
of approximately hydrogenic levels.
In section~\ref{sec:nonperturb}, we 
investigate whether non-perturbative processes,
such as ``bosenova'', could change this picture;
we find that, for $\alpha \lesssim 0.2$, this
seems rather unlikely.

\subsubsection{Growth mechanisms in the presence of self-interactions}

As discussed in section~\ref{secevol}, if 211 is initially
the only state with appreciable occupation number,
then other states $j$ can be built up through processes
of the form
\begin{center}
	\begin{tikzpicture}[scale=0.3]
	\draw (-1,1) -- (1,-1);
	\draw (-1,-1) -- (1,1);
	\node[anchor=east] at (-1,1) {211};
	\node[anchor=east] at (-1,-1) {211};
	\node[anchor=west] at (1,1) {$j$};
	\node[anchor=west] at (1,-1) {BH};
\end{tikzpicture}
\end{center}
Taking $j=322$ gives
the fastest growth rate, since the
forced oscillation damped by the BH has $m=0$
(maximizing the damping rate), and the overlap factors
are large.

If a 322 and 211 abundance are both present, 
then other states can also be built up through 
\begin{center}
	\begin{tikzpicture}[scale=0.3]
	\draw (-1,1) -- (1,-1);
	\draw (-1,-1) -- (1,1);
	\node[anchor=east] at (-1,1) {211};
	\node[anchor=east] at (-1,-1) {211};
	\node[anchor=west] at (1,1) {$j$};
	\node[anchor=west] at (1,-1) {BH};
\end{tikzpicture}
 \, \,
	\begin{tikzpicture}[scale=0.3]
	\draw (-1,1) -- (1,-1);
	\draw (-1,-1) -- (1,1);
	\node[anchor=east] at (-1,1) {211};
	\node[anchor=east] at (-1,-1) {322};
	\node[anchor=west] at (1,1) {$j$};
	\node[anchor=west] at (1,-1) {BH};
\end{tikzpicture}
 \, \,
	\begin{tikzpicture}[scale=0.3]
	\draw (-1,1) -- (1,-1);
	\draw (-1,-1) -- (1,1);
	\node[anchor=east] at (-1,1) {322};
	\node[anchor=east] at (-1,-1) {322};
	\node[anchor=west] at (1,1) {$j$};
	\node[anchor=west] at (1,-1) {BH};
\end{tikzpicture}
\end{center}
However, as well as these processes building up new
states, there are also processes reducing their
abundance;
\begin{center}
	\begin{tikzpicture}[scale=0.3]
	\draw (-1,1) -- (1,-1);
	\draw (-1,-1) -- (1,1);
	\node[anchor=east] at (-1,1) {322};
	\node[anchor=east] at (-1,-1) {$j$};
	\node[anchor=west] at (1,1) {211};
	\node[anchor=west] at (1,-1) {$\infty$};
\end{tikzpicture}
 \, \,
	\begin{tikzpicture}[scale=0.3]
	\draw (-1,1) -- (1,-1);
	\draw (-1,-1) -- (1,1);
	\node[anchor=east] at (-1,1) {211};
	\node[anchor=east] at (-1,-1) {$j$};
	\node[anchor=west] at (1,1) {322};
	\node[anchor=west] at (1,-1) {BH};
\end{tikzpicture}
 \, \,
	\begin{tikzpicture}[scale=0.3]
	\draw (-1,1) -- (1,-1);
	\draw (-1,-1) -- (1,1);
	\node[anchor=east] at (-1,1) {$j$};
	\node[anchor=east] at (-1,-1) {$j$};
	\node[anchor=west] at (1,1) {211};
	\node[anchor=west] at (1,-1) {$\infty$};
\end{tikzpicture}
\,
	\begin{tikzpicture}[scale=0.3]
	\node[anchor=east] at (0,1) {$\dots$};
	\node[anchor=east] at (0,0) {};
\end{tikzpicture}
\end{center}
To determine whether, starting from very small
fluctuations, another level $j$ will start growing,
we can look at the linear-in-$\varepsilon_j$ evolution
terms (i.e. ignore processes such as the last diagram), and see whether the growth rate
is positive or negative.

\subsubsection{\texorpdfstring{$n11$ levels}{$n11$ levels}}
\label{n11}

For a state $j$ with $m=1$, the quartic processes with $j$ in the final
state all have forced oscillations with $m \ge 1$, which are 
growing rather than decaying
(in the parameter space where 211 is superradiant). Consequently, they contribute
a negative term to $j$'s growth rate. Hence,
growth of $j$ can only come about through superradiance.

In the large self-coupling regime, a quasi-equilibrium
for 211 and 322 can be reached with very little effect
on the BH spin, so the superradiance rates
for $m=1$ states are still positive. The fastest
such rates are for the $n11$ states.
The linear-order evolution of the occupation number
is set by
\begin{align}
\frac{\dot \varepsilon_{n11}}{\varepsilon_{n11} } = &\gamma^\text{SR}_{n11}
-(\gamma_{211\times n11}^{322\times \text{BH}}+\gamma_{n11\times 322}^{211\times\infty})\varepsilon_{211}\varepsilon_{322}.
\label{eq:n11evol}
\end{align}
Substituting in the equilibrium values for
$\varepsilon_{211}$ and $\varepsilon_{322}$, we have
\begin{align}
\frac{\dot \varepsilon_{n11}}{\gamma^\text{SR}_{n11}\varepsilon_{n11} }  &\simeq 1
	- \frac{2}{3}\frac{\gamma^\text{SR}_{211}}{\gamma^\text{SR}_{n11}}
	\frac{\gamma_{211\times n11}^{322\times \text{BH}} + \gamma_{n11\times 322}^{211\times\infty}}{\gamma_{211\times 211}^{322\times \text{BH}}} 
\label{eq:n11evoleq}
\end{align}
It is useful to analyse the large-$n$ behaviour 
of this expression. At leading order in small $\alpha$, the
ratio
$\frac{\gamma^\text{SR}_{211}}{\gamma^\text{SR}_{n11}}
\frac{\gamma_{211\times n11}^{322\times \text{BH}}}{\gamma_{211\times 211}^{322\times \text{BH}}}$ is independent of $\alpha$ and
$a_*$; it exceeds 1 for $n \gtrsim 10$, and approaches 1.27 at large $n$ (see App. \ref{app:ndependence} and Fig. \ref{fig:n11plot1}).
As discussed in section~\ref{secboundints}, the most
important finite-$\alpha$ effects on the quartic BH rates
arise via the horizon flux of the associated forced oscillation.
Since they are driven by near-horizon behaviour,
these do not have large effects on \emph{ratios}
of rates (Fig.~\ref{fig1120rate}). Consequently,
the ratio of analytic superradiance rates should
be accurate at the few-percent level, except
close to the superradiance boundary.

The ratio $\frac{\gamma^\text{SR}_{211}}{\gamma^\text{SR}_{n11}}
\frac{\gamma_{n11\times 322}^{211\times\infty}}{\gamma_{211\times 211}^{322\times \text{BH}}}$ scales as $\alpha^{-3}$ at small $\alpha$.
For $n$ large, it approaches
\beq
\label{eq:ratios_for_n11_growth_2}
\frac{2\gamma^\text{SR}_{n11}}{3\gamma^\text{SR}_{211}}
\frac{\gamma_{n11\times 322}^{211\times\infty}}{{\gamma_{211\times 211}^{322\times\text{BH}}}} \rightarrow \left(\frac{0.29}{\alpha \tilde{r}_+^{1/3}}\right)^3, \quad n\rightarrow\infty, \alpha \ll 1
\eeq
(see App. \ref{app:ndependence} and Fig.~\ref{fig:n11plot2}).\par
\begin{figure}[t!]
	\begin{centering}
		\includegraphics[width=\columnwidth]{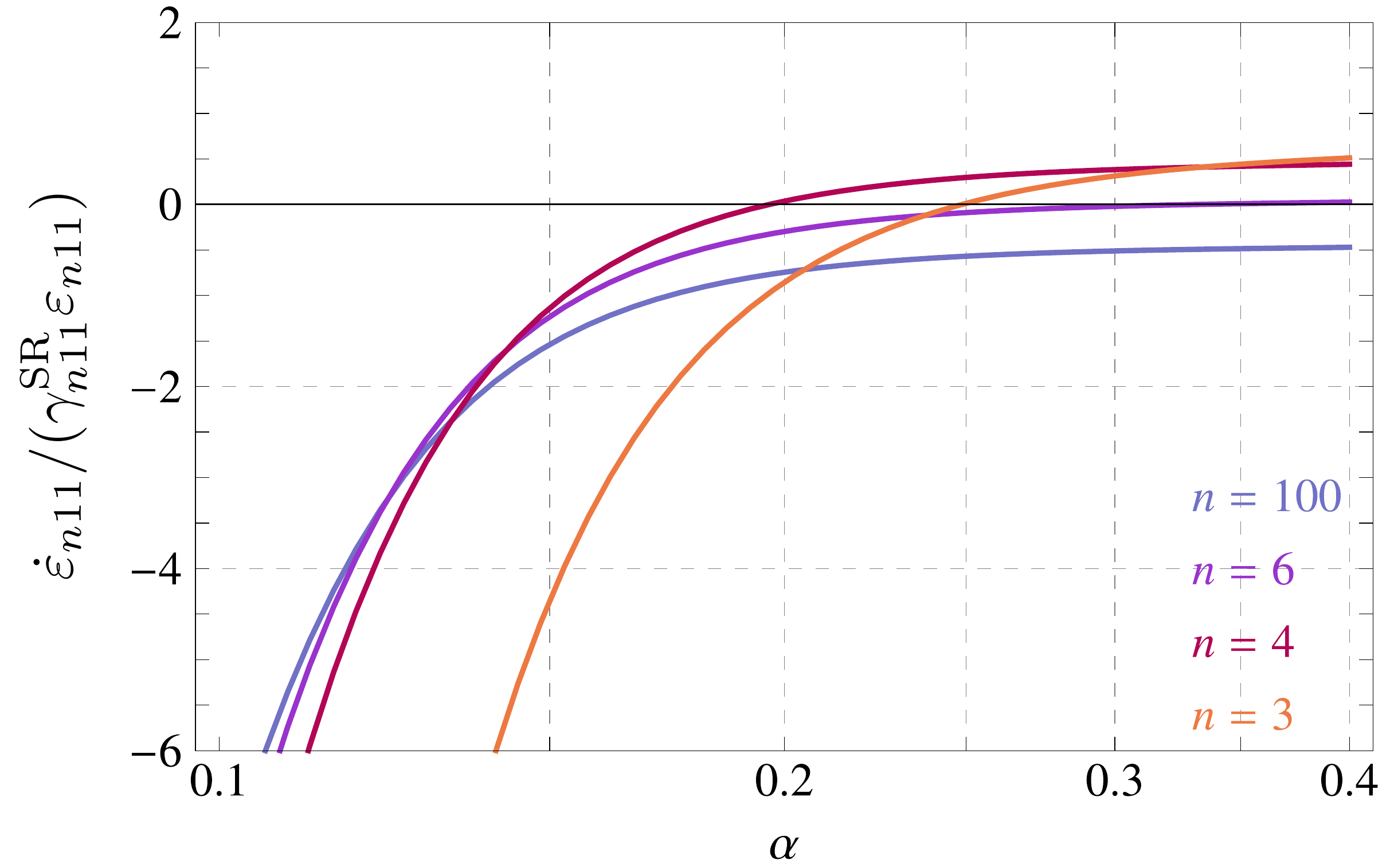}
		\caption{Growth rates of $n11$ levels once 211/322 quasi-equilibrium
		has been reached, relative to their superradiance rates.
At $\alpha\lesssim 0.2$ none of the levels have positive growth rates; levels with $n\gtrsim10$ have negative growth rates for all $\alpha$,
		within our hydrogenic approximation.}
	\label{fig:n11rate}
	\end{centering}
\end{figure}

 
The combination of these negative contributions means
that no $n11$ level with $n\gtrsim 6$ gets populated, at least for $\alpha \tilde r_+ \lesssim 0.3$.\footnote{If
$\alpha$ is large enough that we are in the
``overshoot'' regime, where the maximum occupation
numbers are reached before the equilibrium phase,
the negative contributions to the growth rate
during the overshoot are even larger than
in equilibrium.} For $n=3$, the process $211\times 311 \rightarrow 322 \times \BH$ is resonant, as discussed in section \ref{secboundints};
this makes it more difficult to populate 311.
However, for $\alpha \gtrsim 0.2$, we expect that the 411 level
will grow, given enough time. This is illustrated in Fig.~\ref{fig:n11rate}.

Since the 411 superradiance rate is $\OO(10)$ smaller than
that of 211, the evolution of the 211/322 two-level
system should proceed, at first, without modifications.
Therefore, in the moderate and large self-coupling
regimes we are considering, 211 and 322 will reach
their two-level quasi-equilibrium occupation
numbers, as described in section~\ref{sec:two-level}.
After two-level quasi-equilibrium is reached,
we can initially treat 211 and 322 as constant
sources while 411 grows (since the BH spin-down
timescale is relatively very long).
As a result, 411 grows with an ``effective" superradiance rate which is smaller than its usual superradiance rate,
\begin{align}
\gamma_{411}^{\text{SR-eff}}\equiv\gamma_{411}^{\text{SR}}-(\gamma_{211\times 411}^{322\times \text{BH}}+\gamma_{411\times 322}^{211\times\infty})\varepsilon_{211}^\text{eq}\varepsilon_{322}^\text{eq}
\label{eq:eff411rate}
\end{align}
where the quasi-equilibrium concentrations are given by Eqs. \eqref{eq:quasi_equlibrium_occupations_211} \&
\eqref{eq:quasi_equlibrium_occupations_322}. 

\begin{figure}[t!]
	\begin{centering}
		\includegraphics[width=.49\textwidth]{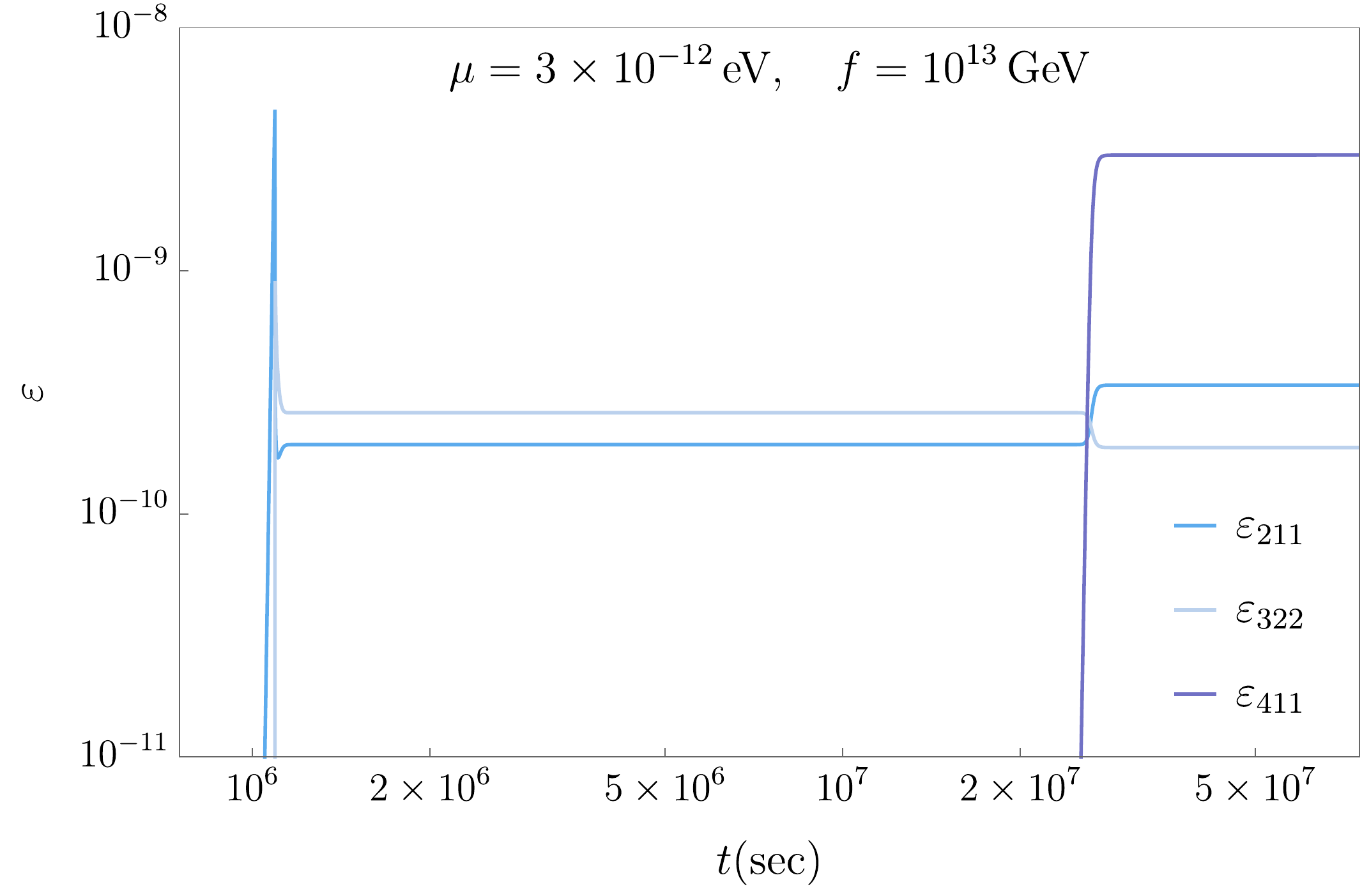}
		\caption{Example of 411 level growth after
		a period of 211/322 quasi-equilibrium.
		This plot 
		assumes a $10\msun$ BH, with $\alpha\simeq0.22$, and an initial BH spin of $0.9$. As discussed in section~\ref{n11},
		the three levels reach a new quasi-equilibrium state,
		in which we expect the 422 level to grow,
		becoming large at later times than those shown here.}
	\label{fig:411sample}
	\end{centering}
\end{figure}

After $\OO(100)$ e-folds, the occupation number
of 411 will become comparable to those of
211 and 322, and the three levels
reach a new quasi-equilibrium. The most striking feature of this
is that the equilibrium 411 occupation number
is significantly higher than the equilibrium occupation numbers
in the two-level 211/322 equilibrium.
The 411 evolution equation is
\begin{align}
\begin{split}
	\frac{\dot \varepsilon_{411}}{\varepsilon_{411}}
	&\simeq \gamma_{411}^{\rm SR} -(\gamma_{211\times 411}^{322\times \text{BH}}+\gamma_{411\times 322}^{211\times\infty})\varepsilon_{211}\varepsilon_{322}\\  & \quad \, - \gamma_{411 \times 411}^{322 \times {\rm BH}}
	\varepsilon_{322} \varepsilon_{411} \nonumber \\
	&= \gamma_{411}^{\rm SR-eff} - \gamma_{411 \times 411}^{322 \times {\rm BH}}
	\varepsilon_{322} \varepsilon_{411}
	\end{split}
\end{align}
Since the numerical
coefficient of the
$\gamma_{411 \times 411}^{322 \times {\rm BH}}$
rate is significantly smaller
than e.g.\ that of $\gamma_{211\times 411}^{322\times {\rm BH}}$
(see Table~\ref{tablebound}),
then unless $\gamma_{411}^{\rm SR-eff}$ is significantly
smaller than the components of Eq.~\eqref{eq:eff411rate},
we need $\varepsilon_{411}^{\rm eq} \gg \varepsilon_{211,322}^{\rm eq}$
to compensate. This is illustrated in
Fig.~\ref{fig:411sample}, which shows
the growth of 411, and development of a new
three-level equilibrium, for $\alpha \simeq 0.22$.
From numerical calculations, 411 grows
to be up to $\sim 50$ times larger than
the benchmark two-level quasi-equilibrium
value of 211 (Eq.~\eqref{eq:quasi_equlibrium_occupations_211}).

Given this enhanced occupation number, it is natural
to ask whether higher-order or non-perturbative processes
could occur, even if they do not
for the two-level system.
As discussed in section~\ref{sec:nonperturb},
the more spread-out wavefunction of the 411
level makes this unlikely.
The emission of scalar radiation
will also be enhanced, as discussed 
in section~\ref{sec:scalarwaves}.

This three-level quasi-equilibrium is unlikely to be the full
story. As we discuss in the next section,
within the two-level equilibrium, we do not expect
$n22$ levels to grow. However, the
large value of $\varepsilon_{411}^{\rm eq}$
can change this conclusion. For example,
the dominant processes building up and depleting
the 422 level, in the presence of equilibrium
211, 322 and 411 occupations, are
\begin{center}
	\begin{tikzpicture}[scale=0.3]
	\draw (-1,1) -- (1,-1);
	\draw (-1,-1) -- (1,1);
	\node[anchor=east] at (-1,1) {411};
	\node[anchor=east] at (-1,-1) {411};
	\node[anchor=west] at (1,1) {422};
	\node[anchor=west] at (1,-1) {BH};
\end{tikzpicture}
 \, \,
	\begin{tikzpicture}[scale=0.3]
	\draw (-1,1) -- (1,-1);
	\draw (-1,-1) -- (1,1);
	\node[anchor=east] at (-1,1) {411};
		\node[anchor=east] at (-1,-1) {422};
	\node[anchor=west] at (1,1) {211};
	\node[anchor=west] at (1,-1) {$\infty$};
\end{tikzpicture}
\end{center}
The first diagram is almost on-shell
for a $400$ forced oscillation, so the
$411 \times 411 \rightarrow 422 \times$ BH process
is ``resonant'', like the $211 \times 311 \rightarrow 322
\times$ BH process discussed in section~\ref{secboundints}.
Consequently, its rate is suppressed by a lower
power of $\alpha$. Along with the large
value of $\varepsilon_{411}$ relative to $\varepsilon_{211}$,
this means that the growth rate of 422 is positive
for the three-level equilibrium occupation numbers.
As a result, after $\OO(100)$ e-folds of
this new growth time, the three-level equilibrium would
be disrupted by the growth of the 422 level.

We leave a more detailed analysis of evolution
in this large-$\alpha$ regime to future work
(as well as the evolution being complicated,
our hydrogenic approximations
are less reliable here). It is possible that further levels
will grow after 422 does, leading to a 
complicated, multi-state superradiant cloud.
In particular, is is possible that 
the cloud could reach large enough field amplitudes
that
higher-order or non-perturbative processes
become important, as we discuss in section~\ref{sec:nonperturb}.

\subsubsection{$n22$ levels}
\label{n22}

$n22$ states grow and are depleted 
similarly to the 322 level, via the processes
\begin{center}
	\begin{tikzpicture}[scale=0.3]
	\draw (-1,1) -- (1,-1);
	\draw (-1,-1) -- (1,1);
	\node[anchor=east] at (-1,1) {211};
	\node[anchor=east] at (-1,-1) {211};
	\node[anchor=west] at (1,1) {$n22$};
	\node[anchor=west] at (1,-1) {BH};
\end{tikzpicture}
 \, \,
	\begin{tikzpicture}[scale=0.3]
	\draw (-1,1) -- (1,-1);
	\draw (-1,-1) -- (1,1);
	\node[anchor=east] at (-1,1) {322};
		\node[anchor=east] at (-1,-1) {$n22$};
	\node[anchor=west] at (1,1) {211};
	\node[anchor=west] at (1,-1) {$\infty$};
\end{tikzpicture}
\end{center}
at linear order in $\varepsilon_{n22}$ (the superradiance rate of $n22$ states is small enough
not to be important, for parameters of interest).
The linear-order growth rate is 
\beq
\dot\varepsilon_{n22} = \gamma_{211\times211}^{n22\times\text{BH}}\left(1-\frac{\gamma_{n22\times322}^{211\times\infty}}{\gamma_{211\times211}^{n22\times\text{BH}}}\eta\right)\varepsilon_{211}^2\varepsilon_{n22},
\label{eq:tdotn22}
\eeq
where
$\eta \equiv \varepsilon_{322}/\varepsilon_{211}.$\par
At early times, $\varepsilon_{322}/\varepsilon_{211} \ll 1$,
and $n22$ is sourced in the same way as 322. However,
since the 322 growth rate is at least $\OO(1)$ larger,
it has an exponentially larger occupation number
than the other $n22$ levels by the time quasi-equilibrium
is established. For example,
\beq
\label{eq:ratio_for_n22_growth_1}
\frac{\gamma_{211\times 211}^{422\times\text{BH}}}{\gamma_{211\times211}^{322\times\text{BH}}}\simeq 0.36;\quad \frac{\gamma_{211\times211}^{n22\times\text{BH}}}{\gamma_{211\times 211}^{322\times\text{BH}}}\propto n^{-3}.
\eeq
(see App.~\ref{app:ndependence} and Fig.~\ref{fig:n22plot} for further details).
For the quasi-equilibrium abundances of 211 and 322, the
negative term in Eq.~\eqref{eq:tdotn22} dominates,
reaching a value of $1.96$ for $n=4$  ($1.69$ for $n\rightarrow \infty$),    \begin{align}
\frac{\gamma_{n22\times322}^{211\times\infty}}{\gamma_{211\times211}^{n22\times\text{BH}}} \eta &\gtrsim
\frac{1}{2}\frac{\kappa_{n22\times322}^{211\times\infty}}{\kappa_{322\times322}^{211\times\infty}}
 \frac{\kappa_{211\times211}^{322\times\text{BH}}}{\kappa_{211\times211}^{n22\times\text{BH}}}\gtrsim 1.69.
\end{align}
Including higher order corrections to the equilibrium ratio of 322 to 211, as well as the superradiance of 322, increases the ratio further. Thus the time derivative of $n22$ becomes negative at leading order in $\alpha$, independently of $\alpha, n,$ and $a_*$.

\subsubsection{$n33$  levels}
\label{n33}

$n33$ states grow and are depleted by
\begin{center}
	\begin{tikzpicture}[scale=0.3]
	\draw (-1,1) -- (1,-1);
	\draw (-1,-1) -- (1,1);
	\node[anchor=east] at (-1,1) {211};
	\node[anchor=east] at (-1,-1) {322};
	\node[anchor=west] at (1,1) {$n33$};
	\node[anchor=west] at (1,-1) {BH};
\end{tikzpicture}
 \, \,
	\begin{tikzpicture}[scale=0.3]
	\draw (-1,1) -- (1,-1);
	\draw (-1,-1) -- (1,1);
	\node[anchor=east] at (-1,1) {322};
		\node[anchor=east] at (-1,-1) {$n33$};
	\node[anchor=west] at (1,1) {211};
	\node[anchor=west] at (1,-1) {$\infty$};
\end{tikzpicture}
\end{center}
giving
\begin{align}
	\dot\varepsilon_{n33} &= (\gamma_{211\times 322}^{n33\times \text{BH}}-\gamma_{322\times n33}^{211\times \infty})\varepsilon_{211}\varepsilon_{322}\varepsilon_{n33} \nonumber \\
	&\simeq
	\left(
	\kappa_{211\times 322}^{n33\times \text{BH}} \tilde r_+\alpha^{11} -\kappa_{322\times n33}^{211\times \infty} \alpha^8 \right)
	\left(\frac{\Mpl}{f}\right)^4
\varepsilon_{211}\varepsilon_{322}\varepsilon_{n33}
\end{align}
at linear order in $\varepsilon_{n33}$.
Due to the different $\alpha$ scaling, the grow rate is negative at small enough $\alpha$.
Quantitatively,
\begin{equation}
\label{eq:ratio_for_n33_growth}
	\left(\frac{\kappa_{322\times n33}^{211\times \infty}}
	{\kappa_{211\times 322}^{n33\times \text{BH}}}\right)^{1/3}
	= \begin{cases} 0.31 & n=4 \\ 0.5 & n \rightarrow \infty 
	\end{cases}
\end{equation}
so at high spin, where $\tilde r_+ \simeq 1$, 
the growth rate is always negative for $\alpha \lesssim 0.3$ (see App. \ref{app:ndependence} and Fig.~\ref{fig:n33plot}).

\subsubsection{$n44$ levels}

For $n44$, we have
\begin{center}
	\begin{tikzpicture}[scale=0.3]
	\draw (-1,1) -- (1,-1);
	\draw (-1,-1) -- (1,1);
	\node[anchor=east] at (-1,1) {322};
	\node[anchor=east] at (-1,-1) {322};
	\node[anchor=west] at (1,1) {$n44$};
	\node[anchor=west] at (1,-1) {BH};
\end{tikzpicture}
 \, \,
	\begin{tikzpicture}[scale=0.3]
	\draw (-1,1) -- (1,-1);
	\draw (-1,-1) -- (1,1);
	\node[anchor=east] at (-1,1) {322};
		\node[anchor=east] at (-1,-1) {$n44$};
	\node[anchor=west] at (1,1) {211};
	\node[anchor=west] at (1,-1) {$\infty$};
\end{tikzpicture}
\end{center}
giving
\begin{align}
	\dot\varepsilon_{n44} &= \left(\gamma_{322\times322}^{n44\times \text{BH}}\frac{\varepsilon_{322}}{\ep}-\gamma_{n44\times 322}^{211\times \infty}\right)\varepsilon_{211}\varepsilon_{322}\varepsilon_{n44}
\nonumber \\
	&\simeq\left(\kappa_{322\times322}^{n44\times \text{BH}}
	\tilde r_+ \alpha^{3} \eta
	-\kappa_{n44\times 322}^{211\times \infty} \right)\alpha^8
	\left (\frac{\Mpl}{f}\right)^4
	\varepsilon_{211}\varepsilon_{322}\varepsilon_{n44}
\end{align}
at linear order in $\varepsilon_{n44}$.

With quasi-equilibrium occupations for 211 and 322, the growth of $n44$ states occurs when $\alpha$ is large enough that
\begin{align}
\label{eq:ratio_for_n44_growth}
&\frac{\kappa_{322\times322}^{n44\times \text{BH}}}{\kappa_{n44\times 322}^{211\times \infty}}\tilde r_+ \alpha^{3}\eta\approx\frac{1}{2}\frac{\kappa_{322\times322}^{n44\times\text{BH}}}{\kappa_{n44\times322}^{211\times\infty}}
\frac{\kappa_{211\times211}^{322\times\text{BH}}}{\kappa_{322\times322}^{211\times\infty}} \alpha^6\tilde r_+^{2} 
\gtrsim 1,
\end{align}
or equivalently
\begin{align}
\alpha\tilde r_+^{1/3} \gtrsim 0.3
\end{align}
where the right hand side is as large as $0.34$ for $n=5$ ($0.3$ for $n\rightarrow\infty$) (see App. \ref{app:ndependence} and Fig.~\ref{fig:n44plot}).

\subsubsection{Other levels}

The $n22,n33$ and $n44$ levels considered 
above are the only ones which can be built up via
quartic processes where the forced oscillation has
$l=m=0$.\footnote{This is not strictly true --- the Kerr potential breaks spherical symmetry, so $l$ is no longer a good quantum number, and e.g.\ $n42$ can also be build up via a $m=0$ forced oscillation. However, in the small-$\alpha$ limit, the overlaps for such processes
are suppressed by more powers of $\alpha$.} To build up other processes via self-interactions,
starting from 211 and 322, we need to use forced oscillations
with $l > 0$, which have a parametrically smaller
flux through the BH horizon.
They therefore stand even less chance of having 
positive growth rates.
For $l \ge 2$, we can often rule out these processes
being relevant on astrophysical timescales, simply by
estimating the magnitude of the growth rate.
For example, for $l=2$,
we have
\begin{align}
    \begin{split}
     \gamma_{322\times 322}^{766\times \text{BH}(2,-2)} (\varepsilon_{322}^\text{eq})^2 \sim{}& 10^{-2}\left(\frac{\alpha}{0.3}\right)^{19}\left(\frac{M_\odot}{M}\right)\Myr^{-1},
    \end{split}
\end{align}
where the superscript $\BH(l,m)$ indicates the angular momentum numbers of the damped leg.\par
Taking an $l=1$ example,
\begin{align}
    \begin{split}
     {}& \dot\varepsilon_{655} = \\{}&\gamma_{322\times 322}^{655\times \BH(1,-1)}\left(1-\frac{\gamma_{655\times 322}^{211\times\infty}}{\gamma_{322\times 322}^{655\times \BH(1,-1)}}\frac{\varepsilon_{211}}{\varepsilon_{322}}\right)\varepsilon_{322}^2\varepsilon_{655}.
    \end{split}
\end{align}
The depletion term dominates at equilibrium as long as 
\begin{align}
    \begin{split}
     \alpha\tilde r_+^{1/9}\lesssim\left(\frac{\kappa_{655\times322}^{211\times\infty}}{\kappa_{322\times 322}^{655\times\BH(1,-1)}}\frac{1}{\eta^B}\right)^{1/9} \approx 0.7.
    \end{split}
\end{align}
Similar checks can be performed for other processes involving mixing
with an $l = 1$ damped state (see App.~\ref{app:ndependence}).
One finds that, for all of them, the depletion process to infinity
dominates over the pumping process for the entire range of $\alpha$ for
which $m=1$ states can be superradiant ($\alpha \lesssim 0.5$).

\section{Non-perturbative behavior}
\label{sec:nonperturb}

So far, our analysis has assumed that the scalar
field is always well-approximated by a combination
of approximately hydrogenic bound states,
and that quartic interactions result in the slow transfer of energy to and from
these bound states.
However, if the field amplitude becomes large enough,
we expect this picture to break down.
Most directly, for a generic potential, higher-order field interactions
can become important. 
In addition, for large enough amplitudes, attractive interactions
would make hydrogenic bound states unstable
to collapse, in a ``bosenova''~\cite{Arvanitaki:2010sy,Yoshino:2012kn,Yoshino:2015nsa}.

As we explored in section~\ref{secperturb},
for large self-couplings, the quartic interactions
lead to the saturation of the cloud to a quasi-equilibrium
configuration (for much of the parameter space of
interest), with field amplitude $\propto f$.
For a potential of the form 
$V(\varphi) \propto g(\varphi / f)$, this means
that the relative importance of higher-dimensional
interactions
becomes independent of $f$ (for small enough $f$).
As we will show below, for small $\alpha$,
the maximum value of $\theta \equiv \varphi/f$ is small, 
and the quartic-driven behaviour we have investigated
should be a good approximation. Similarly,
for small $\alpha$, the cloud is always far from 
the non-perturbative ``bosenova'' regime.
For $\alpha \gtrsim 0.2$, we expect levels
beyond 211 and 322 to grow in the small-$f$ regime,
as discussed in the previous section,
so their behaviour would need to be analysed to draw
conclusions about non-perturbative behaviour.

\subsection{Maximum field amplitude}
\label{sec:theta}

When a single hydrogenic level dominates
the energy stored in the cloud, the dimensionless
field amplitude $\theta = \varphi/f$ is related
to the occupation number of that level by
$|\theta| \propto 
\alpha^{5/2}\sqrt{\varepsilon}\Mpl/f$.
In the small and moderate self-coupling
regimes, where 211 reaches its saturation occupation
number, $|\theta|$ increases $\propto 1/f$ as $f$ decreases.
However, once we are in the large-self-coupling
regime, the occupation numbers reached are
$\propto f^2$, so $\theta$ becomes independent of $f$.

\begin{figure}[t!]
	\begin{centering}
		\includegraphics[width=\columnwidth]{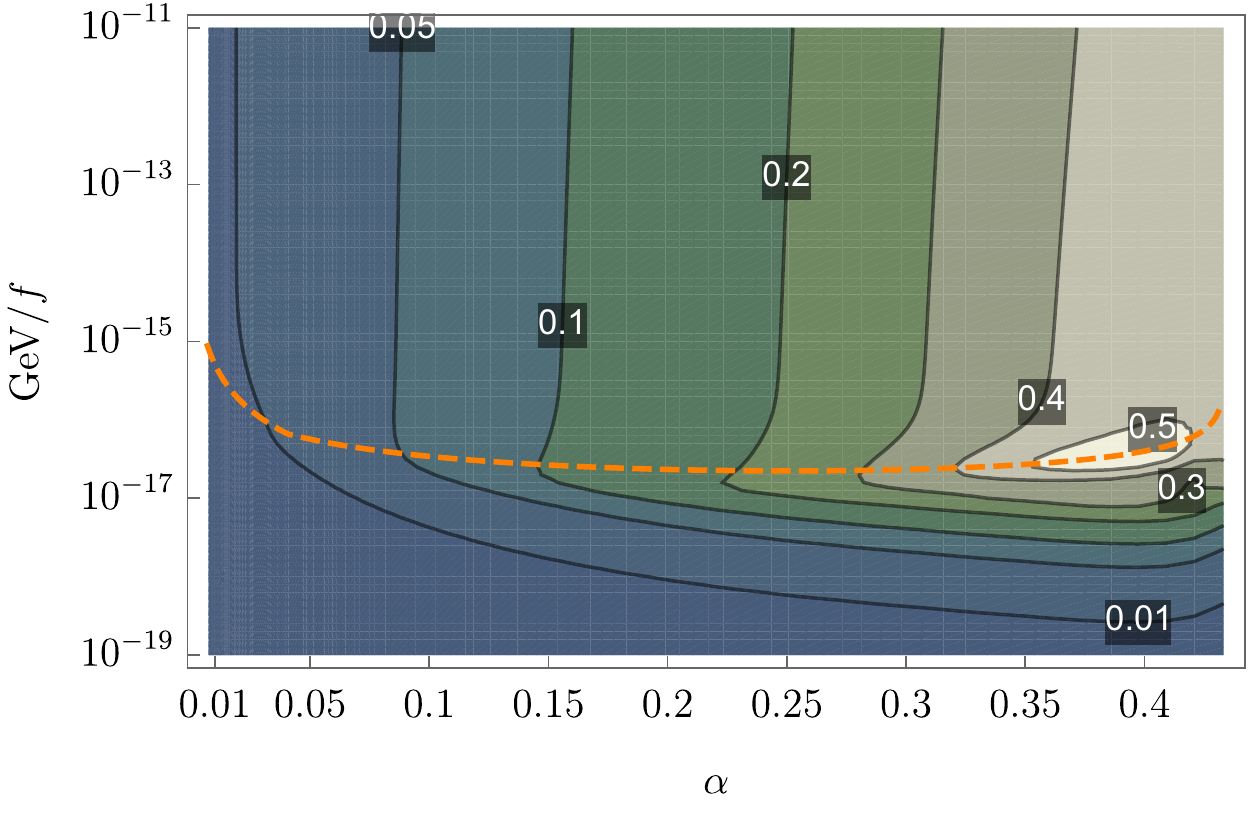}
		\caption{Maximum value of $|\theta| \equiv |\varphi/f|$
		attained during the evolution of the two-level
		211/322 system,
		for a BH with initial 
		spin $a_* = 0.99$ and initial mass $10 \msun$
		(the BH mass only affects this plot via
		the number of $e$-folds $\sim \log (G M^2)$ a level can grow). The dashed
		orange line indicates the boundary between
		the moderate and large self-coupling regimes
		(corresponding to $f_{\rm BC}$ as defined in
		section~\ref{secperturb}).
		$|\theta_{\rm max}|$ is computed by numerically
		solving the evolution equations for
		the 211 and 322 occupation numbers.}
	\label{fig:theta}
	\end{centering}
\end{figure}

If 211 is the dominant level,
then the maximum value of $\theta$ is attained
at $r = 2 a_0$ and $\theta = \pi/2$, with
\begin{align}
|\theta_\text{max}| \approx \alpha^{5/2}\sqrt{\varepsilon_{211}} \left(\frac{\Mpl}{f}\right)\sqrt{\frac{1}{8\pi}}e^{-1}.
\label{eq:thetamax}
\end{align}
As we decrease $f$, this increases until
$f \simeq f_{\rm BC}$ (Eq.~\eqref{eq:f_{BC}}). For $\alpha \gtrsim 0.04$, $f_\text{BC} = f_\text{thresh}$ and
\begin{align}
\begin{split}
\label{eq:bound_on_theta_0}
|\theta_\text{max}(f_\text{BC})| \approx{}& \alpha^{7/4} \left(\frac{\log(GM^2)\kappa^\SR a_*(t_0)}{\kappa^\BH}\right)^{1/4}\frac{e^{-1}}{2\sqrt {\sqrt 2\pi}}\\
\approx{}&0.03\left(\frac{\alpha}{0.05}\right)^{7/4}.
\end{split}
\end{align}
The scalings in \eqref{eq:bound_on_theta_0} are only representative when
$\alpha \ll a_*(t_0)$ (see App. \ref{sec:fa_min_contour}). 
For $\alpha \lesssim 0.04$, $f_\text{BC} = f_\text{eq}$ and the maximum value of $\theta$ is equal to its value at equilibrium:
\begin{align}
\begin{split}
\label{eq:theta_eq}
|\theta^\text{eq}_\text{max}| \approx{}& \alpha \left(\frac{\sqrt{\kappa^\SR a_*(t_0)\kappa^\infty}}{\kappa^\BH}\right)^{1/2} \sqrt{\frac{1}{\sqrt{24}\pi}}e^{-1}\\
\approx{}&0.005 \left(\frac{\alpha}{0.01}\right)\left(\frac{a_*(t_0)}{0.99}\right)^{1/4}.
\end{split}
\end{align}
(again, these scalings are valid when $\alpha
\ll a_*(t_0) $).\footnote{
Although Eq.~\eqref{eq:theta_eq} is valid at equilibrium,
we noted in section~\ref{secllarge} that $\varepsilon_{211}$
can ``overshoot'' its equilibrium value as it evolves
towards equilibrium. 
We have determined numerically that the overshoot estimate of Eq.~\eqref{eq:overshoot2}, or the approximation of Eq.~\eqref{eq:overshoot}, accurately predicts $\varepsilon_{211}^{\max}$
for $\alpha\gtrsim0.05$ with an error less than $1\%$, deep in the
self-interaction regime. Quantitatively, we found numerically that
there is a thin band around the dashed boundary line of Fig.
\ref{fig:theta} (see Eq.~\eqref{eq:fa_min_contour_exact}), with a
width of less than an order of magnitude in $f$, where both the
quasi-equilibrium and the overshoot estimates under-predict
$\varepsilon_{211}^{\max}$ by $\gtrsim5\%$. A significant discrepancy
arises only in the region where $|\theta_\text{max}|$ reaches its
largest value and is $\sim 20\%$. These translate to a $\sim2.5\%$
and $\sim 10\%$ discrepancy in the analytically predicted
$|\theta_\text{max}|$, according to the scaling of Eq.~\eqref{eq:thetamax}.}

These equations suggest that, for small $\alpha$,
the value of $|\theta|$ never becomes large,
so we would generically expect higher-dimensional
interactions to remain unimportant. To see this 
more quantitatively, Fig.~\ref{fig:theta} shows
the maximum value of $|\theta|$ attained
during the evolution of the two-level 211/322 system,
for different values of $\alpha$ and $f$.
This has the expected behaviour, increasing
with decreasing $f$ for $f \gtrsim f_{\rm BC}$,
and reaching a constant value for smaller $f$
(at a given $\alpha$).

As discussed in section~\ref{secperturb},
we expect that, for small $f$ and $\alpha \gtrsim 0.2$,
levels other than 211 and 322 will grow.
At these parameters, the $|\theta_{\rm max}|$ values
in Fig.~\ref{fig:theta} represent a lower bound 
(since the initial 211 overshoot value is still
set by 211/322 dynamics). 
For the 411 level, which
we expect to be the first
to grow after the 211/322 quasi-equilibrium
(section~\ref{n11}),
the maximum occupation reached is 
only around twice the maximum occupation number of 211.
Consequently, the more spread-out wavefunction of
411 means that it does not attain a larger $|\theta|$
value. However, a more careful 
analysis would be required to determine
$|\theta_{\rm max}|$ once other levels grow.

\subsection{Bosenova}
\label{sec:bosenova}

As well as higher-dimensional interactions
becoming important, another possible issue
arising at large occupation numbers is that
the cloud may undergo a sudden collapse
due to attractive self-interactions,
known as a ``bosenova''~\cite{Arvanitaki:2010sy}.
Here, we estimate the occupation number
threshold for a bosenova to occur,
using a variational approach.

The wavefunction for the hydrogenic 211 level is
\begin{equation}
	\psi_{211} = \frac{\sqrt{N_{211}}}{2 \sqrt{6}} a_0^{-5/2} r e^{-r/(2 a_0)} Y_{11}(\theta,\phi)
\end{equation}
where $a_0 \equiv 1/(\alpha \mu)$ is the Bohr radius.
As our variational ansatz, we will take a wavefunction
of this form, but with a modified radius,
\begin{equation}
	\psi = \frac{\sqrt{N}}{2 \sqrt{6}} R^{-5/2} r e^{-r/(2 R)} Y_{11}(\theta,\phi)
	\label{eq:ansatz211}
\end{equation}
For convenience, we will define a dimension-2 wavefunction $\tilde \psi = \sqrt{\mu} \psi$.
Then, the non-relativistic action for $\tilde \psi$ interacting with a gravitational field, sourced both by the central BH and by itself, is given by
\begin{equation}
\begin{split}
\mathcal{S}\simeq\int\di^3\mathbf{}{r}\,\di t\,&\frac{i}{2\mu}\pare{\tilde{\psi}^\ast\partial_t\tilde{\psi}-\tilde{\psi}\partial_t\tilde{\psi}^\ast}-\frac{1}{2\mu^2}|\nabla\tilde{\psi}|^2-\Phi|\tilde{\psi}|^2\\
&+\frac{\lambda}{16\mu^4}|\tilde{\psi}|^4-\frac{1}{8\pi G}|\nabla\Phi|^2-\rho_\text{BH}\Phi
\label{eq:nonrelL}
\end{split}
\end{equation}
The gravitational potential $\Phi$ obeys the Poisson equation,
\begin{align}
\nabla^2\Phi=4\pi G\pare{\rho_\text{BH}+|\tilde{\psi}|^2}
\label{eq:poisson}
\end{align}
where we take
$\rho_\text{BH}=M\delta^{3}(\mathbf{r})$ and $M$ is the mass of the BH.
Using this potential, and integrating the action
of Eq.~\eqref{eq:nonrelL} over space,
we obtain an effective potential for $R$.
Ignoring self-gravity of $\psi$, this is
\begin{equation}
V(\tilde{R})=\frac{\alpha^4\Mpl^2\varepsilon}{\mu}\pare{\frac{1}{8\tilde{R}^2}-\frac{1}{4\tilde{R}}-\frac{3\alpha^3\varepsilon\Mpl^2}{16384\pi \tilde{R}^3f^2}},
\label{eq:effpotential}
\end{equation}
where $\tilde R \equiv R/a_0$.
The first two terms correspond to kinetic and gravitational
energy, and set the radius of small-amplitude hydrogenic levels --- the last
terms arises from attractive self-interactions.
The extrema of the potential $V(\tilde R)$ are at
\begin{align}
\tilde{R}^{\pm}_\text{extrema}= \frac{1}{2}\pm\sqrt{\frac{1}{4}-\frac{9\alpha^3\varepsilon\Mpl^2}{4096\pi f^2}}.
\label{eq:rext}
\end{align}
If we decrease $f$, at some point these extrema will coincide,
and the potential will no longer have a stable minimum.
This leads to a ``bosenova'', with the cloud collapsing.
The critical occupation number for this to occur is
\begin{align}
\varepsilon_\text{crit}=\frac{1024\pi f^2}{9\alpha^3\Mpl^2}
\label{eq:critepsnosg}
\end{align}
Incorporating the effects of self-gravity, this becomes
\begin{align}
\varepsilon_\text{crit}=\frac{32}{711\alpha^2}\sqrt{75840\pi\pare{\frac{f}{M_\text{Pl}}}^2+225\alpha^2}-\frac{160}{237\alpha}
\label{eq:criteps}
\end{align}
which reduces to Eq.~\eqref{eq:critepsnosg} for small $f$, i.e.\ for small clouds.

\begin{figure}[t!]
	\begin{centering}
		\includegraphics[width=\columnwidth]{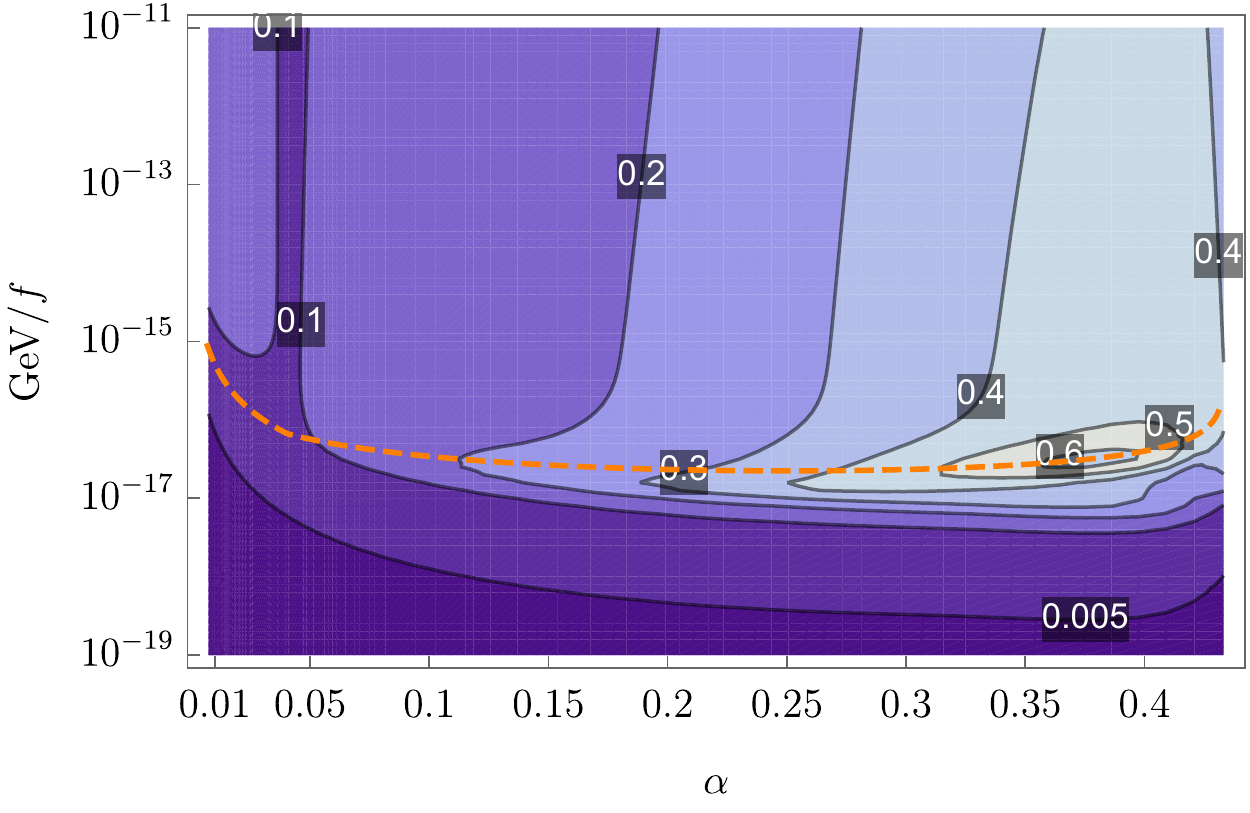}
		\caption{Maximum value of $\varepsilon_{211}/\varepsilon_{211}^{\rm crit}$ attained during the evolution
		of the two-level 211/322 system,
		for a BH with initial spin $a_* = 0.99$.
		$\varepsilon_{211}^{\rm crit}$ is the critical
		occupation number above which a
		rapid collapse of the cloud (a ``bosenova'') is
		expected to occur (section~\ref{sec:bosenova}).
		The dashed orange line indicates the boundary
		between the moderate and large self-coupling
		regimes
		(corresponding to $f_{\rm BC}$ as defined in
		section~\ref{secperturb}).
		$\varepsilon_{211}$ is computed by numerically
		solving the evolution equations for the
		211 and 322 occupation numbers.
		The plot is roughly independent of the BH mass,
		within the range of astrophysical BHs.}
	\label{fig:bosenova}
	\end{centering}
\end{figure}

Given this, we can ask whether the
211 occupation number reaches $\varepsilon_{\rm crit}$
during its perturbative evolution. If it does not, then
our assumption of perturbative evolution can be self-consistent.
Fig.~\ref{fig:bosenova} shows the maximum value
of $\varepsilon_{211}/\varepsilon_{211}^{\rm crit}$ 
attained during the evolution of the two-level 211/322 system.
For $\alpha$ small enough that other levels
do not grow ($\alpha \lesssim 0.2$), we can see
that this ratio is always $\lesssim 0.3$, so we do not expect
a bosenova to occur. This is in contrast
to the conclusions of much of the existing literature.
As emphasized previously, 
other papers neglect the perturbative processes
that lead to energy exchange between hydrogenic levels,
causing the cloud to saturate to a quasi-equilibrium
configuration before its amplitude becomes large enough
for a bosenova.

For $\alpha \gtrsim 0.2$, we expect that levels other than
211 and 322 will grow. This means that the $\varepsilon/\varepsilon_{\rm crit}$ values
in Fig.~\ref{fig:bosenova} represent a lower bound.
As we discussed in the previous subsection,
the more spread-out wavefunction of the 411 level
means that it is unlikely to get closer to 
the critical occupation number than the 211 level;
we leave an analysis of the situation once other levels
have grown to future work.

\subsubsection{Sub-leading effects}

As discussed in App. \ref{app:cloudmass}, superradiance extracts mass
from the BH in addition to angular momentum. As such, the cloud can
actually grow to be somewhat larger than we have assumed so far.
The modified equations for purely gravitational superradiance can
be found in App. \ref{app:cloudmass}. In deriving Fig.
\ref{fig:theta} and \ref{fig:bosenova} we have included the
correction coming from the change of the BH mass or, equivalently,
from the time-dependence of $\alpha$. As expected, we find that 
this correction can become quite large near the superradiance boundary,
as the final spin is slightly modified (see Eq.~\eqref{eq:cmspinsimple}).  However, for strong self-interactions,
where the bosenova might be relevant, there is practically no
significant correction, as the cloud does not grow appreciably and
thus does not extract a significant amount of spin or mass from the
BH. 

One might also ask how the inclusion of another level, say 322, changes
the above picture. Assuming that its fractional occupation number
is small compared to our primary level (e.g.\ 211), we can treat such a
level as a small perturbation and check whether our results are
consistent. In what follows we will neglect self-gravity for clarity
or, equivalently, we will work in the small $f$ (large
self-interactions) limit, where Eq.~\eqref{eq:criteps} coincides with
Eq.~\eqref{eq:critepsnosg}. We add a contribution from 322 to our
variational ansatz
\begin{align}
    \tilde{\psi}\supset \frac{\Mct^{1/2}}{4a_0^{3/2}}\frac{4}{81\sqrt{30}}\pare{\frac{r}{a_0}}^2\exp\pare{-\frac{r}{3a_0}}Y_2^2(\theta,\phi)
    \label{eq:ansatz322}
\end{align}
where $\Mct$ is the mass
of the 322 cloud. Note that we treat 322 as rigid, i.e.\ we do not
allow its radius to change. Following the same procedure as before,
we get an effective potential for 211 with an additional attractive
term, stemming from its interaction with 322
\begin{equation}
\label{eq:potential322}
\begin{split}
&V(\tilde{R})=\\
&\frac{\alpha^4\Mpl^2\varepsilon}{\mu}\pare{\frac{1}{8\tilde{R}^2}-\frac{1}{4\tilde{R}}-\frac{3\alpha^3\varepsilon\Mpl^2}{16384\pi \tilde{R}^3f^2}-\frac{27\tilde{R}^4\alpha^3\varepsilon_2\Mpl^2}{2\pi(3+2\tilde{R})^9f^2}}
\end{split}
\end{equation}
where $\varepsilon_2$ is the fractional occupation number of 322.
Expanding around the critical values as
$\tilde{R}=\frac{1}{2}+\sqrt{\varepsilon_2}\,\delta\tilde{R} $ and
$\varepsilon=\varepsilon_\text{crit}+\varepsilon_2\,\delta\varepsilon$,
we find the correction $\delta\varepsilon=21/16384$, giving
\begin{align}
\frac{\varepsilon_2\,\delta\varepsilon}{\varepsilon}=\frac{21}{16384}\frac{\varepsilon_2}{\varepsilon}\ll 1 \, .
\end{align}
The result is indeed small and, thus, it does not change our
conclusions about the bosenova. In particular, the correction to
$\varepsilon_\text{crit}$ is \textit{positive}. Since the interaction
is attractive, as seen from the potential in Eq.~\eqref{eq:potential322}, the 322 cloud attracts the 211 one and,
since it resides at a larger radius, it effectively dilutes it. 

In Fig.~\ref{fig:bosenova}, we compared
the $\varepsilon_{211}$ value attained during the perturbative
level evolution to $\varepsilon_{\rm crit}$. However,
the rates of the different processes involved
in the evolution were calculated for
the unperturbed hydrogenic wavefunctions. 
Consequently, we should ask whether 
self-interaction-induced
perturbations to the wavefunctions make a significant
difference to the rates, and so the occupation numbers
attained.
From Eq.~\eqref{eq:rext}, we can see that if
$\varepsilon_{211}/\varepsilon_{\rm crit}$ is always small,
then the corrections to the wavefunctions will always
be small, and our calculations should be
self-consistent. Since $\varepsilon_{211}/\varepsilon_{\rm crit}$
only becomes large for larger $\alpha$, where (as 
discussed previously) our perturbative evolution
calculations are already incomplete, we leave a full
analysis to future work.

In plotting $|\theta_\text{max}|$, we have used the field defined
using Eq.~\eqref{eq:ansatz211}, that is, by taking into account the
corrected radius of Eq.~\eqref{eq:rext}. This amounts to multiplying
Eq.~\eqref{eq:thetamax} by a factor of
$\pare{\tilde{R}^+_\text{ext}}^{-3/2}$ (Eq.~\eqref{eq:rext}), giving
\begin{align}
|\theta| \approx \alpha^{5/2}\sqrt{\varepsilon_{211}}\pare{\frac{1}{\tilde{R}^+_\text{ext}}}^{3/2}\left(\frac{\Mpl}{f}\right)\sqrt{\frac{1}{8\pi}}e^{-1}.
\label{eq:thetamaxNP}
\end{align}
We have determined numerically that the radius change is at most $~15\%$ and introduces at most a $~25\%$ change in the region where $|\theta|$ grows to be the largest possible, driving to a value of $\sim0.5$, whereas the change is much smaller everywhere else.

Another possible issue with our variational analysis is that
the evolution is not adiabatic during the last few e-folds before 211
reaches its maximum occupation number. As a result, the cloud might
not trace the minimum of the potential of Eq.~\eqref{eq:effpotential}
but rather oscillate around it, in the manner of an ``excited state".
In this case, the cloud could overcome the barrier at $\tilde{R}^{-}$
(Eq.~\eqref{eq:rext}) and collapse. We note that oscillations
of the radius of the peak seem consistent with the results of ref.
\cite{Yoshino_2014}. The minimum of the potential would need
to be fairly close to critical for this to be an issue,
but 
we leave detailed investigation of this point
to future work.

\subsubsection{Comparison to simulations}

While we expect our hydrogenic ansatz to be a good
approximation, properly understanding the dynamics
of a bosenova requires numerical simulations.
In \cite{Yoshino:2012kn,Yoshino:2015nsa}, the authors numerically simulate
the evolution of a self-interacting scalar field around a high-spin Kerr BH, starting from a hydrogenic bound state profile
with $\theta \sim \OO(1)$.
These simulations effectively operate in the 
large self-coupling regime, taking the cloud's mass
to be very small compared to the BH.
In simulations with $a_* = 0.99$
and $\alpha = 0.3$~\cite{Yoshino:2015nsa},
they find that a 211 bound state with initial amplitude such that $|\theta_{\rm max}|
= 0.4$ does not undergo a bosenova,
but one with $|\theta_{\rm max}| = 0.45$ does.

Comparing these to our variational calculations,
we can convert the critical occupation number \eqref{eq:critepsnosg}
to a field amplitude, giving
the leading-$\alpha$ expression
$|\theta_{\rm max}^{\rm crit}| = \frac{8 \sqrt{2}}{3 e} \alpha
\simeq 0.42 \frac{\alpha}{0.3}$.
This is highly compatible with the threshold behaviour observed
in the simulations.

The simulations in \cite{Yoshino:2012kn,Yoshino:2015nsa} 
were  evolved forward for $t \simeq 2000 r_g$.
This is much shorter than the timescales for
any of the perturbative processes
studied in section~\ref{secperturb}, including
211 superradiance, and the growth of 322 through self-interactions.
A simulation would have to be run for much longer times to
observe these effects.
In particular, the fact that a bosenova was observed
for the initial state 
$|\theta_{\rm max}| = 0.45$ is not evidence that a bosenova
would occur around an astrophysical black hole.
In the latter case, the true initial conditions are at an exponentially
smaller amplitude, and according to our estimates,
the maximum 211 amplitude reached during the evolution
is $|\theta_{\rm max}| \simeq 0.3$ (Fig.~\ref{fig:theta}), at which point
interactions with 322 cut off its growth.

\subsubsection{Repulsive self-interactions}

In \cite{Gruzinov_2016}, it is claimed that if
self-interactions are repulsive, they can completely
suppress the growth of 322, by spreading out
the 211 cloud and 
reducing the rate of the $211 \times 211 \rightarrow
322 \times {\rm BH}$ process. 
We can estimate the effect of repulsive self-interactions
by looking at how they shift the 211 wavefunction radius in 
our variational ansatz. This gives
\begin{equation}
	\tilde R_\text{rep}=\frac{1}{2}\pare{1+\sqrt{1+\frac{\varepsilon_{211}}{\varepsilon_{211}^{\rm crit}}}}
    \label{eq:Rrepulsive}
\end{equation}
with $\varepsilon_{211}^{\rm crit}$ from
Eq.~\eqref{eq:critepsnosg}.
Since the perturbative evolution processes
from section~\ref{secperturb} all depend 
on $\lambda^2$, they are the same for attractive
and repulsive self-interactions.
Consequently, the maximum value of $\varepsilon_{211}$
attained through perturbative evolution
should be the same.
As a result, we expect that, unless
$\varepsilon_{211}/\varepsilon_{211}^{\rm crit}$ becomes
large (which we cannot rule out for $\alpha \gtrsim 0.2$
and small $f$),
the effects of repulsion should be small.


\section{Black hole spin-down}
\label{sec:spindown}

\begin{figure*}[t]
\centering
\includegraphics[width = .991 \columnwidth]{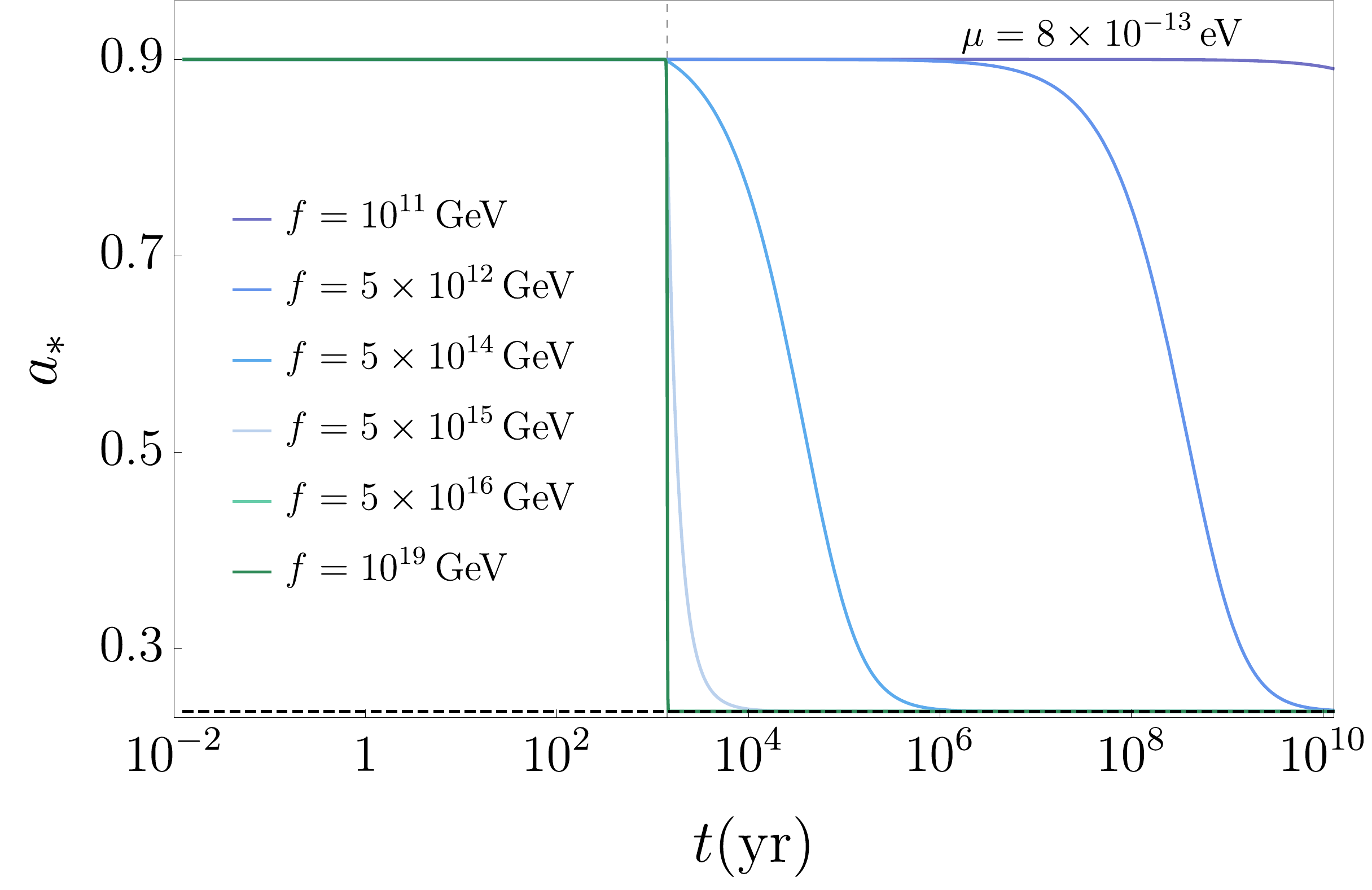}
\centering
\includegraphics[width = .991 \columnwidth]{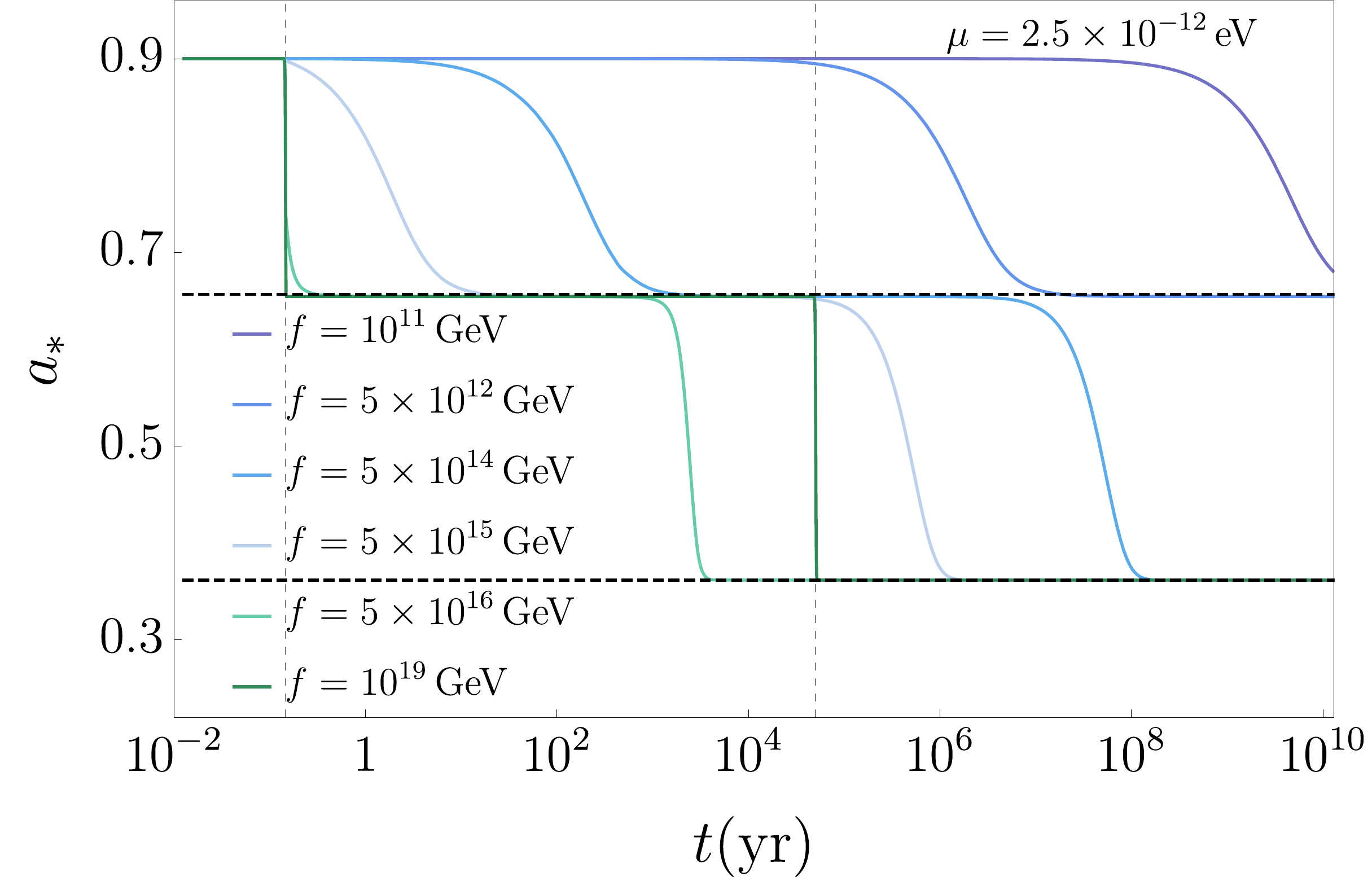}
\caption{Black hole spin-down as a function of time for $\mu = 8\times 10^{-13}\eV$ (\textit{left panel}) and $\mu = 2.5\times10^{-12}\eV$ (\textit{right panel}) for a range of self-interactions strengths, and a $10\msun$ black hole. These axion masses correspond to $\alpha \simeq 0.06$ 
	and $\alpha \simeq 0.19$ respectively. The dashed horizontal lines show the superradiance boundary for levels 211 (upper) and 322 (lower). The dashed vertical lines show the expected spindown time in the limit of no self-interactions for levels 211 (smaller $t$) and 322 (larger $t$). }
\label{fig:BHspindown}
\end{figure*}

One of the observational signatures
of superradiance is the spin-down
of initially fast-spinning
BHs~\cite{Arvanitaki:2009fg,Arvanitaki:2010sy}. 
In
the absence of non-gravitational interactions, if
a BH is born with spin high enough that a mode is
superradiant, and the mode's growth time is much shorter
than the lifetime of the BH, then a superradiant
cloud will form around the BH. This spins down
the BH to the point where the mode is stable,
rather than growing. Consequently, observing a
sufficiently old, sufficiently fast-spinning BH
is good evidence against the existence of a light
boson with such properties. Constraints of this kind have
been placed on spin-0 \cite{Arvanitaki:2014wva,
Cardoso+2018} and spin-1 \cite{Baryakhtar:2017ngi}
particles from measurements of BH spins in X-ray
binaries \cite{Miller:2014aaa,McClintock:2013vwa} 
(higher-spin particles have also been considered~\cite{1304.6725,2002.04055}, though such models encounter theoretical issues, as we discuss
in the conclusions).

In contrast, if self-interactions
are large, then as discussed in section~\ref{secperturb},
the occupation numbers in the
quasi-equilibrium state are suppressed.
Consequently, the rate of energy and angular
momentum extraction from the BH is suppressed, 
and the spin-down constraints described
in the previous paragraph will not apply directly.

Instead, for small enough $f$, the time-averaged
spin extraction rate will be approximately set
by the equilibrium occupation number of the 
211 level (at
least in the case of 211 superradiance), as discussed in section~\ref{secllarge}.
Since $\varepsilon_{211}^{\rm eq} \propto \alpha^{-3} \frac{f^2}{M_{\rm pl}^2}$ (Eq.~\eqref{eq:quasi_equlibrium_occupations_211}),
the time taken to fully spin down the BH (to
the point where 211 superradiance is saturated) scales
$\propto f^{-2}$. Consequently, as reviewed
in section~\ref{seclnosd}, there is some minimum
$f$ below which the BH is not significantly spun
down in the time available.

This behaviour is illustrated, for particular initial
BH parameters, in Fig.~\ref{fig:BHspindown}.
The figure shows how, for $f \lesssim f_{\rm BC}$
(table~\ref{tab:paramsummary}; $f_{\rm BC} \simeq 3 \times 10^{16} 
\GeV$ for the left-hand panel, and
$\simeq 2 \times 10^{17} \GeV$ for the right-hand panel),
spin-down to the $m=1$ superradiance threshold takes 
longer as $f$ is decreased, until it no longer occurs
within the lifetime of the BH for
$f \lesssim f_{\rm CD}$.
The region of $(\mu,f)$ parameter space in
which the BH is spun down to the $m=1$ superradiance
threshold is shown in the bottom-left 
panel of Fig.~\ref{fig:paramspace}.

We have only performed a detailed
analysis (at all $f$) of situations in which 
211 is the first superradiant level to grow,
and levels beyond 211 and 322 do not grow.
From section~\ref{secperturb}, this corresponds to
$\alpha \lesssim 0.2$.
Nevertheless, we can be confident that,
when interactions are weak enough that superradiant
growth of the 322 level is unaffected, the black hole
is spun down as in the purely gravitational
case. This is indicated in the bottom right of 
the lower panels in Fig.~\ref{fig:paramspace}.

\begin{figure}[t!]
	\begin{centering}
		\includegraphics[width=\columnwidth]{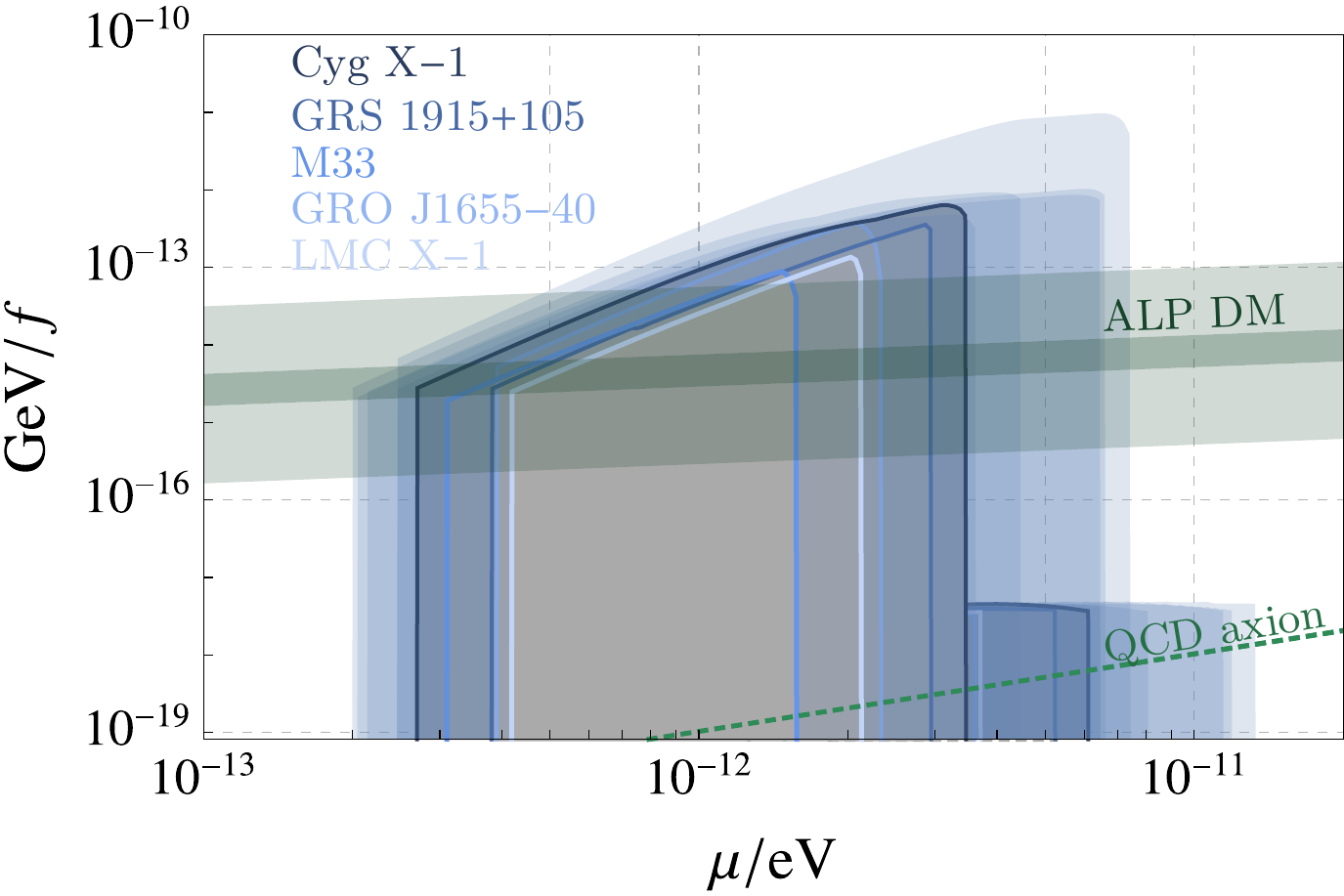}
		\caption{Constraints on axion parameter space from black hole
		spin measurements in X-ray binaries. 
		For each black hole, the region enclosed by the solid
		line of the corresponding color (see key at top left)
		is the intersection of the $m=1$ spin-down regions
		for different BH parameters (mass, spin, lifetime, binary period, and mass of the binary companion) within the observational
		error intervals. This corresponds to the parameter space region
		in which we can be confident that spin-down occurs,
		so is constrained by observations of that BH.
		The light shaded regions of each color
		are the unions of the spin-down regions
		for different BH parameters and could be constrained by improved measurement and analysis of these BHs. Higher axion masses could potentially be constrained using higher-$m$ levels;  we include only the analog of the small and moderate self-coupling regimes A and B (for which self-interactions do not affect the extraction of angular momentum to the level with the largest SR rate) for $m=2$, where the analysis in this work applies. 
		The ``ALP DM'' band
		corresponds to the range of quartic couplings
		that 
		allow the observed DM abundance to be produced by
		the misalignment mechanism.
		The darker middle band corresponds to $\OO(1)$ values
		of the initial misalignment angle ($\theta \in (1,\pi-1)$), while the lighter bands
		above and below correspond to ``tuned'' initial values
		($\theta \in (10^{-1}, \pi - 10^{-6})$).}
	\label{figSpindown1}
	\end{centering}
\end{figure}

Applying this physics to observations of astrophysical
BHs, Fig.~\ref{figSpindown1} shows the regions
in the $(\mu,f)$ plane for which sufficient
spin-down occurs, so that spin measurements from
BHs in X-ray binaries constrain an axion with that
mass and coupling. For each black hole, the solid line
of the corresponding color indicates the region
in which spin-down would occur with high confidence,
given the uncertainties on the measured BH parameters.
The larger shaded regions are those in which spin-down
may occur, given BH parameter values within the confidence
intervals; these represent the regions of parameter
space which may be constrained by future, better observations
of these BHs. Given the uncertainties in our analyses
when $\alpha \gtrsim 0.2$ and $f$ is small, the constraints
in those parts of parameter space should be
treated as estimates requiring further study.

Fig.~\ref{figSpindown1} can be
compared to Fig.\ 11 of~\cite{Arvanitaki:2014wva}.
The latter assumed that the dominant
effect of quartic self-interactions
was to cause periodic bosenova events when 
the cloud became too large;
parametrically, when 
	\begin{equation}
	N\gtrsim 16 \pi \frac{\ell^4}{\alpha}\frac{f^2}{\mu^2}
	\end{equation} 
for an $l,m = \ell$ superradiant level,
as discussed in \cite{Arvanitaki:2010sy}.
From the previous section, we know that,
at small $\alpha$ and small $f$, the critical occupation number
for a bosenova to occur has the same
parametric scaling as the equilibrium 211 occupation 
number, but is numerically larger,
$\varepsilon_{211}^{\rm eq} / \varepsilon_{211}^{\rm crit} \sim 0.1$
(Eq.~\eqref{eq:criteps} and Fig.~\ref{fig:bosenova}).
Consequently, we expect the time-averaged
211 occupation number in our picture to be parametrically
the same as that assumed in~\cite{Arvanitaki:2014wva}.
Numerically, since~\cite{Arvanitaki:2014wva} assumes
that a bosenova completely destroys the cloud,
which then takes $\OO(100)$ e-folds to be rebuilt,
our time-averaged 211 occupation number is actually slightly
larger, for the same parameters, resulting in slightly stronger
spin-down constraints.

The age (or accretion timescale) of the BH limits how small a particle mass $\mu$
can be constrained by spin-down measurements --- 
if $\mu$ is too small, then superradiance
is not fast enough to spin down the BH.
A separate effect is that, for small $\mu$,
the cloud is more dilute,
and can be disrupted by tidal forces
from the companion star~\cite{Baumann+2019}.
These gravitational perturbations
mix superradiant levels with 
decaying ones (e.g.\ 211 with $21-1$),
which can inhibit their growth.
We do not attempt a careful analysis of
the effects on the evolution of the cloud, but adopt the conservative
approach of not placing constraints when the companion
is closer than the maximum radius for the
resonant depletion processes identified
in~\cite{Baumann+2019} (see App.~\ref{app:compmixing}). This sets
the small-$\mu$ boundary of the constrained
region in Fig.~\ref{figSpindown1}.
We are able to constrain axion masses
a factor $\sim 2$ lighter than the limits
from~\cite{Arvanitaki:2014wva}, which included
an unphysical dipole gravitational potential
effect from the companion.

\begin{figure}[t!]
	\begin{centering}
		\includegraphics[width=\columnwidth]{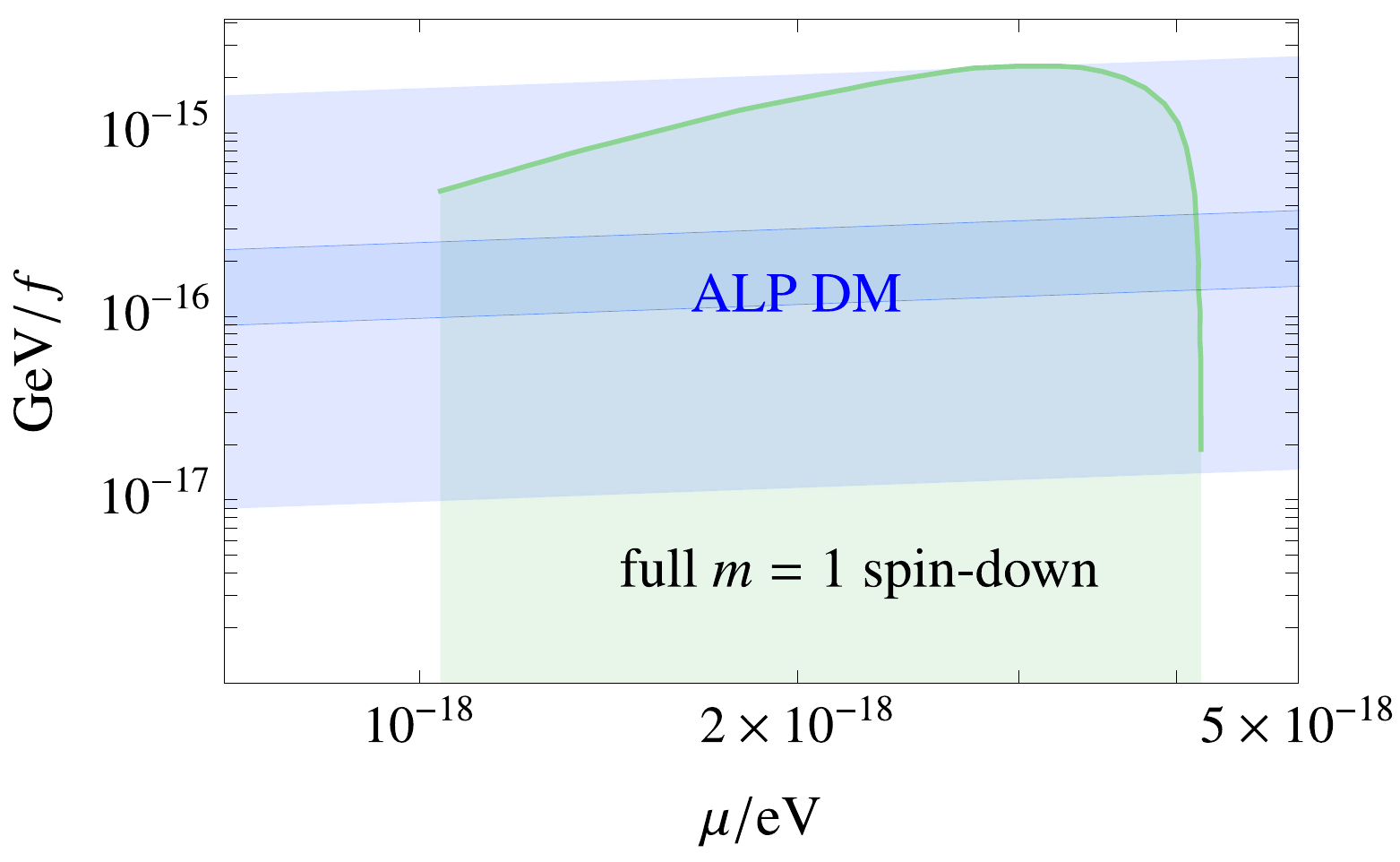}
		\caption{Parameter space for which the 211 level of 
		a supermassive BH ($M_{\rm BH} = 10^7 M_\odot$),
		with initial spin $a_* = 0.9$, spins the BH down
		to saturation within an Eddington accretion timescale,
		$t_{\rm Edd} \simeq 4 \times 10^8 \yr$.
		The ``ALP DM'' band is defined as in Fig.~\ref{figSpindown1}.}
	\label{figsmbh}
	\end{centering}
\end{figure}

In most of this paper, we have taken our nominal BH
mass to be $\OO(10 M_\odot)$. However, our analyses
can be easily rescaled to different BH masses;
the most important dimensionless parameter that changes
is the ratio of the BH lifetime to the light-crossing time.
Fig.~\ref{figsmbh} shows the spin-down parameter space for
a supermassive BH (SMBH), with $M = 10^7 M_\odot$.
This parameter space sits at smaller $\mu$ (due to the larger BH size)
and larger $f$ (due to the smaller $T_{\rm BH} \mu$ parameter) than for a stellar-mass BH. There do exist spin measurements
for some SMBHs~\cite{Reynolds_2013,Reynolds_2013_2,Middleton_2016}, and these could be used
to place constraints on very-low-mass bosons
(see e.g.~\cite{1706.06311,Stott_2018,2009.07206}).
However, the galactic center environments in which
SMBHs live are rather complicated, and understanding
environmental effects on the evolution of a superradiant cloud
(e.g.\ due to the occasional infall of compact objects)
would be necessary to place robust constraints.
We leave such an analysis to future work,
but include Fig.~\ref{figsmbh}
as a guide to the kind of region that might be constrained
by these measurements.

As well as spin measurements for BHs in X-ray
binaries, there are also spin measurements
for $\OO(10 M_\odot)$ BHs from gravitational
wave observations of binary BH mergers at LIGO
and Virgo~\cite{Abbott_2019,Abbott_2019_2,Zackay_2019,Venumadhav_2020,Abbott_2020gyp,Abbott_2020niy}. The statistical
uncertainty of these measurements is generally
much greater than the estimated errors of X-ray binary
spin measurements --- for most of the binary BH mergers
observed so far, the spins of the primary BHs
could lie in an $\OO(1)$ range, and are consistent with zero.
However, there were two events in recent observing runs
for which one of the primary BHs was measured to have high spin
(significantly different from zero); GW190412 and 
GW190517~\cite{Ng_2020ruv}.
The inferred masses of these BHs were $\sim 30 \msun$,
which is significantly heavier than the BHs observed
in X-ray binary systems. Consequently, if one assumes
that the history of the system would have allowed
a superradiant cloud to grow around the BH,
one can constrain smaller boson masses,
in the range $\mu \sim 1.3 \times 10^{-13} \eV$ -- $2.7
\times 10^{-13} \eV$~\cite{Ng_2020ruv}.

Given that we have no reliable information about the
pre-merger history of these BHs, we do not include
them in Fig.~\ref{figSpindown1}. However, with better
understanding of such systems, gravitational wave
observations of binary BH mergers could become
a valuable tool for constraining (or providing evidence
for) light bosons. In addition, while
mergers other than the two mentioned above
do not provide strong evidence regarding
superradiance~\cite{ng2019searching,Ng_2020ruv},\footnote{This
 is in contrast to some works
which claim that earlier GW spin measurements can
put constraints on BH superradiance (e.g.~\cite{Stott_2018}).
These claims appears to be based on a misinterpretation of the spin measurements
presented by the LIGO collaboration. For example, the pre-merger
spin of the primary BH in GW150914 is given as
$0.32^{+0.47}_{-0.29}$, where the errors correspond to a $90\%$ credible
interval~\cite{Abbott_2016}. \cite{Stott_2018} appears
to use the interpretation of spins below $0.32 - 0.29 = 0.03$ as being excluded
at the $90\%$ level, to place constraints on superradiant
processes that would have reduced the spin to below this value.
However, suppose (for example) that we had a uniform prior on $a_* \in [0,1]$,
and that the measurement gave us no information about $a_*$.
Then, $[0.05,0.95]$ would be a $90\%$ interval,
and spins $< 0.05$ would be excluded at the $90\%$ level,
despite obtaining no new information; to set constraints a more complete analysis is needed. }
future data from many such mergers
may provide statistical evidence
for or against superradiant BH spin-down~\cite{Arvanitaki:2016qwi,Brito_2017}.

\subsection{Axion models}
\label{secaxionmodels}

Understanding the parameter space in which spin-down
constraints apply is important in determining the consequences
for motivated particle physics models.
For the QCD axion, Fig.~\ref{figSpindown1} confirms
that, at least 
for 211 and 322 superradiance, self-interactions are small
enough not to affect spin-down constraints.

Another motivated target model is an axion with a
fixed (rather than temperature-dependent) potential.
An initial ``misalignment'' axion field value
in the early universe will lead to a dark matter density
at late times, depending on the axion mass, the shape
of the potential,
and the initial field value.
Consequently, while the mass and self-couplings of
a generic axion can vary independently,
imposing that the misalignment mechanism
must generate the observed DM density 
gives the ``ALP DM'' band in Fig.~\ref{figSpindown1}
(for a cosine potential $V \propto \cos(\varphi/f)$).

The darker central part of this band corresponds
to masses and self-couplings for which a ``generic'', 
$\OO(1)$ misalignment angle, $\theta_{\rm init} = a_{\rm initial}/f
\in (1,\pi-1)$, gives the correct dark matter density.
For the same $\mu$ and $\theta_{\rm init}$, but larger $f$, we would obtain
too large a dark matter density. However, this can
be fixed by ``tuning'' the initial field value
to be close to the bottom of the potential.
Since $\rho_{\rm DM} \propto \mu^{1/2} \theta_{\rm init}^2 f^2$
for small $\theta_{\rm init}$,
the tuning required is simply $\theta_{\rm init} \propto 1/f$.
The lower edge of the band in Fig.~\ref{figSpindown1}
corresponds to $\theta_{\rm init} = 0.1$.

At smaller $f$, we have the opposite problem
of not producing enough DM. For a cosine-type
potential,
this can be solved by tuning the initial field
value to be close to the top of the potential,
so that its transition to matter-like oscillations
around the bottom of the potential is delayed.
This ``large-misalignment mechanism''~\cite{Arvanitaki_2020}
can lead to significant enhancements of dark matter
density perturbations, resulting in a range of
phenomenological signatures. In Fig.~\ref{figSpindown1},
the top edge of the band corresponds to $\theta_{\rm init} = \pi - 10^{-6}$
(see App.~\ref{app:dmestimates} for formulae),
illustrating that, apart from the lower end of the $\mu$
range, BH spin-down constraints still apply to such models.

As well as affecting dark matter in the early universe,
self-interactions could have effects at late times,
leading to DM-DM scattering in halos.
The associated relaxation rate is, parametrically~\cite{Semikoz_1995,Sikivie_2009},
\begin{align}
	\Gamma &\sim \frac{\rho_{\rm DM}^2}{f^4 \mu^3 v^2} \nonumber
	\\
	&\sim 3 \times 10^{-26} {\rm \, yr^{-1}} \left(\frac{\rho_{\rm DM}}{\GeV \cm^{-3}}\right)^2
	\left(\frac{10^{11} \GeV}{f}\right)^4 \nonumber \\ 
	& \,\,\,\,\,\, \times \left(\frac{10^{-12} \eV}{\mu}\right)^3
	\left(\frac{10^{-3}}{v}\right)^2
\end{align}
where $v$ is the halo's virial velocity
(this should be compared to the relaxation
rate $\Gamma \sim \frac{\rho_{\rm DM}^2}{M_{\rm pl}^4 \mu^3 v^6}$
for gravitational interactions~\cite{Hui_2017,Levkov_2018,Bar_Or_2019}).
Consequently, unless DM forms very dense structures,
quartic self-interactions will not be significant in halos,
for the parameter space we have been considering.

\section{Gravitational waves}
\label{sec:gw}

Gravitational waves emitted by the
 superradiant cloud are
a unique signal of ultralight bosons,
turning gravitational wave observatories into 
indirect particle detectors~\cite{Arvanitaki:2009fg,Arvanitaki:2010sy}. The superradiant cloud can grow to up
to several percent of the black hole's mass, 
and sources gravitational waves through its oscillating
stress-energy tensor. These are almost-monochromatic,
coherent, and long-lasting.
Such emission occurs in two parametrically-different
frequency ranges; higher-frequency ``annihilation'' signals,
with $\omega \simeq 2\mu$, and lower-frequency
``transitions'', with $\omega = \omega_j - \omega_{j'}$
set
by the frequency difference between different bound levels.

Conceptually, annihilation signals are sourced by 
the annihilation of two axions into a graviton.
Consequently, they are emitted by any level populated
by a single real scalar field. The timescale over which such emission
lasts is parametrically longer than the superradiant
growth time (Sec.~\ref{seclsmall}), making them promising for detection at
gravitational wave observatories. Up to thousands of potential
annihilation signals could be detectable, from black holes in the
Milky Way, at Advanced LIGO and Virgo \cite{Arvanitaki:2014wva,
Arvanitaki:2016qwi,Brito+2017,Brito+2017a,Zhu:2020tht}.  
Such signals, and their detectability, have been studied
in the context of continuous wave searches~\cite{Arvanitaki:2014wva,
Arvanitaki:2016qwi}, stochastic searches~\cite{Brito+2017,Brito+2017a,Tsukada:2018mbp},  directed searches for
clouds around products of binary mergers
\cite{Arvanitaki:2016qwi,Isi:2018pzk}, and directed searches for
clouds around BHs in  X-ray binaries \cite{Yoshino:2014wwa,
Sun:2019mqb}. Searches with LIGO/Virgo data are ongoing; so far,
no signals have been observed~\cite{Dergachev:2019oyu,Palomba:2019vxe,Zhu:2020tht}, though using this non-observation to constrain superradiance relies on poorly
measured black hole population properties, and may suffer from
down-weighting of the signal \cite{Zhu:2020tht}.
Searches at space-based, lower-frequency gravitational wave detectors such as
LISA will be sensitive to lighter axions~\cite{Arvanitaki:2014wva,Brito+2017,Brito+2017a}, while heavier
axions may be observable with future higher-frequency detectors
\cite{Arvanitaki:2012cn,Aggarwal:2020umq}.

\begin{figure*}[ht!]
	\begin{centering}
		\includegraphics[width=.48\textwidth]{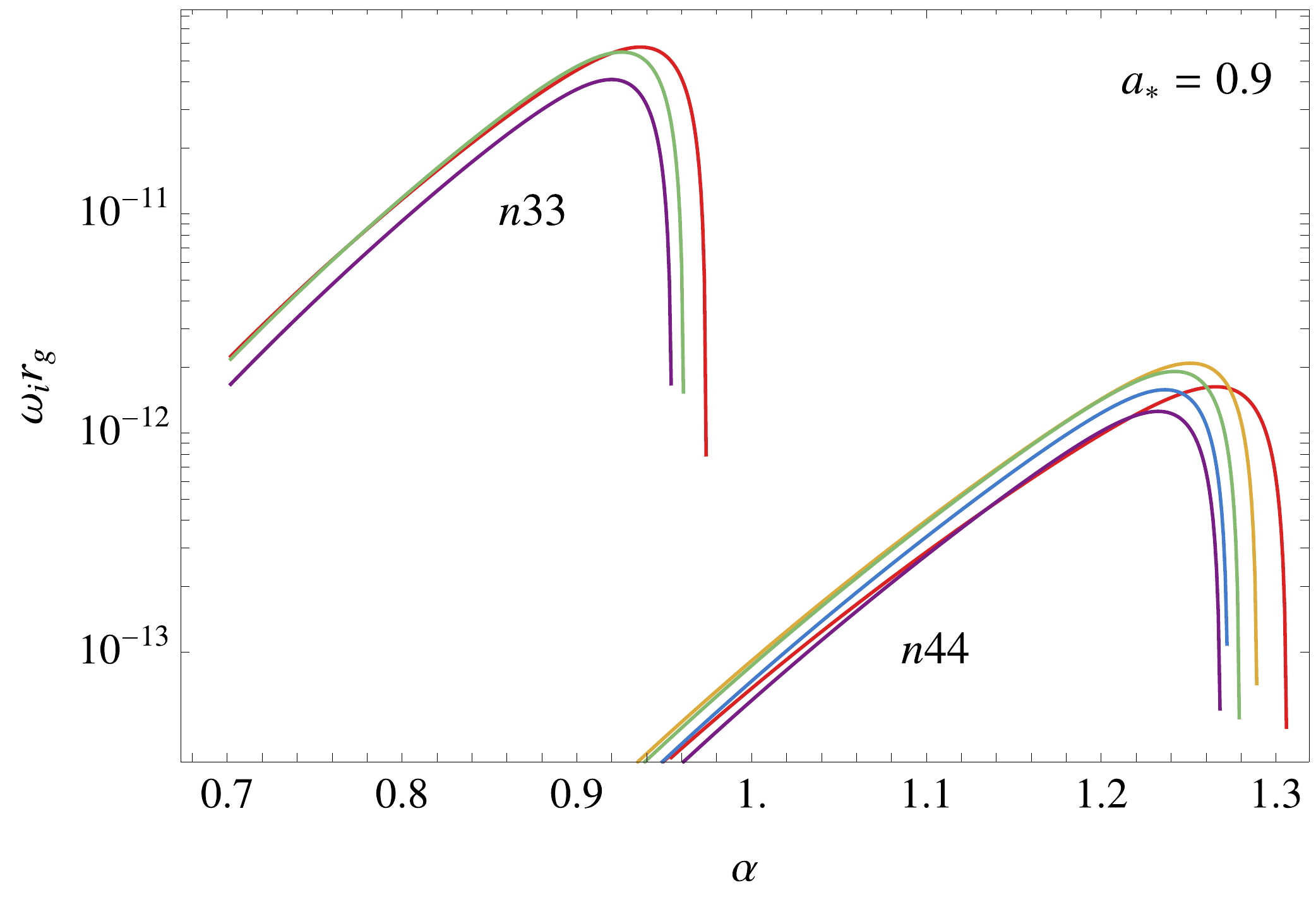}
		\includegraphics[width=.48\textwidth]{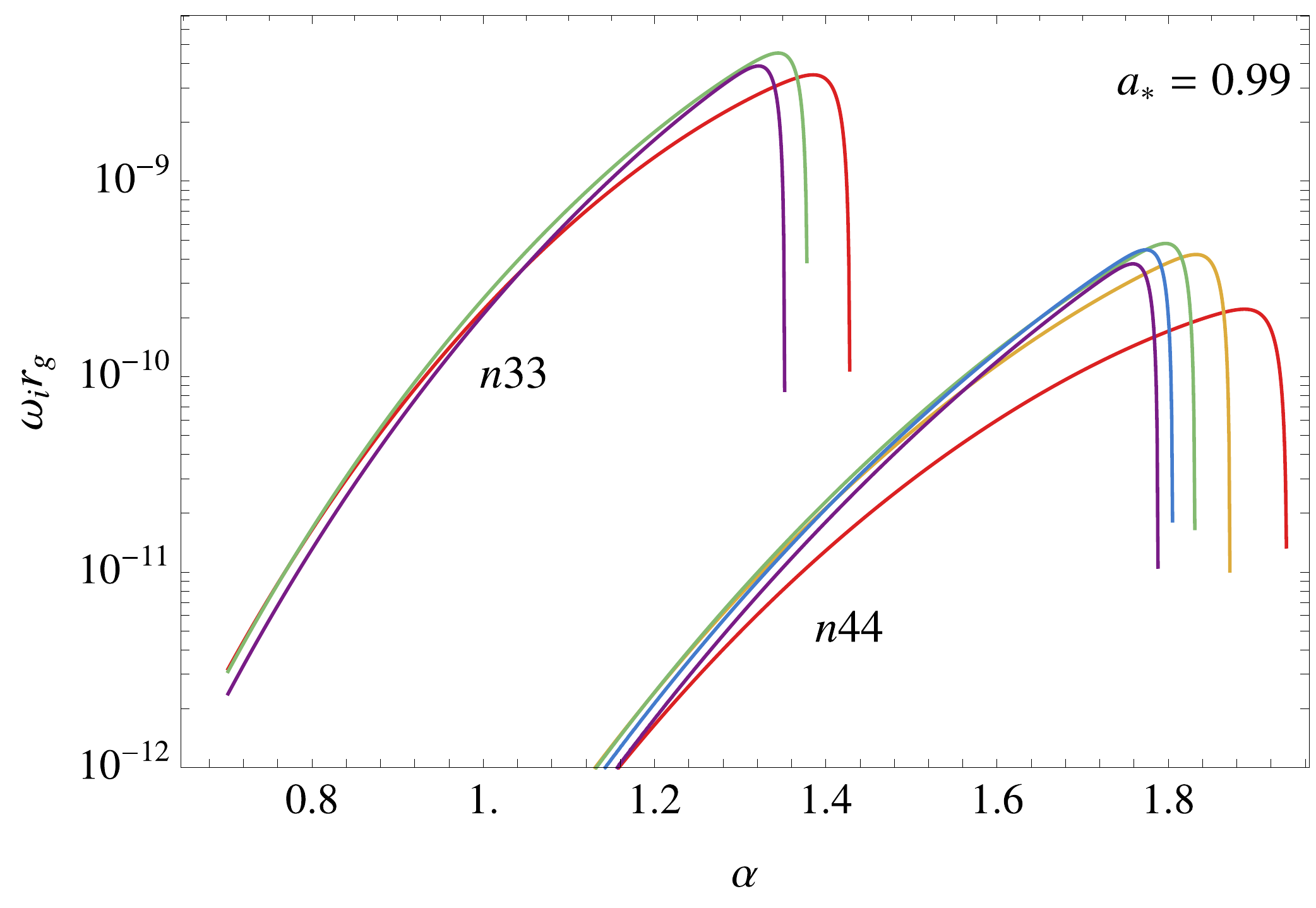}
		\caption{Superradiance rates for the $n33$ and $n44$
		hydrogenic bound states, computed numerically on the full
		Kerr background (using the continued fraction
		method of~\cite{Dolan2007}). The left-hand plot shows rates
		for $a_* = 0.9$, and right-hand plot those for $a_* = 0.99$. 
		The red curves correspond to the levels with smallest
		$n$; levels with larger $n$ have cutoffs at progressively
		smaller $\alpha$.
		These plots illustrate how, at some $\alpha$ parameters,
		different hydrogenic levels can have the same
		superradiance rates.
		As discussed in Sec.~\ref{sec:gw}, this can give rise to gravitational wave
		transition signals.
		}
	\label{figsrm}
	\end{centering}
\end{figure*}

Transition signals correspond to axions dropping into 
a more deeply bound level, emitting gravitational radiation
at the frequency set by the level splitting.
Attaining a significant emission rate
requires both levels to have large occupation numbers
simultaneously. For the case of purely-gravitational
superradiance, these circumstances only arise
for higher-$l$ levels and for short times,
leading to limited observational 
prospects at current gravitational wave
observatories~\cite{Arvanitaki:2014wva}.

More specifically, for a given $m < 3$, the fastest-growing
superradiant level is also the most tightly bound one, so other modes
with the same $m$ have exponentially smaller
occupation numbers. For $m \ge 3$, this is not always the case --- for example,
at large $a_*$ and near-threshold $\alpha$, the growth rate of 433 becomes smaller
than that of 533 and higher levels. This can lead to multiple $m=3$ levels having large
occupation numbers simultaneously. Similar
crossings happen for $m=4$ and higher
levels, as illustrated in Fig.~\ref{figsrm}.

These circumstances allow gravitational wave transition
signals of non-negligible amplitude to occur
around astrophysical BHs. Even so, compared to annihilation
signals, they offer less promising observational prospects.
The total energy released, if the occupation number
of the higher level transitions entirely to the lower
one, is $E = \Delta \omega N \lesssim \alpha^2 \mu
N$, whereas annihilations can emit the entire
energy stored in a cloud, $E \sim \mu N$.
In addition, signal durations for transitions are typically
of order a superradiance time, compared to the parametrically
longer annihilation signals~\cite{Arvanitaki:2014wva}.
Nevertheless, transition signals could probe interesting
parts of parameter space, providing sensitivity to heavier
axions than annihilation signals do (for a given BH mass).

Compared to the purely-gravitational behavior
summarized in the preceding paragraphs, the presence
of self-interactions can have a significant effect
on the gravitational wave signatures of superradiance.
For annihilations, self-interactions suppress
the potential signals due to two main effects:
the gravitational wave power emitted is reduced due to
the smaller cloud size, and the new energy loss mechanisms via
scalar radiation reduce the total energy emitted in GWs.
On the other hand, self-interactions provide a mechanism to populate multiple levels simultaneously,
potentially increasing the parameter space
for transition signals (though the cloud size and scalar radiation
caveats still apply). In the rest of this section,
we discuss annihilation and transition signals and their observational
prospects in more detail. We focus on continuous wave searches for such signals, which are well-suited to louder signals from within our galaxy, and can provide a wealth of information about the detected signal properties. Stochastic searches to look for excess power in a narrow frequency range could potentially be performed more (computationally) cheaply and would also be interesting to study in future work.

\subsection{Annihilations}
\label{sec:annihilations}

In this subsection, we focus on the prospects for observing
annihilation signals from the 211 level, for a range of
self-couplings, at current gravitational wave observatories.  We
also comment briefly on other types of annihilation signals,
including annihilation signals from complex scalar fields.

\begin{figure*}[t!]
	\begin{centering}
		\includegraphics[width=.49\textwidth]{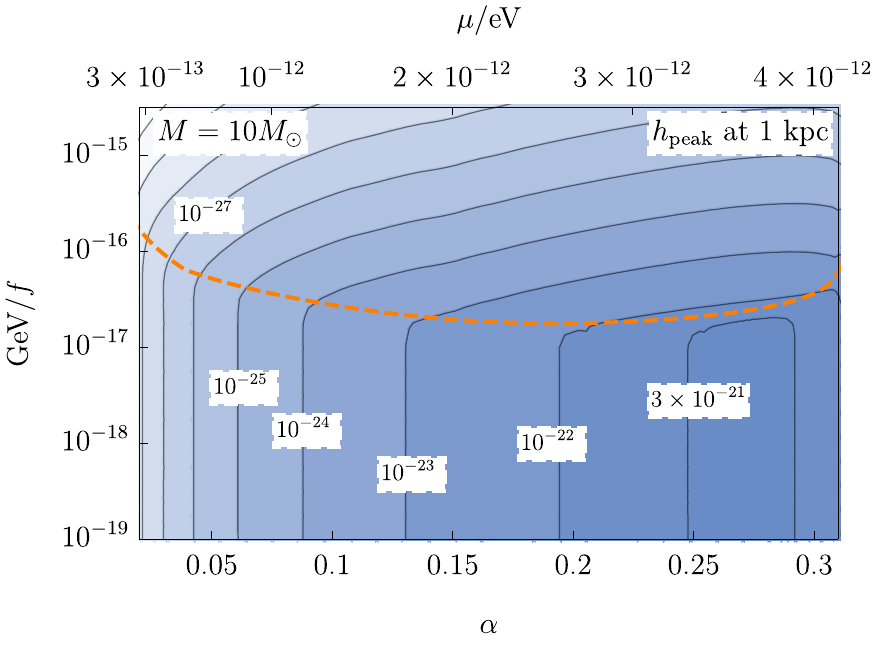}
		\includegraphics[width=.49\textwidth]{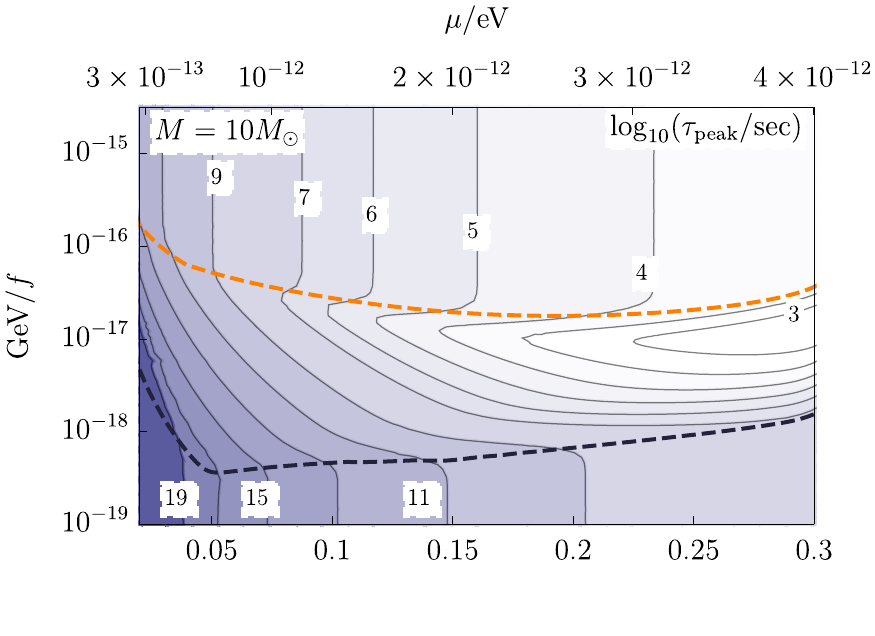}
		\vspace{0.3cm}
		\includegraphics[width=.49\textwidth]{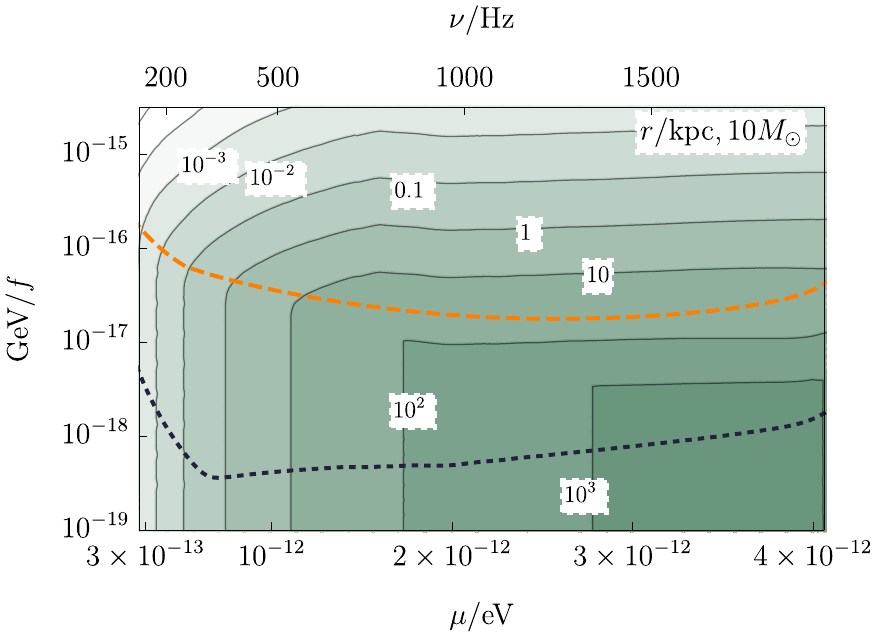}
		\includegraphics[width=.49\textwidth]{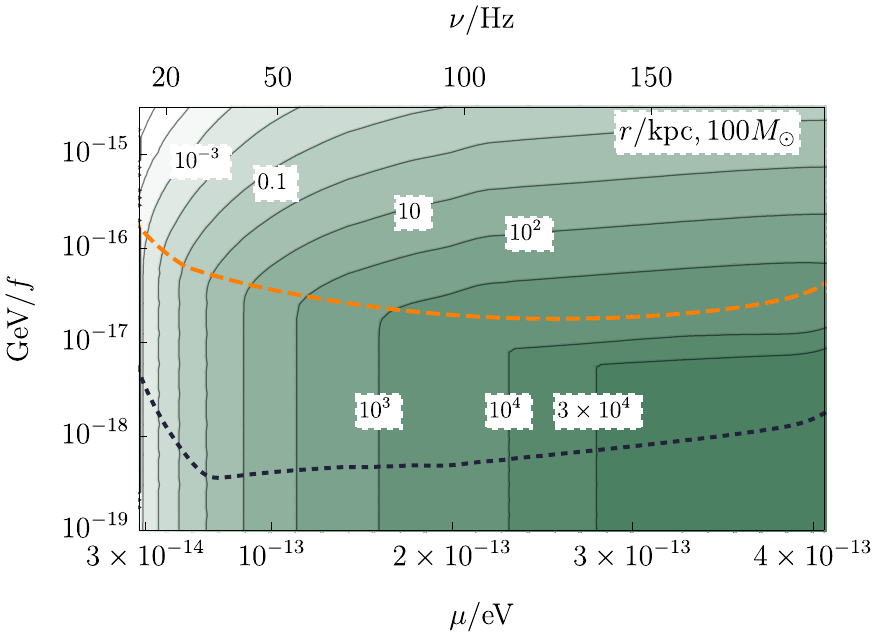}
		\caption{\emph{Upper left:} Peak strain from $211 \times 211 \rightarrow$ GW annihilations for an observer at 1 kpc from a $10 M_{\odot}$ BH, with initial spin 0.9. \emph{Upper~right:}  Typical duration $\tau_{\rm peak}$ of peak signal, $\log_{10}(\tau_{\rm peak}/{\rm sec})$. In the large self-interactions regime, we show the time-scale of the overshoot regime, corresponding to the peak signal strain.  \emph{Lower~left:}  Sensitivity reach in kpc to a 10 solar mass BH, for continuous wave searches at Advanced LIGO design sensitivity~\cite{1411.4547}. \emph{Lower~right:}  Reach in kpc to a 100 solar mass BH. The dashed orange line indicates the boundary between the moderate and large self-coupling regimes (corresponding to $f_{\rm BC}$, Sec.~\ref{secperturb}), while the dotted black line indicates the boundary of the regime in which the 322 level grows appreciably ($f_{\rm AB}$).}	
	\label{fig:annihilations}
	\end{centering}
\end{figure*}

Figure~\ref{fig:annihilations} illustrates
the effects of self-interactions on gravitational
signatures of 211 superradiance, showing the
peak signal amplitude, signal duration
and sensitivity reach for different axion masses and
self-couplings. To estimate the projected reach,
we take the design strain sensitivity
of Advanced LIGO~\cite{1411.4547}, and assume all-sky semi-coherent continuous wave (CW) search strategies,
with coherent integration times of 240 hours,
and sensitivity depth $\mathcal{D}^c(f)\sim50/\sqrt{\mathrm{Hz}}$.
The sensitivity depth is defined by $\mathcal{D}^c(f) \equiv \sqrt{S_h(f)}/h_0^c(f)$,
where $\sqrt{S_h(f)}$ is the noise spectral density and $h_0^c(f)$
is the strain limit at the desired confidence level $c$.
It allows comparisons of different searches,
independently of the data used,
and depends on the detailed search technique,
coherent integration time, total integration time,
etc. \cite{Behnke:2014tma}.
The latest searches with O2 data have used
coherence times of up to $T_{\mathrm{coh}} = 60 \, \mathrm{hrs}$ with 
$N_{\rm seg} = 64$
segments in the first analysis stage \cite{Steltner:2020hfd}, and
have reached sensitivity depths of $\sim 30/\sqrt{\mathrm{Hz}}$
\cite{Pisarski:2019vxw} to $\sim 50/\sqrt{\mathrm{Hz}}$ \cite{Steltner:2020hfd}
for $c=90\%$ exclusion limits.
Since the CW searches assumes a constant signal
amplitude over the entire integration time,
while our signals may change on times
shorter than the coherent search time,
we conservatively penalize our reach by
$\sqrt{\tau_{\mathrm{sig}}/T_{\mathrm{coh}}}$
(though the searches could be improved to take
into account the time dependence of the signal,
alleviating this penalty).

\begin{figure}[t!]
	\begin{centering}
		\includegraphics[width=\columnwidth]{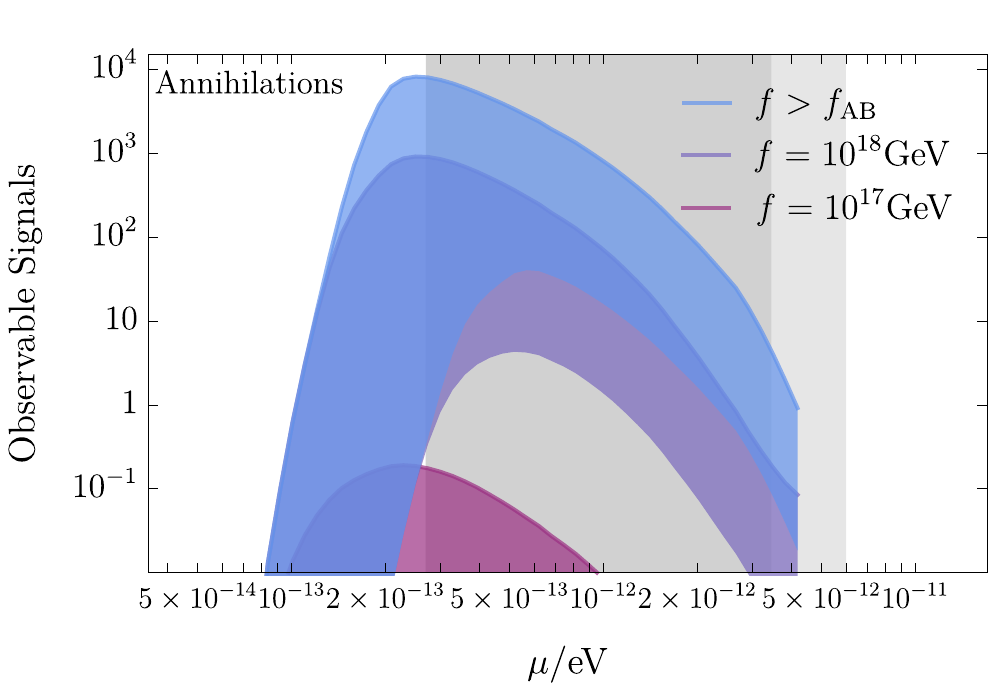}
		\caption{Projections for the number of observable $211 \times 211 \rightarrow$ GW annihilation signals, using continuous wave searches at Advanced LIGO (with design sensitivity), for a range of self-interaction strengths (see text for details). The width of the bands results from varying the BH spin distribution and maximum BH mass as described in the text. The highest number of observable signals is in the small self-interactions regime, which includes gravitational superradiance and QCD axion parameter space.  Increasing self-interactions reduces the number of signals expected. At high masses, the signal frequency falls above the band of typical CW searches ($\nu\gtrsim 2$~kHz). The darker (lighter) shaded regions are disfavored by black hole spin down for initially superradiating levels with $m=1$ ($m=2$) (see Sec.~\ref{sec:spindown}).}
	\label{fig:GWann}
	\end{centering}
\end{figure}

While the sensitivity reach is a useful quantity
for a search targeting a specific BH,
standard CW searches are `blind', and look for
signals from sources anywhere in the sky.
Figure~\ref{fig:GWann} shows the expected number
of events in such a search at Advanced LIGO, given assumptions
about the galactic BH population, for different 
self-couplings.\footnote{It should be noted
that for small axion masses, where there may be multiple
long-duration signals from galactic BHs, stochastic searches
for excess power within a frequency range may be
an advantageous approach. We leave a quantitative comparison
of stochastic and CW searches to future work.} We assume a power-law BH mass distribution,
$dN/dM\propto M^{-2.35}$, with a minimum black hole mass of $5\msun$, and vary the maximum black hole mass from $20$ to $45\msun$\cite{1811.12940}. For the BH spatial distribution,
we take a combination of the disk and bulge distributions
as in~\cite{Zhu:2020tht}, with a total
number of $10^8$ BHs, born at a uniform rate
throughout the age of the galaxy.
We vary the BH spin distribution,
with our extreme cases having $10\%$ and $0.2\%$
of BHs with initial spin $a_*(t_0) 
\ge 0.9$, respectively. The $10\%$ figure is consistent with
spin measurements from X-ray binaries \cite{1302.3260,1408.4145},
and $0.2\%$ with models of rare high spin BHs associated with gamma ray bursts  \cite{Yoon:2005tv,Woosley:2005gy}, making them reasonable upper
and lower bounds.

The shaded bands in Fig~\ref{fig:GWann} correspond to
this range of BH population assumptions. While these
unknowns do give rise to orders of magnitude uncertainty
in the expected event rate, we can see that,
for particle masses just below
the spin-down threshold, even the pessimistic distributions
give a promising number of events for purely-gravitational
superradiance. 
Conversely, the very large number of events
(at design sensitivity) predicted
by the optimistic distributions
means that some of this parameter space is already
ruled out by existing observations; axions with gravitational interactions and mass between $3-7\times 10^{-13}$~eV would yield more than 10 signals  in current LIGO data for all the BH mass and spin distributions considered here; masses between $2\times 10^{-13}-2\times 10^{-12}$~eV would yield 10 or more signals for the most optimistic spin distribution considered here \cite{Zhu:2020tht}. An analysis of existing data taking into account the reduced event rates at larger self-interactions has not been performed and would be very valuable.

Once we incorporate self-interactions,
there are three
different parameter space regimes, with distinct behavior
(as per Sec.~\ref{secperturb}).
In the small self-coupling
regime, $f > f_{\rm AB}$, the 322 level does not grow
through self-interactions, and the dynamics proceeds as in the purely gravitational case.
Consequently, the annihilation signal properties are independent
of the self-coupling, and existing analyses of
gravitational wave signals will apply without modification.  This
regime, which (for stellar mass BHs)
includes $f \sim \Mpl$ as well as QCD axion self-couplings,
can lead to as many as thousands of signals 
at LIGO/Virgo, as shown in Fig.~\ref{fig:GWann}.

In the moderate self-coupling regime, $f_{\rm AB} > f > f_{\rm BC}$, the growth of the 211
level is unaffected, but 322 grows earlier than it would otherwise have done.
The main effect on the annihilation signal is through the addition of another energy loss
process for the cloud, via $322 \times 322 \rightarrow
211 \times \infty$ emission. Consequently, while the peak emission amplitude
is unaffected, the signal duration is reduced. This corresponds to the parameter space
region between the orange and black dashed lines in the upper-right panel of Fig.~\ref{fig:annihilations}.
More specifically, when 211 is primarily depleted through gravitational waves,
the signal strain as a function of time is given by,
\begin{align}
h^\text{GW,ann}(t) = \frac{h_{\rm peak}}{1+t/\tau_{\mathrm{ann}}}
\end{align}
with $\tau_{\mathrm{ann}}$ defined in Eq.~\eqref{eq:t_ann}. However, due to the self-interaction processes, there is additional energy lost from the cloud, changing the time-evolution to that in Eq.~\eqref{eq:scalarevol}, with 
\begin{align}
h^\text{GW,ann}(t) \propto\sqrt{\tau_{\mathrm{scalar}}/t}
\end{align}
at late times, where $\tau_{\mathrm{scalar}} \propto
(f/\Mpl)^4$, Eq.~\eqref{eq:t_scalar}. For $f$
in the moderate self-coupling regime,
$\tau_{\rm scalar}$ can be significantly less
than $\tau_{\rm ann}$.
Given the typical assumptions on black hole formation rates and
distributions, the shortest signals that are likely to be observable
in an all-sky continuous wave search have signal times on the order
of $10^{4}$~years or more \cite{Zhu:2020tht}.

Since, for moderate self-couplings, the peak
signal strain is not affected, the sensitivity reach
of gravitational wave detectors for
signals observed around the optimum time is only moderately
affected, as illustrated in the bottom panels
of Fig.~\ref{fig:annihilations}.
One effect is that,
especially for lighter black holes, the signal
duration can become comparable to the
typical coherent integration times used in
continuous wave searches~(e.g. \cite{Steltner:2020hfd}),
which degrades the signal to noise.

For blind searches, the faster decrease
of signal strain with time leads to less chance 
of seeing a signal, as illustrated in Figure~\ref{fig:GWann}.
The expected number of observable signals
at $f \sim 10^{18} \GeV$, which is in the moderate 
self-interactions regime for $\mu \sim 10^{-12} \eV$,
is around an order of magnitude lower than in the purely 
gravitational case. 
For larger and smaller $\mu$, this value
of $f$ falls back into the weak self-interactions regime,
so the difference is reduced.
At $f \sim 10^{17} \GeV$, which is in the moderate
self-interactions regime for the whole $\mu$ range,
the signal durations are much shorter, and the
expected number of observable signals is less than $1$. As a result,
such signals are unlikely to observed with current detectors,
in a blind search. In addition,
the faster time-evolution can lead to larger frequency
drifts, which could degrade search sensitivity further
(see Sec.~\ref{sec:fdot}).

For strong self-couplings, $f > f_{\rm BC}$, the peak
signal amplitude drops with increasing coupling
as $(f/f_{\rm BC})^2$ (Fig.~\ref{fig:annihilations}).
In particular, this drop-off starts at
larger $f$ than for the suppression of BH spin-down,
since $f_{\rm BC} > f_{\rm CD}$. Consequently,
with current detectors, self-interactions strong enough
to avoid BH spin-down constraints (Sec.~\ref{sec:spindown})
also render GW annihilation signals undetectable,
for any plausible BH spin and mass distributions. 
For $f \lesssim f_{\rm BC}$, i.e. $f \lesssim 10^{16} \GeV$
for stellar-mass BHs, the expected number of
events in a blind search is $\lesssim 10^{-3}$,
while for $f \lesssim 10^{15} \GeV$, where
signal durations become comparable to those in the small
self-interaction regime, signals beyond $10-100 {\rm \, pc}$ are
unlikely to be visible at Advanced LIGO sensitivities.

Nevertheless, it is possible that advanced future
detectors, such as the Cosmic Explorer~\cite{1607.08697,1907.04833}
or Einstein Telescope~\cite{Punturo:2010zz,1012.0908,1206.0331,1912.02622},
may be able to probe this parameter space. The
signal strain in the quasi-equilibrium regime is a factor $\mathcal{O}(1-5)$
below the overshoot peak shown in the left panel, but 
 the quasi-equilibrium regime lasts parametrically longer than in the moderate self-interaction regime,
$\tau_{\mathrm{sig}}\propto(f_{\rm BC}/f)^2$ (see Fig.~\ref{fig:ScalarSignalTime}). If smaller strains come within reach of future detectors, the long-lasting signals would have an increased chance of being observed in the quasi-equilibrium regime.

\emph{Additional annihilation channels.}
In addition to $211 \times 211 \rightarrow$ GW annihilations,
as occur in the purely-gravitational case, the presence of the 322 level allows
$211 \times 322 \rightarrow$ GW and $322 \times 322 \rightarrow$ GW
processes. These GWs will still have frequency $\omega \simeq  2\mu$,
but due to the larger angular momentum of the 322 level,
their rates are suppressed by higher powers of $\alpha$, $P_{\rm GW}^{l,l'} \propto \alpha^{16 + 2 (l + l')}$, where $l$ and $l'$ are the angular momentum numbers of the two levels.
These powers are significantly smaller than the primary $211 \times 211 \rightarrow$ GW annihilation
channel, and are further suppressed by the smaller occupation number of $322$ at small $\alpha$ (App.~\ref{sec:equilibrium_ratio}). For example, the $322 \times 322 \rightarrow$ GW process would lead to signals strains $\mathcal{O}(10^{-4})$ weaker than the primary signal at $\alpha\sim 0.3$. ``Cross-annihilation'' signals between two levels, $211 \times 322 \rightarrow$ GW, may be observable for the closest black holes; further study would require numerical GW power calculations which have not yet been performed for cross-annihilation signals.

\begin{figure*}[t!]
	\begin{centering}
		\includegraphics[width=.49\textwidth]{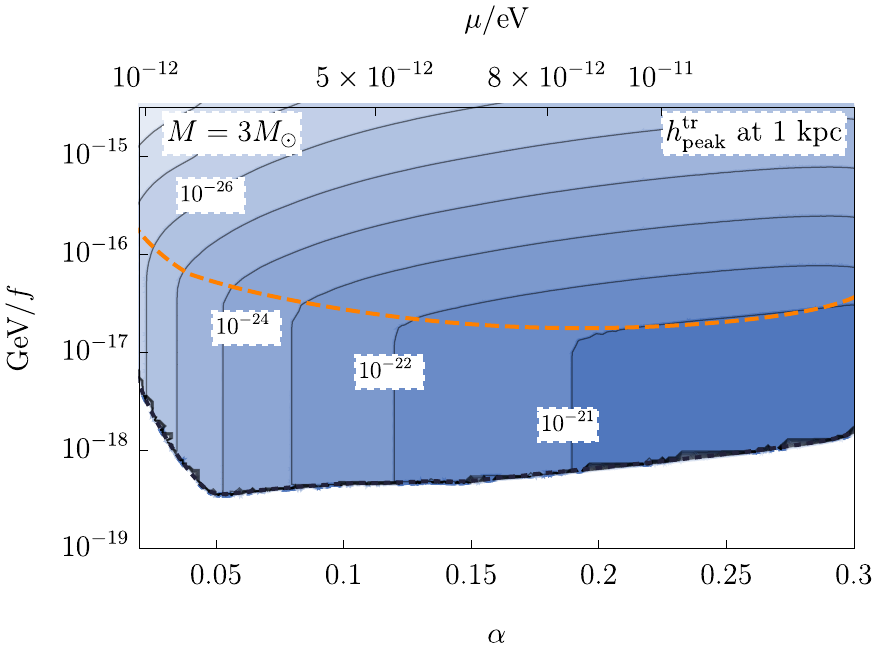}
		\includegraphics[width=.49\textwidth]{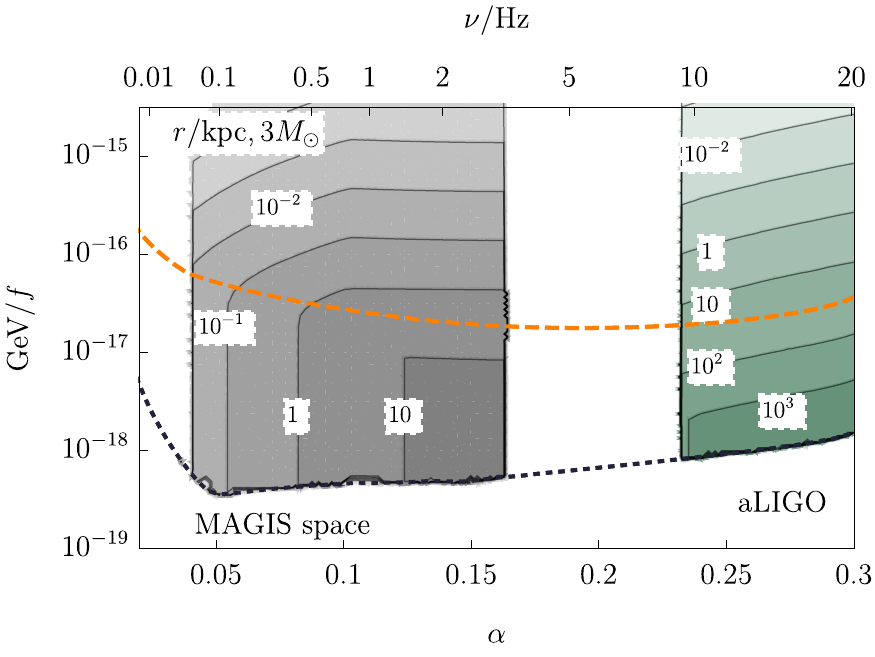}
				\caption{\emph{Left panel:} peak strain of
				the $322 \rightarrow 211 + $ GW transition
				signal at $1 \kpc$ from a BH of mass
				$3 \msun$, as a function of the mass $\mu$ 
				and self-coupling scale $f$ of the scalar particle.
				\emph{Right panel:} sensitivity reach
				for the detection of such signals,
				using the Advanced LIGO detector,
				or with the MAGIS proposal for a future space-based
				atom interferometer~\cite{graham2017midband}.
				The dashed orange and dotted black lines are the $f_{\rm BC}$ and $f_{\rm AB}$ curves, respectively, as in Fig.~\ref{fig:annihilations}.}
		\label{fig:transitions}
	\end{centering}
\end{figure*}

\emph{Annihilation signals from complex fields.}
In this section, and throughout the rest of this paper,
we have considered superradiance of a single, real spin-0 field.
As has been pointed out in a number of papers~\cite{Hod_2012,Herdeiro_2014,Ganchev_2018,Degollado_2018}, for the case of 
two scalar fields of degenerate masses (equivalently, a single
complex scalar field), there are cloud configurations with
a time-independent stress-energy tensor, which consequently do
not emit any gravitational radiation. In complex field terms,
these correspond to all-particle or all-antiparticle 
field configurations, whereas gravitational waves
arise from particle-antiparticle annihilation. This has sometimes been interpreted~\cite{procastar} as
indicating that annihilation radiation,
of the type considered in this section, is not expected
from superradiance of complex fields. 

However, as per the discussion
in Sec.~\ref{secspin0}, the initial conditions for
the growth of superradiant modes are either
vacuum fluctuations, or whatever pre-existing astrophysical
fields are present. In the former case, we can view the
growth of the particle and antiparticle
field modes as effectively separate, and generically,
they will obtain $\OO(1)$-similar occupation numbers.
For pre-existing astrophysical fields, a generic
expectation in many circumstances is 
for $\OO(1)$-similar initial conditions
for particle and antiparticle fields. Consequently,
unless some mechanism drives us to an all-particle
or all-antiparticle state, we expect that the
particle and antiparticle fields generically attain roughly comparable
occupation numbers. Compared to a real scalar field,
this results in a total GW annihilation signal
energy that is only $\OO(1)$ smaller.

\subsection{Transitions}
\label{sec:trans}

For large enough self-interactions (regions B,C,D in
Fig.~\ref{fig:paramspace}), the 322 level grows
earlier than it would have done otherwise,
and both 211 and 322 can have significant occupation numbers
at the same time. This gives rise to GW transition signals.

The transition quadrupole moment
for the $322 \rightarrow 211 + {\rm GW}$ process
vanishes at leading order, so its rate
is suppressed by a larger power of $\alpha$
than other gravitational transition processes
(such as the $644 \rightarrow 544$ process considered
in~\cite{Arvanitaki:2010sy,Arvanitaki:2014wva,Arvanitaki:2016qwi}).
At leading order in $\alpha$, the emitted power,
as a function of polar angle $\theta$, is
\begin{align}
	&\frac{dP}{d\Omega} =
	\frac{G N_{322} N_{211}}{\pi r_g^4} \alpha^{14} \times \\
	&\left(\frac{2^5}{3^6 5^8} (1 - \cos^4\theta)
	+ \frac{(27 + 28 \cos(2 \theta) + 9 \cos(4\theta)) \sin^2 \theta}{2^2 3^6 5^{10} 7^2}
	\right)\nonumber
\end{align}
where the first term corresponds to $l,m=2,1$ emission,
and the second to $l,m=3,1$. This gives a total
emitted power of~\cite{Arvanitaki:2010sy}
\begin{equation}
	P = \frac{2^8 \times 5717}{3^5 5^{11} 7^3} \frac{G N_{322}
	N_{211}}{r_g^4} \alpha^{14}.
\end{equation}
The emitted radiation is at a frequency
$\omega = \omega_{322} - \omega_{211} \simeq \frac{5}{72} \alpha^2 \mu$.
In terms of the normalized occupation numbers, it contributes
a term
\begin{equation}
	\dot \varepsilon_{322} \simeq - 5 \times 10^{-6} \alpha^{10}
	\varepsilon_{211} \varepsilon_{322} + \dots
\end{equation}
to the equations of motion.

Compared to the processes discussed in Sec.~\ref{secperturb},
which drive the evolution of the superradiant cloud,
the effects of GW transitions are always subdominant.
While this does reduce the peak signal amplitude,
it also means that signal timescales can be longer compared to the transitions in the purely gravitational regime, which is helpful for detection. 

Fig.~\ref{fig:transitions} shows projections for the peak signal strain, and
sensitivity
reach, for transition signals from a fairly light BH, $M_{\rm BH} = 3
\msun$. The signal durations (for a given BH mass) are the same as those
for annihilations (Fig.~\ref{fig:annihilations}) in the region where
322 grows, $f<f_{\mathrm{AB}}$, as the two levels evolve together
over time.  Given the lower frequency compared to annihilations, the
signal strains are typically larger (Fig.~\ref{fig:transitions}
left). However, transition signals only occur in the moderate and
large self-interaction regimes, where much of the energy loss
is through scalar radiation. 
Furthermore, for given BH mass,
the frequency decreases $\propto \mu^3$ with decreasing
$\mu$, rapidly falling out of the sensitivity band
of current detectors such as Advanced LIGO. For heavier BHs,
the frequency of transition signals would always be too low
for ground-based GW detectors, due to overwhelming
seismic and gravity-gradient noise.

For a narrow range of axion masses above $10^{-11}$~eV, current detectors could
potentially probe signals in the moderate self-interaction regime
(Fig.~\ref{fig:transitions}, right). Although the reach is poor at small $f$,
there is a roughly order-of-magnitude range in $f$ for which sensitivity to
signals from the galactic centre would be possible. The signal times in this
region last on the order of minutes to hours, and the expected number of signals in a blind search is heavily
dependent on the poorly-measured black hole distribution in the `mass gap' below $5
M_{\odot}$ \cite{Bailyn:1997xt,Ozel:2010su,kreidberg2012mass,belczynski2012missing} (although evidence for  compact objects in this mass range is emerging \cite{Abbott:2020khf,Thompson637,Margalit:2017dij}). Consequently, blind searches with current detectors are unlikely to
lead to observable signals.

However, future space-based detectors such as LISA \cite{1702.00786, 1907.06482} and atom
interferometer missions~\cite{graham2017midband}, could have
promising sensitivity to such signals. For illustration we show the
reach of the MAGIS proposal~\cite{graham2017midband} in the right panel of
Fig.~\ref{fig:transitions}, which can achieve a reach of 10 kpc for
axions around $3 M_{\odot}$ black holes, and up to $10^3$ kpc for
$100 M_{\odot}$ black holes. Some of the more promising signals fall
in the $0.1-10$ Hz range, where future  proposals such as
DECIGO~\cite{2006.13545} could improve transition detection
prospects.

\subsection{Frequency drifts}
\label{sec:fdot}

While the frequency of gravitational wave
annihilation signals is almost constant
at $\nu_\text{ann} \equiv 2\omega/(2\pi) \simeq 2 \mu / (2 \pi)$ (we will use
frequency rather than angular frequency
in this section, to match the GW literature),
the potentially long signal durations
mean that even very small frequency drifts
can be measured. Moreover,
the search algorithms employed in continuous
wave detection analyses can be strongly
affected by these small frequency drifts,
so it is important to quantify them to 
determine the appropriate search strategy and sensitivity~\cite{Wette2012}.

The self-energy of the cloud, from both
gravity and self-interactions, affects
the frequency of the bound axions,
and therefore the frequency of the 
GWs emitted \cite{Arvanitaki:2014wva}.
As the occupation numbers of the
levels evolve, the self-energy contribution to the binding energy $\Delta \omega$
and thus the emitted frequency $\nu$ change over time. 

The gravitational  and self-interaction contributions to the energy of axions in level 211 are, respectively, (see App.~\ref{app:selfgravity} and App.~\ref{app:freqcorr} )
\begin{align}
\label{eqn:domega}
\Delta \omega_g &\simeq -{0.19}\mu \alpha^3 \varepsilon_{211} \\
\Delta \omega_{\lambda} &\simeq -3.5\times10^{-5} \mu \alpha^5 \varepsilon_{211} \left(\frac{M_\text{pl}}{f} \right)^2,
\end{align}
 where the energy is decreased  (increased) in the presence of an attractive (repulsive) self-interaction. These corrections are always small compared to the axion mass, as well as the energy splitting between levels (for occupation numbers below the non-perturbative
 regime --- see Sec.~\ref{sec:nonperturb}).
   
As the cloud is growing through superradiance, the frequency changes
relatively rapidly  as $\propto\mu\alpha \dot{\alpha}$ on the order of the
superradiance time due to the changing BH mass. However this period
is short, and generally does not contribute much
of the detectable signal.
At late times, the
cloud size is depleted over time, and the level's frequency drift is
positive (assuming negligible or attractive
self-interactions). 
This is in contrast to standard astrophysical sources
of continuous gravitational radiation, such as spinning
neutron stars, and may provide a hint that a
detected signal arises from superradiance.
We describe the main contributions to these
frequency drifts, at leading order in $\alpha$, below. For a more
complete discussion of frequency drifts we refer the reader to App.
\ref{app:fdrifts}.

At small self-interactions, the frequency drift is dominated by the depletion of the gravitational self-binding energy through annihilations, resulting in a frequency drift  of order 
\begin{align}
\label{eqn:gfdrift}
    \dot{\nu}_\text{ann}\simeq7\times10^{-15}\frac{\text{Hz}}{\text{s}}\pare{\frac{\alpha}{0.1}}^{17}\pare{\frac{\mu}{10^{-12}\text{ eV}}}^2,
\end{align}
to leading order in $\alpha$. Throughout the small self-interaction regime $f>f_{\mathrm{AB}}$ (see also Fig. \ref{fig:paramspace}), the gravitational frequency drift dominates any contribution from the self-interactions.

As self-interactions increase, the frequency drift from the gravitational binding energy is increased due to the faster depletion of the cloud from axion emission,
\begin{align}
    \dot{\nu}_\text{g}\simeq10^{-10}\frac{\text{Hz}}{\text{s}}\pare{\frac{10^{17}\,\text{GeV}}{f}}^4\pare{\frac{\mu}{10^{-12}\,\text{eV}}}^{2}\pare{\frac{\alpha}{0.1}}^{17},\label{eqn:gravdrift}
\end{align}
and there is an additional frequency drift from the change of self-interaction energy,
\begin{align}
\label{eqn:scalarfdrift}
\dot{\nu}_{\lambda}\simeq 10^{-10}\frac{\text{Hz}}{\text{s}}\pare{\frac{10^{17}\,\text{GeV}}{f}}^6\pare{\frac{\mu}{10^{-12}\,\text{eV}}}^{2}\pare{\frac{\alpha}{0.1}}^{19}.
\end{align}
The latter dominates when $f \lesssim 8.5\times 10^{16}\text{ GeV}(\alpha/0.1)$. Finally, in the strong self-interactions  regime $f<f_{\mathrm{BC}}$, the cloud reaches a long-lived quasi-equilibrium
configuration, and the dominant source of frequency drifts comes from the slow spindown of the BH. 

Gravitational wave signals from $322 \rightarrow 211 +$ GW transitions have frequency
$\nu_{322}-\nu_{211}$, so the changing contributions to the 211 and
322 frequencies partially cancel, making frequency drifts a factor of a few
smaller than for annihilations, and negative in most parts of the parameter space. Similarly
to annihilations, for moderate self-couplings, self-interactions
dominate the frequency drifts for
$f\lesssim10^{17}\mathrm{GeV}(\alpha/0.1)$.

At small $\alpha$, the frequency drift can be small enough
so as to be unobservable. Over a year,
the minimum frequency change
that can be measured is $\sim {\rm yr}^{-1} \sim 3 \times 10^{-8}
\Hz$, so if the frequency drift is
$\lesssim {\rm yr}^{-2} \simeq 10^{-15} \Hz {\rm \, s}^{-1}$,
it has no observational effect.
At the other extreme, too large a frequency
drift can be problematic for the search
algorithms employed.
Current LIGO/Virgo continuous wave searches cover a range of
positive to  negative frequency derivatives of e.g. $2\times
10^{-9}$ Hz/s through $-1\times 10^{-8}$ Hz/s
\cite{Pisarski:2019vxw}.
More sensitive searches, using longer coherent integration
times, may require even smaller frequency drifts~\cite{Dergachev:2019oyu}.
In the small coupling regime, the drift of the signal becomes
larger than this threshold at $\alpha\sim0.25$.   In the moderate
self-interactions regime, both annihilation and transition signals
have drifts large compared to the current search range for
$f\lesssim5\times 10^{16}\mathrm{GeV}(\alpha/0.1)^{17/4}$. 
However, as discussed above, the observational
prospects for GW signals at such small $f$ are not promising,
with current-generation experiments.


\section{Axion waves}
\label{sec:scalarwaves}

As well as emitting gravitational radiation,
the cloud also emits both relativistic (section~\ref{secrelscalar}) and
non-relativistic (section~\ref{secnonrelem}) scalar waves.
If the scalar $\varphi$ has non-gravitational interactions\footnote{If the scalar $\varphi$'s interactions with the SM
are purely gravitational, then its 
interaction rate with matter is $\propto G^2 \sim 1/\Mpl^4$,
whereas for gravitational radiation,
the interaction rate is $\propto G \sim 1/\Mpl^2$. Consequently,
such $\varphi$ radiation would be practically undetectable.} with the SM, such $\varphi$ radiation could be detected in laboratory experiments. 
For an axion-like particle, a natural assumption
is that interactions with the SM are suppressed
by parametrically the same symmetry breaking scale $f$
that sets the axion potential. If this is the case,
then we have the unusual feature that, in the large
self-coupling regime $f < f_\text{BC},$ the signal does not decouple: while the power in axion radiation decreases as the quasi-equilibrium size of the cloud decreases, this is compensated for
by the increased interaction strength from the smaller $f$.
In addition, the BH spin-down time increases
with decreasing $f$, so such signals can last for
very long times, increasing the chance of observing
them.
Consequently, axion waves could be a probe
of the small-$f$ regime, in which both GW and spin-down
signatures are suppressed.

Quantitatively, if we take the 211 and 322 quasi-equilibrium
occupation numbers \eqref{eq:quasi_equlibrium_occupations}, then the emitted power
is dominated by non-relativistic
$322 \times 322 \rightarrow 211 \times \infty$ radiation.
At large distances $r$ from the BH, this radiation has
energy density 
\begin{align}
    \begin{split}
		\rho_{\rm rad} &\sim{} \frac{\mu}{4\pi r^2}\frac{GM^2 \gamma_{322\times 322}^{211\times \infty}(\varepsilon_{322}^\text{eq})^2\varepsilon_{211}^\text{eq}}{v}
	 \\{}&\simeq
         10^{-6}\GeV/\cm^3\left(\frac{\alpha}{0.1}\right)^6\left(\frac{10\kpc}{r}\right)^2\\{}& \quad \times \left(\frac{f}{10^{16}\GeV}\right)^2,
\end{split}
\end{align}
where $v=\alpha/6$ is the velocity of the non-relativistic axions emitted.
The energy density $\rho_{\rm rad}$ depends only
on $\alpha$, and not on $\mu$ and $M_{\rm BH}$ independently.
For given $f$, the emitted power is maximized
when the superradiance rate is largest, at high $a_*$ and $\alpha$.
The corresponding dimensionless amplitude $\theta$ of the axion waves is
\begin{align}
    \begin{split}
        \theta \simeq 10^{-19} \left(\frac{10^{-12}\eV}{\mu}\right)\left(\frac{\alpha}{0.1}\right)^3\left(\frac{10\kpc}{r}\right),\\
    \end{split}
\end{align}
independent of $f$. This is in contrast to GW signals,
for which the amplitude at Earth decreases as $f^2$
in the quasi-equilibrium regime. 
Relativistic axion radiation
from the $3 \rightarrow 1$ process
(section~\ref{secrelscalar}),
and $2 \rightarrow 1$ cubic emission,
also have $f$-independent $\theta$, but are suppressed
by higher powers of $\alpha$, and are smaller
than the non-relativistic radiation
for the parameter space we are interested in.

As we discussed in section~\ref{sec:higherlevels},
for $\alpha \gtrsim 0.2$ and small $f$ we expect additional
hydrogenic levels, other than 211 and 322, to be
populated. While we have not performed a full analysis 
in this regime, a example of the possible
effects can be seen from the 411 build-up
studied in section~\ref{n11}, which for $\alpha$ not too
far above $0.2$ is expected to be the first additional level to grow.
The 211, 322, and 411 levels form a new quasi-equilibrium,
with the 411 level having enhanced occupation number
relative to those of the 211/322 equilibrium.
Consequently, the rate of scalar radiation
during this equilibrium is enhanced; numerically,
we find that $\rho^{\rm rad}_{\rm{3-level}} \sim 25 \rho^{\rm rad}_{\rm{2-level}}$
for $\alpha \simeq 0.3$.
While this equilibrium will be disrupted in turn by the growth
of further levels, this illustrates that,
while the parametric behaviour in $f$ should remain the same,
additional levels may change the numerical factors
affecting the scalar radiation power.
As discussed in section~\ref{sec:nonperturb},
if the growth of additional levels
leads to large enough field amplitudes
in the cloud, then higher-order processes
or a non-perturbative collapse of the cloud may become possible,
significantly altering the behaviour.

Since the axion radiation is non-relativistic
and narrow-bandwidth, its effects on
a laboratory system
are similar to those of axion dark matter at the same 
mass. The masses of interest correspond to rather low
frequencies, e.g.\  
$10^{-12} \eV \simeq 2\pi \times 200 \Hz$.
For this parameter space, the axion-SM couplings
most amenable to laboratory detection experiments
are those to nuclear spins and to photons,
which we discuss below. 

Searches for axion DM via the axion-gluon
coupling 
$\LL_{\rm int} \propto (\varphi/f) G_{\mu\nu} \tilde G^{\mu\nu}$ have 
promising sensitivity reach
at low axion masses~\cite{kimball2017overview}. 
However, if an axion-like particle has the same
$G \tilde G$ coupling, but a smaller mass
than the QCD axion (or equivalently,
a larger $G \tilde G$ coupling for the same mass),
then it is strongly constrained
by its behaviour in dense
environments such as the early universe
and stellar cores~\cite{Blum:2014vsa,Hook_2018}.
For superradiance-sourced signals, $G \tilde G$ couplings
significantly higher than the QCD axion value (for a given
axion mass) are needed to have experimental
sensitivity, and are affected
by these constraints.

\subsection{Nucleon spin coupling}
\label{secnucleonspin}

The axion coupling to fermion spins is
$\LL \supset g_N (\partial_\mu \varphi) \bar \psi \gamma^\mu \gamma^5 \psi$,
where we generically expect $g_N \sim 1/f_a$. For a non-relativistic fermion, this gives an axion-dependent
term in the fermion Hamiltonian,
\begin{equation}
	H \supset g_N \vec \sigma \cdot \left(\nabla \varphi + \dot \varphi \vec v\right)
\end{equation}
where $\vec \sigma$ is the fermion's spin, and $\vec v$ is its
velocity. We will focus on couplings to nucleons, 
which for low axion frequencies are easier
to detect than couplings to electrons.

Since the $322 \times 322 \rightarrow 211 \times \infty$ axion radiation from the BH has
$v \sim \alpha/6$ (Sec.~\ref{secnonrelem}),
while the nucleon velocity changes associated
to low-energy laboratory processes are
much smaller, the ``axion wind'' term
$H_{\rm wind} = g_N \vec \sigma\cdot\nabla\varphi$
dominates. Due to the $\sim \alpha/6$
velocity being significantly larger than the
virial velocity of DM in the galaxy, $\sim
10^{-3}$, and because of the
coherent nature of the emitted radiation, an
experiment searching for the axion wind coupling
will have better sensitivity to BH-sourced
radiation than it would for DM for an equivalent
axion energy density.

\begin{figure}[t!]
	\begin{centering}
		\includegraphics[width=\columnwidth]{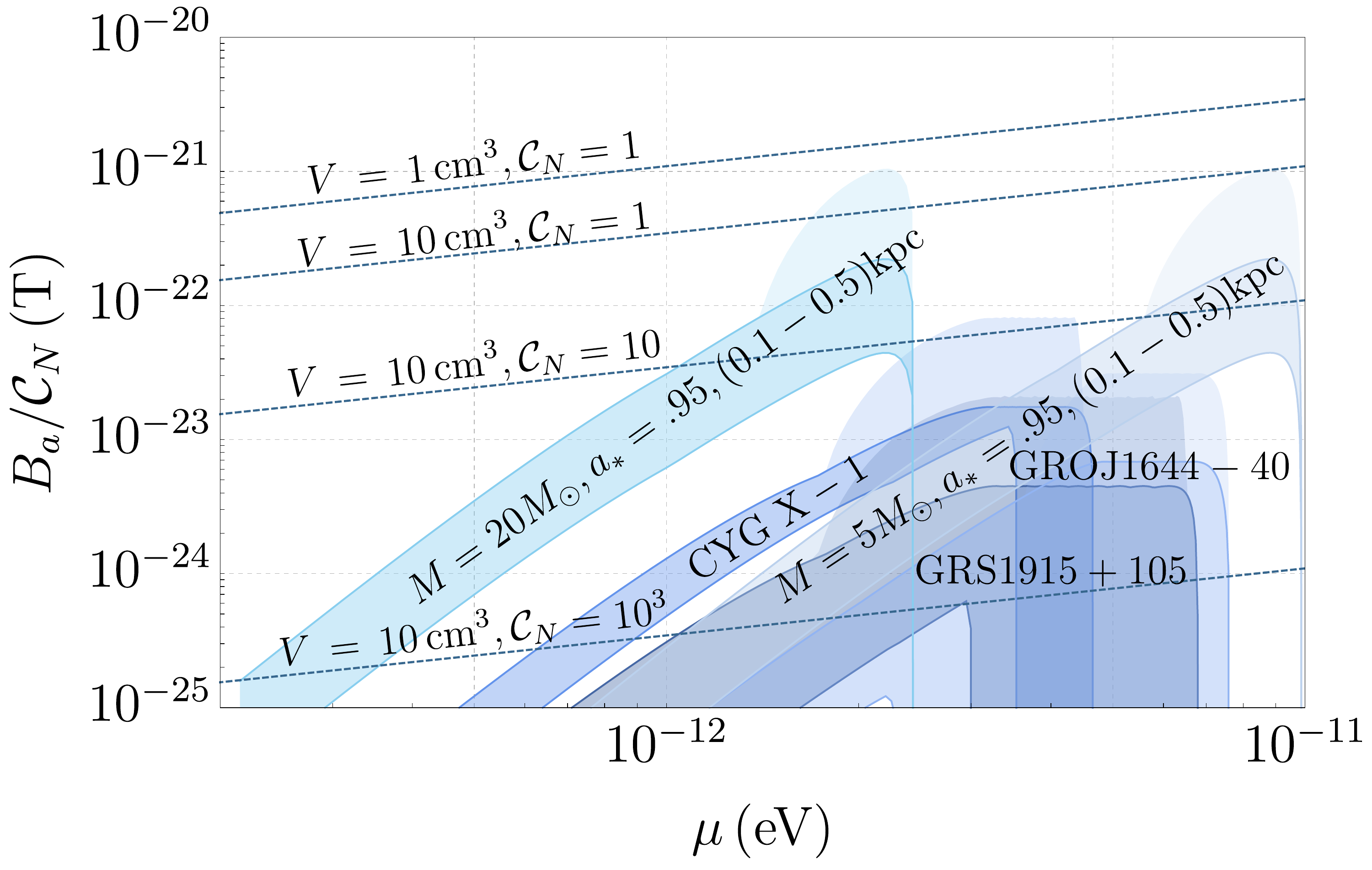}
		\caption{Projected detectability of 
		non-relativistic axion radiation, assuming an axion-nucleon coupling.
		The signal strength is expressed in terms
		of the equivalent pseudo-magnetic field
		felt by nuclei.
		The blue dotted lines correspond to sensitivity
		estimates for NMR axion-wind detection experiments~\cite{kimball2017overview}
		with the indicated parameters.
		The bands correspond to signals
		from three astrophysical BHs and
		two nominal BHs with the indicated parameters.
		The widths of the bands correspond to the uncertainty
		on the BH parameters (for the nominal BHs,
		to the distance range indicated).
		The darker bands bounded by solid contours
		correspond to the signal emitted
		during two-level quasi-equilibrium (Sec.~\ref{secperturb}).
		The lighter-shaded extensions above
		 represent the enhanced signal
		from the three-level equilibrium with 411 (Sec.~\ref{n11}), illustrating the potential range of signals.}
	\label{fig:ScalarSignals}
	\end{centering}
\end{figure}

\begin{figure}[t!]
	\begin{centering}
		\includegraphics[width=\columnwidth]{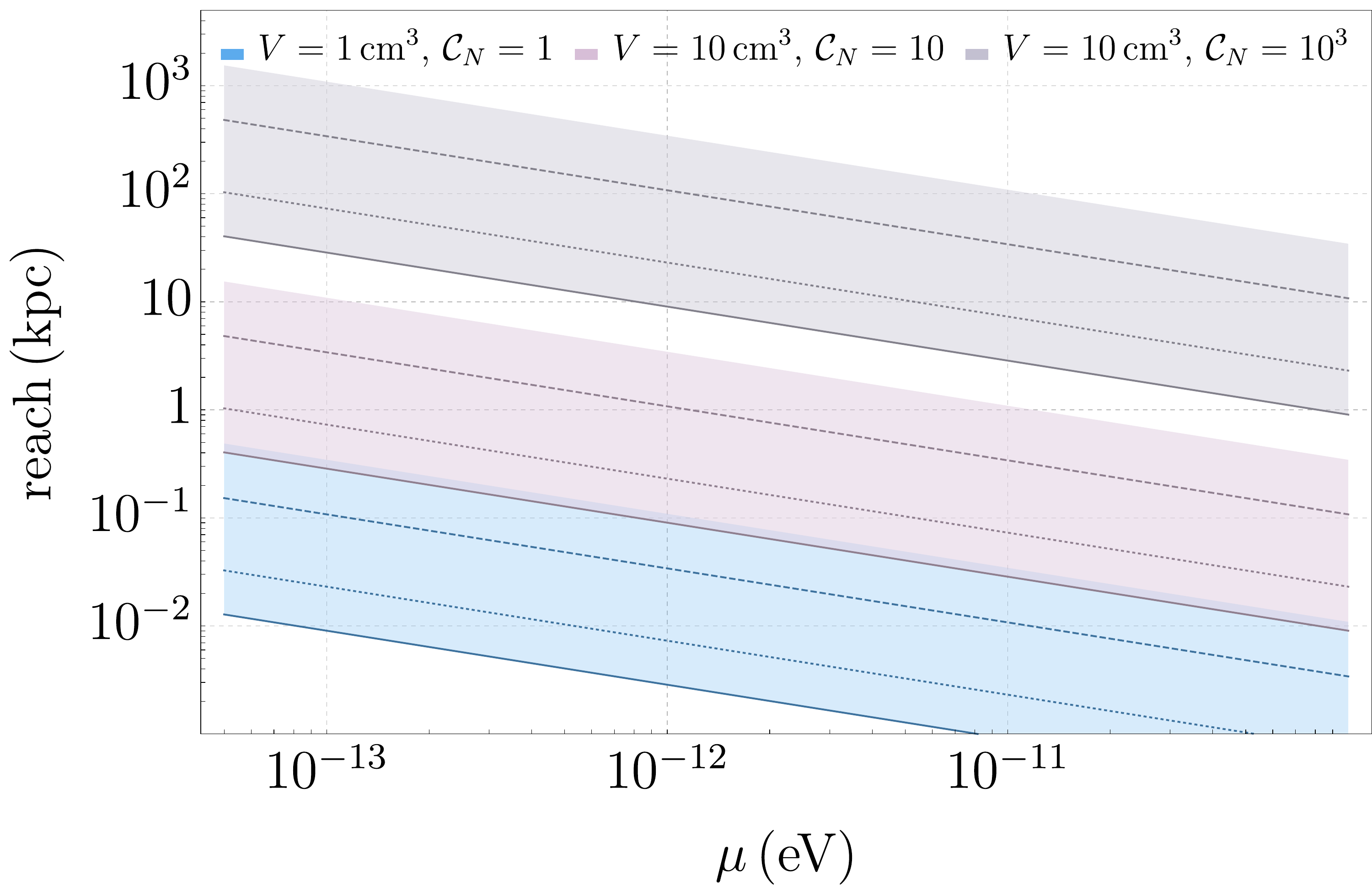}
		\caption{Projected sensitivity reach (SNR = 1) for the detection of non-relativistic axion waves from a BH-cloud system at large self-interactions, $f<f_\text{BC}$, in a NMR-based axion-wind detection experiment. The bands show the reach to a BH-cloud system, ranging from a two-level quasi-equilibrium with parameters $a_* = 0.9$ and $\alpha = 0.2$ (lower edge, solid), to that of a three-level quasi-equilibrium system with $a_* = 0.99$ and $\alpha = \alpha^\text{optimal}(0.99)\approx 0.41$ (upper edge). The reach for a BH-cloud system with $a_* = 0.9$ and $\alpha = \alpha^\text{optimal}(0.9)\approx 0.28$ is also indicated for a two-level equilibrium (dotted line), and a three-level equilibrium (dashed line) system.}
	\label{fig:ScalarReach}
	\end{centering}
\end{figure}

The best-developed experimental proposal aiming to
detect the axion wind coupling is CASPEr-Wind~\cite{kimball2017overview}, which
employs Nuclear Magnetic Resonance (NMR) technologies.
This uses a liquid xenon target, whose nuclear spins
are polarized in a strong magnetic field.
The axion wind coupling acts on the nuclei
like an effective magnetic field,
$H_{\rm wind} = g_N \vec \sigma \cdot \nabla \varphi
\equiv B_a \cdot \vec \mu_n$, 
where $\mu_n$ is the nuclear magnetic moment and $B_a$ is the effective axion `magnetic field'.
If this effective magnetic field oscillates
at close to the Larmor frequency
of the nucleons in the external magnetic field,
then the resulting spin precession
of the nuclei is resonantly enhanced.
This spin precession can then be picked up by a sensitive magnetometer.

In App.~\ref{appcasper}, we review the sensitivity
of such experimental setups to a monochromatic
axion oscillation. If we are uncertain about the axion
mass, and want to experimentally probe an $\OO(1)$
axion mass range around an angular frequency $\omega_0$, then a signal can be detected for
\begin{equation}
	B_a^2 \gtrsim {\rm few} \times \frac{\omega_0}{\mu_n^2 N_n T_{\rm tot}},
\end{equation}
 where $T_{\rm tot}$
is the total experimental running time, and $N_n$ is the number
of aligned spins in our spin-polarized sample.\footnote{This sensitivity estimate is for the detection of a single,
	monochromatic signal.
	As mentioned in section~\ref{sec:gw}, in situations where
	many galactic sources are emitting at any given
	time, it may be more effective to perform
	a ``stochastic'' search, looking for multiple 
	unresolved signals within a given bandwidth.
	We leave analysis of such scenarios to
	future work.}
This is a best-case sensitivity estimate, limited
by the fundamental spin-projection noise
of the sample
--- to achieve it, a well-shielded
sample and a sufficiently sensitive magnetometer
would be required. Experiments capable of sensing
nuclear spin projection noise have been carried out~\cite{Sleator_1985},
and such sensitivities are
a goal for the CASPEr-Wind experimental
program~\cite{kimball2017overview}.

A fully polarized liquid $^{129}$Xe sample
has $\sim 10^{22}$ spins/cm$^3$~\cite{kimball2017overview},
so the sensitivity limit for a relatively small 
target volume is
\begin{equation}
	B_a \gtrsim 10^{-20} {\rm \, T} 
	\sqrt{
		\frac{\nu}{\kHz} \frac{10^{22}}{N_n} \frac{{\rm yr}}{T_{\rm tot}}}
\end{equation}
For comparison, an axion DM signal at the sensitivity threshold
estimated in~\cite{kimball2017overview}, for these
parameters, has an effective
magnetic field of $\sim {\rm few} \times 10^{-20} {\rm \, T}$.
The effective magnetic field from axion radiation emitted
by a superradiant cloud is
\begin{equation}
	B_a \simeq 3 \times 10^{-24} {\rm \, T} \times \mathcal{C}_N \left(\frac{\alpha}{0.1}\right)^{4}\left(\frac{1\kpc}{r}\right),
\end{equation}
for a high-spin BH, where $\mathcal{C}_N \equiv g_N f$.
Consequently, some combination
of 
larger experimental volumes (as planned for
CASPEr-Wind phase II~\cite{kimball2017overview}),
larger $\mathcal{C}_N$, larger $\alpha$ and a closer BH would
enable laboratory experiments to be sensitive to axion waves.


This is illustrated in Fig.~\ref{fig:ScalarSignals},
which shows projected signal strengths for a selection
of astrophysical BHs (both nominal and observed), along
with sensitivity thresholds for different experimental 
configurations.
While $\mathcal{C}_N \sim \OO(1)$ is the `natural' expectation
in many models, larger values of $\mathcal{C}_N$ are possible.
In particular, it is interesting to consider how large a reach
can be obtained in as-yet-unconstrained parameter space,
below the existing astrophysical limits of
$g_N \lesssim ({\rm few} \times 10^8 \GeV)^{-1}$~\cite{1803.00993,1906.11844,1806.07151,1512.07828}. While
much of the axion mass range in Fig.~\ref{fig:ScalarSignals}
is excluded for large $f$ by BH spin measurements
(Fig.~\ref{figSpindown1}), these constraints
do not apply for $f \lesssim 10^{12}-10^{13} \GeV$,
where the BH spin-down is too slow. The astrophysical
bounds translate into $|\mathcal C_N| \lesssim 10^3 (f/10^{12} \GeV)$; the $\mathcal C_N = 10^3$ line in Fig.~\ref{fig:ScalarSignals}
illustrates that 
such couplings can give good detection prospects
for a wide range of BHs and axion masses.

To reflect the uncertain behavior
of the superradiant cloud at $\alpha \gtrsim 0.2$,
Fig.~\ref{fig:ScalarSignals} displays the signal resulting from the radiation
power during the three-level quasi-equilibrium phrase,
as a shaded area above the signal from the two-level equilibrium.
The signal curves illustrate that, with larger-volume
experiments, sensitivity to astrophysical
BHs may be possible for  $\mathcal C_N\sim \mathcal{O}(1)$. They also strongly motivate detailed
numerical analyses of the high-$\alpha$
regime, where the strongest signals would arise.

Fig.~\ref{fig:ScalarReach} displays
the sensitivity reach to an optimal BH
for a given axion mass. Again, we see that for larger
experimental volumes, astrophysically-relevant reaches
--- in particular, to the Galactic Center $\sim 8 {\rm \, kpc}$
away ---
may be possible for fairly natural $\mathcal C_N$ values.

\begin{figure}[t!]
	\begin{centering}
		\includegraphics[width=\columnwidth]{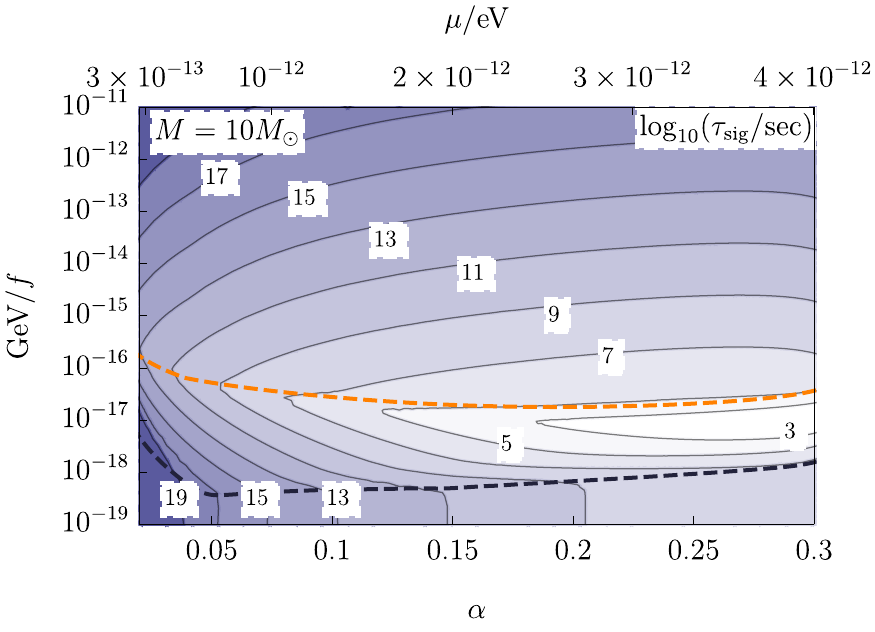}
		\caption{Typical duration $\log_{10}(\tau_{\rm sig}/{\rm sec})$ the axion wave signal  for  a $10 M_{\odot}$ BH with initial spin 0.9. In the large self-interactions regime, we show the time-scale corresponding to  the quasi-equilibrium evolution. For $f \lesssim 10^{12}$~GeV the signals can last longer than the age of the universe (note that $10 {\rm \, Gyr} \simeq 3 \times 10^{17} {\rm \, s}$). The dashed orange and dotted black lines are the $f_{\rm BC}$ and $f_{\rm AB}$ curves, respectively, as in Fig.~\ref{fig:annihilations}.}
	\label{fig:ScalarSignalTime}
	\end{centering}
\end{figure}

If we are interested in the signal from a specific, known
BH, then the sensitivity reach is the most important parameter.
However, as is the case for gravitational wave searches,
many signals are expected to arise from as-yet-unobserved BHs,
and could only be detected via a ``blind'', all-sky search.
In this situation, another important factor is the typical
duration of signals, which affects the probability that
a given BH is still emitting today. Figure~\ref{fig:ScalarSignalTime}
shows the duration of the peak axion signal (which contributes
most of the detectable SNR) from a nominal BH,
as a function of axion mass and coupling. 
Lower $f$ values lead
to slower BH spin-down, and so to longer durations of quasi-equilibrium
signal emission; this is relevant down to $f \sim 10^{11} - 10^{12} \GeV$, 
below which signals can last longer than the age of the universe.

\begin{figure}[t!]
	\begin{centering}
		\includegraphics[width=\columnwidth]{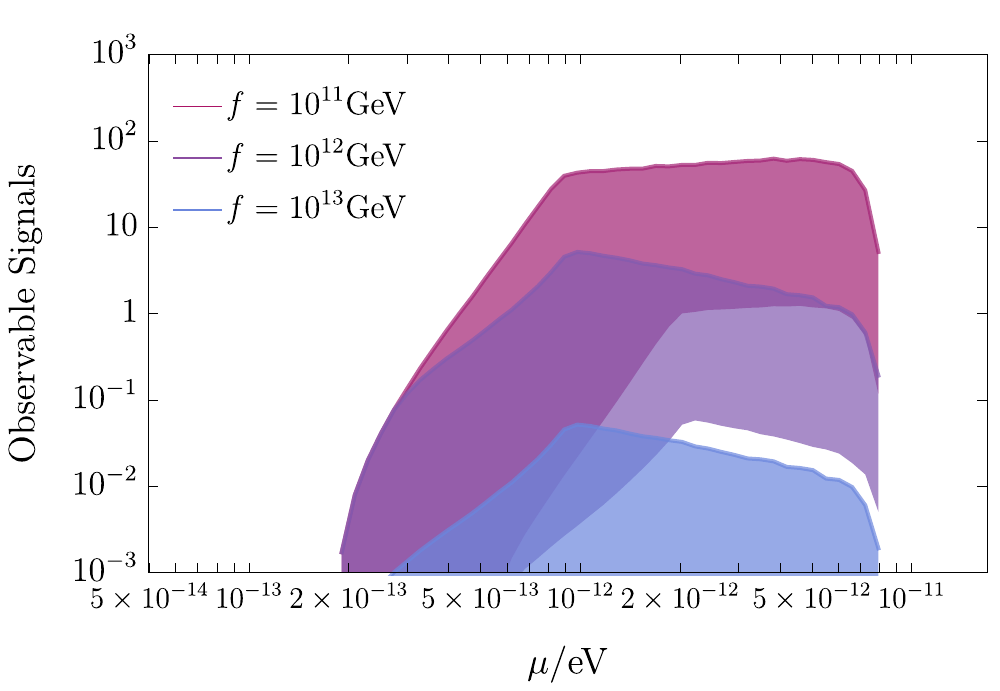}
		\caption{Number of observable signals expected in an NMR axion wind experiment with $V = 10 {\rm \, cm^3}$ and $\mathcal{C}_N=100$, 
		with different bands corresponding to different
		quartic coupling scales $f$. We require observable signals to have SNR $\ge 10$, given the blind search strategy required for these events.
The width of the bands results from varying the assumed BH spin distribution and maximum BH mass (see Section~\ref{sec:annihilations}).
		For a fixed $\mathcal{C}_N$, the number of observable signals increases for smaller $f$, due to longer signal durations, saturating at $f\sim10^{11}$~GeV.}
	\label{fig:ScalarEvents2}
	\end{centering}
\end{figure}

\begin{figure}[t!]
	\begin{centering}
		\includegraphics[width=\columnwidth]{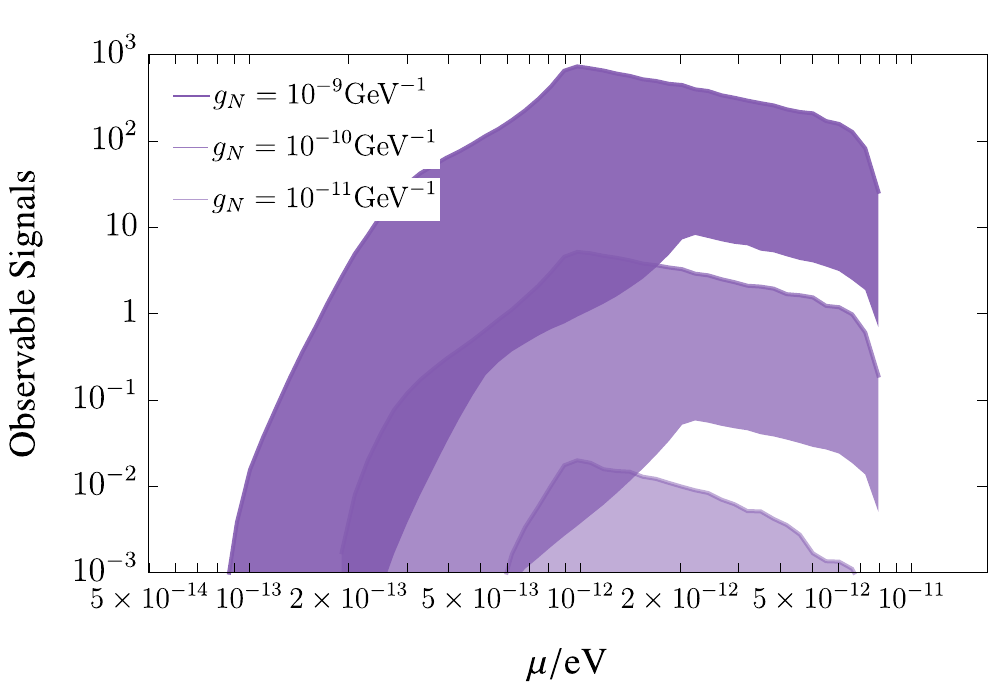}
		\caption{Number of observable signals expected in an NMR axion wind experiment with $V = 10 \cm^3$, for $f=10^{12}$~GeV and different couplings to nuclear spins, as shown. 
		We require observable signals to have SNR $\ge 10$, given the blind search strategy required for these events. The width of the bands results from varying the assumed BH distribution as in Fig.~\ref{fig:ScalarEvents2}. For a fixed self-interaction strength, the highest number of observable signals is for the largest coupling strength to nuclei.}
	\label{fig:ScalarEvents}
	\end{centering}
\end{figure}

Since, in the quasi-equilibrium regime,
the peak signal strength at Earth is independent of $f$ for
fixed $\mathcal{C}_N$, decreasing $f$ down to $\sim 10^{11} \GeV$
increases the expected number of events in a blind
search. This is illustrated in Fig.~\ref{fig:ScalarEvents2}.
If, rather than fixing $\mathcal{C}_N$, we require
that $g_N$ is below the astrophysical bounds, then as shown
in Fig.~\ref{fig:ScalarEvents}, there is a wide range
of axion masses over which we might expect visible signals in
an all-sky search (depending on the mass and spin distribution
of astrophysical BHs).  In both the Fig.~\ref{fig:ScalarEvents2} and Fig.~\ref{fig:ScalarEvents} projections we assume the reach to  the axion waves from the two-level equilibrium, not taking into account the possible enhancements in power from additional levels; on the other hand, the dynamics of additional levels could shorten the signal lifetime at large $\alpha$ values. In the blind search, an analysis similar to the techniques employed by Continuous Waves searches at LIGO/Virgo (Sec.~\ref{sec:annihilations}) would be required, to make use of the extremely long signal coherence times while at the same time taking into account the Doppler shifts from the many relative motions between the experiment and the unknown black hole positions.

Unless $\mathcal{C}_N$ is extremely large, the effects 
of the axion field on spins in the vicinity of the black
hole, and the effect of these spins on the axion field,
are always small. The largest effective magnetic
field obtained in the cloud is $\sim |\mathcal{C}_N| 10^{-6} {\, \rm T}
\frac{\mu}{10^{-12} \eV}$, which would not have
any significant affect on accretion disk behavior.
Similarly, the axion field sourced by a coherent 
nuclear spin density, if any exists in the accretion
disk, is tiny compared to the fields of a superradiant cloud.
For any reasonable nuclear spin response to small
magnetic field perturbations, 
the effect of spin response on the dynamics of quasi-bound
axion levels will be extremely small, 
so the growth of the cloud will not be affected.
Similar considerations apply to the propagation of scalar
waves through interstellar space; these will
be undisturbed to a very good approximation.

\subsection{Photon coupling}
\label{sec:photons}

The axion coupling to photons is $\LL \supset
- \frac{g_{a\gamma\gamma}}{4} \varphi F_{\mu \nu} \tilde F^{\mu\nu}
= g_{a\gamma\gamma} \varphi E \cdot B$. Generically, we expect
the coupling constant to be $g_{a\gamma\gamma} = \mathcal{C}_\gamma
\frac{\alpha_{\rm EM}}{2 \pi f_a}$, where
$\mathcal{C}_\gamma \sim \OO(1)$ is related
to the charged matter content of the UV theory~\cite{Di_Luzio_2017}.

An axion oscillation sources EM fields through
the effective current density $J_a = g_{a\gamma\gamma} (\dot \varphi B
+ \nabla \varphi \times E)$ (and the corresponding effective
charge density $\rho_a = - g_{a\gamma\gamma} \nabla \varphi \cdot B$).
Axion DM, which is non-relativistic,
has $|\dot \varphi| \gg |\nabla \varphi|$, 
so detection experiments use strong magnetic fields
to maximize $J_a$.
Searches for low-frequency ($\sim \kHz$)
axions have been proposed using static background magnetic
fields~\cite{Kahn_2016,1904.05806},
or GHz-frequency fields in superconducting 
cavities~\cite{Lasenby_2020,Berlin_2020,1912.07751,2007.15656}.\footnote{Experiments using optical-frequency fields have also
	been proposed~\cite{DeRocco_2018,Obata_2018,Liu_2019},
	but these have significantly worse theoretical sensitivity.
	}

If they can be realized in the future,
quantum-limited meter-scale experiments could 
probe axion DM couplings
as small as $g_{a \gamma \gamma} \sim 10^{-17} \GeV^{-1}$
at $\sim \kHz$ frequencies (unfortunately, this is still
far from QCD axion sensitivity).
With a monochromatic signal, as opposed
to virialized axion DM, this would correspond
to a sensitivity of $g_{a \gamma \gamma} \sim 10^{-18} \GeV^{-1}
\left(\frac{\rho}{\rho_{\rm DM}}\right)^{-1/2}$.\footnote{The ideal search strategy for monochromatic signals
	may be different from that for
	a virialized axion signal with non-negligible bandwidth.
	For static-field experiments such as
	those proposed in~\cite{1904.05806}, an optimal search
	for monochromatic signals will overcouple
	the amplifier even more strongly to the 
	pickup. However, for $\sim \kHz$ axion frequencies and
	practical temperatures, optimal axion
	DM experiments would already be strongly overcoupled 
	(to the point of having almost $\OO(1)$ fractional
	sensitivity bandwidth~\cite{1904.05806}),
	so there would not be a significant difference
	between the monochromatic and DM search strategies.}
For non-relativistic emission from 
a superradiant cloud, we would obtain a reach of
\begin{equation}
\frac{r}{\kpc} \approx (2\times 10^{-3}) \, |\mathcal C_\gamma|\left(\frac{\mu}{10^{-12}\eV}\right)\left(\frac{\alpha}{0.1}\right)^3,
\end{equation}
Consequently, signals from an superradiant cloud via the axion-photon
coupling could only be seen for an exceptionally close,
fast-spinning BH, and/or in models where $|\mathcal C_\gamma|$ 
is large. 

At the small axion masses we are interested in,
SN1987A observations constrain the axion-photon
coupling to be $|g_{a\gamma\gamma}| \lesssim 5 \times 10^{-12} \GeV^{-1}$~\cite{1410.3747}. This translates
to $|\mathcal C_\gamma| \lesssim 500 \, (f/10^{11} \GeV)$,
which allows for somewhat smaller expected blind-search
event rates than the nucleon-coupling case shown in Fig.~\ref{fig:ScalarEvents}.



Similarly to the case of nucleon couplings, the
effects of astrophysical EM fields on the SR
cloud will be tiny unless $|\mathcal C_\gamma| \gg 1$.
In addition, 
the naive $\varphi \rightarrow \gamma \gamma$ decay rate,
$\Gamma_{\varphi \rightarrow \gamma \gamma} \simeq \frac{g_{a \gamma \gamma}^2 \mu^3}{64 \pi}$, is much longer than the age
of the universe for couplings of interest.
However, in some circumstances
it is possible for parametric resonance
to greatly enhance the photon emission rate~\cite{Hertzberg2018}. Parametrically, in the limit where $g_{a\gamma\gamma}$
is arbitrarily small, and taking $L$ to
be the approximate spatial extent of the axion profile, the total decay rate into a particular mode within the $\sim L^3$ volume is
$\Gamma \sim g_{a\gamma\gamma}^2 \varphi^2 \mu^2 L$,
where $\varphi$ is the typical field amplitude.
Consequently,
the number of photons emitted into that mode, in
the light-crossing time $\sim L$, is 
$\sim \Gamma L \sim g_{a\gamma\gamma}^2 \varphi^2 (\mu L)^2$.
This tells us that for finite $g_{a \gamma \gamma}$, if $\Gamma L \gtrsim 1$, then stimulated emission will become
important; for $\Gamma L \gg 1$, the emission
rate will be exponentially enhanced.

This parametric argument agrees with the conclusions
of~\cite{Hertzberg2018}, which analyses the growth
of electromagnetic perturbations using Floquet theory,
and finds that parametric resonance occurs
if
\begin{equation}
\label{eq:param_res_condition}
	|g_{a \gamma \gamma} \mu \varphi L| \gtrsim {\rm few}.
\end{equation}
Since $g_{a \gamma \gamma} \varphi = \mathcal{C}_\gamma \frac{\alpha}{2\pi}
\theta$, the LHS is maximized (for given $\mathcal{C}_\gamma$)
by maximizing $\theta$. For an axion of mass $\mu$,
this occurs at $f \simeq f_{\rm BC}$ (for the 211 level). Using
Eq.~\eqref{eq:bound_on_theta_0}, we find
that for parametric resonance to occur, we need
\begin{equation}
	|\mathcal{C}_\gamma| \gtrsim (9\times 10^2) \left(\frac{0.1}{\alpha}\right)^{3/4},
\end{equation}
for $a_*(t_0) = 0.99$. Consequently,
if $|\mathcal{C}_{\gamma}| \ll 10^3$, then photon emission
will be unimportant.

It should be noted that the above is a best-case estimate,
which will only hold if the BH is in a sufficiently pristine environment.
The plasma frequency in the interstellar medium is
$\omega_p \sim 10^{-12} - 10^{-10} \eV$, which is comparable
to the mass range for a superradiant axion around a stellar-mass BH.
Moreover, one expects the plasma density in the vicinity 
of the BH to be greater, due to accretion~\cite{Dima_2020rzg}.
Consequently, it is likely that plasma effects suppress 
the parametric resonance process, even at large $|\mathcal C_\gamma|$~\cite{Sen2018}.

\section{Conclusions}
\label{sec:concl}

In this paper, we have investigated some of
the most important consequences of scalar
self-interactions for superradiance around
astrophysical BHs. As we have showed, self-interactions
can result in very rich and complicated dynamics, 
and there are a number of aspects which would benefit from further study. In particular, we have not systematically
treated situations in which the initially fastest-growing
level has $m \ge 2$. While we generally expect
gravitational (and scalar) wave signatures to be
dominated by cases where 211 grows first, 
BH spin-down constraints for higher-mass axions
will depend on higher-$m$ superradiance.

In addition, even for the 211 case, our calculations
have been at the (semi-)analytic level,
and may not be reliable for large enough
$\alpha$. In particular, we found that for
$\alpha \gtrsim 0.2$ and small $f$, levels other than 211 and 322 might
play an important role in the dynamics. One 
route to properly understanding the high-$\alpha$
regime might be to perform numerical simulations
of the (self-interacting) field
equations themselves, rather than of the occupation numbers
of hydrogenic modes. Such approaches have 
been used to study purely-gravitational superradiance
in a number of papers~\cite{1212.0551,1212.1477,1704.04791,1705.01544,1807.00043}. As mentioned in Sec.~\ref{sec:nonperturb},
numerical methods were applied to a self-interacting
scalar field on the Kerr background by~\cite{Yoshino:2015nsa,Yoshino:2012kn}, but they did not evolve the system for long enough to observe the perturbative effects we have studied.
Since the high-$\alpha$ regime is where
observational signatures may be the strongest,
and in which there is the possibility of phenomena
such as bosenova, a fuller treatment would be valuable.

Our analyses focussed on the simplest form
of self-interactions for a spin-0 particle; the lowest-order
(renormalizable) potential terms. In more complicated
hidden sector models, other forms of interactions,
or extra hidden sector states, could affect 
the superradiance behavior. For example,
\cite{Mathur:2020aqv} discusses a model in which
the QCD axion couples to a hidden-sector photon,
and there are hidden-sector fermions which interact
with this photon. Such models illustrate that, while
the minimal DM models we considered in Figures~\ref{figSpindown1}
and~\ref{figsmbh} are often still subject to
BH spin-down constraints, others may not be.

Beyond the spin-0 particle candidates we
considered, superradiance of massive
vectors is also of interest. Vector self-interactions
are somewhat more complicated than those for scalars,
since renormalizable interactions between
vectors must take the form of Yang-Mills theory. 
For abelian theories, ``light-by-light'' scattering could lead to qualitatively similar dynamics to those discussed here, but has to be investigated in the context of a low cutoff and potential production of the charged particles which give rise to the vector self-interaction. Beyond self-interactions, a simple example of both theoretical and
phenomenological interest is a light vector
interacting with the SM via a kinetic
mixing with the SM photon (though plasma
dynamics may make the behavior around astrophysical black holes
very complicated).
A vector may also have interactions with other
hidden sector states --- for example, its mass may come from
a Higgs mechanism, or it may mediate interactions between
hidden sector matter. 
For the purely gravitational story to hold,
such states must be sufficiently heavy, and/or sufficiently
weakly coupled~\cite{Baryakhtar:2017ngi}.
We leave investigations of such
scenarios to future work.

Superradiance of spin-2 particles
has also been investigated in the
literature~\cite{Brito:2013wya,Brito:2020lup}. 
An issue with such models is that an effective
field theory with a spin-2 particle
of mass $\mu$, along with the massless
graviton (a ``bigravity'' theory),
has a cutoff scale at or below $ \Lambda_3= (M_P \mu^2)^{1/3}$~\cite{Bonifacio:2017nnt,Bonifacio:2018aon}.
Here, $M_P \sim \min (\Mpl, \Lambda)$ is an effective mass scale
set by the mass scales $\Mpl$, which suppresses 
massless graviton interactions, and
$\Lambda$, which suppresses massive spin-2 interactions.
At the small masses $\mu$ we are interested
in for BH superradiance, $\Lambda_3 \lesssim 10 \eV \left(\frac{\mu}{
	10^{-12} \eV}\right)^{2/3}$ is small
compared to energy scales of interest. For example,
the energy density in a fully-occupied superradiant
cloud is $\rho \sim (6 \MeV)^4 
\left(\frac{\alpha}{0.2}\right)^5 \left(\frac{\mu}{10^{-12} \eV}\right)^2$.
Consequently, it is unclear whether there are theories
for which reliable calculations can be carried out
in the regimes of interest.

Returning to the topic of spin-0 superradiance;
as well as exploring the new observational signatures
that may arise from self-interactions,
our analyses clarify when self-interactions
are small enough not to affect the usual 
gravitational dynamics of superradiance.
As illustrated in Figures~\ref{figSpindown1} and \ref{figsmbh},
this is important for understanding when constraints
and signatures from motivated models,
such as the QCD axion or misalignment DM, can be trusted. 

As we have demonstrated, adding a simple quartic interaction can dramatically change the dynamics of scalar superradiance. The additional interaction inevitably reduces the efficiency of black hole spindown as well as the strength and timescale of gravitational wave annihilation signals. Nevertheless, the new dynamics can lead to simultaneous population of multiple levels giving rise to gravitational wave transition signals, a narrow range of which may be observable at LIGO/Virgo. Given that the transition signals are at parametrically lower frequencies corresponding to the energy splitting between different levels, signals from scalars around stellar mass black holes generally fall below the LIGO/Virgo sensitivity band in frequency and present new targets for future mid-band detectors.

Perhaps the most novel signature is the emission of particles to infinity: a light, self-coupled axion can extract the energy of rotating  black holes and populate our galaxy with axion waves, without the need for a cosmological abundance or a coupling to Standard Model matter. In the presence of such a coupling, these axion waves could be detected in the lab. While current experiments are not yet sensitive to this population of light axions, this mechanism further motivates the development of light axion direct detection experiments, as well as numerical work on self-interactions in superradiance to better characterize the signal from compact, semi-relativistic axion clouds.


\begin{acknowledgments}
	We thank Asimina Arvanitaki, Savas Dimopoulos, Sergei Dubovsky, Peter Graham,  
Kurt Hinterbichler, Junwu Huang, Ken Van Tilburg,
	and Sylvia Zhu for helpful discussions. We thank Perimeter Institute for warm hospitality during the completion of part of this work. Research at Perimeter Institute is supported by the Government of Canada through Industry Canada and by the Province of Ontario through the Ministry of Economic Development \& Innovation. MB is supported by the James Arthur Postdoctoral
	Fellowship. MG and RL are supported in part by the National Science Foundation under Grant No. PHYS-1720397, and the Gordon and Betty Moore Foundation Grant GBMF7946. OS is supported by the \emph{Fonds de recherche du Qu\'ebec
	Nature et Technologies} and by a DARE Fellowship from the Vice Provost for
	Graduate Education at Stanford University. 
\end{acknowledgments}

\appendix


\section{Parametric oscillator analysis}
\label{app:param}

As discussed in section~\ref{secboundints}, 
a useful way to analyse the growth of bound
levels is to assume that we have some large-amplitude
$\psi_c$, and to treat this as a parametric
forcing in the Gross-Pitaevskii (GP) equation
(Eq.~\eqref{eq:eqgp1}), i.e.\ to solve
\begin{equation}
	(i \partial_t + \MM)\psi = \frac{-3 \lambda}{24 \mu^2}(\psi_c^2 \psi^*
	+ |\psi_c|^2 \psi)
\end{equation}
(here, $\mathcal{M}$ represents the terms
in the non-relativistic Hamiltonian, including
an absorbing term corresponding to the BH horizon).
As compared to the forced oscillation
analysis in section~\ref{secboundints}, we 
ignore back-action for only two of the ``legs'' 
in diagrams such as Eq.~\eqref{eq1120},
rather than for three of them.

To simplify our discussion, we will take
$\psi_c \propto \psi_{211}$ (so we are interested
in processes such as Eq.~\eqref{eq1120}).
It is helpful to extract the time dependence
corresponding to the 211
oscillation, and write $\psi = \Psi e^{- i \tilde \omega_c t}$,
where $\tilde \omega_c \equiv \tilde \omega_{211}$
(for simplicity, we will assume that $\omega_{211}$ is real,
as it is when 211 has reached its saturation value).
Then, if we take a harmonic
ansatz, $\Psi = A e^{- i \hat \omega t} + B
e^{i \hat \omega^* t}$, the GP equation 
\begin{equation}
	\left(i \partial_t + \tilde \omega_c + \MM\right) \Psi 
	=  \tilde \lambda (\Psi_{211}^2 \Psi^*
	+ |\Psi_{211}|^2 \Psi)
\end{equation}
(where $\tilde \lambda \equiv - \frac{3 \lambda}{24 \mu^2}$)
implies that
\begin{equation}
	(\hat \omega + \tilde \omega_c + \MM) A =  
	\tilde \lambda (\Psi_{211}^2 B^* + |\Psi_{211}|^2 A)
	\label{eqgpa}
\end{equation}
and
\begin{equation}
	(-\hat \omega^* + \tilde \omega_c + \MM) B =  
	\tilde \lambda (\Psi_{211}^2 A^* + |\Psi_{211}|^2 B)
	\label{eqgpb}
\end{equation}
If we take the complex conjugate of Eq.~\eqref{eqgpb},
then together with Eq.~\eqref{eqgpa}, we have a linear
eigenvalue problem that we can solve for $\hat \omega$.
For $\lambda = 0$, the solutions
correspond to usual hydrogenic (quasi-bound) states.

For non-Hermitian Hamiltonians, the eigenstates are
generally non-orthogonal~\cite{Sternheim:1972zz}. However, in our case, we
can write $\MM = \MM_R + i \MM_I$, and treat
$\MM_I$ as being diagonal in the basis
of $\MM_R$ eigenstates (that is, we ignore
the detailed dynamics behind the absorption,
since this is outside the regime of the non-relativistic
approximation). In this case, the ($\lambda = 0)$
quasi-bound states $\Psi_k$ are orthogonal~\cite{Sternheim:1972zz},
and we will assume the normalization
$\int dV \Psi_k^* \Psi_j = \delta_{jk}$.

To linear order in $\lambda$, if we start with the unperturbed
solution $A = \Psi_i$, $B = 0$, then we can write
the perturbed solution as $A = \Psi_i + \sum_k \alpha_k \Psi_k$,
$B = \sum_k \beta_k \Psi_k$ (expanding in the unperturbed
basis). Using equations~\eqref{eqgpa} and~\eqref{eqgpb},
\begin{equation}
	(\hat \omega_i - \hat \omega_k) \alpha_k = \tilde \lambda
	\int dV \Psi_k^* |\Psi_{211}|^2 \Psi_i
	\label{eqap1}
\end{equation}
\begin{equation}
	(- \hat \omega_i^* - \hat \omega_k) \beta_k = \tilde \lambda
	\int dV \Psi_k^* \Psi_{211}^2 \Psi_i^*
	\label{eqbp1}
\end{equation}
As well as these perturbations to the wavefunction,
we are interested in finding the perturbation to the frequency
$\hat \omega$ of the state. Writing $\hat \omega = \hat \omega_i + \delta \hat \omega$,
we have 
\begin{equation}
	(\hat \omega_i + \tilde \omega_c + \MM) A
	= - \delta \hat \omega A + \tilde \lambda (\Psi_{211}^2 B^* + |\Psi_{211}|^2 A)
\end{equation}
If we take $A$ to be normalized so that $\int dV \Psi_i^* A = 1$ even for non-zero $\lambda$, then this implies that
\begin{align}
	\delta \hat \omega &= \tilde \lambda \int dV \Psi_i^* 
	(\Psi_{211}^2 B^* + |\Psi_{211}|^2 A) \\
	&= \tilde \lambda \int dV \Psi_i^* |\Psi_{211}|^2 \Psi_i \\
	& - \tilde \lambda^2 \sum_k \frac{1}{\hat \omega_i + \hat \omega_k^*}
	\left|\int dV \Psi_k^* \Psi_{211}^2 \Psi_i^*\right|^2
	\\
	& + \tilde \lambda^2 \sum_k \frac{1}{\hat \omega_i - \hat \omega_k}
	\left|\int dV \Psi_k^* |\Psi_{211}|^2 \Psi_i^*\right|^2
	\label{eqdw2}
\end{align}
The second and fourth lines of this expression
give behaviour similar to standard perturbation theory.
However, the $-1/(\hat \omega_i^* + \hat \omega_k)$ factor
in the second line gives rise to qualitatively different effects.
If the $\Psi_i$ mode is decaying, but
the $\Psi_k$ mode is damped sufficiently
strongly that ${\rm Im}(\hat \omega_i + \hat \omega_k^*) > 0$, then
${\rm Im}\left(\frac{-1}{\hat \omega_i + \hat \omega_k^*}\right) > 0$.
Consequently, the ``mixing'' with the $\Psi_k$ mode
contributes a \emph{growing} term to the perturbed
$\Psi_i$ mode.
In our case, the 211 parametric forcing gives the 322
mode a ``mixing'' with the decaying 100 mode (and the $n00$ modes, etc),
contributing a growing term for 322.
The perturbations to the 322 wavefunction
correspond to the forced oscillation discussed in
section~\ref{secboundints}.
Using Eq.~\eqref{eqdw2}, we obtain
the same 322 growth rate as calculated
from the forced-oscillation picture.

Similarly, mixing with superradiant (rather than decaying)
modes contributes a negative imaginary part to $\delta \hat \omega$.
This is again as we'd expect from the forced oscillation
picture.
Including a growing 211 occupation number,
as is appropriate when 211 is still superradiant,
leads to more complicated expressions. However,
since the superradiant growth timescale is always
much longer than the oscillation period of $\psi_{211}$,
we can separate these timescales, with 322
growth at a particular time being driven by the 211 amplitude
at that time.

\begin{figure}[t!]
	\begin{centering}
		\includegraphics[width=\columnwidth]{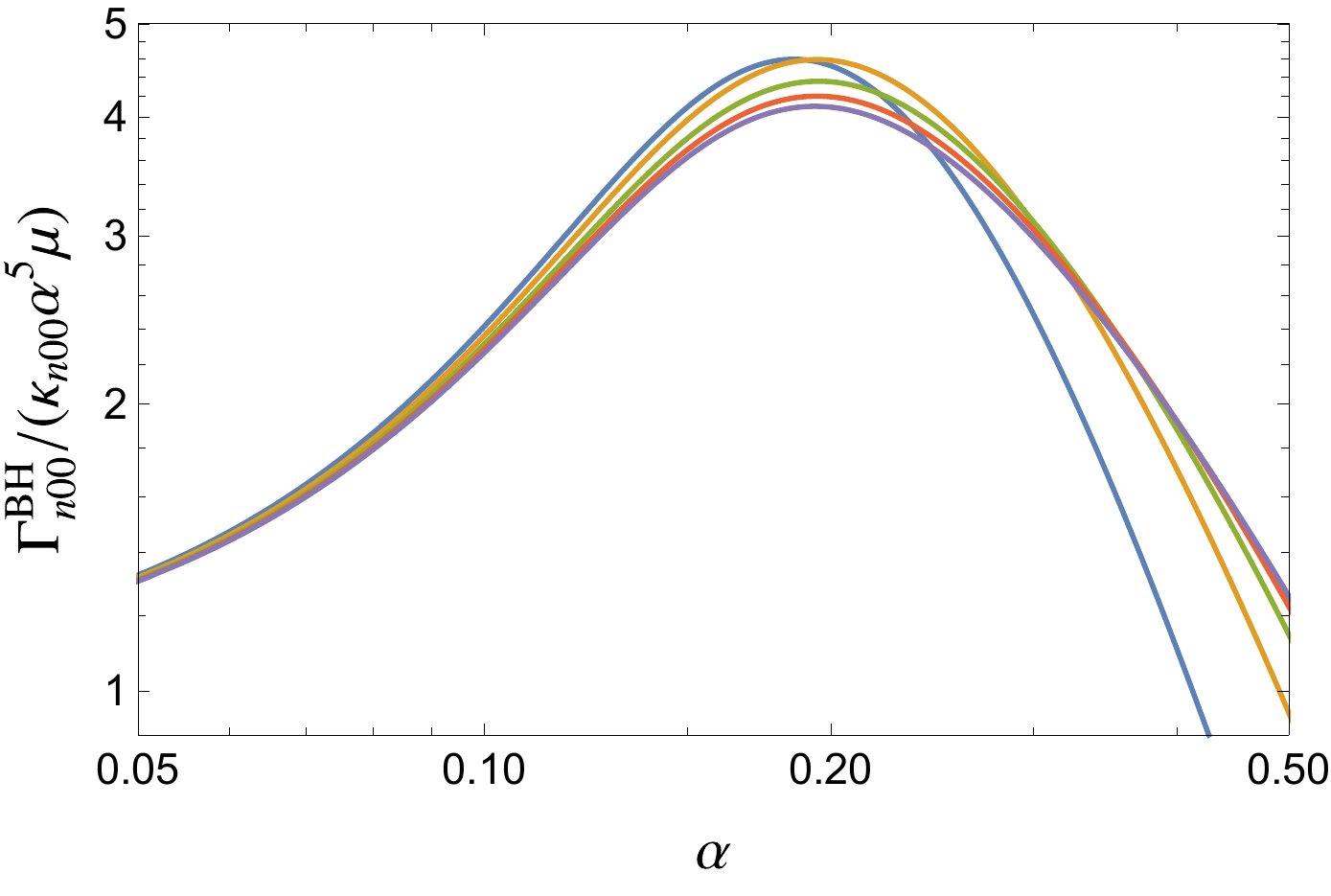}
		\includegraphics[width=\columnwidth]{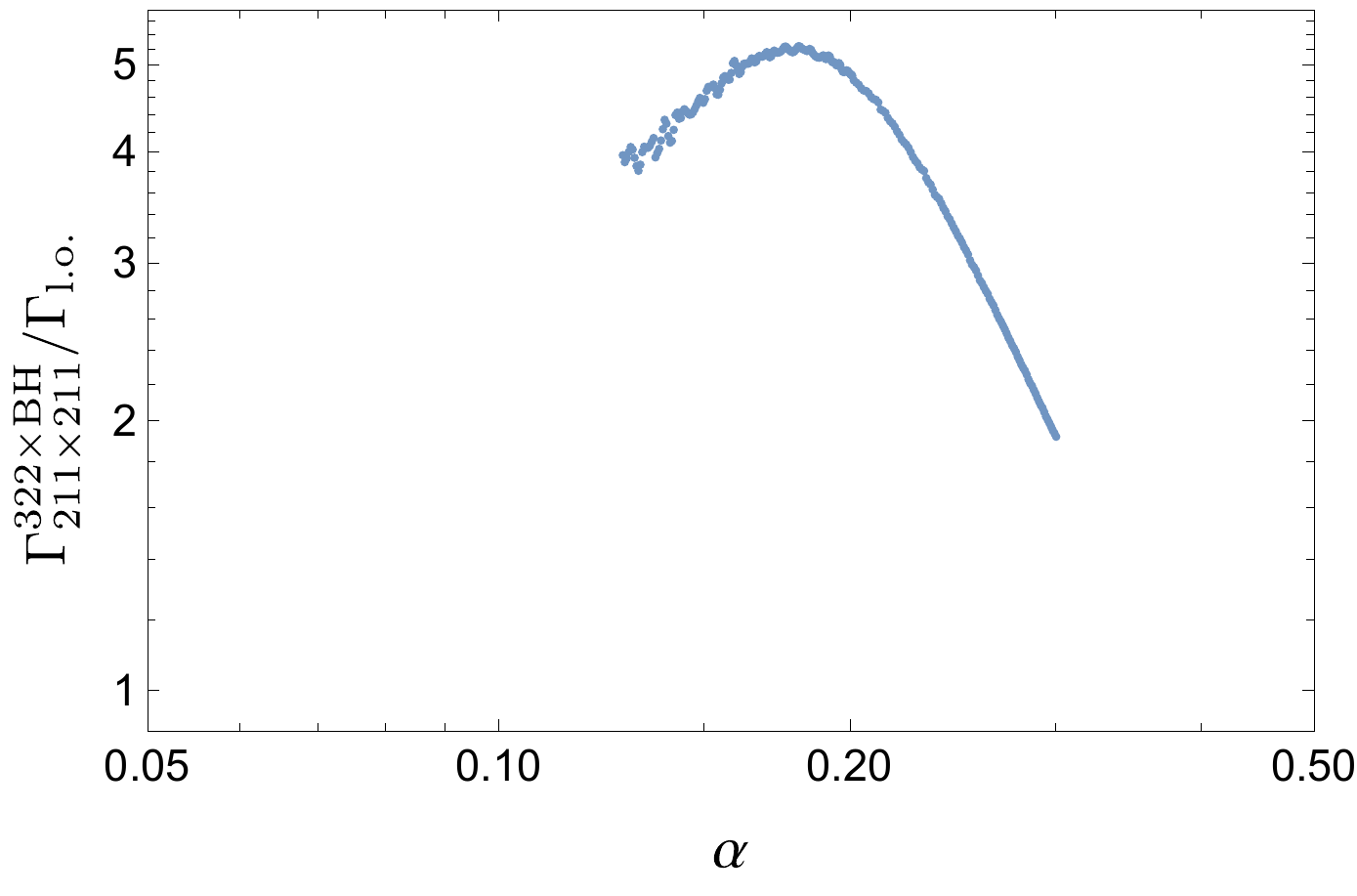}
		\caption{\emph{Top panel:} decay rates of the $n00$ hydrogenic
		levels, for $n=1$ to 5, relative
		to their leading-order power-law behaviour
		as a function of $\alpha$ (for a BH with $a_* = 0.9$).
		\emph{Bottom panel:} rate of the 
		$211 \times 211 \rightarrow 322 \times {\rm BH}$
		process,
		for a BH with $a_* = 0.9$, relative
		to its leading power-law behaviour $\Gamma_{\rm l.o.}$
		as a function of $\alpha$ 
		(see Table~\ref{tablebound}).
		As discussed in section~\ref{secboundints},
		the deviation of this rate from its leading-order
		form is mostly driven by the same short-distance
		effects that modify the $n00$ decay rates.}
	\label{fig1120rate}
	\end{centering}
\end{figure}

This kind of perturbative analysis can be applied
in the hydrogenic approximation (at leading order in $\alpha$),
or using numerical wavefunctions for the bound states,
which will be more accurate at higher $\alpha$.
In Fig.~\ref{fig1120rate}, we plot the 
decay rate of the 100 level, relative to its
leading-$\alpha$ power-law behavior (the $n00$ levels
have very similar behavior).
The lower panel of the figure also shows a numerical approximation
to the rate of the $211 \times 211 \rightarrow 322 \times
{\rm BH}$ process, computed by numerically integrating
the forced equation of motion in the Kerr background
(for practical reasons, over a restricted range
in $\alpha$). The close correspondence between the behaviours
of these two rates illustrates that, for the $211 \times 211 \rightarrow
322 \times {\rm BH}$ process, the most significant
high-$\alpha$ corrections come from short-distance
effects that affect the flux across the horizon;
at long distances from the BH, the forcing term,
and the forced oscillation, are not strongly affected
(at the $\alpha$ of interest).

The parametric forcing analysis above is not specific
to black hole superradiance. In the simplest case, if we had two oscillators,
with an oscillating coupling between them,
\begin{equation}
	\ddot x + \omega_0^2 x = f \cos (2 \omega_c t) y
\end{equation}
\begin{equation}
	\ddot y + \gamma \dot y + \omega_1^2 y = f \cos (2 \omega_c t) x
\end{equation}
then the same kind of analysis would apply.
In the absence of the damping term $\gamma$,
if $\omega_0 + \omega_1$ is detuned from $2 \omega_c$,
then the system is not unstable to growth.
Introducing $\gamma$ leads to the exponential growth of $x$,
as per above.

While the above analyses were at the level of classical
equations, a similar analysis could be done in terms
of quantum master equations. The most important physical difference
is that, while the classical ground state is stationary,
quantum fluctuations are amplified by the instability,
so the ground states evolves into a probability mixture
of coherent states.
This is precisely analogous to the amplification of
quantum fluctuations by superradiance, as discussed
in section~\ref{secspin0}.

From Eq.~\eqref{eqbp1}, our perturbative
treatment breaks down when
\begin{equation}
	\frac{\tilde \lambda \int dV \Psi_k^* \Psi_{211}^2 \Psi_i^*}{\hat \omega_i^* + \hat \omega_k} \gtrsim 1
	\label{eqb6}
\end{equation}
In terms of the physical mode frequencies,
$\hat \omega_i^* + \hat \omega_k = \omega_i^* + \omega_k - 2 {\rm Re} \, \omega_{211}$. 
For generic hydrogenic modes, this 
is $\OO(\alpha^2) \mu$, and in this case,
the LHS of Eq.~\eqref{eqb6}
is parametrically $\sim \alpha^3 \frac{M_{\rm pl}^2}{f^2}
\varepsilon_{211}$, similarly to the self-energy
corrections (Eq.~\eqref{eq:eqcorrect1}). As we
discuss in section~\ref{sec:nonperturb},
the largest value that $\varepsilon_{211}$ attains
decreases as we decrease $f$, and the numerical value
of this quantity is always small.

In special cases, the source term for the $k$ oscillation
can be almost on resonance, and the denominator
can become smaller. We discuss a specific
example in section~\ref{secboundints} (Eq.~\eqref{eq1120res}),
where it is $\OO(\alpha^4) \mu$
for the $211 \times 311 \rightarrow 322 \times$ BH process. However, in this case,
the source term does not appear to grow large enough for
there to be a problem, in most of the parameter space
of interest.

It is also possible to treat emission to infinity,
e.g. through the $322 \times 322 \rightarrow 211 \times \infty$
process discussed in section~\ref{secnonrelem}, 
in terms of a parametric forcing, with loss to infinity acting as
like a damping term. 


\section{Perturbative calculations of frequency shifts and rates}
\label{appcalcs}

In this appendix, we will provide more detailed derivations of the
leading-$\alpha$ rates for quartic self-interaction processes
involving hydrogenic levels.

Up to corrections from self-gravity, the system obeys the classical equation of motion
\begin{equation}
\left(D^2 - \mu^2\right)\varphi = -\frac{\lambda}{6}\varphi^3,
\end{equation}
where $D^2 = D_\nu D^\nu$ and $D_\nu$ is the covariant derivative of the Kerr geometry. Expending $D^2$ to first order in $r_g / r$, this becomes
\begin{equation}
\left(\pdd{}{t}-\vec\nabla^2 + \mu^2\right)\varphi-\frac{2\alpha}{r}\left(\mu+\hat K\right) = \frac{\lambda}{6}\varphi^3.
\end{equation}
The term
\begin{align}
\begin{split}
\hat K ={}& \frac{1}{r}\pd{}{r}-2\left(\mu^2+\pdd{}{t}\right)-\frac{\hat {L}^2}{r^2}
\end{split}
\end{align}
is parametrically suppressed relative to $\mu$ for non-relativistic components of $\varphi$, and we drop $\hat K$ except for calculations  of relativistic emissions. Here $\hat L^2$ denotes the total angular momentum operator: $\hat L^2 Y^m_{l} = l(l+1)Y^m_l$ for the Laplace spherical harmonics $Y^m_l(\theta,\varphi)$. \par
We seek a perturbative solution in the self-interaction parameter $\lambda$,
\begin{align}
\begin{split}
\varphi = \varphi^{(0)} + \lambda \varphi^{(1)} + \dots
\end{split}
\end{align}
At zeroth order, 
\begin{align}
\begin{split}
\left(\pdd{}{t}-\vec\nabla^2 +\mu^2-\frac{2\alpha \mu}{r}\right)\varphi^{(0)}= 0.
\end{split}
\end{align}
This equation admits non-relativistic  (quasi-)bound states
with hydrogenic waveforms and energies which we identify with the superradiant cloud:
\begin{align}
\begin{split}
\label{eq:source}
\varphi^{(0)} \equiv \sum_{nlm}\varphi_{nlm}^{(0)}=\sum_{nlm}\sqrt{\frac{N_{nlm}}{2\mu}}e^{-i\omega_{nlm}t}\psi_{nlm}+\text{c.c.},
\end{split}
\end{align}
up to phases, where $\psi_{nlm}$ are the normalized hydrogenic wavefunctions $\int |\psi_{nlm}|^2d^3\vec r =1.$

To avoid secular terms at the next perturbative order, we must also introduce a perturbation series for the normal frequencies:
\begin{align}
\begin{split}
\omega_{nlm}=\omega_{nlm}^{(0)}+\lambda \omega_{nlm}^{(1)}+\dots,
\end{split}
\end{align}
where 
\begin{align}
\begin{split}
\omega_{nlm}^{(0)} = \omega_n+ i \Gamma^\text{SR}_{nlm},
\end{split}
\end{align}
\begin{align}
\begin{split}
\omega_n \approx \mu\left(1-\frac{\alpha^2}{2n^2}\right),
\end{split}
\end{align}
and  $\Gamma^\text{SR}_{nlm}$ is the superradiance rate. We call the energy corrections $\Delta\omega_{nlm}\equiv\lambda\omega_{nlm}^{(1)}$

At first order in perturbation theory, this gives  a driven massive Coulomb wave equation,
\begin{align}
\begin{split}
\label{eq:first_order_equation}
\left(\pdd{}{t}-\vec\nabla^2 +\mu^2-\frac{2\alpha \mu}{r}\right)\varphi^{(1)}\\
= \frac{1}{6}\left(\varphi^{(0)}\right)^3+\sum_{nlm}2\mu\omega_{nlm}^{(1)}\varphi_{nlm}^{(0)}.
\end{split}
\end{align}
Plugging \eqref{eq:source} into \eqref{eq:first_order_equation}, and expanding the driving term as a sum of harmonic driving terms gives
\begin{align}
\begin{split}
(\varphi^{(0)})^3 \sim \sum_\Omega f(\vec r,\Omega)e^{-i\Omega t}+\text{c.c.}, \quad \Omega > 0.
\end{split}
\end{align}
Since $\varphi^{(0)}\sim a_0^{-3/2}\mu^{-1/2}\sim \alpha^{3/2}\mu$, the source
$f(\vec{r})\sim (\varphi^{(0)})^3$ scales as $\sim\alpha^{9/2}\mu^3.$ The
physical intuition behind that scaling is that a cloud with larger $\alpha$ has
a smaller characteristic size $a_0$ and therefore larger densities, enhancing
the rate of many-body processes.

The physical nature of the process associated to each summand depends on the value of $\Omega$:
\begin{enumerate}
    \item $\Omega -\mu > 0$ corresponds to \emph{free} radiation emitted in the continuum and travelling to infinity either with non-relativistic or relativistic velocities,
    \item $\Omega - \mu < 0$ and $\Omega \neq \omega_n$ for all $n$ is off-resonant driving of  discrete \emph{bound} modes, i.e.  the production of off-shell particles trapped in the gravitational well
    \item $\Omega = \omega_n$ for some $n$ is resonant driving, which either corresponds to resonant (on-shell) production of particles inside the cloud, or to a correction to the frequencies (one-particle energies) and waveforms (one-particle states) of the zeroth-order normal modes.
\end{enumerate}

For clarity, we focus on the source
\begin{align}
\begin{split}
\label{eq:source_explicit}
\varphi^{(0)}(\vec r,t)=\sqrt{\frac{N_{211}}{2\mu}}e^{-i\omega_2 t}\psi_{211}(\vec r)\\
+\sqrt{\frac{N_{322}}{2\mu}}e^{-i\omega_3 t}\psi_{322}(\vec r)+\text{c.c.}
\end{split}
\end{align}
for the remainder of this appendix. The source \eqref{eq:source_explicit} represents the only two levels of the cloud relevant to the intra-cloud dynamics at small enough $\alpha$, as argued in Sec.~\ref{sec:higherlevels} and App.~\ref{app:ndependence}.

\subsection{Frequency corrections}
\label{app:freqcorr}
The source term includes components at the frequency $\omega_2$ of the 211 bound state,
\begin{align}
\frac{1}{6}\left(\varphi^{(0)}\right)^{3} &\supset \frac{e^{-i\omega_2 t}}{(2\mu)^{3/2}}\times\\
&\left(\frac{1}{2}N_{211}^{3/2}\left(\psi_{211}\right)^2\psi_{211}^*+N_{322}\sqrt{N_{211}}\psi_{322}\psi_{322}^*\psi_{211}\right)\nonumber.
\end{align} 
This source contains components in resonance with the normal mode $\psi_{211}e^{-i\omega_2 t}$ 
which would drive $\varphi^{(1)}$ to very large amplitudes, preventing a perturbative treatment. The frequency correction $\omega_{211}^{(1)}$ is therefore determined by demanding that those resonant components be exactly cancelled:

\begin{align}
\label{eq:se211}
\begin{split}
{}&\omega_{211}^{(1)} =-\frac{1}{4\mu^2}\times\\ {}&\int\left(\frac{N_{211}}{2}|\psi_{211}|^4+N_{322}|\psi_{322}|^2|\psi_{211}|^2\right)d^3\vec r .
\end{split}
\end{align}

The two terms in Eq.~\eqref{eq:se211} correspond to self-energy corrections of the level 211 from its
interaction with itself and with 322, respectively. The integral can be
computed analytically by using the explicit form of the hydrogenic waveforms
$\psi_{211}$ and $\psi_{322}.$ 

Since bound state wavefunctions scale as $\psi_{nlm}\propto 1/a_0^{-3/2} \propto (\alpha\mu)^{3/2}$ and only depend on $\vec r$ through $r/a_0$, the frequency correction scales with  $\alpha$ as $\omega_{nlm}^{(1)}\sim \alpha^3\mu$. A denser cloud gives larger frequency corrections.

We calculated the integral of Eq.~\eqref{eq:se211} and the equivalent for $\omega_{322}^{(1)}$ and we found the corrections:

\begin{align}
    \label{eq:secorrections}
    \Delta\omega_{211}&\simeq-\lambda\alpha^3\mu\pare{1.2\times 10^{-4}N_{211}+3.5\times 10^{-5}N_{322}}\nonumber\\
    &=-\alpha^5\mu\pare{\frac{\Mpl}{f}}^2\pare{1.2\times 10^{-4}\varepsilon_{211}+3.5\times 10^{-5}\varepsilon_{322}}\\
    \Delta\omega_{322}&\simeq-\lambda\alpha^3\mu\pare{3.5\times 10^{-5}N_{211}+1.4\times 10^{-5}N_{322}}\nonumber\\
    &=-\alpha^5\mu\pare{\frac{\Mpl}{f}}^2\pare{3.5\times 10^{-5}\varepsilon_{211}+1.4\times 10^{-5}\varepsilon_{322}}
\end{align}

\subsection{$l=0$ damped-driven oscillation}
\label{appdd}
When $\Omega - \mu < \mu$ and $\Omega \neq \omega_n$ for any $n$, the source
generates a forced bound oscillation which is damped by the BH. For example,
when the cloud consists of particles in the 211 and 322 levels
\eqref{eq:source_explicit}, the frequency of the forced oscillation is
$\omega_\text{ind} = 2\omega_2-\omega_3 = \mu(1-7\alpha^2/36)<\mu$, so the
oscillation is bound.

The bound state  $\varphi^{(1)} \supset e^{-i\omega_\text{ind}t}\Psi^{(1)}(\vec r)+\text{c.c}$. satisfies the time-independent equation for the {complex} field $\Psi^{(1)}$,
\begin{align}
\begin{split}
\label{eq:driven_oscillation}
{}&\left(k_\text{ind}^2-\vec\nabla^2 -\frac{2\alpha \mu}{r}\right)\Psi^{(1)}(\vec r)=\frac{1}{2}\frac{N_{211}\sqrt{N_{322}}}{(2\mu)^{3/2}}(\psi_{211})^2\psi^*_{322},
\end{split}
\end{align}
where $k_\text{ind}^2=\mu^2-\omega_\text{ind}^2\approx (7/18)\alpha^2\mu^2$.

We expand $\Psi^{(1)}(\vec r)$ in the complete basis of the hydrogenic differential operator $-\nabla^2 - 2\alpha\mu/r$,
\begin{align}
\begin{split}
\label{eq:basis_decomposition}
\Psi^{(1)}(\vec r)= \sum_{nlm}c_{nlm}\psi_{nlm}+\sum_{lm}\int dk c(k)\psi_{klm},
\end{split}
\end{align}
where the eigenfunctions $nlm$ of the discrete spectrum satisfy
\begin{align}
\begin{split}
\label{eq:discrete_eigenvalue_equation}
{}&\left(-\nabla^2 - \frac{2\alpha\mu}{r}\right)\psi_{nlm}=-k^2_n \psi_{nlm},\quad k^2_n = \frac{\alpha^2\mu^2}{n^2},
\end{split}
\end{align}
with $n$ a positive integer, and eigenfunctions $klm$ of the continuous spectrum obey
\begin{align}
\begin{split}
\label{eq:continuous_eigenvalue_eigenvalue}
{}&\left(-\nabla^2 - \frac{2\alpha\mu}{r}\right)\psi_{klm}=k^2 \psi_{klm},\, \frac{k}{\alpha \mu}\in (0,+\infty).
\end{split}
\end{align}
Moreover, the eigenfunctions obey orthonormality conditions:
\begin{subequations}
\label{eq:orthonormality}
\begin{align}
{}&\int d^3 \vec r \psi^*_{nlm}\psi_{n'l'm'} = \delta_{n,n'}\delta_{m,m'}\delta_{l,l'},\\
{}&\int d^3 \vec r \psi^*_{klm}\psi_{k'l'm'} = 2\pi\delta(k'-k)\delta_{m,m'}\delta_{l,l'},\\
{}&\int d^3 \vec r \psi^*_{klm}\psi_{nl'm'} = 0.
\end{align}
\end{subequations}
Explicitly, the states of the discrete spectrum are the usual bound hydrogenic wavefunctions,
\begin{align}
\begin{split}
{}&\psi_{nlm}(r,\theta,\varphi)= R_{nl}(r)Y^m_l(\theta,\varphi),
\end{split}
\end{align}
with the radial part
\begin{align}
{}&R_{nl}(r)= \sqrt{\left(\frac{2}{na_0}\right)^3\frac{(n-l-1)!}{2n(n+l)!}}\\{}&\times\exp{\left[\frac{-r}{na_0}\right]}\left(\frac{2r}{na_0}\right)^lL_{n-l-1}^{2l+1}\left[\frac{2r}{na_0}\right],\nonumber
\end{align}
where $L^{2l+1}_{n-l-1}(x)$ is the generalized Laguerre polynomial of degree $n-l-1$.

The states of the continuous spectrum are stationary Coulomb waves \cite{Landau1977},\footnote{this is appropriate in the hydrogenic
approximation, where we take into account the
the Newtonian $1/r$ gravitational potential. 
Corrections from the full Kerr potential will
be higher order in $\alpha$.}
\begin{align}
\begin{split}
{}&\psi_{klm}(r,\theta,\varphi)= R_{kl}(r)Y^m_l(\theta,\phi)
\end{split}
\end{align}
with the radial part
\begin{align}
\begin{split}
{}&R_{kl}(r)=
\frac{2ke^{\pi/(2ka_0)}|\Gamma(l+1-i/(ka_0))|}{(2l+1)!}\\{}&\times(2kr)^le^{-ikr}{}_1F_1\left(i/(ka_0)+l+1,2l+2,2ikr\right),
\end{split}
\end{align}
where ${}_{1}F_1$ is the confluent hypergeometric function of the first kind.

To obtain the coefficients $c_{nlm}$, we put \eqref{eq:basis_decomposition} in \eqref{eq:driven_oscillation} and integrate both sides against $\psi_{n'l'm'}^*$. We can then use the Hermiticity of $(-\nabla^2-2\alpha\mu/r)$ (which in this case amounts to integrating by parts, so that $-\nabla^2-2\alpha\mu/r$ acts on $\psi_{n'l'm'}^*$), along with \eqref{eq:discrete_eigenvalue_equation} and \eqref{eq:orthonormality} to find
\begin{align}
\begin{split}
{}&c_{nlm} = 
\frac{1}{k_\text{ind}^2-k^2_n}\times\\{}&\int \frac{1}{2}\frac{N_{211}\sqrt{N_{322}}}{(2\mu)^{3/2}}(\psi_{211})^2\psi^*_{322} \psi_{nlm}^*d^3\vec r.
\end{split}
\end{align}
Similarly, the values of the transform $c(k)$ are obtained by integrating both sides of \eqref{eq:driven_oscillation} against $\psi_{k'l'm'}$. The analogue procedure then yields
\begin{align}
\begin{split}
{}&c(k) = 
\frac{1}{2\pi}\frac{1}{k_\text{ind}^2+k^2}\times\\{}&\int \frac{1}{2}\frac{N_{211}\sqrt{N_{322}}}{(2\mu)^{3/2}}(\psi_{211})^2\psi^*_{322} \psi_{klm}^*d^3\vec r.
\end{split}
\end{align}
It is appropriate to do these integrals in units of the Bohr radius $a_0 =
(\alpha \mu)^{-1}$ to reconstitute the dependence on $\alpha$. The prefactors
of $k_\text{ind}$, $k_n$ and $k$ are naturally in units of $a_0^{-1}$, while
bound state wavefunctions are in units of $a_0^{-3/2}$ and continuum
wavefunctions are in units of $a_0^{-1}$. The $c_{nlm}$'s then have dimension
$a_0^{-1}\mu^{-3/2}$ and $c(k)$ has units of $a_0^{-1/2}\mu^{3/2}$. The
amplitude of the induced oscillation $\varphi^{(1)}$ therefore has units of
$a_0^{-5/2}\mu^{-3/2} = \alpha^{5/2}\mu$.

These overlap integrals are non-vanishing for $l=0,2,4$ and $m=0$. For
$l>0$ however, the angular momentum barrier suppresses 
the field amplitude at the horizon, and therefore the corresponding rates of
absorption are smaller. This in turn leads to
a smaller induced growth rate, as discussed previously.
We therefore focus on $l=m=0$ and ignore the $l=2,4$ terms.

For $l=0$ states, the power absorbed at the horizon in terms of complex field $\Psi^{(1)}$ goes as the square of the norm at the origin:
\begin{align}
\begin{split}
P_\text{abs} \approx 4\alpha^2(1+\sqrt{1-a_*^2})\lambda^2|\Psi(0)^{(1)}|^2.
\end{split}
\end{align}
In terms of particles in the cloud carrying energy $\approx \mu$, this contributes
\begin{align}
\begin{split}
\dot N_\text{c} \supset- \Gamma_\text{damp}N_{211}^2N_{322},
\end{split}
\end{align}
with
\begin{align}
\begin{split}
\Gamma_\text{damp}=\frac{P_\text{abs}}{\mu N_{211}^2N_{322}}.
\end{split}
\end{align}

\subsection{Non-relativistic emission}
\label{appnonrelem}
Generally, the source term $(\varphi^{(0)})^3$ will generate some driving terms
oscillating at the frequency
$\omega_r^\text{NRE}=\omega_n+\omega_{n'}-\omega_{n"}$. When
$\omega_r^\text{NRE} > \mu $, the driven oscillation is free. These free
emissions are non-relativistic because $\omega_\text{r}^\text{NRE} \approx
\mu+\mathcal O (\alpha^2)$ for the constituents of the superradiant cloud. The
superscript NRE (``non-relativistic emissions'') will be suppressed will be
suppressed for the remainder of this section.

Generically, we seek to solve
\begin{align}
\begin{split}
\left(\pdd{}{t}-\vec \nabla^2+\mu^2-\frac{2\alpha \mu}{r}\right)\varphi_r = e^{-i\omega_\text{r}t}f(\vec r)+\text{c.c.},
\end{split}
\end{align}
where $e^{-i\omega_\text{r}t}f(\vec r)$ is a localized source of radiation with harmonic time-dependence, and $\varphi_r\subset \varphi^{(1)}$ is the “radiation" part of the field. The time-averaged differential power per solid angle that such a source emits in the radiation zone at infinity in the direction $(\theta_{\vec{k}},\varphi_{\vec{k}})$ is
\begin{align}
\begin{split}
\label{eq:differential_power_emitted}
\frac{d\langle P\rangle}{d\Omega}\left(\theta_{\vec{k}},\varphi_{\vec{k}}\right) = 2\frac{\omega_r |\vec k|}{(4\pi)^2}\lambda^2|\tilde f(\vec k)|^2,
\end{split}
\end{align}
where $\vec k = (\sqrt{(\omega_r)^2-\mu^2})\hat r$ is the momentum at spatial
infinity, $\hat r$ is a radial unit vector pointing in the direction
$(\theta_k,\varphi_k)$ and $\tilde f(\vec k)$ is the ``Coulomb'' transform 
\begin{align}
\begin{split}
\label{eq:Coulomb_transform}
{}&\tilde f(\vec k) =\sum_{lm}Y^m_l(\theta_k,\varphi_k) \int d^3 \vec r(4\pi)(-i)^l f(\vec r)\frac{\psi_{klm}^*\left(\vec r\right)}{2k}.
\end{split}
\end{align}
This is analogous to the usual Fourier
transform that one would compute for the emission rate in
flat spacetime, with the regular spherical
Bessel functions having been replaced
by the appropriate regular Coulomb
waves.

For non-relativistic emissions, $k\sim \mathcal O(a_0^{-1})$.
It was noted earlier that $f(\vec r)\sim \alpha^{9/2}\mu^3$. Furthermore, since
it is a product of hydrogenic wavefunctions, $f(\vec r)$ depends on $\vec r$
only through the combination $\vec r/a_0$. On the other hand, $\psi_{klm}(\vec
r)/2k$ is dimensionless and depends on $\vec r$ only through the combination $k
r \sim r/a_0$, for non-relativistic $k$. Therefore, all the dependence of
\eqref{eq:Coulomb_transform} on $\alpha$ can be extracted by evaluating the
intgeral in units of the Bohr radius $a_0=(\alpha \mu)^{-1}$. Thus $\tilde
f(\vec k)\sim \alpha^{3/2}$ and $d\langle P\rangle/d\Omega \sim
\lambda^2\alpha^4 \mu^2.$

The total radiated power is determined by integrating \eqref{eq:differential_power_emitted} over solid angles:
\begin{align}
\begin{split}
P_r^\text{NRE} = \int d\Omega \frac{d\langle P \rangle}{d\Omega}
\end{split}
\end{align}
In terms of particles in the cloud, and particles radiated to infinity with energy $\omega_r \approx \mu$, we have
\begin{align}
\begin{split}
\dot N_\text{c}\supset -\Gamma_\text{r}^\text{NRE}N_{nlm}N_{n'l'm'}N_{n"l"m"},
\end{split}
\end{align}
with the rate
\begin{align}
\begin{split}
\Gamma_r^\text{NRE} = \frac{P_r^\text{NRE}}{\mu N_{nlm}N_{n'l'm'}N_{n"l"m"}}.
\end{split}
\end{align}

A particularly important process
is $211 \times 211 \rightarrow 322 \times \infty$. This
is sourced by
\begin{align}
\begin{split}
{}&\frac{1}{6}(\varphi^{(0)})^3 \supset e^{-i(2\omega_3-\omega_2)t}\times\\
{}& \frac{1}{2}\frac{N_{322}\sqrt{N_{211}}}{(2\mu)^{3/2}}(\psi_{322})^2\psi^*_{211}+\text{c.c.}
\end{split}
\end{align}
By substituting 
\begin{align}
\begin{split}
f(\vec r)\rightarrow \frac{1}{2}\frac{N_{322}\sqrt{N_{211}}}{(2\mu)^{3/2}}(\psi_{322})^2\psi^*_{211}
\end{split}
\end{align}
in the above, we obtain the rate in table~\ref{tab:paramsummary}.

\subsection{Relativistic emission}
\label{app:relemission}
The source term $(\varphi^{(0)})^3$ will also contain terms oscillating at the frequency $\omega_r^{RE} = \omega_n + \omega_{n''}+\omega_{n'''}$. When $\omega_r^{RE}>\mu$, the driven oscillation is free. These free emissions are relativistic because $\omega_r \approx 3\mu + \mathcal{O}(\alpha^2)$ for the constituents of the superradiant cloud. Cubic self-interactions would also generate relativistic emissions through $(\varphi^{(0)})^2$ in the equations of motion. In this case $\omega_r^\text{RE} \approx 2\mu + \mathcal{\alpha}^2$.

For the remainder of this section, the superscript RE (``relativistic emissions'') will be suppressed. 
As is the case for non-relativistic emissions, the radiated power is
controlled by the integral \eqref{eq:Coulomb_transform} which projects
the source onto the Coulomb scattering state with outgoing momentum
$\vec k$. The source $f(\vec r)$ is a product of hydrogenic
wavefunctions,
\begin{align}
	f(\vec r) &\simeq \frac{-\lambda/6}{(2 \mu)^{3/2}}
	N_{211}^{3/2}
	\psi_{211}^3 \nonumber \\
	&= \frac{\lambda}{768 \pi \sqrt{70}} \alpha^{3/2}  a_0^{-3} (r/a_0)^3 e^{-3 r/(2 a_0)} Y_{33} N_{211}^{3/2}
\end{align}
For non-relativistic emission, $k \sim a_0^{-1}$
so we need to use the full form of the Coulomb scattering state.
In contrast, for relativistic emission, $k \sim \mu \sim \alpha^{-1} a_0^{-1}$, so $k a_0 \sim \alpha^{-1}$ is a large parameter.
As a result, we can expand the radial part of the Coulomb
wavefunction around its flat-space,
spherical Bessel function form.

It turns out that the contributions to
$\tilde f (\vec k)$ from the spherical Bessel function,
and from the leading-$\alpha$ correction,
are at the same order in $\alpha$. This effectively occurs
due to the contribution from the spherical Bessel function
suffering a ``cancellation'', making it higher-order
in $\alpha$ than a naive guess based on the 
behaviour of $f(\vec r)$ near the origin
would have indicated. The integral against
the leading-$\alpha$ correction term
does not suffer this kind of cancellation,
making the contributions from both of the same order.
This is why our result for the emitted
power (Eq.~\eqref{eqrelpower}) has the same
$\alpha$ dependence as that derived in~\cite{Arvanitaki:2010sy} 
using a flat-space approximation, but has a larger
constant factor (\cite{Arvanitaki:2010sy} also
treats emission as light-like, taking $\omega_r^2 = k^2$
rather than $\omega^2 = k^2 + \mu^2$).

Higher-order corrections, and effects from working in the full
Kerr metric instead of just a $1/r$ potential, all 
contribute to the emitted power at higher order in $\alpha$.

\section{Mixing beyond 211 and 322}
\subsection{Selection rules for mixing with damped states}
As explained in App. \ref{app:param}, in the presence of a quartic self-coupling $\lambda$, one can view a background SR cloud $\varphi_\text{c}(\vec r,t) \sim e^{-i\mu t}\psi_{211}(t) + e^{-i\mu t}\psi_{322}(t) + \text{c.c}$ as providing a time-dependent mixing potential $V_\text{mixing}\sim \lambda e^{-i2\mu t}\left(\psi_{211}(t)^2 +\psi_{211}(t)\psi_{322}(t) +\psi_{322}(t)^2\right)$ between states. In particular, if the mixing matrix element $\bra{\psi_{n'l'm'}}V_\text{mixing}\ket{\psi_{nl m}^*}$ between a superradiant state $\psi_{nl m}$ and a \emph{decaying} state $\psi_{n'l'm'}$ is non-vanishing, then a forced oscillation $\propto \psi_{n'l'm'}$ is sustained and a growth instability is induced for $\psi_{nlm}$. 

We are therefore interested in the selection rules when $V_\text{mixing}\sim\psi_{211}^2 \sim Y^2_2$, $V_\text{mixing}\sim\psi_{211}\psi_{322} \sim Y^3_3$, and $V_\text{mixing}\sim\psi_{322}^2\sim Y^4_4$. In each case, $V_\text{mixing}\sim Y^{m''}_{l''}$ can be viewed as an element of an irreducible tensor operator representation of the rotation group with angular momentum numbers $(l'',m'')$.\footnote{Strictly
speaking, since the Kerr metric breaks spherical symmetry, $l$ is not a good quantum number (though $m$ is, since we still have axial symmetry). However, since the metric terms that break
spherical symmetry are suppressed at large $r/r_g$,
they lead to effects that are suppressed by more powers
of $\alpha$ in the hydrogenic limit.} Considering further that, $\psi_{nl m}^* \propto Y^{-m}_l$, then by the Wigner-Eckart theorem  $\bra{\psi_{n'l'm'}}V_\text{mixing}\ket{\psi_{nl m}^*}\propto(l,l'',-m,m''|
l'm')$, where $(j_1,j_2,m_1,m_2|
J,M)\equiv \langle{j_1,j_2,m_1,m_2}|{j_1,j_2,J,M}\rangle$ is the Clebsch-Gordon (CG) coefficient for the addition of two irreducible angular momentum representations $j_1$ and $j_2$. Furthermore, since the parity of a spherical harmonic $Y^{m}_l$ is $(-1)^{l}$, inserting parity transformations inside the matrix element yields $\bra{\psi_{n'l'm'}}Y^{m''}_{l''}\ket{\psi_{nl m}^*}= (-1)^{l+l'+l''}\bra{\psi_{n'l'm'}}Y^{m''}_{l''}\ket{\psi_{nl m}^*}$. 

From this we get the selection rules for an induced growth instability to develop:
\begin{enumerate}
    \item Mixing with a damped state: $m'\leq 0$,
    \item CG coefficient: $m'' = m' + m$,
    \item CG coefficient: $|l-l''| \leq l' \leq l+l''$,
     \item Invariance under parity: $l+l'+l'' = \text{even}$.
\end{enumerate}
The first rule assumes that the spin in the BH is such that $m\geq 1$ states are SR.

\subsection{Dependence of rates on the quantum numbers}

\subsubsection{Dependence of rates on overtone number $n$}
\label{app:ndependence}
The sources components $\psi_{211}$ and $\psi_{322}$ are peaked within a few Bohr radii, while hydrogenic wavefunctions in general are peaked further and further away from the origin as the quantum numbers are taken to be larger and larger. Thus, the interaction of a level $n\ell m$ with a combination of 211 and 322 will depend on the behavior of $R_{nl}$ near the $a_0$:

\begin{align}
    \begin{split}
      {}& R_{nl}(r\sim a_0) \\ {}&\sim\left(\frac{2}{(n_r+l+1)a_0}\right)^{3/2}\left(\frac{1}{2(n_r+l+1)}\right)^{1/2}\\
      {}&\times\left(\frac{(n_r+2l+1)!}{n_r!}\right)^{1/2}\frac{1}{(2l+1)!} \left(\frac{2r}{(n_r+l+1)a_0}\right)^l,
    \end{split}
\end{align}
where $n_r = n-l-1$ is the radial quantum number.
If $n_r\rightarrow \infty$, while $l$ is held fixed,
\begin{align}
    \begin{split}
        R_{nl}(r \sim a_0) \sim \left(\frac{1}{n_ra_0}\right)^{3/2}\sim \left(\frac{1}{na_0}\right)^{3/2}.
    \end{split}
\end{align}
Thus, any overlap integral with $R_{nl}$ decreases as $\sim n_r^{-3/2}\sim n^{-3/2}$. This is simply saying that as $n_r$ is taken larger, the characteristic volume of the driving wavefunction $\psi_{nl m}$ gets larger as $\sim (na_0)^3$, and so the driving is uniformly diluted by that same factor. A forced oscillation with a $\psi_{nl}$ component as a source term therefore suffers the same suppression.

Rates (whether emission rates or rates of absorption into the BH) depend on the square of the forced oscillation and therefore behave as $\propto n^{-3}$ in the limit of large $n$. This means that ratios of emissions and absorption processes become independent of $n$.\par
\begin{figure}[h]
	\begin{centering}
		\includegraphics[width=\columnwidth]{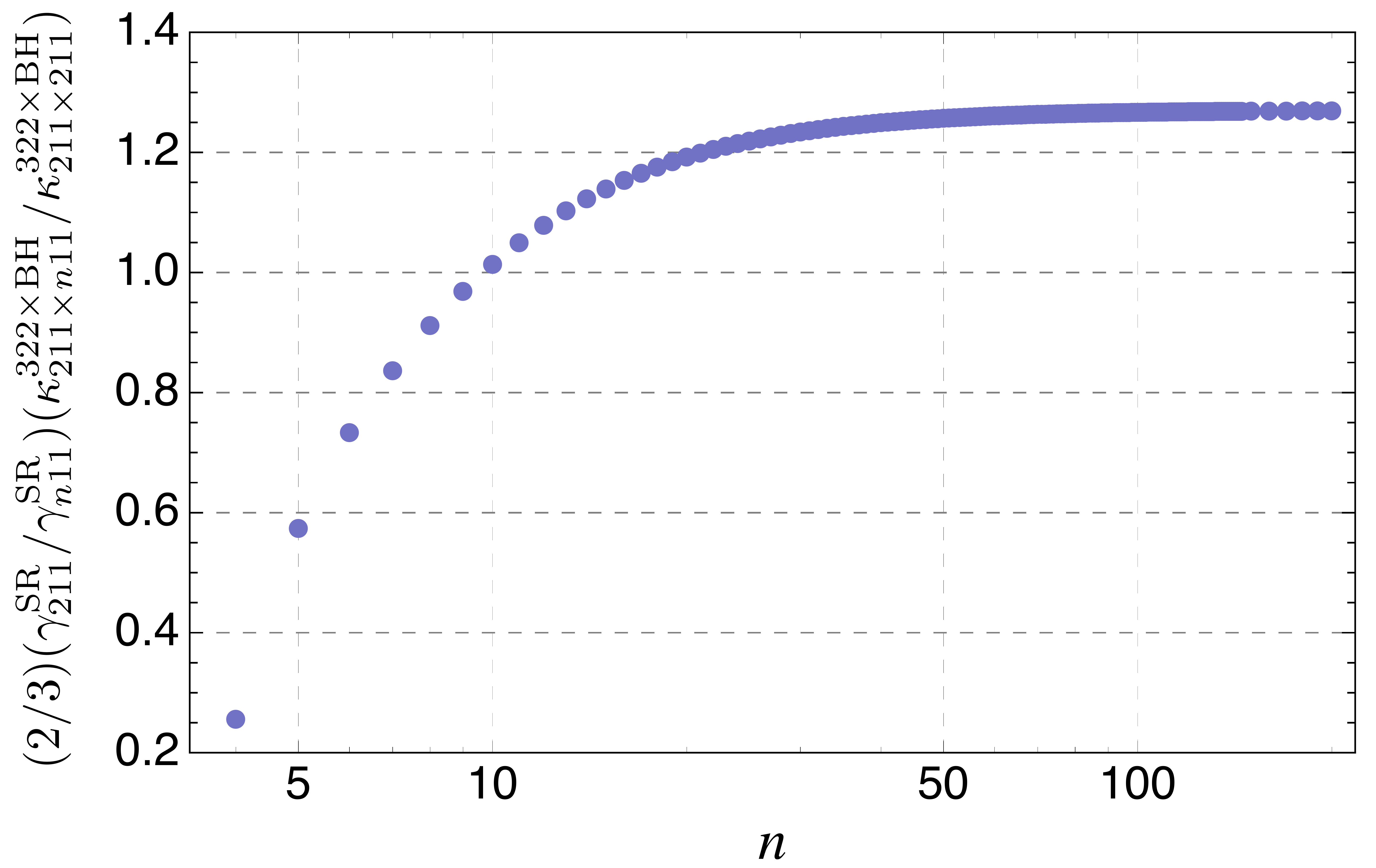}
		\caption{Behavior of the first term in the ratio in \eqref{eq:n11evoleq} as $n\rightarrow \infty$. As discussed in the paragraph below \eqref{eq:n11evoleq}, and as expected from \ref{app:ndependence}, the ratio rapidly becomes independent of $n$ and is $>1$ for $n\gtrsim$10.}
	\label{fig:n11plot1}
	\end{centering}
\end{figure}
\begin{figure}[h]
	\begin{centering}
		\includegraphics[width=\columnwidth]{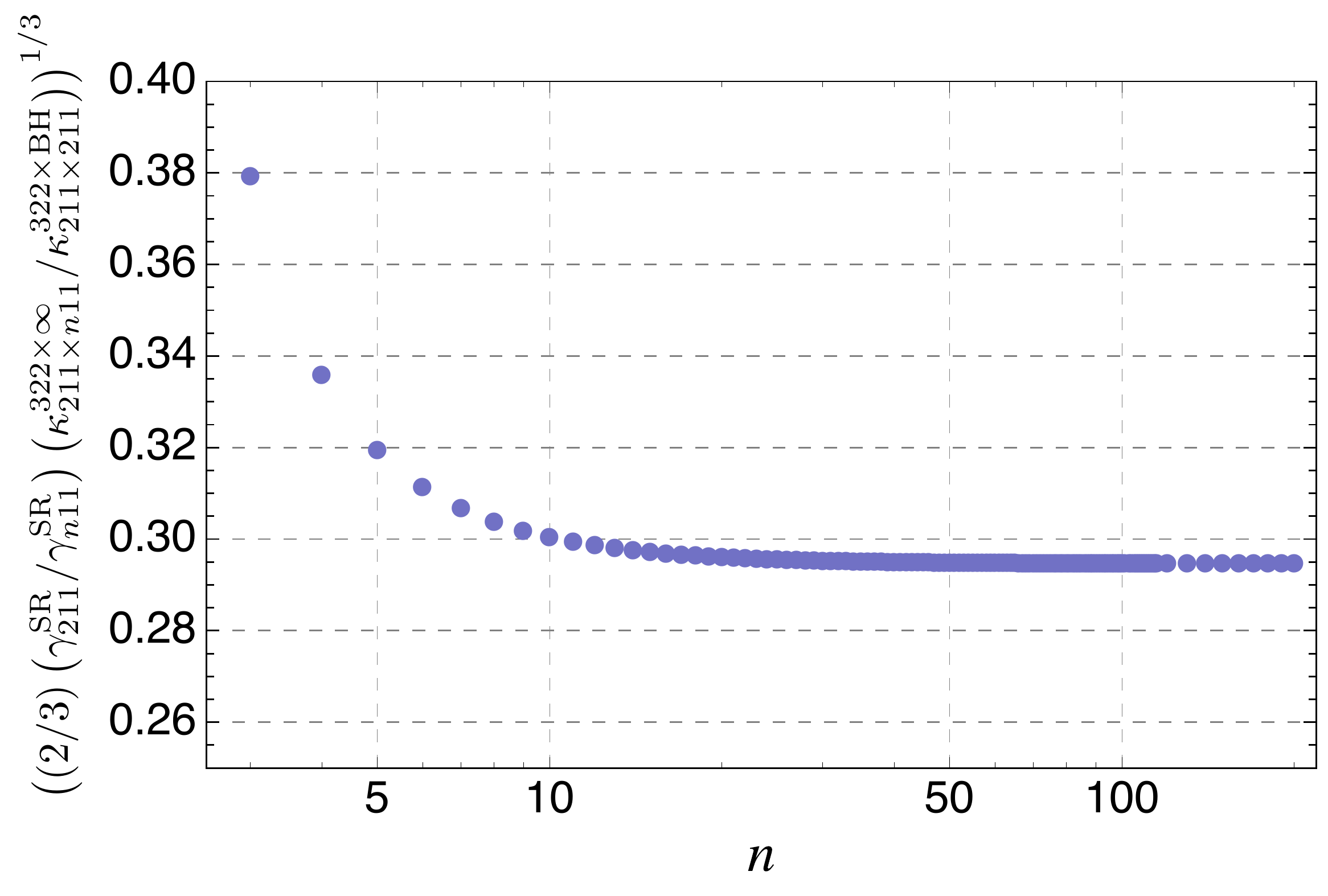}
		\caption{Behavior of the ratio in \eqref{eq:ratios_for_n11_growth_2} as $n\rightarrow \infty$. As expected from \ref{app:ndependence}, the ratio rapidly becomes independent of $n$.}
	\label{fig:n11plot2}
	\end{centering}
\end{figure}
\begin{figure}[h]
	\begin{centering}
		\includegraphics[width=\columnwidth]{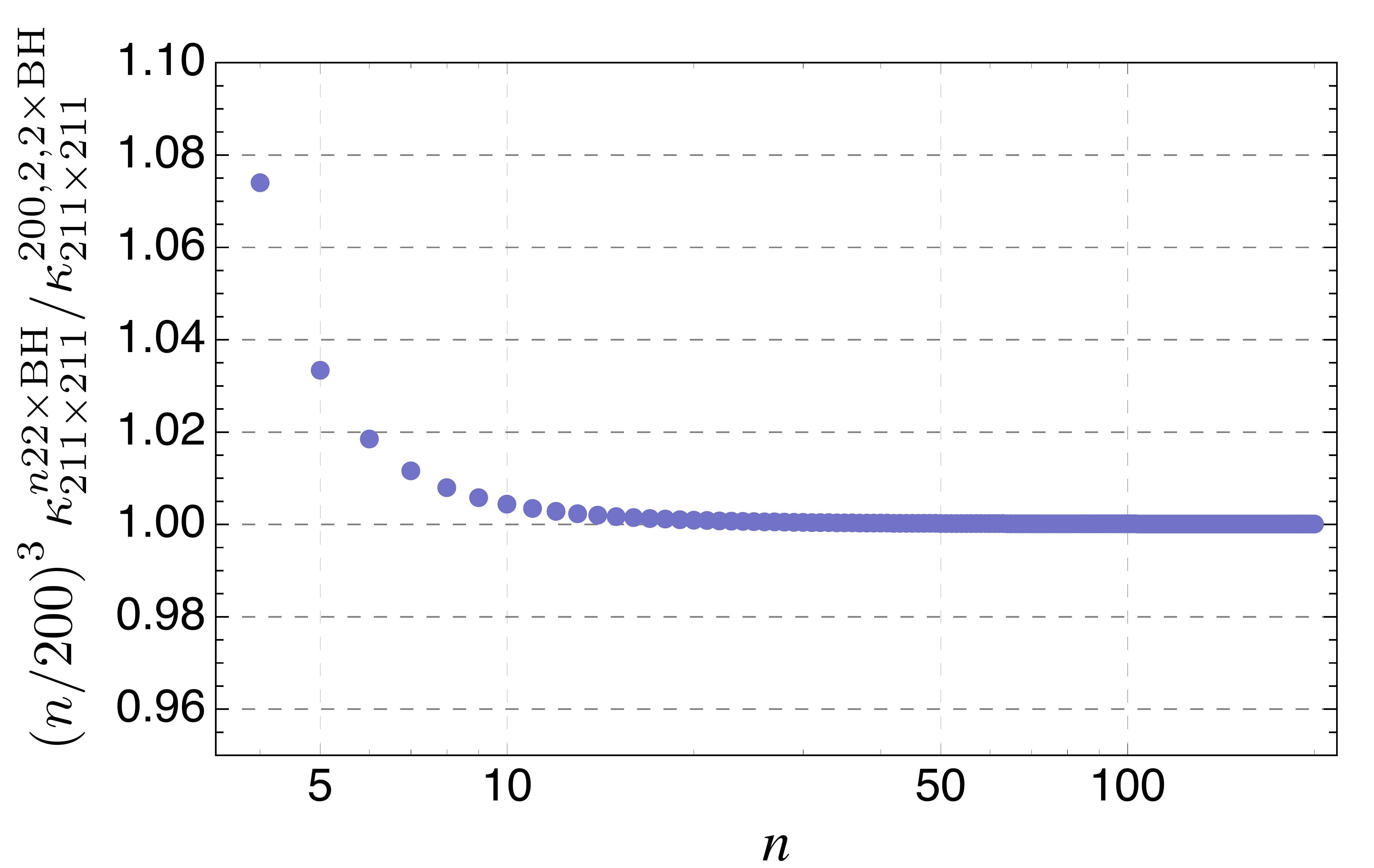}
		\caption{Behavior of the growth ratio $211\times 211\rightarrow n22 \times \BH$ normalized to its value at $n=200$. As stated in \eqref{eq:ratio_for_n22_growth_1}, the ratio scales as $n^{-3}$.}
	\label{fig:n22plot}
	\end{centering}
\end{figure}
\begin{figure}[h]
	\begin{centering}
		\includegraphics[width=\columnwidth]{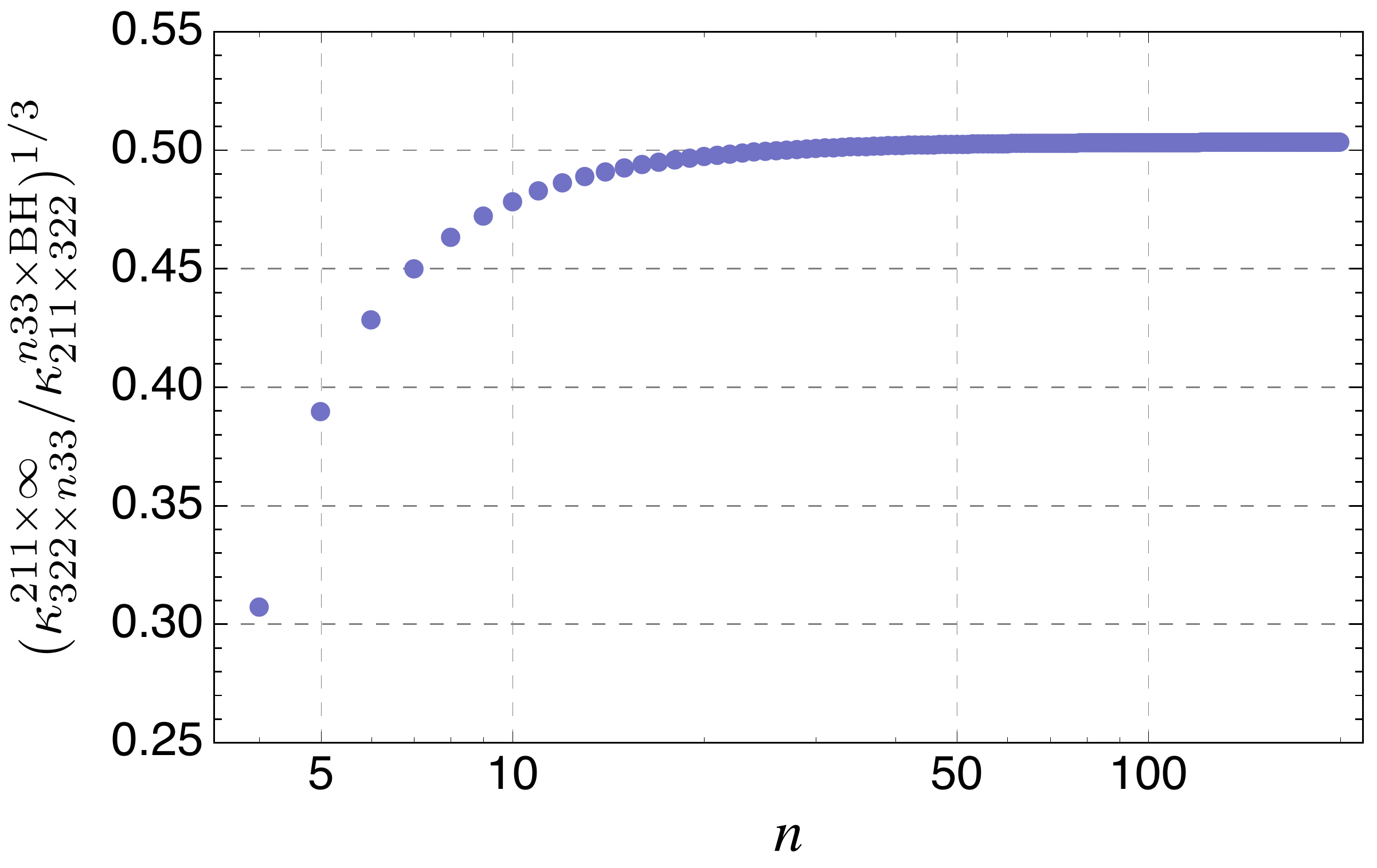}
		\caption{Behavior of the ratio in \eqref{eq:ratio_for_n33_growth} as $n\rightarrow \infty$. As expected from \ref{app:ndependence}, the ratio rapidly becomes independent of $n$.}
	\label{fig:n33plot}
	\end{centering}
\end{figure}
\begin{figure}[h]
	\begin{centering}
		\includegraphics[width=\columnwidth]{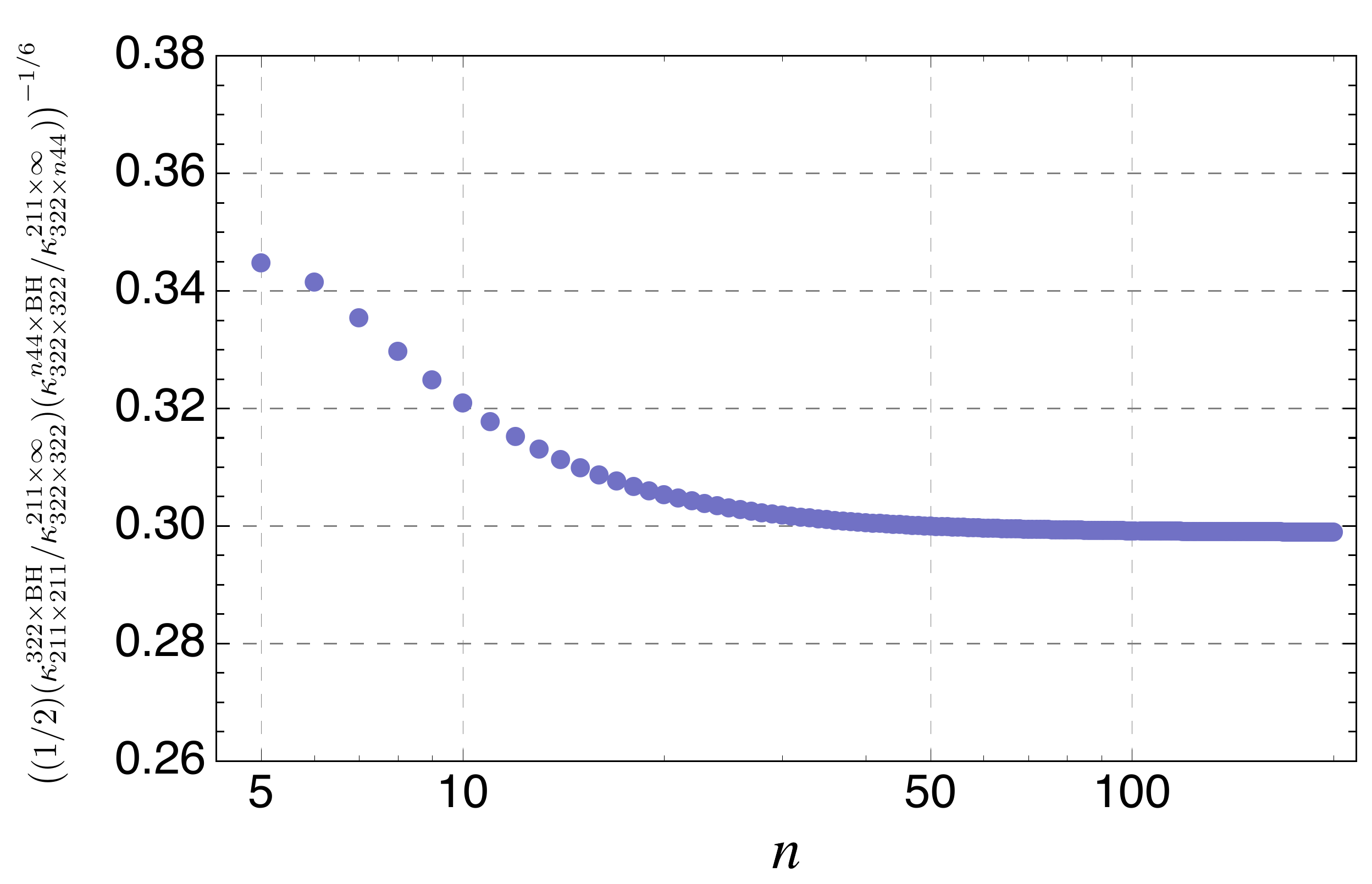}
		\caption{Behavior of the ratio in \eqref{eq:ratio_for_n44_growth}. As expected from \ref{app:ndependence}, the ratio rapidly becomes independent of $n$.}
	\label{fig:n44plot}
	\end{centering}
\end{figure}

The discussion in section \ref{sec:higherlevels} relied on the behavior of various ratios of rates at large $n$. To assess how fast the relevant ratios converge to the expected scaling in $n$, we plot them for the first 200 $n$ (Figs. \ref{fig:n11plot1}, \ref{fig:n11plot2}, \ref{fig:n22plot}, \ref{fig:n33plot}, \ref{fig:n44plot}).

\subsubsection{Mixing with $l'= 0$ damped states}
The analysis of levels that can grow from 211 and 322 mixing with an $l'=0$ forced oscillation is done in the main text (\ref{sec:higherlevels}).

\subsubsection{Mixing with $l'> 0$ damped states}
We give an exhaustive list of the possible processes involving mixing with
$l'=1$ and $l'=2$ damped states. 

For $l'=1$,
\begin{subequations}
\begin{align}
    \begin{split}
     \gamma_{322\times 322}^{655\times \text{BH}(1,-1)} (\varepsilon_{322}^\text{eq})^2 \sim{}& 10^{5}\left(\frac{\alpha}{0.3}\right)^{15}\left(\frac{M_\odot}{M}\right)\Myr^{-1},
    \end{split}
\end{align}
\begin{align}
    \begin{split}
     \gamma_{322\times 322}^{654\times \text{BH}(1,0)} (\varepsilon_{322}^\text{eq})^2 \sim{}& 10^{4}\left(\frac{\alpha}{0.3}\right)^{16}\left(\frac{M_\odot}{M}\right)\Myr^{-1},
    \end{split}
\end{align}
\begin{align}
    \begin{split}
      \gamma_{211\times 322}^{543\times \text{BH}(1,0)} \varepsilon_{211}^\text{eq}\varepsilon_{322}^\text{eq} \sim 10^{6}\left(\frac{\alpha}{0.3}\right)^{13}\left(\frac{M_\odot}{M}\right)\Myr^{-1},
    \end{split}
\end{align}
\begin{align}
    \begin{split}
      \gamma_{211\times 211}^{432\times \text{BH}(1,0)} (\varepsilon_{211}^\text{eq})^2 \sim 10^{7}\left(\frac{\alpha}{0.3}\right)^{10}\left(\frac{M_\odot}{M}\right)\Myr^{-1}.
    \end{split}
\end{align}
\end{subequations}
For $l'=2$,
\begin{subequations}
\begin{align}
    \begin{split}
     \gamma_{322\times 322}^{766\times \text{BH}(2,-2)} (\varepsilon_{322}^\text{eq})^2 \sim{}& 10^{-2}\left(\frac{\alpha}{0.3}\right)^{19}\left(\frac{M_\odot}{M}\right)\Myr^{-1},
    \end{split}
\end{align}
\begin{align}
    \begin{split}
     \gamma_{322\times 322}^{765\times \text{BH}(2,-1)} (\varepsilon_{322}^\text{eq})^2 \sim{}& 10^{-3}\left(\frac{\alpha}{0.3}\right)^{19}\left(\frac{M_\odot}{M}\right)\Myr^{-1},
    \end{split}
\end{align}
\begin{align}
    \begin{split}
     \gamma_{322\times 322}^{764\times \text{BH}(2,0)} (\varepsilon_{322}^\text{eq})^2 \sim{}& 10^{-4}\left(\frac{\alpha}{0.3}\right)^{20}\left(\frac{M_\odot}{M}\right)\Myr^{-1},
    \end{split}
\end{align}
\begin{align}
    \begin{split}
     \gamma_{211\times 322}^{653\times \text{BH}(2,0)} \varepsilon_{322}^\text{eq}\varepsilon_{211}^\text{eq} \sim{}& 10^{-3}\left(\frac{\alpha}{0.3}\right)^{17}\left(\frac{M_\odot}{M}\right)\Myr^{-1},
    \end{split}
\end{align}
\begin{align}
    \begin{split}
     \gamma_{211\times 211}^{542\times \text{BH}(2,0)} \left(\varepsilon_{211}^\text{eq}\right)^2 \sim{}& 10^{-4}\left(\frac{\alpha}{0.3}\right)^{14}\left(\frac{M_\odot}{M}\right)\Myr^{-1}.
    \end{split}
\end{align}
\end{subequations}
Clearly, rates for processes involving $l'\geq 2$ are too small to be
relevant on astrophysical timescales. Rates from mixing with $l'=1$ states
however can become quite large for $\alpha = \mathcal O(0.1)$, but, similarly
to processes with $l'=0$, they should be compared to depletion processes of
the form $nl m\times 322\rightarrow 211\times\infty$.

First,
\begin{align}
    \begin{split}
     {}& \dot\varepsilon_{655} = \\{}&\gamma_{322\times 322}^{655\times \BH(1,-1)}\left(1-\frac{\gamma_{655\times 322}^{211\times\infty}}{\gamma_{322\times 322}^{655\times \BH(1,-1)}}\frac{\varepsilon_{211}}{\varepsilon_{322}}\right)\varepsilon_{322}^2\varepsilon_{655}.
    \end{split}
\end{align}
The depletion term dominates as long as 
\begin{align}
    \begin{split}
     \alpha\tilde r_+^{1/9}\lesssim\left(\frac{\kappa_{655\times322}^{211\times\infty}}{\kappa_{322\times 322}^{655\times\BH(1,-1)}}\frac{2\kappa_{322\times 322}^{211\times\infty}}{\kappa_{211\times211}^{322\times \BH(0,0)}}\right)^{1/9} \approx 0.7.
    \end{split}
\end{align}
Next,
\begin{align}
    \begin{split}
     {}& \dot\varepsilon_{654} = \\{}&\gamma_{322\times 322}^{654\times \BH(1,0)}\left(1-\frac{\gamma_{654\times 322}^{211\times\infty}}{\gamma_{322\times 322}^{654\times \BH(1,0)}}\frac{\varepsilon_{211}}{\varepsilon_{322}}\right)\varepsilon_{322}^2\varepsilon_{654}.
    \end{split}
\end{align}
The depletion term dominates as long as
\begin{align}
    \begin{split}
     \alpha\tilde r_+^{1/10}\lesssim\left(\frac{\kappa_{654\times322}^{211\times\infty}}{\kappa_{322\times 322}^{654\times\BH(1,0)}}\frac{2\kappa_{322\times 322}^{211\times\infty}}{\kappa_{211\times211}^{322\times \BH(0,0)}}\right)^{1/10} \approx 0.7.
    \end{split}
\end{align}
Next,
\begin{align}
    \begin{split}
     {}& \dot\varepsilon_{543} = \\{}&\gamma_{211\times 322}^{543\times \BH(1,0)}\left(1-\frac{\gamma_{543\times 322}^{211\times\infty}}{\gamma_{211\times 322}^{543\times \BH(1,-1)}}\right)\varepsilon_{322}\varepsilon_{211}\varepsilon_{543}.
    \end{split}
\end{align}
The depletion term dominates as long as
\begin{align}
    \begin{split}
     \alpha \tilde r_+^{1/7} \lesssim \left(\frac{\kappa_{543\times322}^{211\times\infty}}{\kappa_{211\times 322}^{543\times \BH(1,0)}}\right)^{1/7} \approx 1.
    \end{split}
\end{align}
Finally,
\begin{align}
    \begin{split}
     {}& \dot\varepsilon_{432} = \\{}&\gamma_{211\times 211}^{432\times \BH(1,0)}\left(1-\frac{\gamma_{432\times 322}^{211\times\infty}}{\gamma_{211\times 211}^{432\times \BH(1,0)}}\frac{\varepsilon_{322}}{\varepsilon_{211}}\right)\varepsilon_{322}\varepsilon_{211}\varepsilon_{432} .
    \end{split}
\end{align}
The depletion term dominates as long as 
\begin{align}
    \begin{split}
     \alpha \lesssim \left( \frac{\kappa_{432\times322}^{211\times\infty}}{\kappa_{211\times211}^{432\times\BH(1,0)}}\frac{1}{2}\frac{\kappa_{211\times211}^{322\times\BH(0,0)}}{\kappa_{322\times322}^{211\times \infty}}\right)^{1/4}\approx 1.
    \end{split}
\end{align}
Since 211 SR stops for $\alpha \geq 0.5$, we conclude that the net growth rate of all four levels (and their radial overtones) is negative over the whole range of relevant parameter space.


\begin{table}
  \begin{center}
    \begin{tabular}{l|l}   
		Process & Rate ($\gamma/\mu$, Eq.~\eqref{eqepsdot})  \\
        \hline
\hline
      $\Gamma^\text{SR}_{211}$ & $  4\times 10^{-2} \alpha^{8} (a_*-2\alpha (1+\sqrt{1-a_*^2}))$  \\
      $\Gamma^\text{SR}_{322}$ &  $  8 \times 10^{-5}\alpha^{12}(a_*-\alpha (1+\sqrt{1-a_*^2}))$ \\
      $\Gamma^\text{SR}_{433}$ &  $  2 \times 10^{-8}\alpha^{16}(a_*-\frac{2}{3}\alpha (1+\sqrt{1-a_*^2}))$ \\
      $\Gamma^\text{SR}_{544}$ &  $  2 \times 10^{-12}\alpha^{20}(a_*-\frac{1}{2}\alpha (1+\sqrt{1-a_*^2}))$ \\      \hline
      $\Gamma^\text{GW,ann}_{211}$ & $1\times 10^{-2} \alpha^{14}$ \\
      $\Gamma^\text{GW,ann}_{322}$ & $3\times 10^{-8} \alpha^{18}$\\
      $\Gamma^\text{GW,tr}_{322\rightarrow 211}$ & $5\times 10^{-6}\alpha^{10}$\\
      \hline                         
     \end{tabular}
		 \caption{Rates for gravitational processes involved in the evolution of the scalar cloud.}
    \label{tablegrav}
  \end{center}
\end{table}

\begin{table}
  \begin{center}
    \begin{tabular}{l|l}   
		Process & Rate ($\gamma/\mu$, Eq.~\eqref{eqepsdot})  \\
        \hline
\hline
      $\Gamma^\text{$322\times$ BH}_{211\times211}$ & $4.3\times 10^{-7} \alpha^{11}\left(\frac{\Mpl}{f}\right)^4(1+\sqrt{1-a_*^2})$\\
      $\Gamma^\text{$422\times$ BH}_{211\times211}$ & $1.5 \times 10^{-7} \alpha^{11}\left(\frac{\Mpl}{f}\right)^4(1+\sqrt{1-a_*^2})$\\
$\Gamma^\text{$322\times$ BH}_{211\times411}$ & $2.5\times 10^{-8} \alpha^{11}\left(\frac{\Mpl}{f}\right)^4(1+\sqrt{1-a_*^2})$\\
$\Gamma^\text{$322\times$ BH}_{411\times411}$ & $9.8\times 10^{-11} \alpha^{11}\left(\frac{\Mpl}{f}\right)^4(1+\sqrt{1-a_*^2})$\\
       $\Gamma^\text{$433\times$BH}_{211\times322}$ & $9.1\times 10^{-8} \alpha^{11}\left(\frac{\Mpl}{f}\right)^4(1+\sqrt{1-a_*^2})$\\
       $\Gamma^\text{$544 \times$ BH}_{322\times322}$& $1.9\times 10^{-9} \alpha^{11}\left(\frac{\Mpl}{f}\right)^4(1+\sqrt{1-a_*^2})$\\
       $\Gamma^\text{$544\times$BH}_{211\times433}$& $1.1\times 10^{-9} \alpha^{11}\left(\frac{\Mpl}{f}\right)^4(1+\sqrt{1-a_*^2})$\\
   $\Gamma^\text{$655\times $ BH}_{322\times433}$& $2.8\times 10^{-10} \alpha^{11}\left(\frac{\Mpl}{f}\right)^4(1+\sqrt{1-a_*^2})$\\       
      $\Gamma^\text{$655\times$BH}_{211\times544}$& $3.6\times 10^{-12} \alpha^{11}\left(\frac{\Mpl}{f}\right)^4(1+\sqrt{1-a_*^2})$\\
      $\Gamma^\text{$766\times$BH}_{433\times433}$& $2.1\times 10^{-10} \alpha^{11}\left(\frac{\Mpl}{f}\right)^4(1+\sqrt{1-a_*^2})$\\
      $\Gamma^\text{$877\times$BH}_{433\times544}$& $5.2\times 10^{-12} \alpha^{11}\left(\frac{\Mpl}{f}\right)^4(1+\sqrt{1-a_*^2})$\\
      $\Gamma^\text{$988\times$BH}_{544\times544}$& $1.6\times 10^{-12} \alpha^{11}\left(\frac{\Mpl}{f}\right)^4(1+\sqrt{1-a_*^2})$\\
     $\Gamma^\text{$1099\times$BH}_{544\times655}$& $5.6\times 10^{-13} \alpha^{11}\left(\frac{\Mpl}{f}\right)^4(1+\sqrt{1-a_*^2})$\\
     $\Gamma^\text{$433\times$200}_{211\times422}$& $1.1\times 10^{-9} \alpha^{7}\left(\frac{\Mpl}{f}\right)^4(1+\sqrt{1-a_*^2})$\\
     \hline                  
     \end{tabular}
		 \caption{Rates for quartic processes involving non-relativistic bound states.}
    \label{tablebound}
  \end{center}
\end{table}

\begin{table}
  \begin{center}
    \begin{tabular}{l|l} 
		Process & Rate ($\gamma/\mu$, Eq.~\eqref{eqepsdot})  \\
        \hline
      \hline
      $\Gamma^{100\times \infty}_{211\times211}$ & $1.3\times10^{-7}\alpha^{8}\left(\frac{\Mpl}{f}\right)^4$\\
       $\Gamma^{100\times \infty}_{211\times322}$ & $8.5\times 10^{-9}\alpha^{8}\left(\frac{\Mpl}{f}\right)^4$\\
      $\Gamma^{100\times \infty}_{322\times322}$ & $1.1\times 10^{-10}\alpha^{8}\left(\frac{\Mpl}{f}\right)^4$\\
      $\Gamma^{211\times \infty}_{322\times411}$ & $3.8\times 10^{-9}\alpha^{8}\left(\frac{\Mpl}{f}\right)^4$\\
            $\Gamma^{211\times \infty}_{322\times322}$ & $1.1\times 10^{-8}\alpha^{8}\left(\frac{\Mpl}{f}\right)^4$\\
      $\Gamma^{211\times\infty}_{322\times433}$ & $2.6\times 10^{-9}\alpha^{8}\left(\frac{\Mpl}{f}\right)^4$\\
      $\Gamma^{211\times \infty}_{433\times433}$& $9.2\times 10^{-11}\alpha^{8}\left(\frac{\Mpl}{f}\right)^4$\\
       $\Gamma^{211\times \infty}_{322\times544}$& $6.1\times 10^{-11}\alpha^{8}\left(\frac{\Mpl}{f}\right)^4$\\
       $\Gamma^{211\times\infty}_{433\times544}$& $1.9\times 10^{-11}\alpha^{8}\left(\frac{\Mpl}{f}\right)^4$\\
       $\Gamma^{211\times\infty}_{544\times544}$& $4.2\times 10^{-13}\alpha^{8}\left(\frac{\Mpl}{f}\right)^4$\\
              $\Gamma^{322\times\infty}_{544\times544}$& $4.4\times 10^{-11}\alpha^{8}\left(\frac{\Mpl}{f}\right)^4$\\
       $\Gamma^{322\times\infty}_{433\times544}$& $7.8\times 10^{-10}\alpha^{8}\left(\frac{\Mpl}{f}\right)^4$\\
      $\Gamma^{21-1\times\infty}_{322\times322}$& $2.3\times10^{-10}\alpha^{8}\left(\frac{\Mpl}{f}\right)^4$\\
       $\Gamma^{211\times\infty}_{655\times322}$& $7.3\times10^{-13}\alpha^{8}\left(\frac{\Mpl}{f}\right)^4$\\
       $\Gamma^{211\times\infty}_{655\times433}$& $4.6\times10^{-13}\alpha^{8}\left(\frac{\Mpl}{f}\right)^4$\\
       $\Gamma^{211\times\infty}_{655\times544}$& $6.9\times10^{-14}\alpha^{8}\left(\frac{\Mpl}{f}\right)^4$\\       
      $\Gamma^{211\times\infty}_{655\times655}$& $1.1\times10^{-15}\alpha^{8}\left(\frac{\Mpl}{f}\right)^4$\\       
     $\Gamma^{322\times\infty}_{655\times433}$& $3.7\times10^{-11}\alpha^{8}\left(\frac{\Mpl}{f}\right)^4$\\
          $\Gamma^{322\times\infty}_{655\times544}$& $1.6\times10^{-11}\alpha^{8}\left(\frac{\Mpl}{f}\right)^4$\\
        $\Gamma^{322\times\infty}_{655\times655}$& $6.2\times10^{-13}\alpha^{8}\left(\frac{\Mpl}{f}\right)^4$\\
       $\Gamma^{433\times\infty}_{766\times766}$& $5.6\times10^{-13}\alpha^{8}\left(\frac{\Mpl}{f}\right)^4$\\
     \end{tabular}
  \end{center}
      \caption{Rates for quartic processes leading to non-relativistic emission.}
    \label{tablenrem}
\end{table}

\begin{table}
  \begin{center}
    \begin{tabular}{l|l} 
		Process & Rate ($\gamma/\mu$, Eq.~\eqref{eqepsdot})  \\
        \hline
      \hline
      $\Gamma^{2\rightarrow1}_{211}(\mathrm{cubic})$ & $1.9 \times 10^{-4} \alpha^{14} |C|^2 \left(\frac{\Mpl}{f}\right)^2$\\ 
	   \hline
      $\Gamma^{3\rightarrow1}_{211}$ & $5\times 10^{-9} \alpha^{21}\left(\frac{\Mpl}{f}\right)^4$\\
      $\Gamma^{3\rightarrow1}_{322}$ & $6\times 10^{-14} \alpha^{27}\left(\frac{\Mpl}{f}\right)^4$\\
     \end{tabular}
  \end{center}
      \caption{Rates for self-interaction induced relativistic emission processes.}
    \label{tab:table2}
\end{table}

\section{Equilibrium ratio for moderate self-interactions}
\label{sec:equilibrium_ratio}
We derive a more precise formula for the value of the time-independent equilibrium ratio by the system of equations \eqref{eq:simplified_evolution}. In terms of the $\gamma$ rates,
\begin{subequations}
\begin{equation}
    \dot\varepsilon_{211} = \gamma_{322\times 322}^{211\times\infty}\varepsilon_{211}\varepsilon_{322}^2-2\gamma_{211\times 211}^{322\times\BH}\varepsilon_{211}^2\varepsilon_{322},
\end{equation}
\begin{equation}
    \dot\varepsilon_{322} = -2\gamma_{322\times 322}^{211\times\infty}\varepsilon_{211}\varepsilon_{322}^2+\gamma_{211\times 211}^{322\times\BH}\varepsilon_{211}^2\varepsilon_{322}.
\end{equation}
\end{subequations}
Therefore,
\begin{align}
\begin{split}
{}&\frac{1}{\varepsilon_{211}\varepsilon_{322}}\frac{d}{dt}\left(\frac{\varepsilon_{322}}{\varepsilon_{211}}\right) =\\ {}&\gamma_{211\times 211}^{322\times \BH}+2\left(\gamma_{211\times 211}^{322\times \BH}-\gamma_{322\times 322}^{211\times \infty}\right)\left(\frac{\varepsilon_{322}}{\varepsilon_{211}}\right)\\
{}&-\gamma_{322\times 322}^{211\times \infty}\left(\frac{\varepsilon_{322}}{\varepsilon_{211}}\right)^2.
\end{split}
\end{align}
The zeros of the right-hand side are
\begin{align}
\begin{split}
{}&\eta^B = \frac{1}{\gamma_{322\times 322}^{211\times \infty}}\Big(\gamma_{211\times211}^{322\times\text{BH}}-\gamma_{322\times322}^{211\times\infty}\\
{}&\pm\sqrt{(\gamma_{211\times 211}^{322\times\text{BH}})^2-\gamma_{211\times 211}^{322\times\text{BH}}\gamma_{322\times322}^{211\times\infty}+(\gamma_{322\times322}^{211\times\infty})^2}\Big).
\end{split}
\end{align}
Since the right-hand side is an inverted parabola, the ``+'' solution is dynamically stable (attractive), while the ``$-$'' solution is unstable. Parametrically, $\gamma_{211\times211}^{322\times\text{BH}}\propto \alpha^{11}$ and $\gamma_{322\times 322}^{211\times \infty} \propto \alpha^8 $. Therefore at small $\alpha$, $\gamma_{211\times 211}^{322\times \text{BH}} < \gamma_{322\times 322}^{211\times\infty}$, and so the ``$-$'' root is negative. Moreover, the "+" root is
\begin{align}
\begin{split}
(\eta^B)_\text{small $\alpha$} \approx \frac{1}{2} \frac{\gamma_{211\times 211}^{322\times \text{BH}}}{\gamma_{322\times 322}^{211\times\infty}} = \frac{\varepsilon_{322}^\text{eq}}{\varepsilon_{211}^\text{eq}}.
\end{split}
\end{align}

\section{Boundary of the regime of early equilibrium}
\label{sec:fa_min_contour}
We derive a more precise formula for the value of $f/\Mpl$ such that the SR growth of 211 is halted before $\mathcal O$(1) of the spin is extracted. At early times, if we neglect the dependence of $\gamma_{211}^\SR$ on the BH spin $a_*$,
\beq
\label{eq:approx_211}
\varepsilon_{211}(t)\approx \frac{1}{GM^2}e^{\gamma^\SR_{211}t}.
\eeq
We use this into
\beq
\dot\varepsilon_{322} = \gamma_{211\times 211}^{322\times \BH}\varepsilon_{211}^2\varepsilon_{322},
\eeq
where we neglect the dependence of $\gamma_{211\times 211}^{322\times \BH}$ on $a_*$. Therefore
\beq
\label{eq:approx_322}
\varepsilon_{322}(t)\approx \frac{1}{GM^2}\exp\left[\frac{\gamma_{211\times 211}^{322\times\BH}}{2\gamma^\SR_{211}}\frac{1}{G^2M^4}\left(e^{2\gamma_{211}^\SR t}-1\right)\right].
\eeq
The condition for SR to be impeded is that
\beq
\label{eq:condition}
\gamma^\SR_{211}\simeq2 \gamma_{211\times211}^{322\times\BH}\varepsilon_{211}(t)\varepsilon_{322}(t).
\eeq
Using the approximations \eqref{eq:approx_211} and \eqref{eq:approx_322}, one finds that \eqref{eq:condition} is satisfied at the time $t_\text{eq}$ such that
\beq
\gamma^\SR_{211} t_\text{eq} \approx \frac{1+4\beta\log\beta-2\beta W(\frac12 \beta e^{1/2\beta})}{4\beta},
\label{eq:overshoot2}
\eeq
where 
\beq
\beta \equiv G^2M^4\frac{\gamma^\SR_{211}}{2\gamma_{211\times 211}^{322\times \BH}},
\eeq
and $W(z)$ is the product logarithm (sometimes called the Lambert W function).

When $\gamma_{211}^\SR t \simeq \log(GM^2\Delta a_*)$, then SR has happened completely. So, in order for \eqref{eq:condition} to be obtained before SR has run its course, we must have
\beq
\label{eq:fa_min_contour_exact}
\frac{1+4\beta\log\beta-2\beta W(\frac12 \beta e^{1/2\beta})}{4\beta} \lesssim \log\left(GM^2\varepsilon_{211}^\text{max}\right).
\eeq
\eqref{eq:fa_min_contour_exact} implicitly defines $f_\text{thresh}$. Note that since $M \gg \Mpl$, $\beta \gg 1$ for much of parameter space. One can then approximate $W(z)$ with the leading terms of its expansion around a large argument: $W(z)\rightarrow \log z - \log(\log z)$ as $z\rightarrow+\infty$. In this approximation, the left-hand side of \eqref{eq:fa_min_contour_exact} becomes $\approx \log\sqrt{2\beta \log \beta}$, and the condition for SR to be halted early simplifies to
\begin{align}
\begin{split}
2\gamma^\text{SR}_\text{211}\log(GM^2) \lesssim \gamma^{322\times\BH}_{211\times 211} (\varepsilon_{211}^\text{max})^2.
\end{split}
\end{align}


\section{Cloud mass}
\label{app:cloudmass}

Here we calculate the mass of the cloud in the case $f\to\infty$, i.e. in the purely gravitational case. We will do the computation for the 211 level for clarity, but it is straightforward to generalize the formalism to any $n\ell m$ level. To simplify notation, we drop the level subscripts for the rest of our discussion here. The cloud parameters are referring to 211, unless stated otherwise.

Since the BH loses $<0.1\%$ of its mass due to SR, we usually treat its mass to be constant, or, equivalent, that $\alpha$ is just a parameter. In the case of self-interactions, in particular, the cloud tends to grow to a smaller occupation number, which strengthens this assumption. A further simplification comes from setting $\omega\simeq\mu$. By noting that $\varepsilon=-\dot{a}_\ast$ in this regime, we get that $\varepsilon_{\max}=\Delta a_\ast$. The final $a_\ast$ can be found by setting the SR rate equal to zero. Eventually, the maximum occupation number one gets is

\begin{equation}
    \label{eq:maxsimple}
    \varepsilon_{\max}=a_\ast(0)-\frac{4\alpha}{1+4\alpha^2}.
\end{equation}

In general, the equations we need to solve are:
\begin{subequations}
\label{eq:clmass}
\begin{align}
       \dot{N}&=\gsr N \\
       \dot{M}&=-\omega_{211}\gsr N\\
       \dot{J}&=-\gsr N\\
       a_\ast&\equiv\frac{J}{G M^2}
       \end{align}

where
       
       \begin{align}
       \label{eq:cmenergy}
       \omega_{211}=\mu\left(1-\frac{\alpha^2}{8}\right)
       \end{align}
\end{subequations}
       
We define the $\varepsilon$ with respect to the initial BH mass $\MBH^i$, i.e. $\varepsilon=N/G \pare{\MBH^i}^2$ and Eqs.~\eqref{eq:clmass} become:

\begin{align}
       \dot{\varepsilon}&=\gsr \varepsilon\label{eq:cm1}\\
         \dot{\alpha}&=-\alpha_i^2\left(1-\frac{\alpha^2}{8}\right)\dot{\varepsilon}\label{eq:cm2}\\
         \dot{a}_\ast&=-\frac{\dot{\alpha}}{\alpha}\parea{2a_\ast-\frac{1}{\alpha\pare{1-\alpha^2/8}}}\label{eq:cm3}
       \end{align}
       
\noindent where $\alpha_i=\alpha(t=0)$, given by the initial BH mass. The usual treatment is to expand these equations for small $\alpha$, which is equivalent to neglecting terms of order $\mathcal{O}(\alpha)$. This reduces Eq.~\eqref{eq:cm2} to $\dot{\alpha}=0$. However, the expansion in Eq.~\eqref{eq:cm3} has to be taken more carefully because the denominator is also small in this limit. By substituting Eq.~\eqref{eq:cm2} it becomes evident that the first term is of order $\mathcal{O}(\alpha)$, whereas the second is independent of $\alpha$. Therefore, we can neglect the former, which gives the standard result $\dot{a}_*=-\dot{\varepsilon}$.

Eq.~\eqref{eq:cm2} has the following solution
\begin{align}
\label{eq:cmalpha}
       \alpha=-2\sqrt{2}\,\text{tanh}\left[\frac{1}{4}\left(\sqrt{2}\alpha_i^2\varepsilon-4\,\text{arctanh}\left[\frac{\alpha_i}{2\sqrt{2}}\right]\right)\right]
       \end{align}
Now, Eq.~\eqref{eq:cm3} can also be solved analytically. The result is
\begin{equation}
  \label{eq:cmspin}
       \begin{split}
a_\ast=\frac{1}{\alpha^2}\parea{a_0\alpha_i^2-2\sqrt{2}\text{arctanh}\pare{\frac{2\sqrt{2}(\alpha-\alpha_i)}{-8+\alpha\alpha_i}}}
     \end{split}
           \end{equation}
The final spin of the BH is that which saturates the SR condition $\omega-m\Omega_H$  is
\begin{equation}
\label{eq:finspin}
       a_\ast^\text{fin}=-\frac{8(-8\alpha_\text{fin}+\alpha_\text{fin}^3)}{16+64\alpha_\text{fin}^2-16\alpha_\text{fin}^4+\alpha_\text{fin}^6},
       \end{equation}
where the ``fin'' superscript denotes final quantities, after the 211 cloud has been saturated and the BH has spun down.

Now we can use Eqs. \eqref{eq:cmalpha}, \eqref{eq:cmspin} and \eqref{eq:finspin} to numerically solve for $\varepsilon_\text{max}$, the final occupation number of the cloud. The mass of the cloud is then $\Mc=\varepsilon_\text{max} G \pare{\MBH^i}^2\omega$. 

By neglecting the $\alpha^2$ term in Eq.~\eqref{eq:cmenergy}, i.e. by approximating $\omega\simeq\mu$, we can get a simpler analytic result for the final BH mass. In this case, the equivalents of Eqs. \eqref{eq:cmalpha}, \eqref{eq:cmspin} are
  \begin{align}
    \alpha&=\alpha_i\left(1-\alpha_i\varepsilon\right)\\
     a_\ast&=\frac{a_0-\varepsilon}{(1-\alpha_i\varepsilon)^2}  \label{eq:cmspinsimple}
       \end{align}
which can be used along with Eq.~\eqref{eq:finspin}, truncated to $\mathcal{O}(\alpha^2)$, to give the final occupation number of the cloud. We find that
 \begin{equation}
   \label{eq:cmmaxsimple}
       \varepsilon_{\max}=\frac{1-8\alpha_i^2+8\alpha_i^3 a_0-\sqrt{1-16\alpha_i^2+32a_0\alpha_i^3-16a_0^2\alpha_i^4}}{8(-\alpha_i^3+a_0\alpha_i^4)}
       \end{equation}
where $a_0=a_\ast(t=0)$.
       
\begin{figure}[h]
\centering
\includegraphics[width=0.48\textwidth]{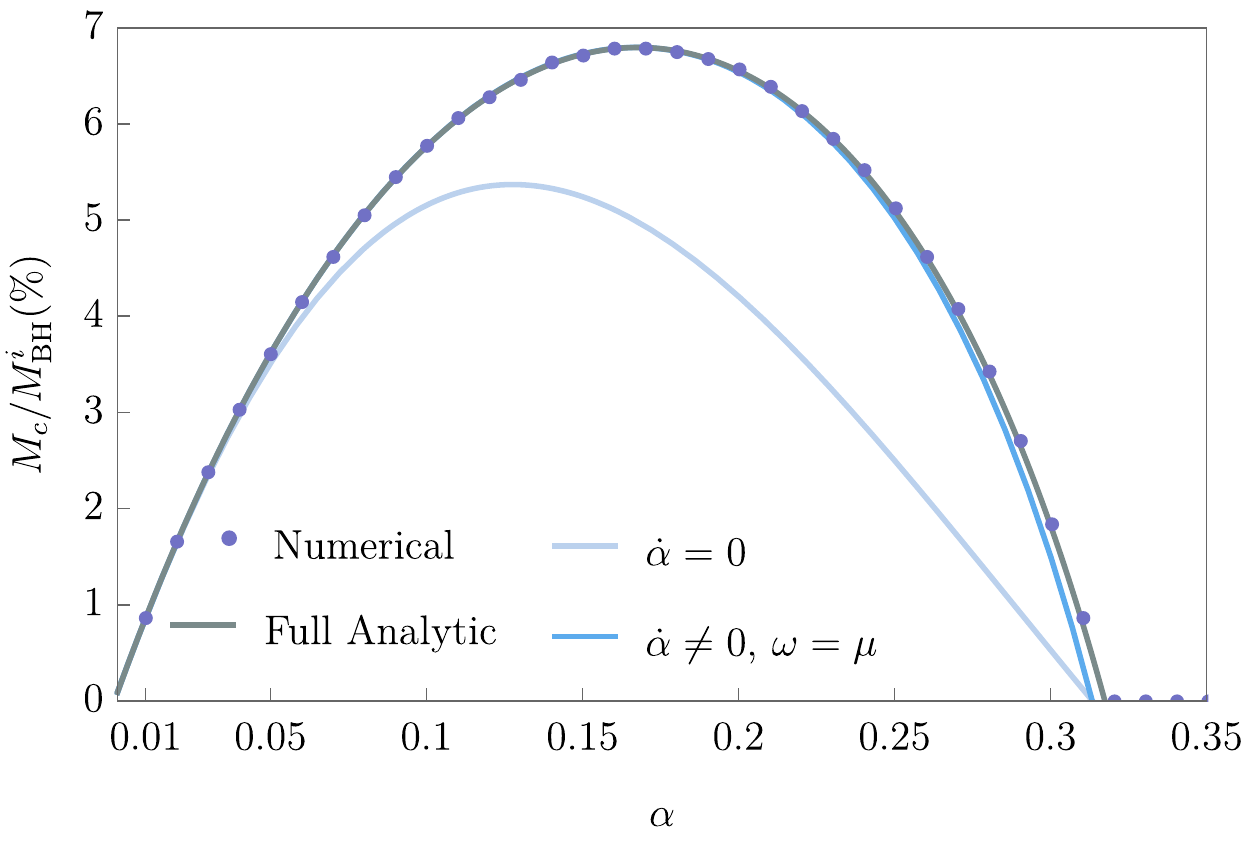}
\caption{Ratio of the final mass of the cloud to the initial BH mass. We plot points from the numerical evolution of Eqs. (\eqref{eq:cm1}-\eqref{eq:cm3}), the full analytic result of Eqs. (\eqref{eq:cmalpha}-\eqref{eq:finspin}), as well as the $\dot{\alpha}=0$ approximation of Eq.~\eqref{eq:maxsimple} and the  $\omega_{211}\simeq\mu,\,\dot{\alpha}\neq0$ approximation of \eqref{eq:cmmaxsimple}. The cloud can grow to have a mass of up to $7\%$ of the initial BH mass. This plot assumes an initial spin $a_*(t_0)=0.9$.}
 \label{fig:cmass}
\end{figure}

In Fig. \ref{fig:cmass} we plot the ratio of the final cloud mass over the initial BH mass. We solve numerically Eq.~\eqref{eq:finspin} with respect to $\varepsilon_{\max}$ and compare it to the numerical evolution of Eqs.~\eqref{eq:cm1}-\eqref{eq:cm3}. We also plot the results of Eq.~\eqref{eq:maxsimple} and Eq.~\eqref{eq:cmmaxsimple} for comparison. We find that the mass of the cloud can grow up to $7\%$ of the initial BH mass.

\section{Self-gravity energy corrections}
\label{app:selfgravity}

The Poisson equation for the gravitational potential sourced by the cloud is

\begin{equation}
    \label{eq:sgpoisson}
    \nabla^2\Phi_\text{SG}=4\pi G\mu\abs{\psi}^2
\end{equation}

where $\psi$ is the wavefunction of the cloud, i.e.

\begin{equation}
    \label{eq:sgwavef}
    \psi(\mathbf{r})=\sum_{nlm}\sqrt{N_{nlm}}\psi_{nlm}
\end{equation}

where $N_{nlm}$ are the occupation numbers of the levels and $\psi_{nlm}$ the hydrogenic wavefunctions. Treating $\Phi_\text{SG}$ as a small perturbation, the energy correction of the $(n,l,m)$ level is

\begin{equation}
\begin{split}
    \label{eq:sgencorr}
    \Delta \omega_{nlm}&=\bra{nlm}\mu\Phi_\text{SG}\ket{nlm}\\
    &=-G\mu^2\int\abs{\psi_{nlm}(\mathbf{r})}^2\int\frac{\abs{\psi(\mathbf{r}')}^2}{|\mathbf{r}-\mathbf{r'}|}\,\di^3\mathbf{r'}\di^3\mathbf{r}
    \end{split}
\end{equation}

Expanding $1/|\mathbf{r}-\mathbf{r'}|$ in spherical harmonics we get

\begin{equation}
\begin{split}
    \label{eq:sgjackson}
    \frac{1}{|\mathbf{r}-\mathbf{r'}|}=4\pi\sum_{l'=0}^\infty\sum_{m'=-l'}^{l'}\frac{1}{2l'+1}\frac{r_<^{l'}}{r_>^{l'+1}}Y_{l'}^{m'\ast}(\theta',\phi')Y_{l'}^{m'}(\theta,\phi),
    \end{split}
\end{equation}

where $r_{<(>)}$ is the smallest (largest) of $r$ and $r'$. We can perform the integration over $\theta$ and $\phi$, since $\psi_{nlm}\propto Y_l^m$.  By the selection rules of the spherical harmonics we can write

\begin{equation}
\begin{split}
    \label{eq:shadd}
    Y_l^m Y_l^{m\ast}=\sum_{k=0}^l c_{k,lm}Y_{2k}^0,\quad  c_{k,lm}=\int \abs{Y_l^m}^2 Y_{2k}^{0\ast}\di\Omega
    \end{split}
\end{equation}

Therefore, the integral over $\theta$ and $\phi$ selects $m'=0$ and $l'=2k$, giving

\begin{equation}
\begin{split}
    \label{eq:sgencorr2}
    \Delta \omega_{nlm}&=-4\pi G\mu^2\sum_{k=0}^l\frac{c_{k,lm}}{4k+1}\int R_{nl}(r) r^2 \\
    &\times\int\abs{\psi(\mathbf{r'})}^2\frac{r_<^{2k}}{r_>^{2k+1}}Y_{2k}^{0\ast}(\theta',\phi')\di^3\mathbf{r}'\di r
    \end{split}
\end{equation}

where $R_{nl}$ are the hydrogenic radial wavefunctions. We will now make the simplifying assumption that the $\psi$ given by Eq.~\eqref{eq:sgwavef} is a sum of levels such that $(n,l,m)=(l+1,l,l)$, which is the case treated in this work. Since $|\psi|^2$ is integrated against $Y_{2k}^0$, only the terms consisting of products of complex conjugates will survive. Thus, we can substitute the integrand as follows:

\begin{equation}
\begin{split}
    \label{eq:sgintsub}
    \abs{\psi(\mathbf{r}')}&Y_{2k}^{0\ast}(\theta',\phi')\to\\
    &\ \sum_{l'=0}\abs{N_{l'+1,l',l'}^{1/2}R_{l'+1,l'}(r')}^2\abs{Y_{l'}^{l'}(\theta',\phi')}^2Y_{2k}^{0\ast}(\theta',\phi').
    \end{split}
\end{equation}

Then the integral over $\theta'$ and $\phi'$ is just $c_{k,l'l'}$, as defined in Eq.~\eqref{eq:shadd}. Note that this integral is non-zero only for $l'>k$. Thus, we can re-write the sum $\sum_{l'=0}\to\sum_{l'=k}$. The coefficients $c_{k,l'l'}$ have a simple analytic form

\begin{equation}
\begin{split}
    \label{eq:ckll}
    c_{k,l'l'}=(-1)^k\frac{2l'+1}{\sqrt{4\pi}}\sqrt{4k+1}&\frac{(2l')!(2k)!(l'+k)!}{(k!)^2(2l'+2k+1)!(l'-k)!},\\
    &\text{for }k<l'.
    \end{split}
\end{equation}

The energy corrections are then

\begin{equation}
\begin{split}
    \label{eq:sgencorr3}
    \Delta \omega_{nlm}&=-4\pi G\mu^2\sum_{k=0}^l\frac{c_{k,lm}}{4k+1}\sum_{l'=k}N_{l'+1,l',l'}c_{k,l',l'}I_{nl}^{kl'}
    \end{split}
\end{equation}

where the last quantity is the radial integral given by

\begin{equation}
\begin{split}
    \label{eq:sgri}
    I_{nl}^{kl'}=\int R^2_{nl}(r)\int R^2_{l'+1,l'}(r')\frac{r_<^{2k}}{r_>^{2k+1}} r'^2r^2\,\di r'\di r,
    \end{split}
\end{equation}

which can be calculated analytically. Assuming a simultaneous occupation of just 211 and 322, the corrections are

\begin{align}
    \label{eq:sgcorrections}
   \Delta \omega_{211}&\simeq-\frac{\alpha^3\mu}{G \MBH^2}\pare{0.19 N_{211}+0.11N_{322}}\\
   \Delta \omega_{322}&\simeq-\frac{\alpha^3\mu}{G \MBH^2}\pare{0.11 N_{211}+0.09N_{322}}
\end{align}


\section{Frequency drifts}
\label{app:fdrifts}

The corrections to the energy of the 211 and 322 levels from self-interactions and
self-gravity were calculated in Apps. \ref{app:freqcorr} and \ref{app:selfgravity} respectively. The angular frequency of a particle occupying 211 or 322 is:

\begin{subequations}
\label{eq:eqcorrect}
\begin{align}
	\omega_{211}=&\mu\pare{1-\frac{\alpha^2}{8}}-\mu\alpha^5\pare{\frac{\Mpl}{f}}^2\pare{\kinto\eone+\kintt\etwo}\nonumber \\ 
&-\mu\alpha^3\pare{\kgro\eone+\kgrt\etwo},	\label{eq:eqcorrect1}\\
		\omega_{322}=&\mu\pare{1-\frac{\alpha^2}{36}}-\mu\alpha^5\pare{\frac{\Mpl}{f}}^2\pare{\kintop\eone+\kinttp\etwo} \nonumber\\
&-\mu\alpha^3\pare{\kgrop\eone+\kgrtp\etwo},\label{eq:eqcorrect2}
\end{align}
\end{subequations}

where $\alpha=\mu \MBH \Mpl^{-2}$ and $\kinto=1.2\times 10^{-4}$, $\kintt=3.5\times 10^{-5}$, $\kgro=0.19$, $\kgrt=0.11$, $\kintop=3.5\times 10^{-5}$, $\kinttp=1.4\times 10^{-5}$, $\kgrop=0.11$ and $\kgrtp=0.09$ are numerical coefficients.

In what follows, we define the frequency $\nu$ as
\begin{align}
    \nu\equiv\frac{\omega}{2\pi}.
\end{align}
So the frequency drifts $\dot{\nu}$ are given by,

\begin{subequations}
\label{eq:fdrift}
\begin{align}
\dot{\nu}_{211}=&-\frac{\mu\alpha^2}{2\pi}\bigg[\frac{1}{4}\frac{\dot{\alpha}}{\alpha}+\alpha^2\rinv^2\label{eq:fdrift1}\\
&\times\parea{\alpha\pare{\kinto\eoned+\kintt\etwod}+5\pare{\kinto\eone+\kintt\etwo}\dot{\alpha}}+\nonumber\\
&+\parea{\alpha\pare{\kgro\eoned+\kgrt\etwod}+3\pare{\kgro\eone+\kgrt\etwo}\dot{\alpha}}\bigg]\nonumber\\
\dot{\nu}_{322}=&-\frac{\mu\alpha^2}{2\pi}\bigg[\frac{1}{18}\frac{\dot{\alpha}}{\alpha}+\alpha^2\rinv^2\label{eq:fdrift2}\\
&\times\parea{\alpha\pare{\kintop\eoned+\kinttp\etwod}+5\pare{\kintop\eone+\kinttp\etwo}\dot{\alpha}}+\nonumber\\
&+\parea{\alpha\pare{\kgrop\eoned+\kgrtp\etwod}+3\pare{\kgrop\eone+\kgrtp\etwo}\dot{\alpha}}\bigg]\nonumber
\end{align}
\end{subequations}
to leading order in $\alpha$ for every term.

The mass of the BH evolves according to \eqref{eq:massevol}, which can be written equivalently as an equation for $\alpha$ as

\begin{equation}
\label{eq:masschange}
\dot{\alpha}\simeq-\alpha^2\pare{\gsro\eone+\gsrt\etwo-\gamma_{211\times211}^{322\times\text{BH}}\eone^2\etwo},
\end{equation}

As a result, the last terms in the second and third row of Eqs. \eqref{eq:fdrift1} and \eqref{eq:fdrift2} are parametrically suppressed by an additional power of $\alpha$ and $\varepsilon_i$ compared to the respective first term and thus will be neglected in what follows. In addition, all drifts are given to leading order in $\alpha$ and are the maximum possible for each individual regime.

In what follows we calculate the frequency drifts of the GWs coming from annihilations of two 211 particles and from transitions from 322 to 211. These are given by the relations $\dot{\nu}_\text{ann}\equiv2\dot{\nu}_{211}$ and $\dot{\nu}_\text{tr}\equiv\dot{\nu}_{322}-\dot{\nu}_{211}$ We separate the sources of frequency drifts in the following categories:
\begin{enumerate}
    \item Due to the change of the mass of the BH, given by the first terms of \eqref{eq:fdrift1} and \eqref{eq:fdrift2}, denoted as $\nu^\alpha$.
    \item Due to the change in the self-interaction energy, given by the second term of \eqref{eq:fdrift1} and \eqref{eq:fdrift2}, denoted as $\nu^\lambda$.
    \item Due to the change in the self-gravitational energy, given by the third term of \eqref{eq:fdrift1} and \eqref{eq:fdrift2}, denoted as $\nu^\text{gr}$.
\end{enumerate}

In the regime of small self-interactions we treat the depletion due to gravitational radiation (annihilations and transitions) separately for points 2 and 3 above, and we denote by the superscript ``GW''.

We also note that there is an additional source of frequency drift coming from the change of the radial velocity of the BH to the observer, but for isolated black holes it is $  \dot{\nu}_\text{Doppler}<10^{-19}\mathrm{ Hz/s}$ \cite{Zhu:2020tht}, which is negligible.

For reference, LIGO/Virgo continuous wave searches currently cover a range of positive to  negative frequency derivatives of \cite{Pisarski:2019vxw}
\begin{align}
    \label{eq:ligo}
    2\times 10^{-9} \text{ Hz/s}\quad \text{through}\quad\! -1\times 10^{-8} \text{ Hz/s}.
\end{align}

All drift calculations carried out here are to leading approximation in $\alpha$ (which is accurate only for $\alpha\ll a_\ast(0)$) but the formalism includes in principle all higher-order corrections. At higher $\alpha$ the calculations can be carried out numerically using the full expressions and the numerical rates, but at $\alpha\gtrsim0.2$ the approximation of the two-level system essentially breaks down. We have verified that, for our purposes, the leading order approximation gives accurate results.

\subsection{Small self-coupling}

Here we revisit the frequency drifts from purely-gravitational interactions, i.e. $f\to\infty$  as described in \cite{Arvanitaki:2014wva}, which corresponds to region (A) of Fig. \ref{fig:paramspace}. There is a clear separation of times when different levels grow, so whenever a higher level gets populated, the lower ones have already fallen back into the BH, as their SR rates have become negative. In what follows, we will consider only 211, from which comes the stronger signal.

The interesting region for signatures is when the BH has spun down, the level has saturated and slowly gets depleted by radiating GWs. The only source of a frequency drift then comes from the gravitational self-energy of the cloud, given by the last line of Eqs. \eqref{eq:fdrift1} and  \eqref{eq:fdrift2}. In particular, the last term is exactly zero, since $\dot{\alpha}=0$.

The 211 cloud obeys the equation $\eoned=-2\gamma_{211\times211}^{\text{GW}}\alpha^{14}\eone^2$. The maximum drift comes about when $\varepsilon_{211}=\varepsilon^{\max}_{211}\simeq\Delta a_\ast$ (for a better estimate, see App. \ref{app:cloudmass}), when SR shuts just off. For $a_\ast(0)=0.9$ we find the drifts to be


\begin{align}
    \label{eq:pureGWdriftsi}
    \dot{\nu}_{\text{ann}}^{\lambda,\text{GW}}\simeq4\times 10^{-22}\frac{\text{Hz}}{\text{sec}}\pare{\frac{\alpha}{0.075}}^{19}\pare{\frac{\mu}{10^{-12}\text{eV}}}^2\pare{\frac{10^{19}\text{GeV}}{f}}^2
\end{align}

\begin{align}
    \label{eq:pureGWdriftgr}
    \dot{\nu}_\text{ann}^{\text{gr,GW}}\simeq8\times10^{-17}\frac{\text{Hz}}{\text{sec}}\pare{\frac{\alpha}{0.075}}^{17}\pare{\frac{\mu}{10^{-12}\text{ eV}}}^2
\end{align}

In the small self-interactions regime, the drift coming from self-interactions is always subdominant to that of self-gravity in the parameter space of interest.

The drift can become larger than the range LIGO/Virgo cover (Eq. \eqref{eq:ligo}) only for $\alpha$ around $0.27$, taking higher order $\alpha$ contributions into account.

\subsection{Moderate self-coupling}

Here we are interested in the region where both levels are occupied and they drift away
slowly, which corresponds to region (B) of Fig. \ref{fig:paramspace}. In this regime 211 reaches its maximum occupation $\Delta a_\ast$ and we can use Eq.~\eqref{eq:equilibrium_ratio} to relate the $\etwo$ to $\eone$. Note that even though the BH has spun down due to the growth of 211, $\dot{\alpha}\neq0$, since particles fall back into the BH, as described by the last term of Eq.~\eqref{eq:masschange}. The resulting frequency drifts are as follows:\\

\noindent\emph{Due to the change of the BH mass:}
\begin{subequations}
\label{eq:dram}
\begin{align}
&\dot{\nu}_{\text{ann}}^{\alpha}\simeq- 10^{-11}\frac{\text{Hz}}{\text{sec}}\pare{\frac{10^{17}\,\text{GeV}}{f}}^4\pare{\frac{\mu}{10^{-12}\,\text{eV}}}^{2}\pare{\frac{\alpha}{0.075}}^{17}\label{eq:annam}\\
&\dot{\nu}_{\text{tr}}^{\alpha}\simeq 3\times10^{-12}\frac{\text{Hz}}{\text{sec}}\pare{\frac{10^{17}\,\text{GeV}}{f}}^4\pare{\frac{\mu}{10^{-12}\,\text{eV}}}^{2}\pare{\frac{\alpha}{0.075}}^{17}\label{eq:tram}
\end{align}
\end{subequations}
The negative sign in Eq.~\eqref{eq:annam} comes from the fact that the SR rates are zero, so the BH is actually gaining mass by the depletion of 211, from the last term of Eq.~\eqref{eq:masschange}.\\

\noindent\emph{Due to self-interactions:}

\begin{subequations}
\label{eq:drlm}
\begin{align}
&\dot{\nu}_{\text{ann}}^{\lambda}\simeq 6\times 10^{-13}\frac{\text{Hz}}{\text{sec}}\pare{\frac{10^{17}\,\text{GeV}}{f}}^6\pare{\frac{\mu}{10^{-12}\,\text{eV}}}^{2}\pare{\frac{\alpha}{0.075}}^{19}\label{eq:annlm}\\
&\dot{\nu}_{\text{tr}}^{\lambda}\simeq -2\times 10^{-13}\frac{\text{Hz}}{\text{sec}}\pare{\frac{10^{17}\,\text{GeV}}{f}}^6\pare{\frac{\mu}{10^{-12}\,\text{eV}}}^{2}\pare{\frac{\alpha}{0.075}}^{19}\label{eq:trlm}
\end{align}
\end{subequations}

\noindent\emph{Due to self-gravity:}

\begin{subequations}
\label{eq:drgm}
\begin{align}
&\dot{\nu}_\text{ann}^\text{gr}\simeq 10^{-11}\frac{\text{Hz}}{\text{sec}}\pare{\frac{10^{17}\,\text{GeV}}{f}}^4\pare{\frac{\mu}{10^{-12}\,\text{eV}}}^{2}\pare{\frac{\alpha}{0.075}}^{17}\label{eq:anngm}\\
&\dot{\nu}_\text{tr}^\text{gr}\simeq -2\times10^{-12}\frac{\text{Hz}}{\text{sec}}\pare{\frac{10^{17}\,\text{GeV}}{f}}^4\pare{\frac{\mu}{10^{-12}\,\text{eV}}}^{2}\pare{\frac{\alpha}{0.075}}^{17}\label{eq:trgm}
\end{align}
\end{subequations}
These are calculated for $a_\ast(0)=0.9$. Note that $\alpha$ scalings of Eqs. \eqref{eq:dram} and \eqref{eq:drgm} are the same, which comes from the fact that SR has shut off and the scalings in both $\dot{\alpha}$ and $\dot{\varepsilon}_i$ of Eqs. \eqref{eq:masschange} and of \eqref{eq:fdrift} are set by the same term, i.e. $\gbh \eone^2\etwo$. This is why the numerical coefficients of both the annihilation and transition drifts are very close. In particular, for the annihilation drift we find more precisely that 

\begin{align}
    &\dot{\nu}_\text{ann}^\alpha+\dot{\nu}_\text{ann}^\text{gr}\label{eq:addedfrdr}\\
    &\simeq 1.4\times10^{-12}\frac{\text{Hz}}{\text{sec}}\pare{\frac{10^{17}\,\text{GeV}}{f}}^4\pare{\frac{\mu}{10^{-12}\,\text{eV}}}^{2}\pare{\frac{\alpha}{0.075}}^{17}\nonumber
\end{align}

\begin{figure}[h]
\centering
\includegraphics[width=0.48\textwidth]{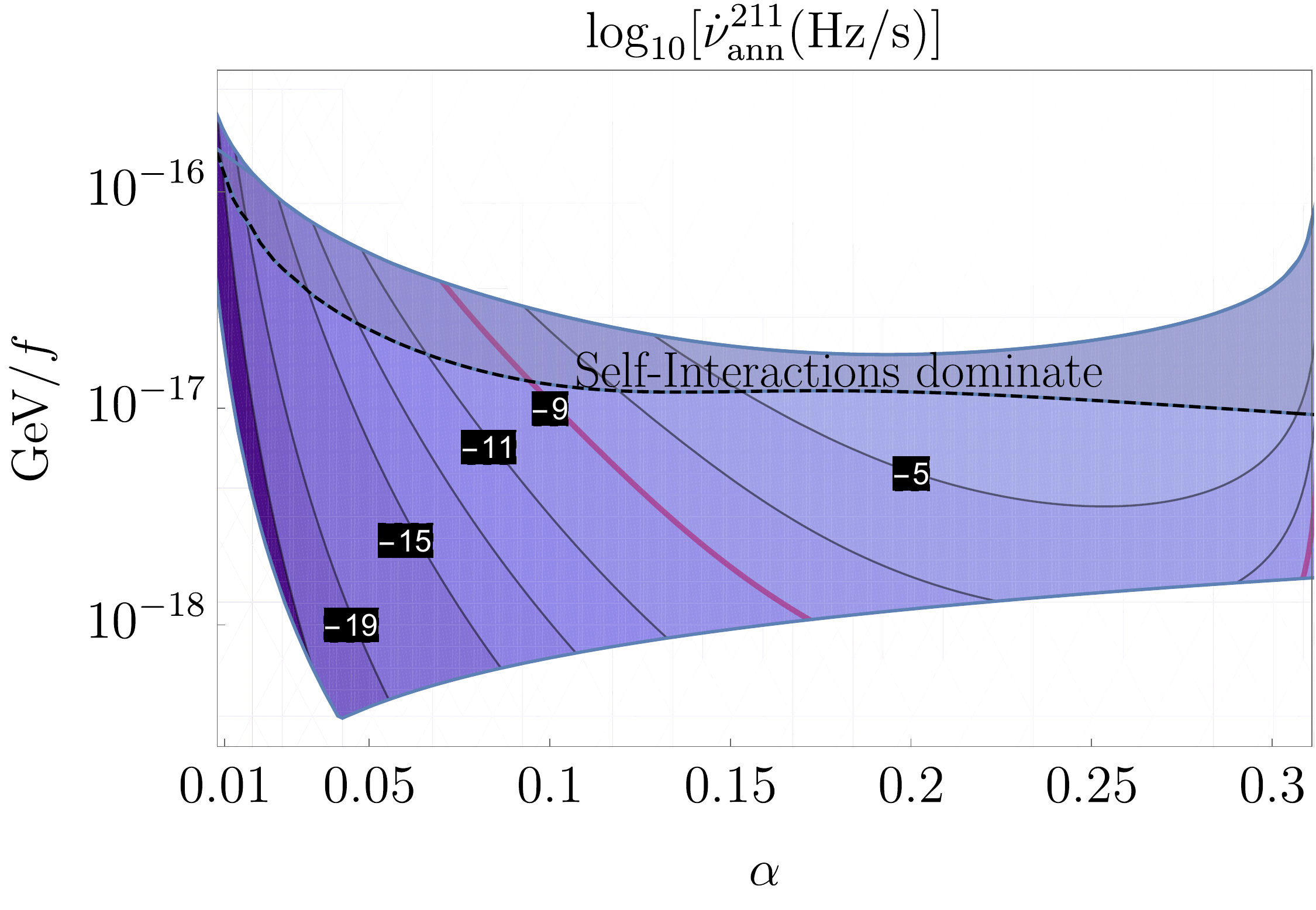}
\caption{Frequency drift contours for annihilations of axions to GWs, given by twice the quantity in Eq. \eqref{eq:fdrift1}, in the moderate self-coupling regime. The gray shaded region above the dashed black contour is where the drift due to self-interactions (second line of Eq. \eqref{eq:fdrift1}) dominates. The red contour corresponds to the largest positive drift covered by LIGO/Virgo continuous searches, taken here to be $2\times10^{-9}$ Hz/s \cite{Pisarski:2019vxw}.}
\label{fig:anndrifts}
\end{figure}

Within the moderate self-interactions regime, we find that self-interactions are the dominant source of frequency drift for $f\lesssim8.5\times 10^{16}(\alpha/0.1)\text{ GeV}$. 
The drift can become larger than the range LIGO/Virgo cover (Eq. \eqref{eq:ligo}) for $f\lesssim 5.6\times10^{16}(\alpha/0.1)^{17/4}\text{ GeV}$. In Fig. \ref{fig:anndrifts} we plot the full annihilation frequency drift stemming from Eq. \eqref{eq:fdrift1} in this regime.

\begin{figure}[h]
\centering
\includegraphics[width=0.48\textwidth]{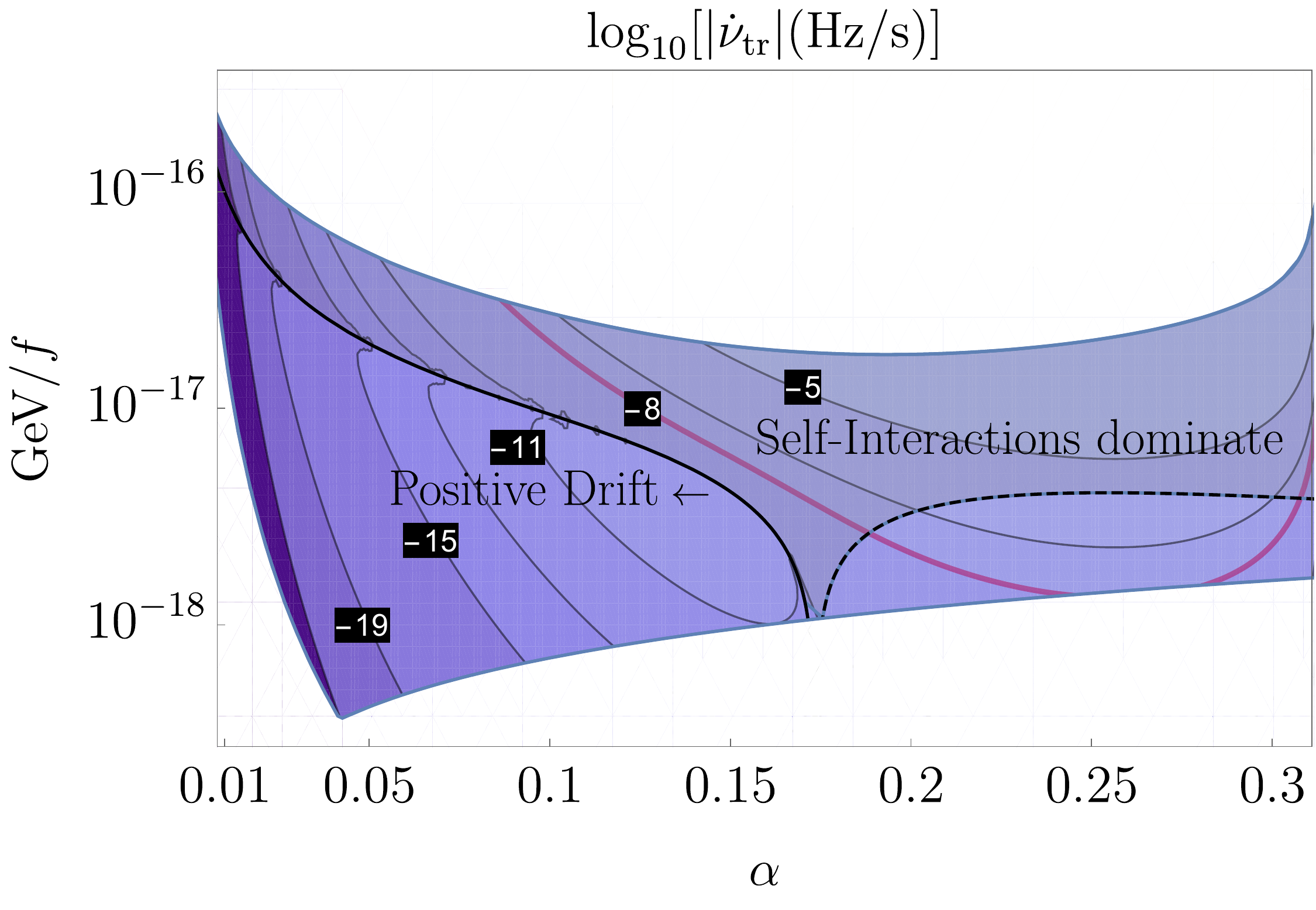}
\caption{Frequency drift contours for GWs sourced by axion transitions from 322 to 211, given by the difference of Eq. \eqref{eq:fdrift1} and Eq. \eqref{eq:fdrift2}, in the moderate self-coupling regime. The gray shaded region above the black contour (solid and dashed) is where the drift due to self-interactions (second line of Eqs. \eqref{eq:fdrift}) dominates. The frequency drift is negative to the right (i.e. to the large-$\alpha$ side) of the solid black line. Note that here we are plotting the absolute value of the frequency drift. The red contour corresponds to the largest \emph{negative} drift covered by LIGO/Virgo continuous searches, taken here to be $-1\times10^{-8}$ Hz/s \cite{Pisarski:2019vxw}.}
\label{fig:trdrifts}
\end{figure}

Analogously, for transitions,
self-interactions are the dominant source of frequency drift for $f\lesssim10^{17}(\alpha/0.1)\text{ GeV}$.  The drift can become larger than the range LIGO/Virgo cover (Eq. \eqref{eq:ligo}) for $f\lesssim 4\times10^{16}(\alpha/0.1)^{17/4}\text{ GeV}$. In Fig. \ref{fig:trdrifts} we plot the full annihilation frequency drift stemming from Eqs. \eqref{eq:fdrift1} \& \eqref{eq:fdrift2}, in this regime. 

\subsection{Large self-coupling}

We are interested in the part of the evolution where the levels have reached
their equilibrium values, given by Eqs. \eqref{eq:quasi_equlibrium_occupations_211} and \eqref{eq:quasi_equlibrium_occupations_322}, which corresponds to region (C) in Fig. \ref{fig:paramspace}.
These are slowly drifting because of the slow spin-down of the BH and the
change of its mass. Neglecting the SR of $\etwo$, which is subdominant, the
spin evolves according to
\begin{equation}
\dot{a}_\ast=-\gsro\eone^\text{eq},
\end{equation}
and its mass changes according to Eq.~\eqref{eq:masschange}. By plugging in the equilibrium values of Eq.~\eqref{eq:quasi_equlibrium_occupations} we get

\begin{equation}
\begin{split}
\dot{\alpha}=-\frac{2\alpha^2\gsro}{3\sqrt{3}}\frac{\sqrt{\gsro\ginf}}{\gbh}.
\end{split}
\end{equation}

Then, the equilibrium values evolve according to

\begin{equation}
\begin{split}
\dot{\varepsilon}&=\dot{\varepsilon}^{\text{eq}}\\
&=\frac{\partial\varepsilon^{\text{eq}}}{\partial a_\ast}\dot{a}_\ast+\frac{\partial\varepsilon^{\text{eq}}}{\partial \alpha}\dot{\alpha}.
\label{eq:epsilon_ast_drift}
\end{split}
\end{equation}

The second term of Eq.~\eqref{eq:epsilon_ast_drift} gives a subdominant contribution and is further suppressed by another power of $\alpha$ compared to the first term. The signal is maximum at the beginning when $a_\ast\simeq a_\ast(0)$. The resulting drifts are given below.\\

\noindent\emph{Due to the change of the BH mass:}

\begin{subequations}
\begin{align}
&\dot{\nu}_{\text{ann}}^{\alpha}\simeq2\times10^{-13}\frac{\text{Hz}}{\text{sec}}\pare{\frac{f}{10^{15}\,\text{GeV}}}^2\pare{\frac{\mu}{10^{-12}\,\text{eV}}}^{2}\pare{\frac{\alpha}{0.075}}^8\\
&\dot{\nu}_{\text{tr}}^{\alpha}\simeq-6\times10^{-14}\frac{\text{Hz}}{\text{sec}}\pare{\frac{f}{10^{15}\,\text{GeV}}}^2\pare{\frac{\mu}{10^{-12}\,\text{eV}}}^{2}\pare{\frac{\alpha}{0.075}}^8
\end{align}
\end{subequations}

\noindent\emph{Due to self-interactions:}

\begin{subequations}
\begin{align}
&\dot{\nu}_{\text{ann}}^{\lambda}\simeq3\times10^{-13}\frac{\text{Hz}}{\text{sec}}\pare{\frac{f}{10^{15}\,\text{GeV}}}^2\pare{\frac{\mu}{10^{-12}\,\text{eV}}}^2\pare{\frac{\alpha}{0.075}}^7\\
&\dot{\nu}_{\text{tr}}^{\lambda}\simeq-10^{-13}\frac{\text{Hz}}{\text{sec}}\pare{\frac{f}{10^{15}\,\text{GeV}}}^2\pare{\frac{\mu}{10^{-12}\,\text{eV}}}^2\pare{\frac{\alpha}{0.075}}^7
\end{align}
\end{subequations}

\noindent\emph{Due to the self-gravity:}

\begin{subequations}
\begin{align}
&\dot{\nu}_\text{ann}^\text{gr}\simeq5\times 10^{-16}\frac{\text{Hz}}{\text{sec}}\pare{\frac{f}{10^{15}\,\text{GeV}}}^4\pare{\frac{\mu}{10^{-12}\,\text{eV}}}^2\pare{\frac{\alpha}{0.075}}^5\\
&\dot{\nu}_\text{tr}^\text{gr}\simeq-10^{-16}\frac{\text{Hz}}{\text{sec}}\pare{\frac{f}{10^{15}\,\text{GeV}}}^4\pare{\frac{\mu}{10^{-12}\,\text{eV}}}^2\pare{\frac{\alpha}{0.075}}^5
\end{align}
\end{subequations}

These are calculated for $a_\ast(0)=0.9$ as well.

In the large self-interactions regime, for $\alpha\gtrsim0.1$ the change of the mass of the BH is the dominant source of frequency drift for annihilations. For $\alpha\lesssim0.1$ self-interactions are dominant.
The drift can become larger than the range LIGO/Virgo cover (Eq. \eqref{eq:ligo})  for $f\gtrsim 3\times10^{16}(\alpha/0.1)^{-4}\text{ GeV}$, which is relevant above $\alpha\simeq0.1$.

Analogously for transitions, for $\alpha\gtrsim0.13$, the change of the mass of the BH dominates and for $\alpha\lesssim0.13$ self-interactions are dominant. 
The drift can become larger than the range LIGO/Virgo cover (Eq. \eqref{eq:ligo}) for $f\gtrsim 2.5\times10^{16}(\alpha/0.15)^{-4}\text{ GeV}$, which is relevant above $\alpha\simeq0.13$.\\

\section{Perturbations from BH companion}
\label{app:compmixing}

When the primary BH has a companion, the perturbation in the gravitational potential induces mixing of different levels. In particular, SR levels can mix with non-SR ones, resulting in the depletion of the cloud. According to \cite{Baumann+2019}, the perturbation $\delta V_\text{gr}$ mixes the levels $\psi_i$ and $\psi_j$ according to
\begin{equation}
    \label{eq:mixing}
    \bra{\psi_j}\delta V_\text{c}\ket{\psi_i}=-\frac{\alpha M_\text{c}}{\Mbh}\sum_{l\geq2}
    \sum_{|m|\leq l}\frac{4\pi}{2l+1}\frac{Y_l^{m*}(\theta_c,\phi_c)}{R_c^{l+1}}I_{\bar{r}}I_\Omega
\end{equation}
where the subscript $M_\text{c}$ is the mass of the companion, $\theta_c$, $\phi_c$ its angular coordinates and $R_\text{c}$ its distance from the primary BH of mass $\Mbh$, whereas the constant $\alpha=G\mu \Mbh$. We have also defined
\begin{align}
    I_r&\equiv\int_0^\infty \di r\, r^{2+l}R_{n_j l_j}(r)R_{n_i l_i}(r),\label{eq:mixingr}\\
    I_\Omega&\equiv \int\di\Omega \,Y_{l_j}^{m_j*}(\theta,\phi) Y_{l_i}^{m_i}(\theta,\phi) Y_{l}^{m}(\theta,\phi)\label{eq:mixingOmega}
\end{align}
where $R_{nl}$ is the radial part of the hydrogenic wavefunction and $Y_m^l$ are the spherical harmonics.\\

Note that the first sum in Eq.~\eqref{eq:mixing} starts from $l=2$, which demonstrates the fact that the first non-zero correction from gravity comes from the quadrupole term, as expected from the equivalence principle.\footnote{In \cite{Arvanitaki:2014wva} it was incorrectly assumed that the leading order contribution came from a dipole term. See \cite{Baumann+2019} for an explanation.}

We are interested in the mixing of the 211 level with non-SR levels of
the BH, which can lead, in principle, to the depletion of our cloud.
The dominant contribution comes from $n=2,l=1,m=-1$, and it is largest when
the companion lies on the plane perpendicular to the spin of the
primary BH, i.e.\ when $\theta_c=\pi/2$. 

The horizon flux becomes \textit{positive}, i.e. more axions fall back into the BH than are extracted due to SR \cite{Arvanitaki:2014wva}, when 
\begin{equation}
    \label{eq:fluxpositive}
    \abs{\frac{\Gamma_{\text{dump}}^{j}}{\Gamma_{i}}}^{1/2}\abs{\frac{\bra{\psi_j}\delta V_\text{c}\ket{\psi_i}}{\Delta\omega_{ji}}}>1
\end{equation}
where ``dump'' denotes the non-SR level that mixes with the SR one,
$\Gamma$ are the superradiance rates, and
$\Delta\omega_{ji}$ is the difference of the energies between the two
levels, which are given by \cite{Baumann+2019}
\begin{equation}
\begin{split}
    \label{eq:baumannenergies}
    \omega_{nlm}&=\mu\left(1-\frac{\alpha^2}{2n^2}\right.\\
    &\left.-\frac{\alpha^4}{8n^4}+\frac{(2l-3n+1)\alpha^4}{n^4(l+1/2)}+\frac{2a_\ast m\alpha^5}{n^3l(l+1/2)(l+1)}\right)
\end{split}
\end{equation}

The physical quantities measured for BH binaries are the BH masses, their spins and the orbital period. We assume that the companion is far away (which is where Eq.~\eqref{eq:mixing} is valid), so we relate the distance to the orbital period using Kepler's 3rd Law: $R^3/T^2=G(\Mbh+M_\text{c})/(4\pi^2)$, where $T$ is the orbital period. Then, the condition \eqref{eq:fluxpositive} becomes parametrically:
\begin{equation}
    \label{eq:condition1}
    \frac{M_\text{c}}{\Mbh}\frac{144\pi^2\sqrt{3}}{a_\ast\alpha^7\pare{1+\frac{M_\text{c}}{\Mbh}}(\mu T)^2}\gtrsim1
\end{equation}
where we have omitted an $\mathcal{O}(1)$ factor in the $\alpha$ region of interest. 

The cloud may also be depleted by resonances that can occur when the period of the companion hits the energy difference between two levels, as shown in \cite{Baumann+2019}. To estimate when this happens, we can compare the period to the energy splitting of the two mixing levels. As the companion spirals closer to the primary BH, its orbital period increases. When it crosses the value $\Delta\omega_{ji}^{-1}$, we expect that the cloud will be significantly depleted. A more careful analysis can be found in \cite{Baumann+2019}. The condition, therefore, is
\begin{equation}
\label{eq:condition2}
    \frac{1}{6}a_\ast \alpha^5 \mu T\simeq 1
\end{equation}
       
       In deriving the BH spin bounds in Sec. \ref{sec:spindown}, we take into account both Eqs. \eqref{eq:condition1} and \eqref{eq:condition2}.
       

\section{Axion wind sensitivity projections}
\label{appcasper}

As discussed in section~\ref{secnucleonspin}, given an
axion coupling to nucleon spins,
an axion oscillation $\varphi(t) = \varphi_0 \cos \omega t$ will
act on nuclei as an effective magnetic field
$B_a(t) = B_a \cos \omega t$. For nuclei
which are spin-polarized
in an external magnetic field,
with Larmor frequency $\omega_0 \simeq \omega$, a transverse
$B_a$ will induce a transverse magnetic moment
\begin{equation}
	\mu_a \simeq \mu_n^2 N_n B_a \frac{\omega_0}{\omega^2 - \omega_0^2 + i \omega \gamma}
	\label{eqwindresponse}
\end{equation}
where $\mu_n$ is the nuclear magnetic moment,
$N_n$ is the total number of nuclei,
and $\gamma$ is the damping rate
(in terms of the spin coherence time $T_2$,
$\gamma = 2/T_2$~\cite{Sleator_1985}).

In the absence of an axion forcing, the fluctuation
spectrum for the transverse magnetic momentum is
\begin{equation}
	S_{\mu\mu} \simeq \frac{\mu_n^2 N_n}{\gamma}\frac{1}{1 + T_2^2 (\omega-\omega_0)^2}
	\label{eqsmm}
\end{equation}
which is related to the response function (Eq.~\eqref{eqwindresponse})
by the fluctuation-dissipation relation~\cite{Sleator_1985}. 

If we read out the transverse magnetic moment using a 
sufficiently sensitive
magnetometer (e.g. a SQUID~\cite{Sleator_1985,kimball2017overview}), 
then it is possible to detect fluctuations as small
as the quantum fluctuations from Eq.~\eqref{eqsmm}.
With a sensor that is bounded by the Standard Quantum Limit~\cite{Kampel:2017gze,Clerk_2010},
this is possible over a bandwidth $\sim 1/T_2$.
Consequently, for an integration time of $T \gtrsim T_2$, we need
\begin{equation}
	\frac{\mu_a^2}{S_{\mu\mu}/T} \simeq \frac{1}{2} \mu_n^2 N_n B_a^2 T T_2 \gtrsim {\rm few}
	\label{eqsnr2}
\end{equation}
in order to reliably detect an axion signal.

To cover an $\OO(1)$ axion mass range, we need
to operate in $\sim \omega_0 T_2$ different resonant
configurations (we will not be careful about
constant factors). Consequently,
if our total experimental time is $T_{\rm tot}$, the time we spend in 
each configuration is $T \sim T_{\rm tot} / (\omega_0 T_2)$,
and our sensitivity limit
is~\footnote{If $T \ll T_2$, then Eq.~\eqref{eqsnr2} will
not apply, since the response
signal will not have time to ring up fully (equivalently,
we cannot resolve the bandwidth of the response function).
For the experimental parameters of interest, we will
not be in this regime.}
\begin{equation}
	B_a^2 \gtrsim {\rm few} \times \frac{\omega_0}{\mu_n^2 N_n T_{\rm tot}}
\end{equation}
Note that, while this naive form does not
depend on $T_2$, the signal amplitude from Eq.~\eqref{eqwindresponse} is $\propto T_2$; consequently, achieving a sensitive
enough magnetometer may be easier for larger $T_2$.
As discussed in section~\ref{secnucleonspin}, 
the CASPEr-Wind project aims to achieve
spin-noise-limited sensitivities at frequencies 
in the $\kHz - 30 \kHz$ range~\cite{kimball2017overview}.


\section{Dark matter abundance}
\label{app:dmestimates}

In section~\ref{secaxionmodels}, we reviewed models in
which an axion dark matter abundance is generated via
the early-universe misalignment mechanism. 
For attractive potentials, if the
initial value of the axion field is tuned close to the
top of its potential, then the generated dark matter
abundance can be enhanced
through the ``large-misalignment mechanism'' \cite{Arvanitaki_2020}.
In this appendix, we give formulae for the DM density obtained
in this way.

For a general cosine potential of the form $V(\varphi)=m^2f^2\parea{1-\cos(\varphi/f)}$, the enhanced final density for a large initial misalignment is given by
\begin{align}
    \label{eq:dmdensity}
    \frac{\rho}{\rho_{\pi/2}}&\simeq0.2\parea{t_\mu^\text{osc}+4\log t_\mu^\text{osc}}^2\\
    t_\mu^\text{osc}&\equiv\log\parea{\frac{1}{\pi-|\theta_0|}\frac{2^{1/4}\pi^{1/2}}{\Gamma(5/4)}}
\end{align}
where $\rho_{\pi/2}$ is the final density when the initial amplitude of the field is $\theta_0=\phi_0/f=\pi/2$, and $t_\mu^\text{osc}$ marks the onset of the oscillation in units of $\mu$.

\begin{figure}[h]
\centering
\includegraphics[width=0.48\textwidth]{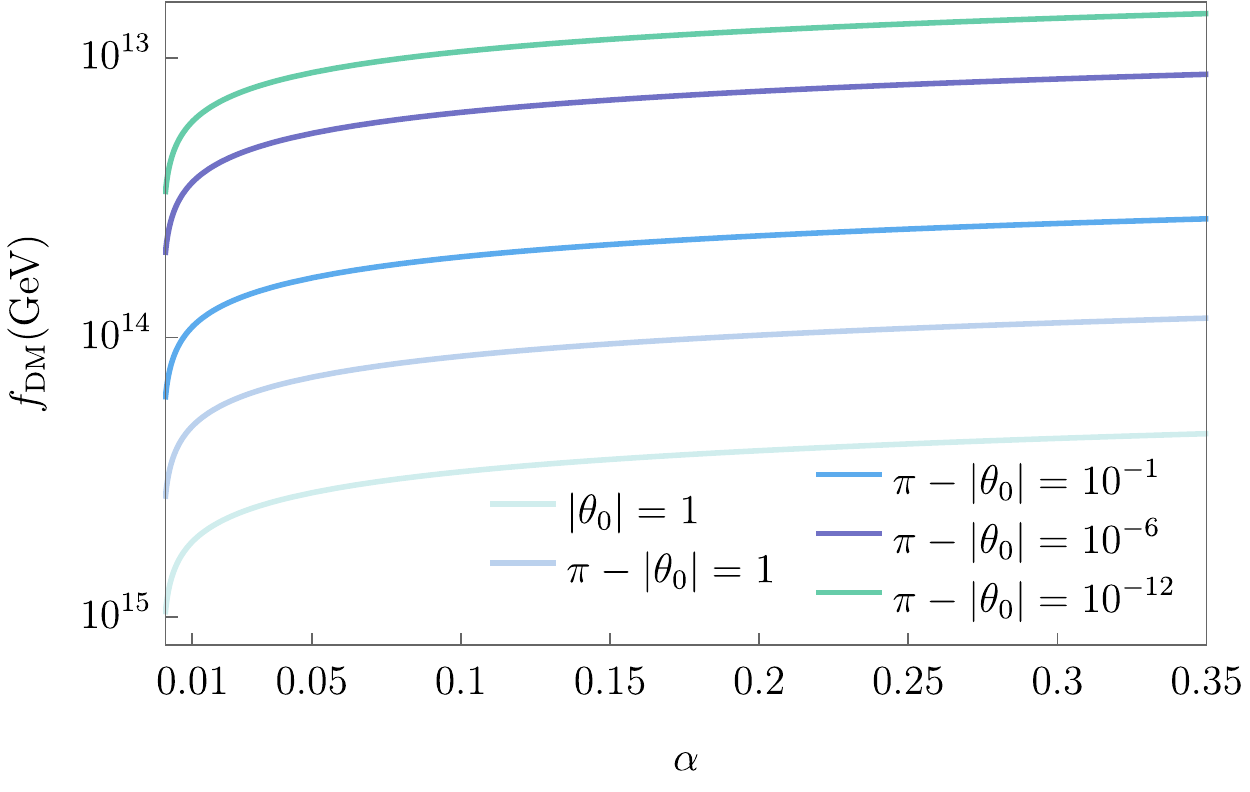}
\caption{The decay constant $f$ of Eq.~\eqref{eq:DMf} that gives the observed DM abundance today, as a function of $\alpha=G\MBH\mu$ for a $10\msun$ BH. Note that the vertical axis is reversed. We are assuming a general cosine potential of the form $V(\varphi)=\mu^2f^2\pare{1-\cos(\varphi/f)}$ and plot for different large initial misalignments from the top of the potential, following the results of \cite{Arvanitaki_2020}. We also plot for usual misalignment values of $|\theta_0|=1,\,\pi-1$.}
\label{fig:plotdm}
\end{figure}

Fixing the final density to be the observed DM abundance today, we arrive at the relation \cite{Arvanitaki_2020}
\begin{equation}
    \label{eq:DMf}
    \frac{f_\text{DM}}{m_\text{pl}}\simeq\frac{3^{1/2}}{2^{5/4}C_{\pi/2}^{1/2}}\pare{\frac{\rho}{\rho_{\pi/2}}}^{-1/2}\pare{\frac{H_\text{eq}}{\mu}}^{1/4}
\end{equation}
where $C_{\pi/2}\simeq 1.15$, $H_\text{eq}$ is the Hubble parameter at matter-radiation equality and $m_\text{pl}$ is the reduced Planck mass. We plot Eq.~\eqref{eq:DMf} for different initial misalignments in Fig. \ref{fig:plotdm}, as a function of $\alpha=G\MBH\mu$, for a $10\msun$ BH.

Somewhat separately, we can compare the energy density in
a superradiant cloud to the
DM energy density.
The energy density of the cloud is $\rho_\text{c}\sim \theta^2 f^2\mu^2$, up to an $\mathcal{O}(1)$ prefactor, which can be found to be
\begin{equation}
\begin{split}
    \label{eq:rho}
    \rho_\text{c}&=\frac{1}{2}\dot{\varphi}^2+\frac{1}{2}(\nabla\varphi)^2+\frac{1}{2}\mu^2+\frac{1}{4!}\frac{\mu^2}{f^2}\varphi^4\\
    &\sim \pare{1+\frac{\alpha^2}{2(\tilde{R}^+)^2}+\frac{\theta^2}{4!}}\theta^2 (f\mu)^2
    \end{split}
\end{equation}
where $\tilde{R}^+$ is given by Eq.~\eqref{eq:rext}. We estimate it to be
\begin{equation}
\begin{split}
    \label{eq:DMestimate}
        \rho_\text{c}&\sim 2\times10^{28}\frac{\text{GeV}} {\text{cm}^{3}}\pare{\frac{\mu}{10^{-12}\text{ eV}}}^2\pare{\frac{f}{10^{16}\text{ GeV}}}^2\pare{\frac{\theta}{0.04}}^2\\
        &~\sim 2\times10^{28}\frac{\text{GeV}} {\text{cm}^{3}}\pare{\frac{\MBH}{10 \msun}}^{-2}\pare{\frac{\alpha}{0.07}}^2\times\\
        &{\color{white} ~\sim\frac{\text{GeV}} {\text{cm}^{3}}\pare{\frac{\MBH}{10 \msun}}^{-2}}\times\pare{\frac{f}{10^{16}\text{ GeV}}}^2\pare{\frac{\theta}{0.04}}^2
            \end{split}
\end{equation}

Even for the smallest $f$ and $\alpha$ we show
in our plots, this density is far larger than astrophysical DM
densities. For example, in the SMBH parameter
space shown in Fig.~\ref{figsmbh}, $\rho_c \gtrsim 10^{14}
\GeV \cm^{-3}$.

\newpage
\bibliography{biblio}

\end{document}